\documentclass[12pt]{iopart}
\pdfoutput=1 

\usepackage{iopams}
\usepackage{amstext} 
\usepackage{amssymb} 
\usepackage{bbold} 
\usepackage{bbm} 
\usepackage{amsthm} 
\usepackage{stmaryrd} 
\usepackage{stix}
\usepackage{etoolbox} 
\usepackage{xparse}
\usepackage{pgffor}

\usepackage[utf8]{inputenc} 
\usepackage{textgreek} 
\usepackage{relsize} 
\usepackage{microtype} 
\usepackage{ulem} 
\usepackage[compress]{cite} 
\usepackage{csquotes} 
\usepackage{setspace} 
\usepackage{scalerel}

\usepackage{graphicx} 
\usepackage[usenames,dvipsnames]{xcolor} 
\usepackage{tikzit} 
\usepackage{stackengine}
\usepackage{circuitikz} 
\usetikzlibrary{
    arrows,
    shapes,
    decorations,
    intersections,
    backgrounds,
    positioning,
    circuits.ee.IEC
    }

\tikzstyle{inline text}=[text height=1.5ex, text depth=0.25ex, yshift=0.5mm]
\tikzstyle{upground}=[circuit ee IEC, thick, ground, rotate=90, scale=2]
\tikzstyle{downground}=[circuit ee IEC, thick, ground, rotate=-90, scale=1.5]
\tikzstyle{point}=[regular polygon, regular polygon sides=3, draw, scale=0.75, inner sep=-0.5pt, minimum width=9mm, fill=white, regular polygon rotate=180, tikzit fill={rgb,255: red,242; green,255; blue,92}]
\tikzstyle{wide copoint}=[fill=white, draw, shape=isosceles triangle, shape border rotate=90, isosceles triangle stretches=true, inner sep=0pt, minimum width=1.5cm, minimum height=6.12mm]
\tikzstyle{wide point}=[fill=white, draw, shape=isosceles triangle, shape border rotate=-90, isosceles triangle stretches=true, inner sep=0pt, minimum width=1.5cm, minimum height=6.12mm, yshift=-0.0mm]
\tikzstyle{wide dpoint}=[wide point, doubled]
\tikzstyle{copoint}=[regular polygon, regular polygon sides=3, draw, scale=0.75, inner sep=-0.5pt, minimum width=9mm, fill=white, tikzit fill={rgb,255: red,255; green,128; blue,0}, tikzit draw={rgb,255: red,255; green,128; blue,0}]
\tikzstyle{dot}=[inner sep=0mm, minimum width=2mm, minimum height=2mm, draw, shape=circle]
\tikzstyle{black dot}=[dot, fill={gray!30}, text depth=-0.2mm]
\tikzstyle{white dot}=[dot, fill=white, text depth=-0.2mm]
\tikzstyle{small box}=[rectangle, inline text, fill=white, draw, minimum height=5mm, yshift=-0.5mm, minimum width=5mm, font={\small}]
\tikzstyle{small gray box}=[small box, fill={gray!30}]
\tikzstyle{medium box}=[rectangle, inline text, fill=white, draw, minimum height=5mm, yshift=-0.5mm, minimum width=10mm, font={\small}]
\tikzstyle{square box}=[small box]
\tikzstyle{medium gray box}=[small box, fill={gray!30}]
\tikzstyle{semilarge box}=[rectangle, inline text, fill=white, draw, minimum height=5mm, yshift=-0.5mm, minimum width=12.5mm, font={\small}]
\tikzstyle{large box}=[rectangle, inline text, fill=white, draw, minimum height=5mm, yshift=-0.5mm, minimum width=15mm, font={\small}]
\tikzstyle{large gray box}=[small box, fill={gray!30}]
\tikzstyle{dpoint}=[point, doubled]
\tikzstyle{dcopoint}=[copoint, doubled]
\tikzstyle{boldedge}=[doubled, shorten <=-0.17mm, shorten >=-0.17mm]
\tikzstyle{normal}=[line width=0.9pt]
\tikzstyle{doubled}=[line width=1pt]
\tikzstyle{boldedge}=[doubled, shorten <=-0.17mm, shorten >=-0.17mm]
\tikzstyle{small dbox}=[small box, doubled]
\tikzstyle{white ddot}=[white dot, doubled]
\tikzstyle{black ddot}=[black dot, doubled, tikzit fill=black]
\tikzstyle{map}=[draw, shape=NEbox, inner sep=2pt, minimum height=6mm, fill=white]
\tikzstyle{box}=[draw, shape=rectangle, inner sep=2pt, minimum height=6mm, minimum width=6mm, fill=white]
\tikzstyle{dbox}=[draw, doubled, shape=rectangle, inner sep=2pt, minimum height=6mm, minimum width=6mm, fill=white]
\tikzstyle{dmap}=[draw, doubled, shape=NEbox, inner sep=2pt, minimum height=6mm, fill=white]
\tikzstyle{dmapdag}=[draw, doubled, shape=SEbox, inner sep=2pt, minimum height=6mm, fill=white]
\tikzstyle{dmapadj}=[draw, doubled, shape=SEbox, inner sep=2pt, minimum height=6mm, fill=white]
\tikzstyle{dmaptrans}=[draw, doubled, shape=SWbox, inner sep=2pt, minimum height=6mm, fill=white]
\tikzstyle{dmapconj}=[draw, doubled, shape=NWbox, inner sep=2pt, minimum height=6mm, fill=white]
\tikzstyle{map}=[draw, shape=NEbox, inner sep=2pt, minimum height=6mm, fill=white]
\tikzstyle{dashedmap}=[draw, dashed, shape=NEbox, inner sep=2pt, minimum height=6mm, fill=white]
\tikzstyle{mapdag}=[draw, shape=SEbox, inner sep=2pt, minimum height=6mm, fill=white]
\tikzstyle{mapadj}=[draw, shape=SEbox, inner sep=2pt, minimum height=6mm, fill=white]
\tikzstyle{maptrans}=[draw, shape=SWbox, inner sep=2pt, minimum height=6mm, fill=white]
\tikzstyle{mapconj}=[draw, shape=NWbox, inner sep=2pt, minimum height=6mm, fill=white]
\tikzstyle{semilarge map}=[draw, shape=NEbox, inner sep=2pt, minimum height=6mm, fill=white, minimum width=9.5mm]
\tikzstyle{semilarge dmap}=[draw, doubled, shape=NEbox, inner sep=2pt, minimum height=6mm, fill=white, minimum width=9.5mm]
\tikzstyle{kpointdag}=[kpoint adjoint]
\tikzstyle{kpointadj}=[kpoint adjoint]
\tikzstyle{kpointconj}=[kpoint conjugate]
\tikzstyle{kpointtrans}=[kpoint transpose]
\tikzstyle{kpoint common}=[draw, fill=white, inner sep=1pt, minimum height=4mm]
\tikzstyle{kpoint sc}=[shape=cornerpoint, kpoint common]
\tikzstyle{kpoint adjoint sc}=[shape=cornercopoint, kpoint common]
\tikzstyle{kpoint}=[shape=cornerpoint, shorten left=5pt, kpoint common, tikzit fill={rgb,255: red,255; green,128; blue,0}]
\tikzstyle{kpoint adjoint}=[shape=cornercopoint, shorten left=5pt, kpoint common, tikzit fill={rgb,255: red,255; green,128; blue,0}]
\tikzstyle{kpoint conjugate}=[shape=cornerpoint, shorten right=5pt, kpoint common]
\tikzstyle{kpoint transpose}=[shape=cornercopoint, shorten right=5pt, kpoint common]
\tikzstyle{kpoint symm}=[shape=cornerpoint, shorten left=5pt, shorten right=5pt, kpoint common]
\tikzstyle{wide kpoint}=[kpoint, minimum width=1 cm, inner sep=2pt]
\tikzstyle{wide kpointdag}=[kpointdag, minimum width=1 cm, inner sep=2pt]
\tikzstyle{wide kpointconj}=[kpointconj, minimum width=1 cm, inner sep=2pt]
\tikzstyle{wide kpointtrans}=[kpointtrans, minimum width=1 cm, inner sep=2pt]
\tikzstyle{wider kpoint}=[kpoint, minimum width=1.25 cm, inner sep=2pt]
\tikzstyle{wider kpointdag}=[kpointdag, minimum width=1.25 cm, inner sep=2pt]
\tikzstyle{wider kpointconj}=[kpointconj, minimum width=1.25 cm, inner sep=2pt]
\tikzstyle{wider kpointtrans}=[kpointtrans, minimum width=1.25 cm, inner sep=2pt]
\tikzstyle{dkpoint}=[kpoint, doubled, tikzit fill={rgb,255: red,255; green,85; blue,210}]
\tikzstyle{wide dkpoint}=[wide kpoint, doubled, tikzit fill={rgb,255: red,68; green,255; blue,0}]
\tikzstyle{dkpointdag}=[kpoint adjoint, doubled]
\tikzstyle{wide dkpointdag}=[wide kpointdag, doubled]
\tikzstyle{label}=[fill=white, draw=white, shape=circle, tikzit draw={rgb,255: red,10; green,26; blue,255}, tikzit fill={rgb,255: red,0; green,12; blue,255}, font={\small}]
\tikzstyle{squarelabel}=[fill=white, draw=white, shape=rectangle, tikzit draw=black]
\tikzstyle{eslabel}=[tikzit draw={rgb,255: red,255; green,191; blue,191}, tikzit fill={rgb,255: red,255; green,191; blue,191}, font={\tiny}]
\tikzstyle{large dmap}=[draw, doubled, shape=NEbox, inner sep=2pt, minimum height=6mm, fill=white, minimum width=12mm]
\tikzstyle{gray point}=[point, fill={gray!40!white}]
\tikzstyle{gray dpoint}=[gray point, doubled, tikzit draw={rgb,255: red,128; green,128; blue,128}, tikzit fill={rgb,255: red,128; green,128; blue,128}]
\tikzstyle{gray copoint}=[copoint, fill={gray!40!white}, tikzit fill={rgb,255: red,128; green,128; blue,128}]
\tikzstyle{gray dcopoint}=[gray copoint, doubled, tikzit fill={rgb,255: red,128; green,128; blue,128}]
\tikzstyle{circlenew}=[draw=black, shape=circle, inner sep=1pt]
\tikzstyle{blue label}=[text=NavyBlue, tikzit draw={rgb,255: red,0; green,96; blue,167}, tikzit fill={rgb,255: red,35; green,68; blue,255}]
\tikzstyle{big dot}=[fill=white, draw=black, shape=circle, minimum width=6mm, minimum height=6mm]
\tikzstyle{3d box}=[fill=white, draw=black, shape=trapezium, trapezium left angle=-70, trapezium right angle=70, rotate=10]
\tikzstyle{slant red box}=[fill={rgb,255: red,191; green,0; blue,64}, draw={rgb,255: red,191; green,0; blue,64}, shape=rectangle, xslant=0.5, font={\tiny}, text={rgb,255: red,191; green,0; blue,64}, fill opacity=0.5, line width=1pt]
\tikzstyle{slant point}=[regular polygon, regular polygon sides=3, draw, scale=0.75, inner sep=-0.5pt, minimum width=9mm, fill white, regular polygon rotate=180, yslant=-0.3]
\tikzstyle{tiny orange label}=[font={\tiny}, text={rgb,255: red,255; green,128; blue,0}, tikzit draw={rgb,255: red,255; green,128; blue,0}]
\tikzstyle{tiny red label}=[font={\tiny}, text={rgb,255: red,191; green,0; blue,64}, tikzit draw={rgb,255: red,191; green,0; blue,64}, draw=none]
\tikzstyle{red label}=[text={rgb,255: red,191; green,0; blue,64}, tikzit draw={rgb,255: red,191; green,0; blue,64}]
\tikzstyle{slant label black}=[font={\tiny}, xslant=0.5, tikzit draw=black]
\tikzstyle{slant label red}=[font={\tiny}, xslant=0.5, text={rgb,255: red,191; green,0; blue,64}, tikzit draw={rgb,255: red,191; green,0; blue,64}]
\tikzstyle{slant label orange}=[font={\tiny}, xslant=0.5, text={rgb,255: red,255; green,128; blue,0}, tikzit draw={rgb,255: red,255; green,128; blue,0}]
\tikzstyle{slanted point}=[fill={rgb,255: red,191; green,0; blue,64}, draw={rgb,255: red,191; green,0; blue,64}, shape=triangle, regular polygon, regular polygon sides=3, scale=0.75, inner sep=-0.5pt, minimum width=5mm, regular polygon rotate=90, xslant=0.5, fill opacity=0.5, font={\tiny}, line width=1pt, text={rgb,255: red,191; green,0; blue,64}]
\tikzstyle{slanted point black}=[draw=black, shape=triangle, regular polygon, regular polygon sides=3, scale=0.75, inner sep=-0.5pt, minimum width=5mm, regular polygon rotate=90, xslant=0.5, font={\tiny}, line width=0.2pt, text=black, fill=white, tikzit fill=white]
\tikzstyle{red dot}=[fill={rgb,255: red,191; green,0; blue,64}, draw={rgb,255: red,191; green,0; blue,64}, shape=circle, inner sep=0, minimum width=1.5mm, minimum height=1.5mm]
\tikzstyle{black dot}=[fill=black, draw=black, shape=circle, inner sep=0, minimum width=1.5mm, minimum height=1.5mm]
\tikzstyle{orange dot}=[fill={rgb,255: red,255; green,128; blue,0}, draw={rgb,255: red,255; green,128; blue,0}, shape=circle, inner sep=0, minimum width=1.5mm, minimum height=1.5mm]
\tikzstyle{blue dot}=[fill={rgb,255: red,0; green,0; blue,228}, draw={rgb,255: red,0; green,0; blue,228}, shape=circle, inner sep=0, minimum width=1.5mm, minimum height=1.5mm]
\tikzstyle{slant white}=[fill=white, draw=black, shape=rectangle, xslant=0.5, font={\tiny}, line width=1pt]
\tikzstyle{slant small map}=[fill=white, draw=black, xslant=0.5, shape=rectangle, font={\tiny}, line width=1pt, inner sep=0.6mm]
\tikzstyle{slanted copoint black}=[draw=black, shape=triangle, regular polygon, regular polygon sides=3, scale=0.75, inner sep=-0.5pt, minimum width=5mm, regular polygon rotate=-90, xslant=0.5, font={\tiny}, line width=0.2pt, text=black, fill=white, tikzit fill=white]
\tikzstyle{purple dot}=[fill={rgb,255: red,128; green,0; blue,128}, draw={rgb,255: red,128; green,0; blue,128}, shape=circle, inner sep=0, minimum width=1.5mm, minimum height=1.5mm]
\tikzstyle{white dot 2}=[fill=white, draw=black, shape=circle]
\tikzstyle{horizontal point}=[style=point, rotate=-90, tikzit shape=rectangle, tikzit fill={rgb,255: red,191; green,128; blue,64}]
\tikzstyle{pslant orange}=[style=slanted point black, fill={rgb,255: red,255; green,128; blue,0}, draw={rgb,255: red,255; green,128; blue,0}, tikzit fill={rgb,255: red,255; green,128; blue,0}, tikzit draw={rgb,255: red,255; green,128; blue,0}]
\tikzstyle{upground horizontal}=[style=upground, rotate=-90]
\tikzstyle{double horizontal point}=[style=horizontal point, line width=1pt]
\tikzstyle{double point}=[style=point, line width=1pt]
\tikzstyle{double copoint}=[style=copoint, line width=1pt]
\tikzstyle{horizontal copoint}=[style=double copoint, rotate=-90]
\tikzstyle{slant label purple}=[style=slant label black, tikzit draw={rgb,255: red,128; green,0; blue,128}, text={rgb,255: red,128; green,0; blue,128}]
\tikzstyle{orange copoint}=[style=pslant orange, rotate=-180, tikzit fill={rgb,255: red,255; green,128; blue,0}]
\tikzstyle{new style 0}=[style=slant white, draw={rgb,255: red,0; green,0; blue,228}, fill={rgb,255: red,0; green,0; blue,228}, fill opacity=0.5, shape=rectangle]
\tikzstyle{wide slanted point}=[style=wide point, xslant=0.5, fill=white, rotate=-90, minimum width=0.8cm, fill={rgb,255: red,128; green,128; blue,128}, fill opacity=0.5, line width=1pt]
\tikzstyle{black dot white}=[style=black dot, text=white, draw=none, tikzit draw={rgb,255: red,191; green,255; blue,0}, shape=circle]

\tikzstyle{new edge style 1}=[-, line width=1pt, shorten <=-0.17mm, shorten >=-0.17mm, tikzit draw={rgb,255: red,204; green,0; blue,3}]
\tikzstyle{diredge}=[-, postaction=decorate, decoration={markings, mark=at position 0.55 with \edgearrow}]
\tikzstyle{bold diredge}=[-, diredge, line width=1pt, tikzit draw={rgb,255: red,128; green,0; blue,128}]
\tikzstyle{grey}=[-, draw={rgb,255: red,188; green,188; blue,188}]
\tikzstyle{classical}=[-, dashed, tikzit draw={rgb,255: red,255; green,128; blue,0}]
\tikzstyle{reddashed}=[-, dashed, draw={rgb,255: red,0; green,128; blue,128}, postaction=decorate, decoration={markings, mark=at position 0.55 with \edgearrow}]
\tikzstyle{reddahednoarrow}=[-, dashed, draw={rgb,255: red,179; green,40; blue,40}]
\tikzstyle{arrow edge}=[-, ->, draw={rgb,255: red,191; green,191; blue,191}, tikzit draw={rgb,255: red,191; green,191; blue,191}, ultra thick]
\tikzstyle{tarrow edge}=[-, ->, draw={rgb,255: red,191; green,191; blue,191}, tikzit draw={rgb,255: red,191; green,191; blue,191}]
\tikzstyle{gray edge}=[-, draw={rgb,255: red,191; green,191; blue,191}, tikzit draw={rgb,255: red,191; green,191; blue,191}, ultra thick]
\tikzstyle{lightgrayedge}=[-, draw={rgb,255: red,207; green,207; blue,207}]
\tikzstyle{green edge}=[-, tikzit draw={rgb,255: red,128; green,128; blue,0}, draw={rgb,255: red,128; green,128; blue,0}]
\tikzstyle{red edge}=[-, draw={rgb,255: red,191; green,0; blue,64}, tikzit draw={rgb,255: red,191; green,0; blue,64}]
\tikzstyle{arrow edge black}=[-, ->]
\tikzstyle{solid blue}=[-, draw={rgb,255: red,0; green,96; blue,167}, tikzit draw={rgb,255: red,0; green,96; blue,167}]
\tikzstyle{classical blue}=[-, draw={rgb,255: red,0; green,96; blue,167}, tikzit draw={rgb,255: red,0; green,96; blue,167}, dashed]
\tikzstyle{fill gray}=[-, fill=gray]
\tikzstyle{bold gray}=[-, line width=1pt, tikzit draw={rgb,255: red,128; green,128; blue,128}]
\tikzstyle{fill pink}=[-, fill={rgb,255: red,193; green,100; blue,94}, fill opacity=0.5, draw={rgb,255: red,134; green,68; blue,65}, line width=1pt, tikzit draw={rgb,255: red,134; green,68; blue,65}, tikzit fill={rgb,255: red,193; green,100; blue,94}]
\tikzstyle{fill carta da zucchero}=[-, fill={rgb,255: red,129; green,158; blue,219}, fill opacity=0.5, line width=0.4mm]
\tikzstyle{fill white}=[-, fill=white]
\tikzstyle{fill purple}=[-, fill={rgb,255: red,113; green,69; blue,128}, fill opacity=0.5, draw={rgb,255: red,79; green,48; blue,90}, tikzit fill={rgb,255: red,113; green,69; blue,128}, tikzit draw={rgb,255: red,79; green,48; blue,90}, line width=1pt]
\tikzstyle{fill green}=[-, fill={rgb,255: red,62; green,128; blue,120}, fill opacity=0.5, draw={rgb,255: red,33; green,68; blue,63}, tikzit fill={rgb,255: red,62; green,128; blue,120}, tikzit draw={rgb,255: red,33; green,68; blue,63}, line width=1pt]
\tikzstyle{bold orange}=[-, draw={rgb,255: red,255; green,128; blue,0}, fill=none, line width=1pt]
\tikzstyle{bold black}=[-, line width=1pt, draw=black, fill=none, tikzit draw=black]
\tikzstyle{bold red}=[-, draw={rgb,255: red,191; green,0; blue,64}, fill=none, line width=1pt]
\tikzstyle{fill light green}=[-, fill={rgb,255: red,166; green,166; blue,112}, fill opacity=0.5, draw={rgb,255: red,121; green,121; blue,81}, line width=1pt]
\tikzstyle{new edge style 0}=[-, fill=yellow, fill opacity=0.5, draw={rgb,255: red,146; green,146; blue,0}, tikzit fill=yellow, tikzit draw={rgb,255: red,146; green,146; blue,0}]
\tikzstyle{bold dashed red}=[-, draw={rgb,255: red,191; green,0; blue,64}, fill=none, line width=1pt, dashed]
\tikzstyle{bold dashed orange}=[-, draw={rgb,255: red,255; green,128; blue,0}, dashed, line width=1pt]
\tikzstyle{bold blue}=[-, draw={rgb,255: red,0; green,0; blue,228}, line width=1pt]
\tikzstyle{arrow red}=[draw={rgb,255: red,191; green,0; blue,64}, ->, line width=1pt]
\tikzstyle{new edge style 2}=[-, draw={rgb,255: red,191; green,0; blue,64}, line width=1pt]
\tikzstyle{boldish}=[-, line width=0.6mm, fill=cyan]
\tikzstyle{white edge}=[-, draw=white]
\tikzstyle{purple edge}=[-, draw={rgb,255: red,128; green,0; blue,128}, line width=1pt]
\tikzstyle{light gray}=[-, fill={rgb,255: red,191; green,191; blue,191}, draw={rgb,255: red,191; green,191; blue,191}, tikzit fill={rgb,255: red,191; green,191; blue,191}, tikzit draw={rgb,255: red,191; green,191; blue,191}, fill opacity=0.3]
\tikzstyle{invisible edge}=[-, fill opacity=0, fill=none]
\tikzstyle{carta da zucchero thin}=[-, style=fill carta da zucchero, line width=0.1pt, fill={rgb,255: red,129; green,158; blue,219}, tikzit fill={rgb,255: red,129; green,158; blue,219}]
\tikzstyle{pink thin}=[-, style=fill pink, line width=0.1pt, fill={rgb,255: red,193; green,100; blue,94}]
\tikzstyle{fill green thin edge}=[-, style=fill green, tikzit fill={rgb,255: red,62; green,128; blue,120}, line width=0.1pt]

\definecolor{evred}{rgb}{0.996, 0.403, 0.537}
\definecolor{evgreen}{rgb}{0.501, 1.0, 0.505}
\definecolor{evblue}{rgb}{0.2, 0.588, 1.0}

\usepackage[frozencache]{minted} 

\setminted[python]{
bgcolor=lightgray!10,
xleftmargin=2mm,
fontsize=\footnotesize,
frame=lines,
linenos,
python3=true}
\setminted[text]{
bgcolor=lightgray!10,
xleftmargin=2mm,
fontsize=\footnotesize}

\theoremstyle{definition}
\newtheorem{definition}{Definition}[section]
\newtheorem{remark}{Remark}[section]
\newtheorem{theorem}{Theorem}[section]
\newtheorem{proposition}[theorem]{Proposition}
\newtheorem{lemma}[theorem]{Lemma}
\newtheorem{observation}[theorem]{Observation}
\newtheorem{corollary}[theorem]{Corollary}

\usepackage{xifthen}

\newcommand\pig[1]{\scalerel*[6pt]{\big#1}{%
  \ensurestackMath{\addstackgap[1.5pt]{\big#1}}}}
\newcommand\pigl[1]{\mathopen{\pig{#1}}}
\newcommand\pigr[1]{\mathclose{\pig{#1}}}

\newcommand{\opapp}[2]{\ensuremath{#1\left(#2\right)}} 
\newcommand{\opapptxt}[2]{\ensuremath{\text{#1}\left(#2\right)}} 

\newcommand{\reals}{\ensuremath{\mathbb{R}}}

\newcommand{\tsuchthat}[2]{\ensuremath{\left\{#1\middle|#2\right\}}} 
\newcommand{\suchthat}[2]{\tsuchthat{\,#1\,}{\,#2\,}} 

\newcommand{\downset}[1]{\ensuremath{#1\!\downarrow}}

\newcommand{\domSym}{\text{dom}}
\newcommand{\dom}[1]{\opapp{\domSym}{#1}}
\newcommand{\suppSym}{\text{supp}}
\newcommand{\supp}[1]{\opapp{\suppSym}{#1}}

\newcommand{\restrict}[2]{#1|_{#2}}
\newcommand{\Subsets}[1]{\opapp{\mathcal{P}\!}{#1}} 
\newcommand{\PFun}[1]{\opapptxt{PFun}{#1}} 
\newcommand{\id}[1]{\text{id}_{#1}}

\newcommand{\ev}[1]{\text{#1}} 
\newcommand{\discrete}[1]{\opapptxt{discrete}{#1}} 
\newcommand{\indiscrete}[1]{\opapptxt{indiscrete}{#1}} 
\newcommand{\total}[1]{\opapptxt{total}{#1}} 
\newcommand{\seqcomposeSym}{\rightsquigarrow}

\newcommand{\LsetsSym}{\Lambda} 
\newcommand{\Lsets}[1]{\opapp{\LsetsSym}{#1}} 
\newcommand{\Hist}[1]{\opapptxt{Hist}{#1}} 
\newcommand{\ExtHist}[1]{\opapptxt{ExtHist}{#1}} 
\newcommand{\ExtSym}{\text{Ext}} 
\newcommand{\Ext}[1]{\opapptxt{Ext}{#1}} 
\newcommand{\Prime}[1]{\opapptxt{Prime}{#1}} 
\newcommand{\allJoinsSym}{\;\dot{\vee}\;}
\newcommand{\tips}[2]{\opapp{\text{tips}_{#1}}{#2}} 
\newcommand{\tip}[2]{\opapp{\text{tip}_{#1}}{#2}} 
\newcommand{\Events}[1]{{E}^{#1}} 
\newcommand{\Inputs}[1]{{I}^{#1}} 
\newcommand{\SpacesFC}[1]{\opapp{\text{Spaces}_{\text{FC}}}{#1}} 
\newcommand{\CSwitchSpaces}[1]{\opapptxt{CSwitchSpaces}{#1}} 
\newcommand{\CausFun}[1]{\opapptxt{CausFun}{#1}} 
\newcommand{\ExtFun}[1]{\opapptxt{ExtFun}{#1}} 
\newcommand{\ExtCausFun}[1]{\opapptxt{ExtCausFun}{#1}} 
\newcommand{\CausFunInj}[4]{i_{#1, #2; #3, #4}}
\newcommand{\TipHists}[2]{\opapp{\text{TipHists}_{#1}}{#2}} 
\newcommand{\histconstrSym}[1]{\sim_{#1}}
\newcommand{\histconstr}[3]{#2\!\histconstrSym{#1}\!\!#3}
\newcommand{\nothistconstr}[3]{#2\!\not\histconstrSym{#1}\!\!#3}
\newcommand{\histconstreqcls}[2]{\ensuremath{\left[#1\right]_{\histconstrSym{#2}}}}
\newcommand{\TipEqCls}[2]{\opapp{\text{TipEq}_{#1}}{#2}}
\newcommand{\DistSym}{\mathcal{D}}
\newcommand{\Dist}[1]{\opapp{\DistSym}{#1}}
\newcommand{\CausDist}[1]{\opapptxt{CausDist}{#1}} 
\newcommand{\StdCov}[1]{\opapptxt{StdCov}{#1}} 
\newcommand{\SolCov}[1]{\opapptxt{SolCov}{#1}} 
\newcommand{\ClsCov}[1]{\opapptxt{ClsCov}{#1}} 
\newcommand{\Covers}[1]{\opapptxt{Covers}{#1}} 
\newcommand{\EmpModels}[1]{\opapptxt{EmpMod}{#1}} 

\newcommand{\hist}[1]{
    \ensuremath{
        \left\{
            \foreach \i\j [count=\idx] in {#1}{%
                \ifnum\idx=1%
                    \ev{\i}\!:\!\j%
                \else%
                    ,\,\ev{\i}\!:\!\j%
                \fi%
            }
        \right\}
    }
}
\newcommand{\evset}[1]{
    \ensuremath{
        \left\{
            \foreach \i [count=\idx] in {#1}{%
                \ifnum\idx=1%
                    \ev{\i}%
                \else%
                    ,\ev{\i}%
                \fi%
            }
        \right\}
    }
}

\begin{document}

\title{The Topology of Causality}

\author{Stefano Gogioso$^{1,2}$ and Nicola Pinzani$^{1,3}$}

\address{$^1$Hashberg Ltd, London, UK}
\address{$^2$Department of Computer Science, University of Oxford, Oxford, UK}
\address{$^3$QuIC, Universit\'{e} Libre de Bruxelles, Brussels, BE}
\ead{$^1$stefano.gogioso@cs.ox.ac.uk, $^2$nicola.pinzani@ulb.be}
\vspace{10pt}

\begin{abstract}
    We provide a unified operational framework for the study of causality, non-locality and contextuality, in a fully device-independent and theory-independent setting.
    Our work has its roots in the sheaf-theoretic framework for contextuality by Abramsky and Brandenburger, which it extends to include arbitrary causal orders---be they definite, dynamical or indefinite.
    This paper is the second instalment in a trilogy: spaces of input histories, our dynamical generalisation of causal orders, were introduced in ``The Combinatorics of Causality'', while the polytopes formed by empirical models will be studied in ``The Geometry of Causality''.

    We define a notion of causal function for arbitrary spaces of input histories, and we show that the explicit imposition of causal constraints on joint outputs is equivalent to the free assignment of local outputs to the tip events of input histories.
    We prove factorisation results for causal functions over parallel, sequential, and conditional sequential compositions of the underlying spaces.

    We prove that causality is equivalent to continuity with respect to the lowerset topology on the underlying spaces, and we show that partial causal functions defined on open sub-spaces can be bundled into a presheaf.
    In a striking departure from the Abramsky-Brandenburger setting, however, we show that causal functions fail, under certain circumstances, to form a sheaf.

    We define empirical models as compatible families in the presheaf of probability distributions on causal functions, for arbitrary open covers of the underlying space of input histories.
    We show the existence of causally-induced contextuality---a phenomenon arising when the causal constraints themselves become context-dependent---and we prove a no-go result for non-locality on total orders, both static and dynamical.
\end{abstract}

\maketitle








\section{Introduction}
\label{section:introduction}

In classical causal modelling, causality is understood as a fundamental deterministic structure, relating the different aspects of a phenomenon in a net of causes and effects.
Probabilistic descriptions classically arise as ignorance of latent variables, but this poses issues on data resulting from the interaction with quantum systems, where it becomes problematic to uphold the view that probabilistic behaviour is merely the obfuscation of some causal ontic structure.
The intrinsic stochasticity of quantum effects---which overrides the traditional explanatory mechanisms of causes and effects as delegated to functions and probabilities---should not be seen as a theoretical limitation, but rather as enabling the understanding of a new notion of causality which ``excludes analysis on classical lines'' \cite{kalckar1996philosophy}, only rendered manifest when one appeals to the unification of complementary---i.e. contextual---perspectives.

As Niels Bohr puts it, we are here concerned with ``new uniformities which cannot be framed into the frame of ordinary causal description'' \cite{bohr1937causality}, the exploration of ``harmonies which cannot be comprehended in the pictorial conceptions adapted to the account of more limited fields of physical experience'' \cite{kalckar1996philosophy}: harmonies not amenable to standard causal understanding, but nevertheless governed by precise rules and regularities, within a new domain of applicability.
Bohr uses these words only two years after the EPR proposal---decades before the Bell and Kochen-Specker arguments---but he is already expressing the position that there is something truly essential in the role that contextuality plays in the causality of quantum theory.
Our work is then inspired by the belief that current obstacles to analysing quantum processes with the tools of classical causal discovery may be a manifestation Bohr's original position---the unavoidability of synthesising quantum phenomena by ``combined use of contrasting pictures'' \cite{bohr1996discussion}---a position which is further corroborated by recent works on fine tuning and contextuality \cite{pearl2021classical,cavalcanti2018classical}.

In a nutshell, the objective of our work is the creation of a mathematical framework to understand processes in spacetime, where complementary choices of parameters settings are allowed to influence the causal structure of events.
We achieve this by encoding operational assumptions and causal constraints into suitably defined topological spaces, describing how to attach causal data to open sets in a way which leads to a presheaf of causal distributions: within this setting, we are able to reconcile causality and contextuality from an operational perspective, in a fully theory-independent way.
This completes the unification program started in seminal 2011 work by Abramsky and Brandenburger \cite{abramsky2011sheaf}: contextuality becomes the impossibility of explaining an empirical model in terms of causal classical functions, those which are compatible with a given---definite, dynamical or indefinite---causal order.

This relative definition of contextuality, where an ambient causal order is imposed, is a crucial feature of our extension: a given empirical model can be compatible with multiple causal orders, and its degree of contextuality varies with available causal functions.
In particular, it is known \cite{abbott2016multipartite} that processes which don't violate causal inequalities are always local when we allow their decomposition within a mixture of fixed or dynamical \textit{total} orders, a result which we also obtain within our sheaf-theoretic framework.
To further substantiate our relative perspective, we discuss an empirical model which is local with respect to a fixed causal order, but maximally non-local with respect to a finer causal order involving dynamical causal constraints.

This new interplay between contextuality and causality does not stop at a re-definition of the former: it leads to new phenomena, where the two notions are allowed to influence one another.
In a radical departure from the Abramsky-Brandenburger framework, for example, we discover that causal functions don't always form a sheaf: a certain class of dynamical causal orders admits context-dependent causal constraints, making it possible to define strongly contextual deterministic empirical models on a suitably fine open cover.
Furthermore, these examples of ``causally-induced'' contextuality are far from exceptional: around 2/3 of dynamical causal orders on 3 events with binary inputs---1767 out of 2644, to be precise---exhibit it.

In the companion work ``The Geometry of Causality'' \cite{gogioso2022geometry}, we define a similarly relative notion of causal inseparability, as the failure for empirical models to admit a definite or dynamical causal description subject to some hidden variable and constrained by a given---definite, dynamical or indefinite---causal order.
By exploring novel examples involving entangled and contextually controlled quantum switches, we provide clear evidence for a new phenomenon of ``contextual causality'', where contextuality of the empirical model strongly correlates to---and, in certain examples, coincides with---their causally inseparable fraction.
In the most complicated example, the calculation of contextual fraction would naively involve billions of causal functions with 4096-dimensional empirical models: exploiting causal constraints and factorisation results allows us to restrict the computation to 131072 causal functions, making it possible to obtain detailed landscapes with reasonable computational resources.

\subsection{Background on contextuality}

Before being reconciled as different facets of the same mathematical notion in seminal work by Abramsky and Brandenburger \cite{abramsky2011sheaf}, the study of contextuality and non-locality followed parallel and independent trajectories.
Almost simultaneously to Bell's proof of inadequacy for local realism \cite{bell1964on}, Kochen and Specker \cite{kochen1968problem} tackled the classical-quantum dichotomy without relying on the causal structure of a specific protocol, by considering the possibility of assigning deterministic values to measurements in a way which adheres to our classical intuition.
The impossibility of such deterministic assignments initiated the study of quantum contextuality: much like non-locality, contextuality exposes the limits of the extent to which quantum phenomena can be classically explained.

The aim of Kochen and Specker's work differs from that of Bell's proof: where the latter provides a viable experiment to violate local causality, the former proves a mathematical statement specific to quantum theory.
Indeed, Kochen and Specker's argument directly relies on assumptions about the algebraic structure of quantum observables, leaving open the possibility---at least in principle---of a non-contextual explanation within a different theory.
A violation of Bell's inequality, on the other hand, certifies the inadequacy of local explanation as a descriptive principle of nature, independently on the theory used to describe the correlations.

Because of the intrinsic theory-dependence of the notion, attempts to experimentally ``verify'' contextuality are conceptually controversial \cite{kent1999noncontextual,barrett2004non,mermin1999kochenspecker,simonl2001hidden}.
The original aim of contextuality is to exclude classical descriptions within standard quantum formalism, but what happens if we assume as little as possible about the theory which generated an empirical model?
What are the structural assumptions that one needs to impose in order to consistently speak about a contextuality of raw experimental data?
Barrett and Kent express a strong opinion about this:
\begin{quotation}
\noindent We only wish to note that the class of such alternatives is not merely as general and natural as the class of locally causal theories. So far as the project of verifying the contextuality of Nature (as opposed to the contextuality of hidden variable interpretations of standard quantum formalism) is concerned, the question is of rather limited relevance and interest.\\
\phantom{.}\hfill---Jonathan Barrett and Adrian Kent, \cite{barrett2004non}, p.19
\end{quotation}

Contextuality in quantum theory has been investigated using a plethora of different mathematical presentations \cite{abramsky2011sheaf,cabello2021converting,dzhafarov2016contextuality,spekkens2005contextuality,staton2018effect,schmid2021characterization}.
Amongst the most popular, in recent years, is Spekkens's approach \cite{spekkens2005contextuality}, which provides solid operational foundations to what Budroni and co-authors refer to as the ``effect perspective'' on contextuality \cite{budroni2021quantum}.
Spekkens's framework has seen continuous development across the years \cite{kunjwal2015from,mazurek2016experimental,schmid2018all} includes several discussions of experimental verifiability, as well as a crossover with process theories \cite{schmid2020structure,schmid2021characterization}, but is of little relevance to our work.

The Abramsky-Brandenburger sheaf-theoretic framework \cite{abramsky2011sheaf} shows that the non-locality of Bell---focused on correlations between space-like separated events---and the contextuality of Kochen-Specker---dealing with ideal measurements that do not disturb the outcome statistics of compatible observables---are instances of the same mathematical problem: finding a global section for a compatible family in a certain presheaf of distributions.
Not only does the sheaf-theoretic approach provide a powerful unifying framework, but it also clarifies the standing of related notions from previous literature on hidden variables, such as ``parameter independence'', ``$\lambda$-independence'', ``factorisability'' and ``determinism''.
The core of the sheaf theoretic contextuality has grown steadily through the years, with the addition of operational semantics \cite{abramsky2014operational}, All-vs-Nothing arguments \cite{abramsky2015contextuality}, logical characterisations \cite{abramsky2016possibilities}, continuous variable models \cite{barbosa2019continuous}, reduction from more general general classes of contextuality\cite{mansfield2014extendability}, and the development of additional categorical structure for empirical models \cite{abramsky2017quantum,karvonen2018categories,abramsky2019comonadic,karvonen2021neither}.
Our adoption of the same sheaf-theoretic machinery means that---once topological spaces and presheaves of distributions suitable to our needs are defined---many important definitions and results from the breadth of existing literature will straightforwardly generalise from no-signalling to causality.

One of the salient features of the Abramsky-Brandenburger approach is the possibility to discuss contextuality and non-locality not just qualitatively, but also quantitatively: for example, Figure 1(a) of \cite{abramsky2017contextual} shows the degree of non-locality exhibited by quantum instruments on the Bell state, where each agent is allowed to choose between two arbitrary qubit measurements in the $XY$ plane (both agents have the same two measurements to choose from, parametrised by two angles $\phi_1$ and $\phi_2$).
The notion of ``contextual fraction''---of which ``non-local fraction'' arises as a special case---quantifies the largest component of an empirical model that can be explained classically, as a convex combination of local functions with probabilities specified by some hidden latent variable.
The quantitative aspects of the sheaf-theoretic framework play an important role in our work, where the correlation between contextual fraction and causally inseparable fraction is brought as evidence for the phenomenon of contextual causality.

The vast majority of established literature on quantum contextuality deals with a static setting, involving either a single system or a spacelike separated collection of systems.
A notable exception is 1985 work on macro-realism by Leggett and Garg \cite{leggett1985quantum}, describing inequalities which rule out macro-realist explanations subject to the assumptions of ``macroscopic realism'' and ``non-invasive measurability''.
Exactly what the Leggett-Garg inequalities show has been the subject of some controversy in recent years: for example, it has been argued \cite{maroney2014quantum,bacciagaluppi2014leggett} that the formal similarity in approach to Bell's inequalities is only superficial, and that Leggett-Garg inequalities cannot be used to rule out macro-realism in a suitable theory-independent manner.
In our work, we show that violation of the Leggett-Garg inequalities is actually ascribable to a failure of no-signalling, and not to a generalised notion of contextuality as has been argued in the past.

Another work on contextuality in a non-static setting is Mansfield and Kashefi's \cite{mansfield2018quantum} proof that ``contextuality in time'' is necessary and sufficient to get advantage for the deterministic computation of non-linear functions; in their definition, contextuality is the failure for empirical data obtained by sequential transformations to be reproduced by analogous modular transformations of the underlying ontic states.
Finally, aspects of the interaction between contextuality and indefinite causality have been investigated in \cite{shrapnel2018causation}, where it is shown that no ontological theory which is both instrument and process non-contextual can explain all protocols described by process matrices.
Neither work is directly relevant to our efforts, which are not concerned with ontological descriptions. 

\section{The topology of causality}
\label{section:topology-causality}

The ultimate purpose of this Section is to provide a definition for ``empirical models'', probability distributions on joint outputs conditional to joint inputs which respect given causal constraints.
Contrary to most works on causality, where such constraints take the form of linear equations and inequalities, the definition of empirical models given in this paper is purely topological, in a setting which is natural for the study of non-locality and contextuality.
In the companion work ``The Geometry of Causality'' \cite{gogioso2022geometry}, we show how a geometric description in terms of causal polytopes arises naturally from our topological definition.

Empirical models are defined as ``compatible families'' for a ``presheaf'' of probability distributions over causal functions, the latter being the deterministic mappings of input histories to local outputs which respect causal the causal structure.
The underlying space of input histories is enriched from a combinatorial object to a topological one, where a choice of open cover corresponds to the possible ``contexts'' over which probability distributions are guaranteed to be simultaneously definable.
The hierarchy formed by open covers under refinement then corresponds to all possible kinds and degrees of contextuality, including:
\begin{itemize}
    \item The ``standard cover'', accommodating generic causal distributions on joint outputs conditional to the maximal extended input histories.
    It models settings where it is, at the very least, possible to define conditional distributions when all events are taken together.
    \item The ``classical cover'' is the coarsest cover, lying at the top of the hierarchy.
    It models settings admitting a deterministic causal hidden variable explanation.
    \item The ``fully solipsistic cover'' is the finest cover, lying at the bottom of the hierarchy.
    It models settings more restrictive than those modelled by the standard cover, where it might only be possible to define distributions over the events in the past of some event, which ``witnesses'' their existence, so to speak.
    That is, the fully solipsistic cover accommodates all causal distributions on joint outputs conditional to the maximal input histories.
\end{itemize}

\subsection{Causal Functions (tight causally complete spaces)}
\label{subsection:topology-causality-cfun-tightcc}

Many of the choices made in the definition of spaces of input histories are dictated by our desire to accommodate a certain notion of ``causal function'', mapping inputs to outputs in a way which respects a given causal structure.
To better connect with previous literature, we start by defining our functions on the joint inputs of an operational scenario, independently of any causal constraints.

\begin{definition}
\label{definition:joint-io-function}
A \emph{joint input-output (IO) function} for an operational scenario $(E, \underline{I}, \underline{O})$ is any function $F: \prod_{\omega \in E}I_\omega \rightarrow \prod_{\omega \in E} O_\omega$ which maps joint inputs to joint outputs.
For every choice $k \in \prod_{\omega \in E}I_\omega$ of joint inputs and every event $\omega \in E$, we refer to the component $F(k)_{\omega} \in O_\omega$ as the \emph{output of $F$ at $\omega$}.
\end{definition}

\begin{remark}
Recall that we can formulate a more general definition of operational scenarios with ``dependent'' outputs $\underline{O} = (O_{\omega, i})_{\omega \in E, i \in I_\omega}$.
In this case, joint IO functions for the scenario become ``dependent'' functions, where the output at each event is allowed to depend on the input at that event: 
\[
F \in \prod_{i \in \prod_{\omega \in E}I_\omega} \prod_{\omega \in E} O_{\omega, i_\omega}
\]
When $O_{\omega, i}$ is independent of $i \in I_\omega$, the product type above reduces to the ordinary function type $\prod_{\omega \in E}I_\omega \rightarrow \prod_{\omega \in E} O_\omega$.
\end{remark}

In previous discussion \cite{gogioso2022combinatorics}, we have conceptualised spaces of input histories as defining causal constraints, by specifying the possible input histories upon which the output at any given event $\omega$ is allowed to depend---namely, those having $\omega$ as (one of) their tip event(s).
That conceptualisation is now made concrete in the following definition of a ``causal'' function.

\begin{definition}
\label{definition:causal-function-whole}
Let $F: \prod_{\omega \in E} I_\omega \rightarrow \prod_{\omega \in E} O_\omega$ be a joint IO function for an operational scenario $(E, \underline{I}, \underline{O})$ and let $\Theta$ be a space of input histories such that $\underline{\Inputs{\Theta}} = \underline{I}$.
We say that $F$ is \emph{causal} for $\Theta$ if the output of $F$ at any given event $\omega$ only depends on the data specified by the input histories having $\omega$ as a tip event.
In other words, for all input histories $h \in \Theta$, the outputs $\restrict{F(k)}{\tips{\Theta}{h}}$ at the tips of $h$ are the same for all joint inputs $k \in \prod_{\omega \in E} I_\omega$ such that $h \leq k$.
\end{definition}

As a concrete example, we look at space $\Theta_{33}$ from \cite{gogioso2022combinatorics} Figure~5 (p.~42).
Recall that $\Theta_{33}$ is a tight space, induced by causal order $\total{\ev{A},\ev{B}}\vee\discrete{\ev{C}}$ with a choice of binary inputs for all events.
\begin{center}
    \begin{tabular}{cc}
    \includegraphics[height=3.5cm]{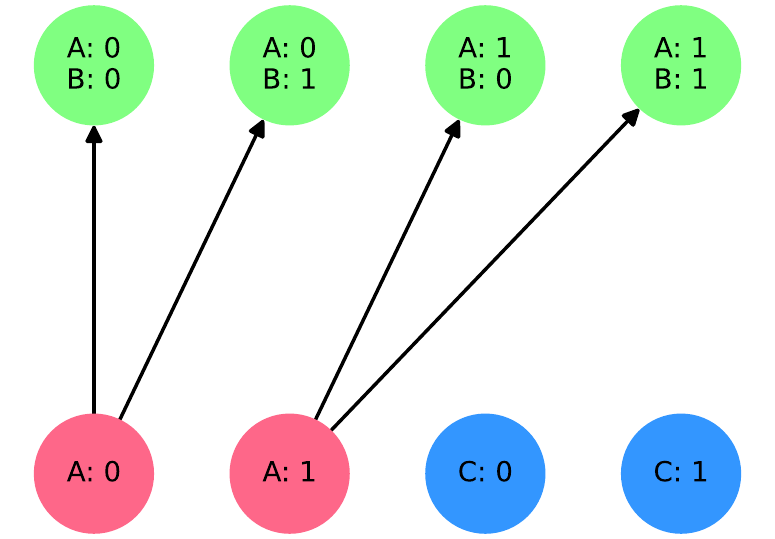}
    &
    \includegraphics[height=3.5cm]{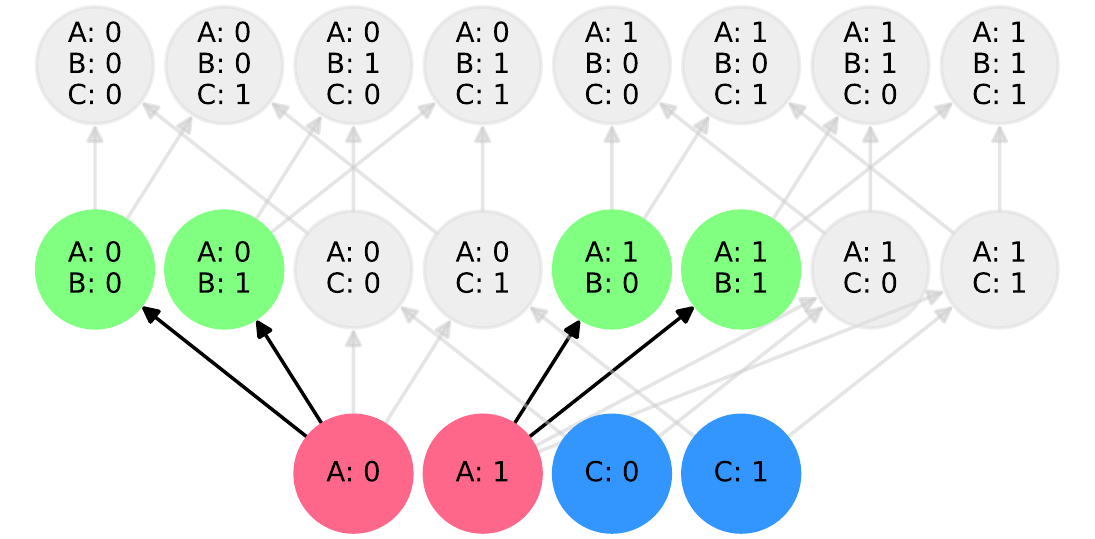}
    \\
    $\Theta_{33}$
    &
    $\Ext{\Theta_{33}}$
    \end{tabular}
\end{center}
A generic function $F: \{0,1\}^{3}\!\rightarrow\!\{0,1\}^{3}$ with binary outputs takes following form: 
\[
F(k_\ev{A}, k_\ev{B}, k_\ev{C})
=
\left(
\begin{array}{c}
F_\ev{A}(k_\ev{A}, k_\ev{B}, k_\ev{C})\\
F_\ev{B}(k_\ev{A}, k_\ev{B}, k_\ev{C})\\
F_\ev{C}(k_\ev{A}, k_\ev{B}, k_\ev{C})\\
\end{array}
\right)
\]
where the binary functions $F_\omega: \{0,1\}^{\evset{A,B,C}} \rightarrow \{0,1\}$ produce the outputs at the three events.
In the case of causal order $\total{\ev{A},\ev{B}}\vee\discrete{\ev{C}}$, we expect joint IO functions $F$ which are causal for $\Theta_{33}$ to take the following simplified form, for generic $G_{\ev{A}}$, $G_{\ev{B}}$ and $G_{\ev{C}}$:
\[
F(k_\ev{A}, k_\ev{B}, k_\ev{C})
=
\left(
\begin{array}{c}
G_\ev{A}(k_\ev{A})\\
G_\ev{B}(k_\ev{A}, k_\ev{B})\\
G_\ev{C}(k_\ev{C})\\
\end{array}
\right)
\]
Indeed, we show that Definition \ref{definition:causal-function-whole} implies the above form for $F(k)$:
\begin{itemize}
    \item The component $F_\ev{A}(k)$ must have the same value for all $k \in \prod_{\omega \in E} I_\omega$ such that $\hist{A/k_\ev{A}} \leq k$, for each choice of $k_\ev{A} \in \{0,1\}$:
    \[
        \begin{array}{rl}
         &F_\ev{A}\left(\hist{A/k_\ev{A},B/0,C/0}\right)
         =F_\ev{A}\left(\hist{A/k_\ev{A},B/0,C/1}\right)\\
        =&F_\ev{A}\left(\hist{A/k_\ev{A},B/1,C/0}\right)
         =F_\ev{A}\left(\hist{A/k_\ev{A},B/1,C/1}\right)
        \end{array}
    \]
    This means that $F_\ev{A}(k_\ev{A}, k_\ev{B}, k_\ev{C}) = G_\ev{A}(k_\ev{A})$ for a generic function $G_\ev{A}: \{0,1\} \rightarrow \{0,1\}$.
    \item The component $F_\ev{B}(k)$ must have the same value for all $k \in \prod_{\omega \in E} I_\omega$ such that $\hist{A/k_\ev{A}, B/k_\ev{B}} \leq k$, for each choice of $k_\ev{A}, k_\ev{B} \in \{0,1\}$:
    \[
         F_\ev{B}\left(\hist{A/k_\ev{A},B/k_\ev{B},C/0}\right)
        =F_\ev{B}\left(\hist{A/k_\ev{A},B/k_\ev{B},C/1}\right)
    \]
    This means that $F_\ev{B}(k_\ev{A}, k_\ev{B}, k_\ev{C}) = G_\ev{B}(k_\ev{A}, k_\ev{B})$ for a generic function $G_\ev{B}: \{0,1\}^2 \rightarrow \{0,1\}$.
    \item The component $F_\ev{C}(k)$ must have the same value for all $k \in \prod_{\omega \in E} I_\omega$ such that $\hist{C/k_\ev{C}} \leq k$, for each choice of $k_\ev{C} \in \{0,1\}$:
    \[
        \begin{array}{rl}
         &F_\ev{C}\left(\hist{A/0,B/0,C/k_\ev{C}}\right)
         =F_\ev{C}\left(\hist{A/0,B/1,C/k_\ev{C}}\right)\\
        =&F_\ev{C}\left(\hist{A/1,B/0,C/k_\ev{C}}\right)
         =F_\ev{C}\left(\hist{A/1,B/1,C/k_\ev{C}}\right)
        \end{array}
    \]
    This means that $F_\ev{C}(k_\ev{A}, k_\ev{B}, k_\ev{C}) = G_\ev{C}(k_\ev{C})$ for a generic function $G_\ev{C}: \{0,1\} \rightarrow \{0,1\}$.
\end{itemize}
Analogous reasoning can be used to prove a general result about the structure of joint IO functions which are causal for spaces induced by causal orders (possibly indefinite).

\begin{proposition}
\label{proposition:causal-function-traditional}
Let $\Theta = \Hist{\Omega, \underline{I}}$ be a space of input histories induced by a causal order $\Omega$ and let $\underline{O} = (O_\omega)_{\omega \in \Omega}$ be a family of non-empty output sets.
The joint IO functions $F: \prod_{\omega \in \Omega}I_\omega \rightarrow \prod_{\omega \in \Omega} O_\omega$ which are causal for $\Theta$ take the following form, for a generic choice of functions $G_\omega: \{0,1\}^{\downset{\omega}} \rightarrow \{0,1\}$:
\[
F\left(k\right)_\omega
=
G_\omega\left(k|_{\downset{\omega}}\right)
\]
\end{proposition}
\begin{proof}
See \ref{proof:proposition:causal-function-traditional}.
\end{proof}

The structure of causal functions derived in Proposition \ref{proposition:causal-function-traditional} is a mainstay of classical causality, so it is legitimate to ask why we didn't use it directly as part of Definition \ref{definition:causal-function-whole}.
The reason is simple: the characterisation of causality as independence of output components from certain input components only works for those (few) spaces which are induced by causal orders.
As an example not fitting this mold, consider space $\Theta_{101}$ from \cite{gogioso2022combinatorics} Figure~5 (p.~42):
\begin{center}
    \begin{tabular}{cc}
    \includegraphics[height=3.5cm]{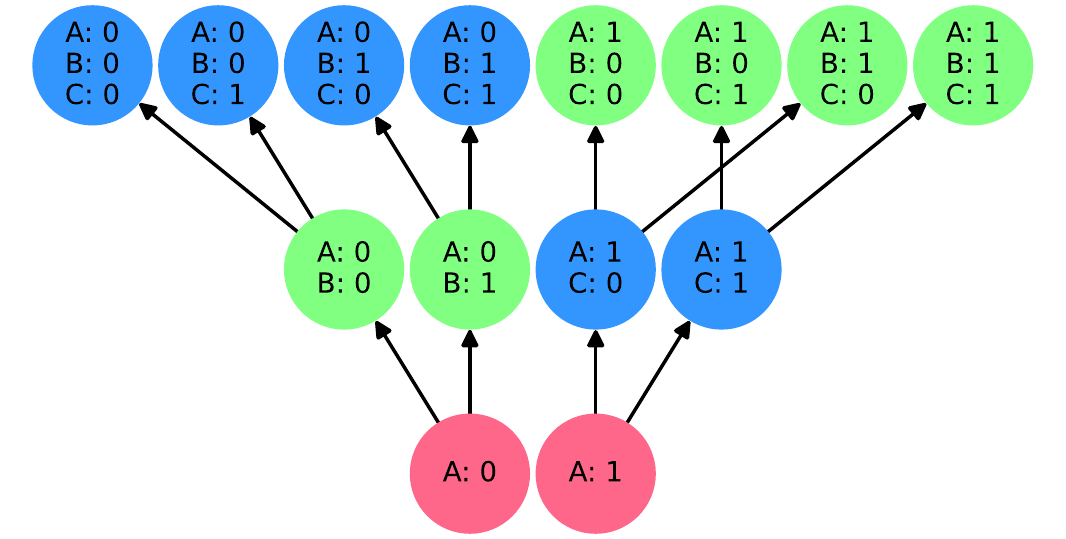}
    &
    \includegraphics[height=3.5cm]{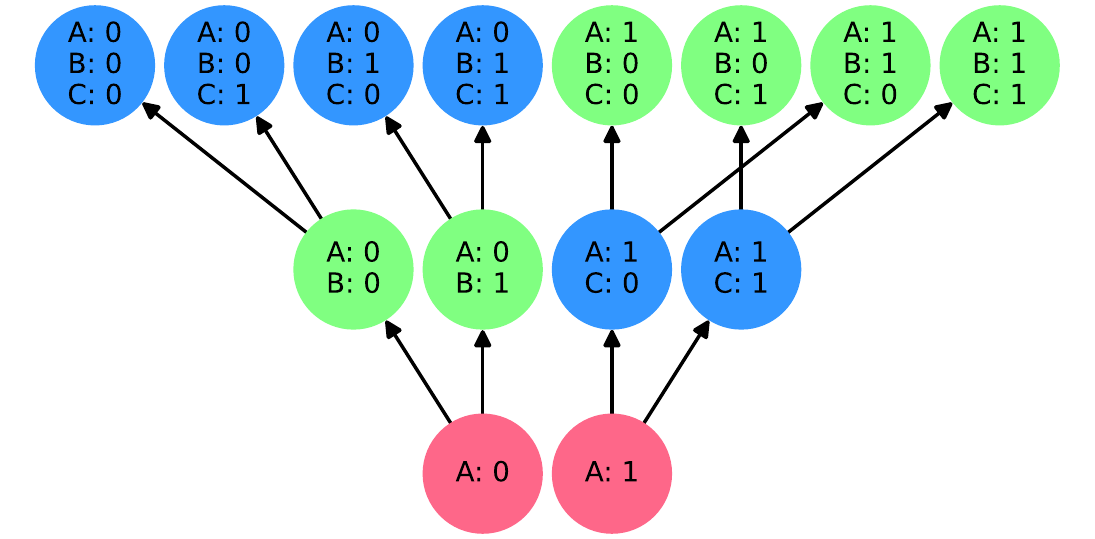}
    \\
    $\Theta_{101}$
    &
    $\Ext{\Theta_{101}}$
    \end{tabular}
\end{center}
Recall that this space is tight, but not order-induced: the definite causal order between events \ev{B} and \ev{C} depends on the input choice at event \ev{A}.
Lets consider again a generic function $F: \{0,1\}^{3}\!\rightarrow\!\{0,1\}^{3}$ with binary outputs: 
\[
F(k_\ev{A}, k_\ev{B}, k_\ev{C})
=
\left(
\begin{array}{c}
F_\ev{A}(k_\ev{A}, k_\ev{B}, k_\ev{C})\\
F_\ev{B}(k_\ev{A}, k_\ev{B}, k_\ev{C})\\
F_\ev{C}(k_\ev{A}, k_\ev{B}, k_\ev{C})\\
\end{array}
\right)
\]
Definition \ref{definition:causal-function-whole} gives the following constraints on a joint IO function $F$ which is causal for $\Theta_{101}$:
\begin{itemize}
    \item The component $F_\ev{A}(k)$ must have the same value for all $k \in \prod_{\omega \in E} I_\omega$ such that $\hist{A/k_\ev{A}} \leq k$, for each choice of $k_\ev{A} \in \{0,1\}$:
    \[
        \begin{array}{rl}
         &F_\ev{A}\left(\hist{A/k_\ev{A},B/0,C/0}\right)
         =F_\ev{A}\left(\hist{A/k_\ev{A},B/0,C/1}\right)\\
        =&F_\ev{A}\left(\hist{A/k_\ev{A},B/1,C/0}\right)
         =F_\ev{A}\left(\hist{A/k_\ev{A},B/1,C/1}\right)
        \end{array}
    \]
    This means that $F_\ev{A}(k_\ev{A}, k_\ev{B}, k_\ev{C}) = G_\ev{A}(k_\ev{A})$ for a generic function $G_\ev{A}: \{0,1\} \rightarrow \{0,1\}$.
    \item The component $F_\ev{B}(k)$ must have the same value for all $k \in \prod_{\omega \in E} I_\omega$ such that $\hist{A/0, B/k_\ev{B}} \leq k$, for each choice of $k_\ev{B} \in \{0,1\}$:
    \[
         F_\ev{B}\left(\hist{A/0,B/k_\ev{B},C/0}\right)
        =F_\ev{B}\left(\hist{A/0,B/k_\ev{B},C/1}\right)
    \]
    This means that $F_\ev{B}(0, k_\ev{B}, k_\ev{C}) = G_{\ev{B},0}(k_\ev{B})$ for a generic function $G_{\ev{B},0}: \{0,1\} \rightarrow \{0,1\}$.
    \item The component $F_\ev{B}(k)$ must have the same value for all $k \in \prod_{\omega \in E} I_\omega$ such that $\hist{A/1,B/k_\ev{B},C/k_\ev{C}} \leq k$, for each choice of $k_\ev{B},k_\ev{C} \in \{0,1\}$.
    This doesn't impose any constraints, as the only such $k$ is $k=\hist{A/1,B/k_\ev{B},C/k_\ev{C}}$ itself.
    This means that $F_\ev{B}(1, k_\ev{B}, k_\ev{C}) = G_{\ev{B},1}(k_\ev{B}, k_\ev{C})$ for a generic function $G_{\ev{B},1}: \{0,1\}^2 \rightarrow \{0,1\}$.
    \item The component $F_\ev{C}(k)$ must have the same value for all $k \in \prod_{\omega \in E} I_\omega$ such that $\hist{A/1, C/k_\ev{C}} \leq k$, for each choice of $k_\ev{C} \in \{0,1\}$:
    \[
         F_\ev{C}\left(\hist{A/0,B/0,C/k_\ev{C}}\right)
        =F_\ev{C}\left(\hist{A/0,B/1,C/k_\ev{C}}\right)
    \]
    This means that $F_\ev{C}(1, k_\ev{B}, k_\ev{C}) = G_{\ev{C},1}(k_\ev{C})$ for a generic function $G_{\ev{C},1}: \{0,1\} \rightarrow \{0,1\}$.
    \item The component $F_\ev{C}(k)$ must have the same value for all $k \in \prod_{\omega \in E} I_\omega$ such that $\hist{A/0,B/k_\ev{B},C/k_\ev{C}} \leq k$, for each choice of $k_\ev{B},k_\ev{C} \in \{0,1\}$.
    This doesn't impose any constraints, as the only such $k$ is $k=\hist{A/0,B/k_\ev{B},C/k_\ev{C}}$ itself.
    This means that $F_\ev{C}(0, k_\ev{B}, k_\ev{C}) = G_{\ev{B},0}(k_\ev{B}, k_\ev{C})$ for a generic function $G_{\ev{C},0}: \{0,1\}^2 \rightarrow \{0,1\}$.
\end{itemize}
Putting all constraints above together, we get the following characterisation of a generic $F$ which is causal for $\Theta_{101}$, for generic functions $G_\ev{A}, G_{\ev{B}, 0}, G_{\ev{B}, 1}, G_{\ev{C}, 0}, G_{\ev{C}, 1}$:
\[
F(0, k_\ev{B}, k_\ev{C})
=
\left(
\begin{array}{c}
G_\ev{A}(0)\\
G_{\ev{B}, 0}(k_\ev{B})\\
G_{\ev{C}, 0}(k_\ev{B}, k_\ev{C})\\
\end{array}
\right)
\hspace{1cm}
F(1, k_\ev{B}, k_\ev{C})
=
\left(
\begin{array}{c}
G_\ev{A}(1)\\
G_{\ev{B}, 1}(k_\ev{B}, k_\ev{C})\\
G_{\ev{C}, 1}(k_\ev{C})\\
\end{array}
\right)
\]

Definitions \ref{definition:joint-io-function} and \ref{definition:causal-function-whole} have the advantage of talking about causality as a property of functions which are themselves defined independently of the causal space.
They allow us to start with a joint IO function $F$ and ask the question ``Is $F$ causal for space $\Theta$?'', or to take a joint IO function $F$ which we already know to be causal for some space $\Theta$ and ask the question ``Is $F$ also causal for space $\Theta'$?''.
They act as a bridge between the causality constraints for different spaces on the given events and inputs.

The same definitions, however, have a major disadvantage: they talk about causality as a set of constraints that a generic function must satisfy, rather than focusing on the causal data which underlies the functions.
Given a joint IO function $F$ for an operational scenario, they make it straightforward to check whether $F$ is causal for a space $\Theta$; given $\Theta$, on the other hand, they don't make it easy to characterise the set of joint IO functions which are causal for $\Theta$.

As it turns out, input histories are the key to understanding the structure of causal functions: up to a minor complication arising from lack of tightness, the joint IO functions which are causal for a causally complete space $\Theta$ are in exact correspondence with functions mapping input histories in $\Theta$ to outputs at their tip events.
We use this as the basis to define ``causal functions'' for a space of input histories $\Theta$, and proceed to show that this definition of causality agrees with that of Definitions \ref{definition:joint-io-function} and \ref{definition:causal-function-whole}. 
We start from the special case of tight causally complete spaces---where things work right off the bat---then generalise to arbitrary tight spaces, and finally generalise to arbitrary spaces of input histories. 

\begin{definition}
\label{definition:causal-function-tight-cc}
Let $\Theta$ be a tight causally complete space and let $\underline{O} = (O_\omega)_{\omega \in \Events{\Theta}}$ be a family of non-empty sets of outputs.
The \emph{causal functions} $\CausFun{\Theta, \underline{O}}$ for space $\Theta$ and outputs $\underline{O}$ are the functions mapping each history in $\Theta$ to the output value for its tip event:
\begin{equation}
    \CausFun{\Theta, \underline{O}}
    :=
    \prod_{h \in \Theta}
    O_{\tip{\Theta}{h}}
\end{equation}
In the special case where the output sets $\underline{O} = (O)_{\omega \in \Events{\Theta}}$ are the same $O$ for all events $\omega$---e.g. the binary case $O=\{0,1\}$---the definition takes the simplified form of functions from the space $\Theta$ to the output set $O$:
\[
    \CausFun{\Theta, O}
    =
    \Theta \rightarrow O
\]
In such cases, $\CausFun{\Theta, O}$ is shorthand for $\CausFun{\Theta, (O)_{\omega \in \Events{\Theta}}}$.
\end{definition}

As an example, consider again the tight causally complete space $\Theta_{33}$, in the binary case $\underline{O} = (\{0,1\})_{\omega \in \evset{A,B,C}}$.
According to Definition \ref{definition:causal-function-tight-cc} above, the functions in $\CausFun{\Theta_{33}, \{0, 1\}}$ are exactly the $2^8=256$ functions $f: \Theta_{33} \rightarrow \{0,1\}$, mapping each one of the 8 histories $h \in \Theta_{33}$ to a binary output value.

We now wish to relate the causal functions from Definition \ref{definition:causal-function-tight-cc} to the joint IO function from Definitions \ref{definition:joint-io-function} and \ref{definition:causal-function-whole}.
We do so by turning the causal functions $f \in \CausFun{\Theta, \underline{O}}$ into corresponding ``extended causal functions'' $\Ext{f}$: the latter are defined on all extended input histories, assigning to each $k \in \Ext{\Theta}$ the output values specified by $f$ on all events in $\dom{k}$.

\begin{definition}
\label{definition:space-extended-function}
Let $\Theta$ be a space of input histories and let $\underline{O} = (O_\omega)_{\omega \in \Events{\Theta}}$ be a family of non-empty output sets.
The \emph{extended functions} on $\Theta$ (with output $\underline{O}$) are those of the following type:
\begin{equation}
    \ExtFun{\Theta, \underline{O}}
    :=
    \prod_{k \in \Ext{\Theta}}\prod_{\omega \in \dom{k}}O_\omega
\end{equation}
\end{definition}

\begin{definition}
\label{definition:ext-f-tight-cc}
Let $\Theta$ be a tight causally complete space and let $\underline{O} = (O_\omega)_{\omega \in \Events{\Theta}}$ be a family of non-empty sets of outputs.
For each causal function $f \in \CausFun{\Theta, \underline{O}}$, define the corresponding \emph{extended causal function} $\Ext{f} \in \ExtFun{\Theta, \underline{O}}$ as follows:
\begin{equation}
    \Ext{f}(k)
    :=
    \left(f\left(h_{k,\omega}\right)\right)_{\omega \in \dom{k}}
    \text{ for all }
    k \in \Ext{\Theta}
\end{equation}
where $h_{k,\omega}$ is the unique input history $h \in \Theta$ such that $h \leq k$ and $\tip{\Theta}{h} = \omega$ (by definition of a tight space).
We refer to $\Ext{f}(k)$ as the \emph{extended output history} corresponding to extended input history $k$.
We say that an extended function $\hat{F} \in \ExtFun{\Theta, \underline{O}}$ is \emph{causal} if it is an extended causal function, i.e. if takes the form $\hat{F} = \Ext{f}$ for some $f \in \CausFun{\Theta, \underline{O}}$.
We write $\ExtCausFun{\Theta, \underline{O}}$ for the subset of $\ExtFun{\Theta, \underline{O}}$ consisting of the extended causal functions.
\end{definition}

An alternative way to describe the extended causal function $\Ext{f}$ is as a ``gluing''---by means of compatible joins---of the output values of $f$ over compatible input histories.
Such ``gluing'' characterisation is made precise by the following definitions and result.

\begin{definition}
\label{definition:consistency-condition}
Let $\Theta$ be a space of input histories and let $\hat{F} \in \ExtFun{\Theta, \underline{O}}$ be an extended function.
We say that $\hat{F}$ satisfies the \emph{consistency condition} if it assigns consistent output histories to compatible extended input histories: $\hat{F}(k') \leq \hat{F}(k)$ for all $k, k' \in \Ext{\Theta}$ such that $k' \leq k$.
We say that $\hat{F}$ satisfies the \emph{gluing condition} if it respects compatible joins: $\hat{F}(k)$ and $\hat{F}(k')$ are compatible for all compatible $k, k' \in \Ext{\Theta}$, and we have $\hat{F}(k\vee k') = \hat{F}(k) \vee \hat{F}(k')$.
\end{definition}

\begin{proposition}
\label{proposition:consistency-gluing-conditions}
In Definition \ref{definition:consistency-condition}, the consistency condition is equivalent to the gluing condition.
\end{proposition}
\begin{proof}
See \ref{proof:proposition:consistency-gluing-conditions}
\end{proof}

\begin{theorem}
\label{theorem:ext-f-characterisation-tight-cc}
Let $\Theta$ be a tight causally complete space, let $\underline{O} = (O_\omega)_{\omega \in \Events{\Theta}}$ be a family of non-empty sets of outputs.
The extended functions $\hat{F} \in \ExtFun{\Theta, \underline{O}}$ which are causal are exactly those which satisfy the consistency condition.
Indeed, the following defines a causal function $\Prime{\hat{F}} \in \CausFun{\Theta, \underline{O}}$ such that $\Ext{\Prime{\hat{F}}} = \hat{F}$:
\begin{equation}
    \Prime{\hat{F}} := h \mapsto \hat{F}(h)_{\tip{\Theta}{h}}
\end{equation}
Furthermore, the above causal function is the unique $f \in \CausFun{\Theta, \underline{O}}$ such that $\Ext{f} = \hat{F}$, because of of the following equation:
\begin{equation}
    \Prime{\Ext{f}} = f
\end{equation}
\end{theorem}
\begin{proof}
See \ref{proof:theorem:ext-f-characterisation-tight-cc}
\end{proof}

Since causally complete spaces $\Theta$ satisfy the free-choice condition, we can restrict $\Ext{f}$ to the maximal extended input histories and obtain a joint IO function which is causal for $\Theta$.
Conversely, every joint IO function which is causal for $\Theta$ arises this way, for a unique choice of $f \in \CausFun{\Theta, \underline{O}}$.
The following result formalises these statements, putting the two definitions of causal functions in exact correspondence for the case of tight causally complete spaces.

\begin{proposition}
\label{proposition:causal-function-joint-function-tight-cc}
Let $\Theta$ be a tight causally complete space and let $\underline{O} = (O_\omega)_{\omega \in \Events{\Theta}}$ be a family of non-empty sets of outputs.
For every $f \in \CausFun{\Theta, \underline{O}}$, the restriction of $\Ext{f}$ to the maximal extended input histories is a joint IO function for the operational scenario $(\Events{\Theta}, \underline{\Inputs{\Theta}}, \underline{O})$ which is causal for $\Theta$.
Conversely, any joint IO function $F$ for $(\Events{\Theta}, \underline{\Inputs{\Theta}}, \underline{O})$ which is causal for $\Theta$ arises as the restriction of an extended causal function $\Ext{f}$ to maximal extended input histories, where $f \in \CausFun{\Theta, \underline{O}}$ can be defined as follows:
\begin{equation}
\label{equation:causal-fun-from-extended-fun-tight-cc}
    f(h)
    :=
    F(k)_{\tip{\Theta}{h}}
    \text{ for any maximal ext. input history $k$ s.t. } h \leq k    
\end{equation}
\end{proposition}
\begin{proof}
See \ref{proof:proposition:causal-function-joint-function-tight-cc}
\end{proof}

As an example, consider once more the tight causally complete space $\Theta_{33}$, in the binary case $\underline{O} = (\{0,1\})_{\omega \in \evset{A,B,C}}$.
Recall that the joint IO functions $F$ which are causal for $\Theta_{33}$ take the following form, for generic $G_{\ev{A}}$, $G_{\ev{B}}$ and $G_{\ev{C}}$:
\[
F(k_\ev{A}, k_\ev{B}, k_\ev{C})
=
\left(
\begin{array}{c}
G_\ev{A}(k_\ev{A})\\
G_\ev{B}(k_\ev{A}, k_\ev{B})\\
G_\ev{C}(k_\ev{C})\\
\end{array}
\right)
\]
Given one such joint IO function $F$, the causal function $f \in \CausFun{\Theta_{33}, \{0,1\}}$ defined by Equation \ref{equation:causal-fun-from-extended-fun-tight-cc} takes the following form:
\[
\begin{array}{rcl}
\hist{A/k_\ev{A}}
&\stackrel{f}{\mapsto}&
G_\ev{A}(k_\ev{A})
\\
\hist{A/k_\ev{A},B/k_\ev{B}}
&\stackrel{f}{\mapsto}&
G_\ev{B}(k_\ev{A}, k_\ev{B})
\\
\hist{C/k_\ev{C}}
&\stackrel{f}{\mapsto}&
G_\ev{C}(k_\ev{C})
\\
\end{array}
\]
The extended causal function $\Ext{f}$ then takes the following form:
\[
\begin{array}{rcl}
\hist{A/k_\ev{A}}
&\stackrel{\Ext{f}}{\mapsto}&
\hist{A/G_\ev{A}(k_\ev{A})}
\\
\hist{C/k_\ev{C}}
&\stackrel{\Ext{f}}{\mapsto}&
\hist{C/G_\ev{C}(k_\ev{C})}
\\
\hist{A/k_\ev{A},B/k_\ev{B}}
&\stackrel{\Ext{f}}{\mapsto}&
\hist{A/{G_\ev{A}(k_\ev{A})},B/{G_\ev{B}(k_\ev{A}, k_\ev{B})}}
\\
\hist{A/k_\ev{A},C/k_\ev{C}}
&\stackrel{\Ext{f}}{\mapsto}&
\hist{A/{G_\ev{A}(k_\ev{A})},C/{G_\ev{C}(k_\ev{C})}}
\\
\hist{A/k_\ev{A},B/k_\ev{B},C/k_\ev{C}}
&\stackrel{\Ext{f}}{\mapsto}&
\hist{A/{G_\ev{A}(k_\ev{A})},B/{G_\ev{B}(k_\ev{A}, k_\ev{B})},C/{G_\ev{C}(k_\ev{C})}}
\\
\end{array}
\]
The last line of the definition of $\Ext{f}$ above is its restriction to the maximal extended input histories, which coincides with the original definition of $F$.

As another example, consider once more the tight causally complete space $\Theta_{101}$, in the binary case $\underline{O} = (\{0,1\})_{\omega \in \evset{A,B,C}}$.
Recall that the joint IO functions $F$ which are causal for $\Theta_{101}$ take the following simplified form, for generic $G_\ev{A}, G_{\ev{B}, 0}, G_{\ev{B}, 1}, G_{\ev{C}, 0}, G_{\ev{C}, 1}$:
\[
F(0, k_\ev{B}, k_\ev{C})
=
\left(
\begin{array}{c}
G_\ev{A}(0)\\
G_{\ev{B}, 0}(k_\ev{B})\\
G_{\ev{C}, 0}(k_\ev{B}, k_\ev{C})\\
\end{array}
\right)
\hspace{1cm}
F(1, k_\ev{B}, k_\ev{C})
=
\left(
\begin{array}{c}
G_\ev{A}(1)\\
G_{\ev{B}, 1}(k_\ev{B}, k_\ev{C})\\
G_{\ev{C}, 1}(k_\ev{C})\\
\end{array}
\right)
\]
Given one such joint IO function $F$, the causal function $f \in \CausFun{\Theta_{33}, \{0,1\}}$ defined by Equation \ref{equation:causal-fun-from-extended-fun-tight-cc} takes the following form:
\[
\begin{array}{rcl}
\hist{A/k_\ev{A}}
&\stackrel{f}{\mapsto}&
G_\ev{A}(k_\ev{A})
\\
\hist{A/0,B/k_\ev{B}}
&\stackrel{f}{\mapsto}&
G_{\ev{B},0}(k_\ev{B})
\\
\hist{A/1,C/k_\ev{C}}
&\stackrel{f}{\mapsto}&
G_{\ev{C},1}(k_\ev{C})
\\
\hist{A/0,B/k_\ev{B},C/k_\ev{C}}
&\stackrel{f}{\mapsto}&
G_{\ev{C},0}(k_\ev{B}, k_\ev{C})
\\
\hist{A/1,B/k_\ev{B},C/k_\ev{C}}
&\stackrel{f}{\mapsto}&
G_{\ev{B},1}(k_\ev{B}, k_\ev{C})
\\
\end{array}
\]
The extended causal function $\Ext{f}$ then takes the following form:
\[
\begin{array}{rcl}
\hist{A/k_\ev{A}}
&\stackrel{\Ext{f}}{\mapsto}&
\hist{A/G_\ev{A}(k_\ev{A})}
\\
\hist{A/0,B/k_\ev{B}}
&\stackrel{\Ext{f}}{\mapsto}&
\hist{A/{G_\ev{A}(0)},B/{G_{\ev{B},0}(k_\ev{B})}}
\\
\hist{A/1,C/k_\ev{C}}
&\stackrel{\Ext{f}}{\mapsto}&
\hist{A/{G_\ev{A}(1)},C/{G_{\ev{C},1}(k_\ev{C})}}
\\
\hist{A/0,B/k_\ev{B},C/k_\ev{C}}
&\stackrel{\Ext{f}}{\mapsto}&
\hist{A/{G_\ev{A}(0)},B/{G_{\ev{B},0}(k_\ev{B})},C/{G_{\ev{C},0}(k_\ev{B},k_\ev{C})}}
\\
\hist{A/1,B/k_\ev{B},C/k_\ev{C}}
&\stackrel{\Ext{f}}{\mapsto}&
\hist{A/{G_\ev{A}(1)},B/{G_{\ev{B},1}(k_\ev{B},k_\ev{C})},C/{G_{\ev{C},1}(k_\ev{B})}}
\\
\end{array}
\]
The last two lines of the definition of $\Ext{f}$ above are its restriction to the maximal extended input histories, which coincides with the original definition of $F$.

\subsection{Causal Functions (general case)}
\label{subsection:topology-causality-cfun}

We now proceed to generalise the definition of causal functions to all spaces of input histories.
We do so in several steps:
\begin{enumerate}
    \item We drop the causal completeness requirement in Definition \ref{definition:causal-function-tight-cc}.
    \item We discuss causal inseparability, an inconvenient property of certain causal functions on causally incomplete spaces.
    \item We discuss the additional requirements imposed on causal functions in the non-tight case.
    \item Inspired by the previous discussion, we drop the tightness requirement in Definition \ref{definition:causal-function-tight-cc}.
    \item We extend Propositions \ref{theorem:ext-f-characterisation-tight-cc} and \ref{proposition:causal-function-joint-function-tight-cc} to generic spaces of input histories.
\end{enumerate}
Along the way, we give some definitions and observations that don't depend on whether the spaces involved are causally complete or tight: to avoid pointless repetition further on, we phrase these in full generality from the start, even though the definition of causal functions for arbitrary spaces of input histories will be yet to come.

\begin{definition}
\label{definition:causal-function-tight}
Let $\Theta$ be a tight space and let $\underline{O} = (O_\omega)_{\omega \in \Events{\Theta}}$ be a family of non-empty sets of outputs.
The \emph{causal functions} $\CausFun{\Theta, \underline{O}}$ for space $\Theta$ and outputs $\underline{O}$ are the functions mapping each history in $\Theta$ to the output values for its tip events:
\begin{equation}
    \CausFun{\Theta, \underline{O}}
    :=
    \prod_{h \in \Theta}
    \prod_{\omega \in \tips{\Theta}{h}}
    O_{\omega}
\end{equation}
\end{definition}

\begin{remark}
When $\Theta$ is a causally complete space, we have $\tips{\Theta}{h} = \{\tip{\Theta}{h}\}$ for all $h$, so that Definition \ref{definition:causal-function-tight} is equivalent to Definition \ref{definition:causal-function-tight-cc}:
\[
\prod_{\omega \in \tips{\Theta}{h}} O_{\omega}
=
\prod_{\omega \in \{\tip{\Theta}{h}\}} O_{\omega}
\simeq
O_{\tip{\Theta}{h}}
\]
\end{remark}

Causal functions on a causally incomplete space include all causal functions for its causal completions: because of this observation, it might be tempting to think of causal incompleteness as specifying indefinite causal order to be made precise by a causally complete subspace.
However, not all causal functions on a causally incomplete space arise this way: some of them are ``inseparable'', requiring multiparty signalling and leading to delocalised events.

As a concrete example, consider the tight space $\Theta = \Hist{\Omega, \{0,1\}}$ of in input histories induced by the indefinite causal order $\Omega = \total{\ev{A}, \evset{B,C}}$ with binary inputs.
The space satisfies the free-choice condition, and the 8 maximal extended input histories have $\evset{B,C}$ as their tip events.
\begin{center}
    \includegraphics[height=2.5cm]{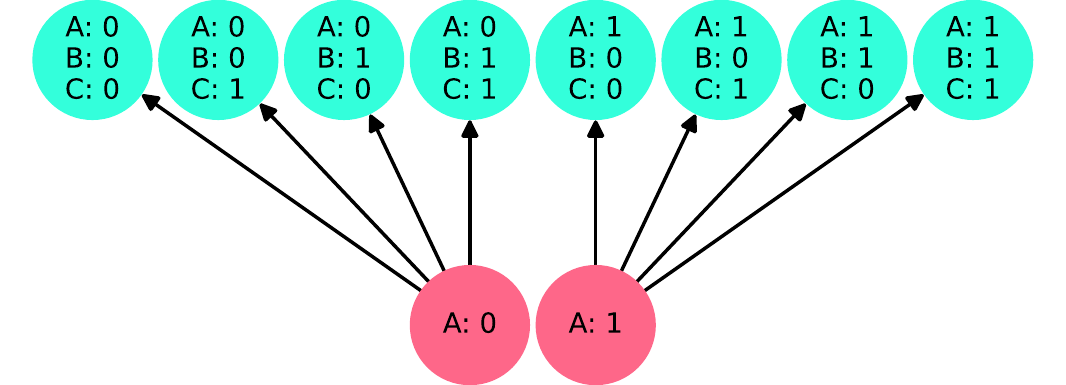}
\end{center}
Recall that there are four causal completions for this space: two where $\ev{B}$ and $\ev{C}$ are unconditionally totally ordered, and two where they are totally ordered conditionally to the input at $\ev{A}$.
\begin{center}
\begin{tabular}{cc}
    \includegraphics[height=3cm]{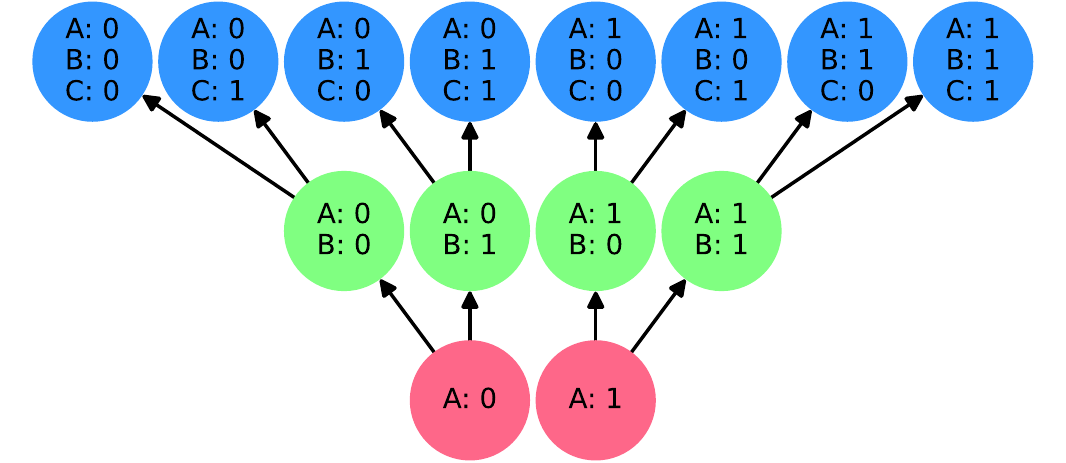}
    &
    \includegraphics[height=3cm]{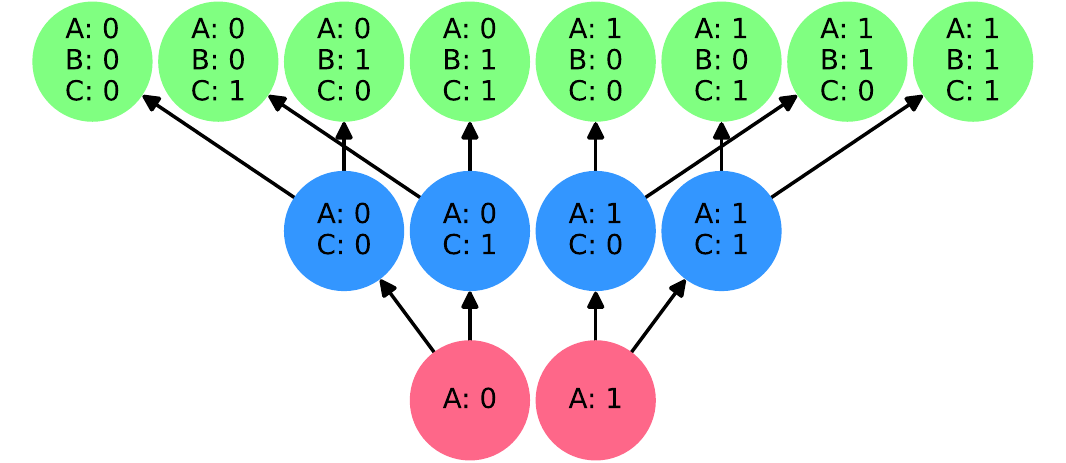}
    \\
    \includegraphics[height=3cm]{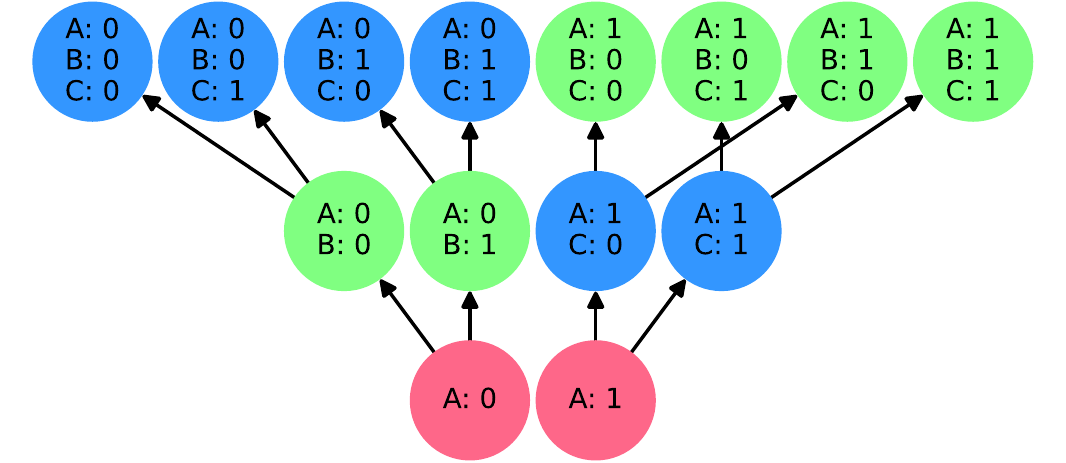}
    &
    \includegraphics[height=3cm]{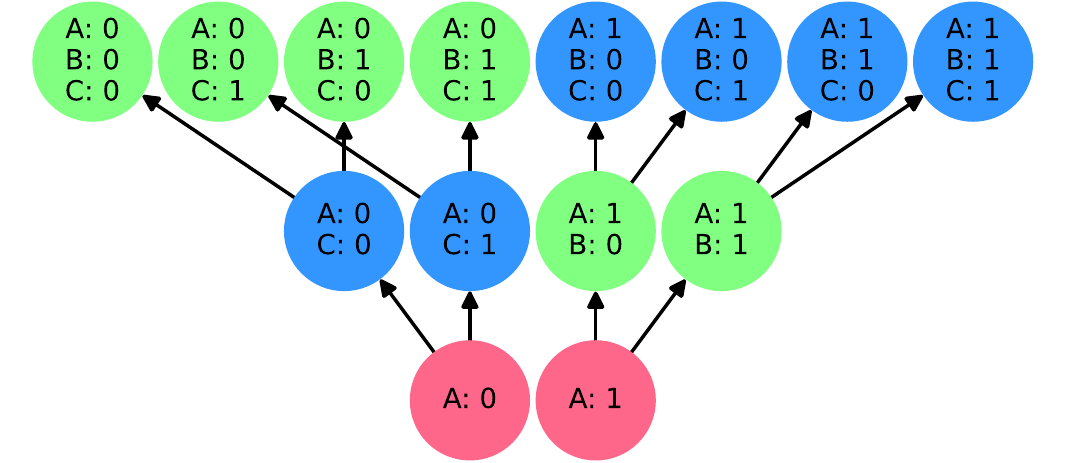}
\end{tabular}
\end{center}
The causally incomplete space $\Theta$ has the following number of causal functions:
\[
\prod_{h \in \Theta} 2^{|\tips{\Theta}{h}|}
=
2^{\sum_{h \in \Theta}|\tips{\Theta}{h}|}
=
2^{2\cdot 1 + 8 \cdot 2}
=
2^{18}
=
262144
\]
Each of the four causal completions has the following number of causal functions:
\[
\prod_{h \in \Theta} 2^{|\tips{\Theta}{h}|}
=
2^{\sum_{h \in \Theta} 1}
=
2^{|\Theta|}
=
2^{14}
=
16384
\]
However, some causal functions are common to multiple causal completions, so only 50176 of the 262144 causal functions on $\Theta$ arise from one of its completions, according to the following definition.

\begin{definition}
Let $\Theta, \Theta'$ be spaces of input histories such that $\Theta' \leq \Theta$.
Let $f \in \CausFun{\Theta, \underline{O}}$ be a causal function on $\Theta$ and let $f' \in \CausFun{\Theta', \underline{O}'}$ be a causal function on $\Theta'$.
We say that $f$ \emph{arises from} $f'$ if the extended causal function $\Ext{f'}$ restricts to  the extended causal function $\Ext{f}$:
\[
f \text{ arises from } f'
\Leftrightarrow
\restrict{\Ext{f'}}{\Ext{\Theta}} = \Ext{f}
\]
where we have used the fact that $\Theta' \leq \Theta$ is defined as $\Ext{\Theta'} \supseteq \Ext{\Theta}$.
\end{definition}

\begin{proposition}
\label{proposition:caus-fun-subspaces}
Let $\Theta', \Theta \in \SpacesFC{\underline{I}}$ be such that $\Theta' \leq \Theta$; define $E := \Events{\Theta} = \Events{\Theta'}$.
Let $\underline{O} = (O_\omega)_{\omega \in E}$ and $\underline{O'} = (O'_\omega)_{\omega \in E}$ be families of non-empty output sets such that $O'_\omega \subseteq O_\omega$ for all $\omega \in E$.
Then the following is an injection:
\[
\begin{array}{rrcl}
\CausFunInj{\Theta'}{\underline{O}'}{\Theta}{\underline{O}}:
&\CausFun{\Theta', \underline{O}'}
&\hookrightarrow
&\CausFun{\Theta, \underline{O}}
\\
&f'
&\mapsto
&\Prime{\restrict{\Ext{f'}}{\Ext{\Theta}}}
\end{array}
\]
We can use the injection above to identify the causal functions for $\Theta'$ and $\underline{O}'$ with a subset of the causal functions for $\Theta$ and $\underline{O}$.
This is safe, because the injections are stable under composition:
\[
\CausFunInj{\Theta'}{\underline{O}'}{\Theta}{\underline{O}}
\circ
\CausFunInj{\Theta''}{\underline{O}''}{\Theta'}{\underline{O}'}
=
\CausFunInj{\Theta''}{\underline{O}''}{\Theta}{\underline{O}}
\]
\end{proposition}
\begin{proof}
See \ref{proof:proposition:caus-fun-subspaces}
\end{proof}

There are examples of spaces where causal functions all arise from those of some sub-space.
For the 3-event case with binary inputs, such spaces are indicated by thick blue arrows in \cite{gogioso2022combinatorics} Figure~5 (p.42), and they are explicitly listed below.
\begin{itemize}
    \item The causal functions on spaces in equivalence classes 1, 2, 3, 6, 7, 9, 10 and 13 all arise from causal functions on the discrete space, the only space in equivalence class 0.
    \item The causal functions on each space in class 11 all arise from causal functions on one of its refinements in class 4.
    \item The causal functions on each space in classes 12, 14, 27, 16 and 19 all arise from causal functions on one of its refinements in class 5.
    \item The causal functions on each space in classes 31, 44, 51 and 59 all arise from causal functions on one of its refinements in class 18.
    \item The causal functions on each space in class 38 all arise from causal functions on one of its refinements in class 25.
    \item The causal functions on each space in class 35 all arise from causal functions on one of its refinements in class 26.
    \item The causal functions on each space in class 47 all arise from causal functions on one of its refinements in class 30.
    \item The causal functions on each space in class 52 all arise from causal functions on one of its refinements in class 32.
    \item The causal functions on each space in classes 61 and 76 all arise from causal functions on one of its refinements in class 45.
    \item The causal functions on each space in class 71 all arise from causal functions on one of its refinements in class 48.
\end{itemize}
Further to the above, there are 21 equivalence classes of spaces which don't introduce new causal functions compared to their refinements, i.e. spaces where each causal function arises from a causal function on one of the their refinements (but not all causal functions arise from the same refinement): 8, 15, 20, 22, 23, 24, 26, 34, 39, 42, 43, 46, 49, 50, 53, 54, 57, 66, 67, 70 and 84. In \cite{gogioso2022combinatorics} Figure~5 (p.~42), these are indicated by a grey node. 

The 211968 causal functions on $\Theta = \Hist{\total{\ev{A}, \evset{B,C}}, \{0,1\}}$ that don't arise from one of its causal completions---and hence don't arise from any one of its causally-complete sub-spaces---are ``inseparable'': the joint output produced at tip events $\evset{B,C}$ on one of the maximal extended input histories functionally depend on the inputs at both events $\ev{B}$ and $\ev{C}$, making it incompatible with any causally complete subspace.
As a simple example, consider the following ``controlled swap'' causal function on $\Theta$:
\[
\begin{array}{rcl}
\hist{A/k_\ev{A}}
&\stackrel{\text{cswap}}{\mapsto}&
\hist{A/k_\ev{A}}
\\
\hist{A/0,B/k_\ev{B},C/k_\ev{C}}
&\stackrel{\text{cswap}}{\mapsto}&
\hist{B/k_\ev{B},C/k_\ev{C}}
\\
\hist{A/1,B/k_\ev{B},C/k_\ev{C}}
&\stackrel{\text{cswap}}{\mapsto}&
\hist{B/k_\ev{C},C/k_\ev{B}}
\\
\end{array}
\]
The controlled swap function requires true bipartite signalling, where events \ev{B} and \ev{C} are delocalised even conditional to the input at \ev{A}.
Indeed, when the input at \ev{A} is 1:
\begin{enumerate}
    \item the output at \ev{B} depends on the input at \ev{C}, which must therefore be in \ev{B}'s past;
    \item the output at \ev{C} depends of the input at \ev{B}, which must therefore be in \ev{C}'s past.
\end{enumerate}
As a consequence, the controlled swap function cannot arise from any causal function $f$ on any causally complete subspace $\Theta' \leq \Theta$ (i.e. we cannot have $\restrict{\Ext{f}}{\Ext{\Theta}} = \Ext{\text{cswap}}$).
To see why, consider the extended input history \hist{A/1,B/0,C/0}, with tips $\evset{B,C}$.
Any causal completion $\Theta'$ of our space $\Theta$ must necessarily include as an extended input history one of the following partial functions:
\begin{enumerate}
    \item \hist{A/1,C/0}, obtained by removing $\ev{B}$ from the domain of \hist{A/1,B/0,C/0}
    \item \hist{A/1,B/0}, obtained by removing $\ev{C}$ from the domain of \hist{A/1,B/0,C/0}
\end{enumerate}
In either case, the causal function $\text{cswap} \in \CausFun{\Theta, \{0,1\}}$ cannot arise from any $f' \in \CausFun{\Theta', \{0,1\}}$, because $\Ext{f'}$ cannot satisfy the consistency condition.
If $\hist{A/1,C/0} \in \Ext{\Theta'}$, we are forced to make the following inconsistent assignments:
\begin{itemize}
    \item from $\hist{A/1,C/0} \leq \hist{A/1,B/0,C/0}$, we must have:
    \[
        \Ext{f'}_{\ev{C}}\left(\hist{A/1,C/0}\right)
        =
        \Ext{f'}_{\ev{C}}\left(\hist{A/1,B/0,C/0}\right)=0
    \]
    \item from $\hist{A/1,C/0} \leq \hist{A/1,B/1,C/0}$, we must have:
    \[
        \Ext{f'}_{\ev{C}}\left(\hist{A/1,C/0}\right)
        =
        \Ext{f'}_{\ev{C}}\left(\hist{A/1,B/1,C/0}\right)=1
    \]
\end{itemize}
If instead $\hist{A/1,B/0} \in \Ext{\Theta'}$, we are forced to make the following inconsistent assignments:
\begin{itemize}
    \item from $\hist{A/1,B/0} \leq \hist{A/1,B/0,C/0}$, we must have:
    \[
        \Ext{f'}_{\ev{B}}\left(\hist{A/1,B/0}\right)
        =
        \Ext{f'}_{\ev{B}}\left(\hist{A/1,B/0,C/0}\right)=0
    \]
    \item from $\hist{A/1,B/0} \leq \hist{A/1,B/0,C/1}$, we must have:
    \[
        \Ext{f'}_{\ev{B}}\left(\hist{A/1,B/0}\right)
        =
        \Ext{f'}_{\ev{B}}\left(\hist{A/1,B/0,C/1}\right)=1
    \]
\end{itemize}
The information above, proving that $\text{cswap}$ is inseparable, can be summarised as follows. There is an extended input history $k=\hist{A/1,B/0,C/0}$ such that for all events $\omega \in \dom{k}$ the function $\Ext{\text{cswap}}$ could not be extended to $\restrict{k}{\dom{k}\backslash\{\omega\}}$:
\begin{enumerate}
    \item if $\omega = \ev{B}$, $\restrict{k}{\dom{k}\backslash\{\omega\}} = \hist{A/1,C/0}$
    \item if $\omega = \ev{C}$, $\restrict{k}{\dom{k}\backslash\{\omega\}} = \hist{A/1,B/0}$
\end{enumerate}
This is because for each $\omega$ there is an extended input history $k'_\omega \in \Ext{\Theta}$ with $\restrict{k}{\dom{k}\backslash\{\omega\}} \leq k'_\omega$ and an event $\xi_\omega \in \dom{k}\backslash\{\omega\}$ such that $\Ext{\text{cswap}}(k)_{\xi_\omega} \neq \Ext{\text{cswap}}(k'_\omega)_{\xi_\omega}$:
\begin{enumerate}
    \item if $\omega = \ev{B}$, we can choose $k'_\omega = \hist{A/1,B/1,C/0}$ and $\xi_\omega = \ev{C}$
    \item if $\omega = \ev{C}$, we can choose $k'_\omega = \hist{A/1,B/0,C/1}$ and $\xi_\omega = \ev{B}$
\end{enumerate}
We refer to such a triple $\left(k, (k'_\omega)_\omega, (\xi_\omega)_\omega\right)$ as an ``inseparability witness'', as it proves that the controlled swap cannot arise from any causal function $f'$ on any causally complete subspace $\Theta' \leq \Theta$.
We will now generalise and formalise our discussion thus far, and show that inseparable functions are exactly those with (at least) one such inseparability witness.

\begin{definition}
Let $\Theta$ be a space of input histories and let $\underline{O} = (O_\omega)_{\omega \in \Events{\Theta}}$ be a family of non-empty output sets.
A causal function $f \in \CausFun{\Theta, \underline{O}}$ is said to be \emph{separable} if it arises from $f' \in \CausFun{\Theta', \underline{O}}$ for some causally complete $\Theta' \leq \Theta$, and \emph{inseparable} otherwise.
We refer to the causal functions which are (in)separable as \emph{(in)separable functions}, for short.
\end{definition}

\begin{observation}
It is always enough to consider the causal completions $\Theta' \leq \Theta$ of $\Theta$, which satisfy $\Events{\Theta'} = \Events{\Theta}$ and $\underline{\Inputs{\Theta'}} = \underline{\Inputs{\Theta}}$.
\end{observation}

\begin{definition}
Let $\Theta$ be a space of input histories and let $\underline{O} = (O_\omega)_{\omega \in \Events{\Theta}}$ be a family of non-empty output sets.
An \emph{inseparability witness} for a causal function $f \in \CausFun{\Theta, \underline{O}}$ is a triple $(k, (k'_\omega)_{\omega \in \dom{k}}, (\xi_\omega)_{\omega \in \dom{k}})$ where: 
\begin{itemize}
    \item $k \in \Ext{\Theta}$ is an extended input history;
    \item for every $\omega \in \dom{k}$, $k'_\omega \in \Ext{\Theta}$ is such that $\restrict{k}{\dom{k}\backslash\{\omega\}} \leq k'_\omega$;
    \item for every $\omega \in \dom{k}$, $\xi_\omega \in \dom{k}\backslash\{\omega\}$ is such that $\Ext{f}(k)_{\xi_{\omega}} \neq \Ext{f}(k'_\omega)_{\xi_{\omega}}$.
\end{itemize}
\end{definition}

\begin{lemma}
\label{lemma:inseparability-witness-1}
Let $\Theta$ be a space of input histories and let $\underline{O} = (O_\omega)_{\omega \in \Events{\Theta}}$ be a family of non-empty output sets.
If $f \in \CausFun{\Theta, \underline{O}}$ is a causal function and $(k, \underline{k'}, \underline{\xi})$ is an inseparability witness for $f$, then $|\dom{k}| \geq 2$ and for all $\omega \in \dom{k}$ the partial function $h_\omega:=\restrict{k}{\dom{k}\backslash\{\omega\}}$ is not an extended input history for $\Theta$.
\end{lemma}
\begin{proof}
See \ref{proof:lemma:inseparability-witness-1}
\end{proof}

\begin{lemma}
\label{lemma:inseparability-witness-2}
Let $\Theta$ be a space of input histories and let $\underline{O} = (O_\omega)_{\omega \in \Events{\Theta}}$ be a family of non-empty output sets.
Let $f \in \CausFun{\Theta, \underline{O}}$ be a causal function and $(k, \underline{k'}, \underline{\xi})$ be an inseparability witness for $f$.
If $f$ arises from $f' \in \CausFun{\Theta', \underline{O}'}$ for some $\Theta' \leq \Theta$, then $(k, \underline{k'}, \underline{\xi})$ is an inseparability witness for $f'$.
\end{lemma}
\begin{proof}
See \ref{proof:lemma:inseparability-witness-2}
\end{proof}

\begin{theorem}
\label{proposition:inseparability-witness}
Let $\Theta$ be a space of input histories and let $\underline{O} = (O_\omega)_{\omega \in \Events{\Theta}}$ be a family of non-empty output sets.
A causal function $f \in \CausFun{\Theta, \underline{O}}$ is inseparable if and only if it has an inseparability witness $(k, \underline{k'}, \underline{\xi})$.
\end{theorem}
\begin{proof}
See \ref{proof:proposition:inseparability-witness}
\end{proof}

Having investigated the effects of causal incompleteness in the definition of causal functions, we now proceed to relax the tightness assumption: for each extended input history $k \in \Ext{\Theta}$ and $\omega \in \dom{k}$, we are no longer guaranteed that the input history $h \leq k$ with $\omega \in \tips{\Theta}{h}$ will be unique.
When $\Theta$ is not tight, this will impose additional constraints to the definition of causal functions $f \in \CausFun{\Theta}$: if $h, h' \leq k$ are distinct input histories such that $\omega \in \tips{\Theta}{h} \cap \tips{\Theta}{h'}$, then the outputs $f(h)_\omega$ and $f(h')_\omega$ at event $\omega$ must coincide.

As a concrete example, consider the causally complete non-tight space $\Theta_{21}$ from \cite{gogioso2022combinatorics} Figure~5 (p.~42).
There are two violations of tightness in this space:
\begin{itemize}
    \item The extended input history \hist{A/0,B/1,C/0} has two sub-histories with event $\ev{B}$ as their tip event: \hist{A/0,B/1} and \hist{B/1,C/0}.
    \item The extended input history \hist{A/1,B/0,C/1} has two sub-histories with event $\ev{C}$ as their tip event: \hist{A/1,C/1} and \hist{B/0,C/1}.
\end{itemize}
The two histories are highlighted in $\Ext{\Theta_{21}}$ below, with a border of the same colour as the corresponding tip event conflicts.
\begin{center}
    \begin{tabular}{cc}
    \includegraphics[height=3cm]{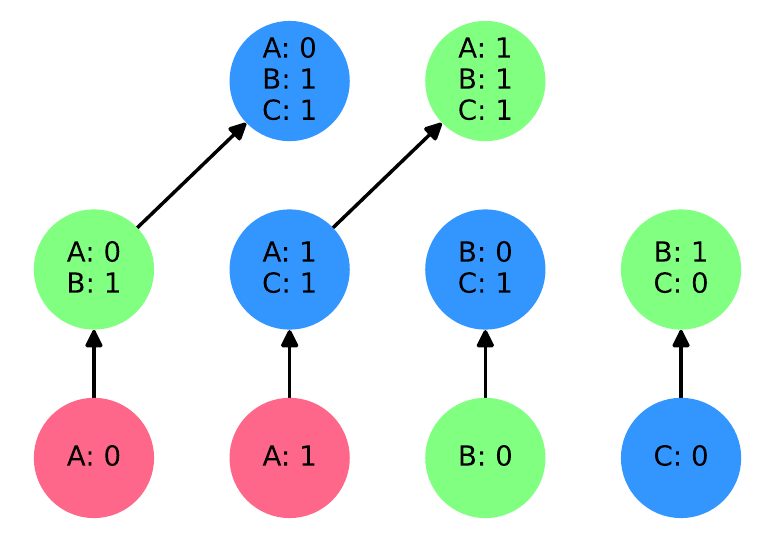}
    &
    \includegraphics[height=3.5cm]{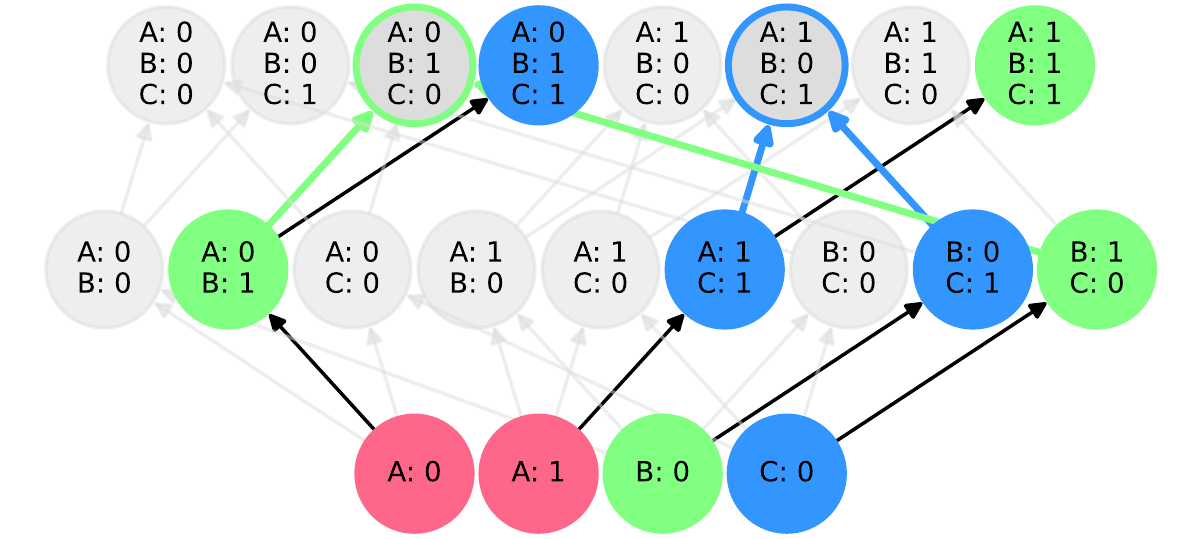}
    \\
    $\Theta_{21}$
    &
    $\Ext{\Theta_{21}}$ with highlights
    \end{tabular}
\end{center}
In order for the extended function $\Ext{f}_{\ev{B}}$ to be well-defined on the (maximal) extended input history \hist{A/0,B/1,C/0}, the outputs associated by a causal function $f$ to the two input histories \hist{A/0,B/1} and \hist{B/1,C/0} must coincide:
\[
\begin{array}{rcl}
\Ext{f}_{\ev{B}}\left(\hist{A/0,B/1,C/0}\right)
&=&
f\left(\hist{A/0,B/1}\right)
\\
&=&
f\left(\hist{B/1,C/0}\right)
\end{array}
\]
Analogously, in order for $\Ext{f}_{\ev{C}}$ to be well-defined on the (maximal) extended input history \hist{A/1,B/0,C/1}, the outputs associated by a causal function $f$ to the two input histories \hist{A/1,C/1} and \hist{B/0,C/1} must coincide:
\[
\begin{array}{rcl}
\Ext{f}_{\ev{B}}\left(\hist{A/1,B/0,C/1}\right)
&=&
f\left(\hist{A/1,C/1}\right)
\\
&=&
f\left(\hist{B/0,C/1}\right)
\end{array}
\]
At first sight, such constrains make is seem like the definition of causal functions is no longer ``free'', but this is not actually the case: instead of a constrained mapping of individual input histories to outputs at their tip event(s), we can think of a causal function on a non-tight space as a free mapping of equivalence classes of input histories to outputs on a common tip event.
In the case of space $\Theta_{21}$ above, we have 10 input histories arranged into 8 pairs of an equivalence class and a common tip event for the histories therein:
\begin{enumerate}
    \item the singleton $\left\{\hist{A/1,B/1,C/1}\right\}$ with tip event $\ev{B}$
    \item the singleton $\left\{\hist{A/0,B/1,C/1}\right\}$ with tip event $\ev{C}$
    \item the pair $\left\{\hist{A/0,B/1},\hist{B/1,C/0}\right\}$ with common tip event $\ev{B}$
    \item the pair $\left\{\hist{A/1,C/1},\hist{B/0,C/1}\right\}$ with common tip event $\ev{C}$
    \item the singleton $\left\{\hist{A/0}\right\}$ with tip event $\ev{A}$
    \item the singleton $\left\{\hist{A/1}\right\}$ with tip event $\ev{A}$
    \item the singleton $\left\{\hist{B/0}\right\}$ with tip event $\ev{B}$
    \item the singleton $\left\{\hist{C/0}\right\}$ with tip event $\ev{C}$
\end{enumerate}
Causal functions on $\Theta_{21}$ are then given by a free choice of output for each equivalence class: for example, the binary case $\CausFun{\Theta_{21}, \{0,1\}}$ features $2^8=256$ causal functions.

For a more complicated example of how to build such equivalence classes, we consider the case of causally complete space $\Theta_7$ (see \cite{gogioso2022classification} for the full hierarchy of causally complete spaces on 3 events with binary inputs).
There are four violations of tightness:
\begin{itemize}
    \item The extended input history \hist{A/0,B/1,C/0} has two sub-histories with event $\ev{B}$ as their tip event: \hist{A/0,B/1} and \hist{B/1,C/0}.
    \item The extended input history \hist{A/0,B/1,C/1} has two sub-histories with event $\ev{B}$ as their tip event: \hist{A/0,B/1} and \hist{B/1,C/1}.
    \item The extended input history \hist{A/1,B/0,C/0} has two sub-histories with event $\ev{A}$ as their tip event: \hist{A/1,B/0} and \hist{A/1,C/0}.
    \item The extended input history \hist{A/1,B/0,C/1} has two sub-histories with event $\ev{A}$ as their tip event: \hist{A/1,B/0} and \hist{A/1,C/1}.
\end{itemize}
The four histories are highlighted in $\Ext{\Theta_{7}}$ below, with a border of the same colour as the corresponding tip event conflicts.
\begin{center}
    \begin{tabular}{cc}
    \includegraphics[height=3cm]{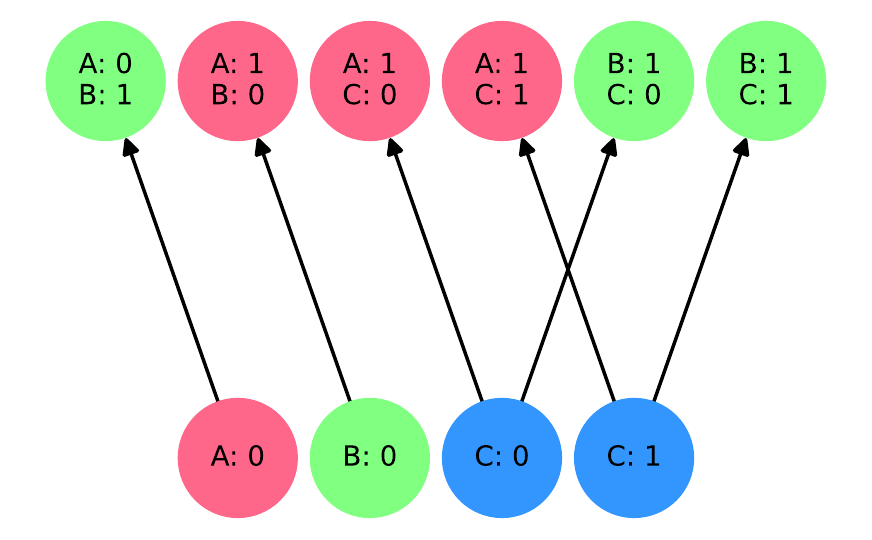}
    &
    \includegraphics[height=3.5cm]{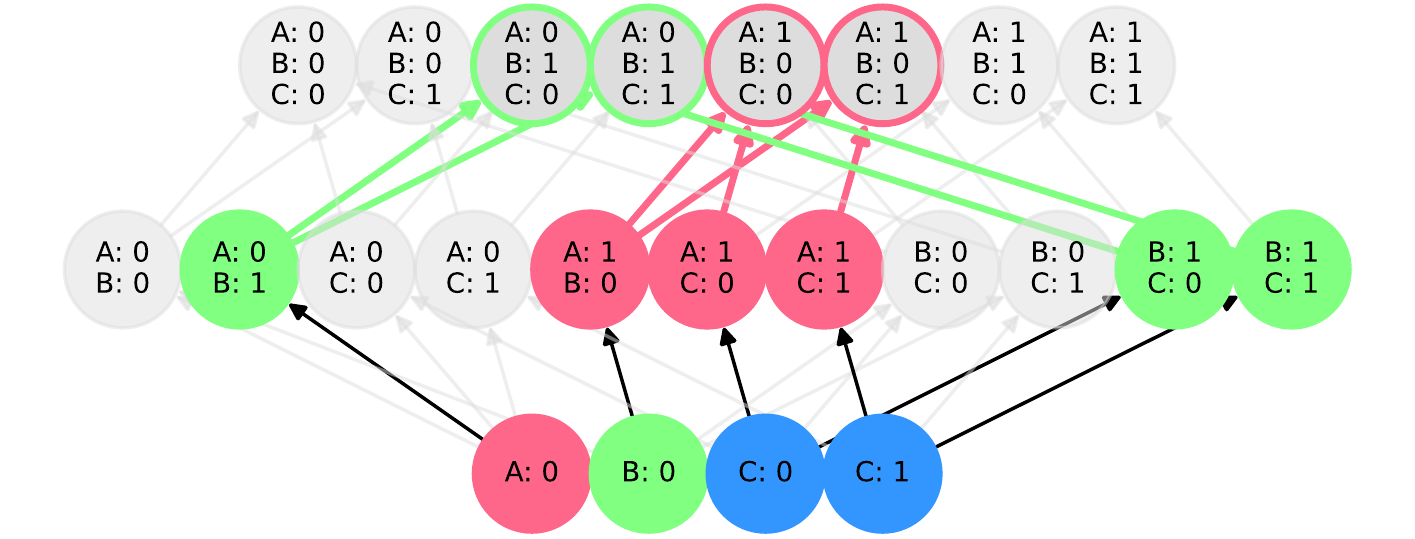}
    \\
    $\Theta_{7}$
    &
    $\Ext{\Theta_{7}}$ with highlights
    \end{tabular}
\end{center}
Well-definition of $\Ext{f}$ on the four (maximal) extended input histories listed above imposes the following constraints on any causal function $f$ for $\Theta_7$:
\[
\begin{array}{rcl}
f\left(\hist{A/0,B/1}\right)
&=&
f\left(\hist{B/1,C/0}\right)
\\
f\left(\hist{A/0,B/1}\right)
&=&
f\left(\hist{B/1,C/1}\right)
\\
f\left(\hist{A/1,B/0}\right)
&=&
f\left(\hist{A/1,C/0}\right)
\\
f\left(\hist{A/1,B/0}\right)
&=&
f\left(\hist{A/1,C/1}\right)
\\
\end{array}
\]
In the case of $\Theta_{21}$, the two constraints were independent of each other.
In the case of $\Theta_{7}$, instead, the four constraints interact because of common histories, fusing into two larger constraints:
\[
\begin{array}{rcccl}
f\left(\hist{B/1,C/0}\right)
&=&
f\left(\hist{A/0,B/1}\right)
&=&
f\left(\hist{B/1,C/1}\right)
\\
f\left(\hist{A/1,C/0}\right)
&=&
f\left(\hist{A/1,B/0}\right)
&=&
f\left(\hist{A/1,C/1}\right)
\\
\end{array}
\]
There are 10 input histories in $\Theta_{7}$ arranged into 6 pairs of an equivalence class and a common tip event for the histories therein:
\begin{enumerate}
    \item the triple $\left\{\hist{B/1,C/0},\hist{A/0,B/1},\hist{B/1,C/1}\right\}$ with common tip event $\ev{B}$
    \item the triple $\left\{\hist{A/1,C/0},\hist{A/1,B/0},\hist{A/1,C/1}\right\}$ with common tip event $\ev{A}$
    \item the singleton $\left\{\hist{A/0}\right\}$ with tip event $\ev{A}$
    \item the singleton $\left\{\hist{B/0}\right\}$ with tip event $\ev{B}$
    \item the singleton $\left\{\hist{C/0}\right\}$ with tip event $\ev{C}$
    \item the singleton $\left\{\hist{C/1}\right\}$ with tip event $\ev{C}$
\end{enumerate}
Causal functions on $\Theta_{7}$ are then given by a free choice of output for each equivalence class: for example, the binary case $\CausFun{\Theta_{7}, \{0,1\}}$ features $2^6=64$ causal functions.

We now formalise the discussion thus far into a definition of causal functions valid for arbitrary spaces of input histories, and generalise the results of Propositions \ref{theorem:ext-f-characterisation-tight-cc} and \ref{proposition:causal-function-joint-function-tight-cc}.
We start by defining the machinery necessary to formulate the constraints associated with lack of tightness, provide a constrained definition on input histories and then prove it equivalent to a free definition.

\begin{definition}
Let $\Theta$ be a space of input histories.
For each $\omega \in \Events{\Theta}$, the \emph{tip histories} for $\omega$ are the input histories which have $\omega$ as a tip event:
\begin{equation}
    \TipHists{\Theta}{\omega}
    :=
    \suchthat{h \in \Theta}{\omega \in \tips{\Theta}{h}}
\end{equation}
\end{definition}

\begin{definition}
\label{definition:hist-constrained-at-event}
Let $\Theta$ be a space of input histories.
For any $\omega \in \Events{\Theta}$, we say that two histories $h, h' \in \Theta$ are \emph{constrained at $\omega$}, written $\histconstr{\omega}{h}{h'}$, if they both have $\omega$ as a tip event and the consistency condition from Definition \ref{definition:consistency-condition} forces all extended functions $\hat{F} \in \ExtFun{\Theta, \{0,1\}}$ to output the same value for $\omega$ on $h$ and $h'$:
\[
    \hat{F}(h)_\omega = \hat{F}(h')_\omega
\]
\end{definition}

\begin{remark}
In Definition \ref{definition:hist-constrained-at-event} above, we could have replaced $\{0,1\}$ with any family of non-empty output sets $\underline{O}$ such that $|O_\omega| \geq 2$.
Equivalently, we could have universally quantified over all $\hat{F} \in \ExtFun{\Theta, \underline{O}}$ for all $\underline{O}$, but this would have made it unnecessarily harder to apply the definition. 
\end{remark}

\begin{proposition}
\label{proposition:histconstr-eqrel}
Let $\Theta$ be a space of input histories.
For any $\omega \in \Events{\Theta}$, the relation $\histconstrSym{\omega}$ is an equivalence relation, obtained as the transitive closure $\bar{R}_\omega$ of the following reflexive symmetric binary relation:
\[
    R_\omega := \suchthat{
        (h, h') \in \Theta^2
    }
    {
        h, h' \in \TipHists{\Theta}{\omega}
        \text{ and }
        \exists k \in \Ext{\Theta}
        \text{ s.t. }
        h, h' \leq k
    }
\]
\end{proposition}
\begin{proof}
See \ref{proof:proposition:histconstr-eqrel}
\end{proof}

\begin{corollary}
\label{corollary:tightness-constrained-histories}
Let $\Theta$ be a space of input histories.
$\Theta$ is tight if and only if $\histconstr{\omega}{h}{h'}$ always implies $h=h'$.
\end{corollary}
\begin{proof}
See \ref{proof:corollary:tightness-constrained-histories}
\end{proof}

\begin{definition}
\label{definition:causal-function}
Let $\Theta$ be a space of input histories and let $\underline{O} = (O_\omega)_{\omega \in \Events{\Theta}}$ be a family of non-empty sets of outputs.
The \emph{causal functions} $\CausFun{\Theta, \underline{O}}$ for space $\Theta$ and outputs $\underline{O}$ are the functions mapping each history in $\Theta$ to the output values for its tip events, subject to the the additional requirement that $f(h)_\omega = f(h')_\omega$ for any input histories $h,h'$ which are constrained at an event $\omega$:
\begin{equation}
    \CausFun{\Theta, \underline{O}}
    :=
    \suchthat{\!\!
        f \in \prod_{h \in \Theta}
        \prod_{\omega \in \tips{\Theta}{h}}
        \!\!\!O_{\omega}
    }
    {
       \histconstr{\omega}{h}{h'} \Rightarrow f(h)_\omega\!=\!f(h')_\omega
    \!\!}
\end{equation}
\end{definition}

\begin{observation}
\label{observation:caus-fun-bijection-tipeqcls}
Let $\Theta$ be a space of input histories and let $\underline{O} = (O_\omega)_{\omega \in \Events{\Theta}}$ be a family of non-empty sets of outputs.
For any $\omega \in \Events{\Theta}$, let $\TipEqCls{\Theta}{\omega}$ be the set of equivalence classes for $\histconstrSym{\omega}$:
\begin{equation}
    \TipEqCls{\Theta}{\omega}
    :=
    \TipHists{\Theta}{\omega}\!/\!\histconstrSym{\omega}
    =
    \suchthat{\histconstreqcls{h}{\omega}}{h \in \TipHists{\Theta}{\omega}}
\end{equation}
There is a bijection between the causal functions in $\CausFun{\Theta, \underline{O}}$ and the functions freely mapping each event $\omega \in \Events{\Theta}$ and each equivalence class $\histconstreqcls{h}{\omega} \in \TipEqCls{\Theta}{\omega}$ to the common output value at $\omega$ for all input histories in $\histconstreqcls{h}{\omega}$:
\begin{equation}
\begin{array}{rcl}
    \CausFun{\Theta, \underline{O}}
    &\longleftrightarrow&
    \prod\limits_{\omega \in \Events{\Theta}} \left(O_\omega\right)^{\TipEqCls{\Theta}{\omega}}
    \\
    f
    &\mapsto&
    \left(
        \left(\omega, \histconstreqcls{h}{\omega}\right) \mapsto f(h)_\omega
    \right)
    \\
    \left(
    h \mapsto
    \left(g\left(\omega,\histconstreqcls{h}{\omega}\right)\right)_{\omega \in \tips{\Theta}{h}}
    \right)
    &\mapsfrom&
    g
\end{array}
\end{equation}
\end{observation}

With Definition \ref{definition:causal-function} in hand, we are finally in a position to fully generalise Definition \ref{definition:ext-f-tight-cc}, Theorem \ref{theorem:ext-f-characterisation-tight-cc} and Proposition \ref{proposition:causal-function-joint-function-tight-cc}.
The changes necessary to achieve this are small: in causally incomplete spaces, we must account for the possibility that causal functions will produce output values for multiple tip events, while in non-tight spaces we must account for the non-uniqueness of the $h \in \TipHists{\Theta}{\omega}$ such that $h \leq k$, used by the definition of extended causal functions.
Both changes cause no trouble to our original proofs, which go through essentially unchanged.

\begin{definition}
\label{definition:ext-f}
Let $\Theta$ be a space of input histories and let $\underline{O} = (O_\omega)_{\omega \in \Events{\Theta}}$ be a family of non-empty sets of outputs.
For each causal function $f \in \CausFun{\Theta, \underline{O}}$, define the corresponding \emph{extended causal function} $\Ext{f} \in \ExtFun{\Theta, \underline{O}}$ as follows:
\begin{equation}
    \Ext{f}(k)
    :=
    \left(f\left(h_{k,\omega}\right)_\omega\right)_{\omega \in \dom{k}}
    \text{ for all }
    k \in \Ext{\Theta}
\end{equation}
where $h_{k,\omega}$ is any input history $h \in \TipHists{\Theta}{\omega}$ such that $h \leq k$.
We refer to $\Ext{f}(k)$ as the \emph{extended output history} corresponding to extended input history $k$.
We write $\ExtCausFun{\Theta, \underline{O}}$ for the subset of $\ExtFun{\Theta, \underline{O}}$ consisting of the extended causal functions.
\end{definition}

\begin{observation}
The function $\Ext{f}$ is well-defined because the definition of the causal function $f$ implies that $f\left(h_{k, \omega}\right)_\omega$ is the same for any choice of $h_{k, \omega}$.
\end{observation}

\begin{theorem}
\label{theorem:ext-f-characterisation}
Let $\Theta$ be a space of input histories, let $\underline{O} = (O_\omega)_{\omega \in \Events{\Theta}}$ be a family of non-empty sets of outputs.
The extended functions $\hat{F} \in \ExtFun{\Theta, \underline{O}}$ which are causal are exactly those which satisfy the consistency condition.
Indeed, the following defines a causal function $\Prime{\hat{F}} \in \CausFun{\Theta, \underline{O}}$ such that $\Ext{\Prime{\hat{F}}} = \hat{F}$:
\begin{equation}
    \Prime{\hat{F}} := h \mapsto \left(\hat{F}(h)_\omega\right)_{\omega \in \tips{\Theta}{h}}
\end{equation}
Furthermore, the above causal function is the unique $f \in \CausFun{\Theta, \underline{O}}$ such that $\Ext{f} = \hat{F}$, because of of the following equation:
\begin{equation}
    \Prime{\Ext{f}} = f
\end{equation}
\end{theorem}
\begin{proof}
See \ref{proof:theorem:ext-f-characterisation}
\end{proof}

We can finally express extended causal functions as ``gluings'' of the output values of causal functions, via compatible joins.

\begin{proposition}
\label{proposition:ext-f-gluing}
Let $\Theta$ be a space of input histories and let $\underline{O} = (O_\omega)_{\omega \in \Events{\Theta}}$ be a family of non-empty sets of outputs.
For every $f \in \CausFun{\Theta, \underline{O}}$, we have:
\begin{equation}
    \Ext{f}
    =
    k \mapsto 
    \bigvee_{h \in \downset{k} \cap \Theta}
    f(h)
\end{equation}
\end{proposition}
\begin{proof}
See \ref{proof:proposition:ext-f-gluing}
\end{proof}

\begin{proposition}
\label{proposition:causal-function-joint-function}
Let $\Theta$ be a space of input histories satisfying the free-choice condition and let $\underline{O} = (O_\omega)_{\omega \in \Events{\Theta}}$ be a family of non-empty sets of outputs.
For every $f \in \CausFun{\Theta, \underline{O}}$, the restriction of $\Ext{f}$ to the maximal extended input histories is a joint IO function for the operational scenario $(\Events{\Theta}, \underline{\Inputs{\Theta}}, \underline{O})$ which is causal for $\Theta$.
Conversely, any joint IO function $F$ which is causal for $\Theta$ arises as the restriction of $\Ext{f}$ to maximal extended input histories, where $f \in \CausFun{\Theta, \underline{O}}$ can be defined as follows for all $\omega \in \tips{\Theta}{h}$:
\begin{equation}
\label{equation:causal-fun-from-extended-fun}
    f(h)_\omega
    :=
    F(k)_\omega
    \text{ for any maximal ext. input history $k$ s.t. } h \leq k    
\end{equation}
\end{proposition}
\begin{proof}
See \ref{proof:proposition:causal-function-joint-function}
\end{proof}

To conclude this subsection, we prove factorisation results for causal functions in the presence of parallel composition, sequential composition and conditional sequential composition of spaces of input histories.

\begin{theorem}
\label{theorem:caus-fun-parallel-composition-factorisation}
Let $\Theta, \Theta'$ be spaces of input histories such that $\Events{\Theta} \cap \Events{\Theta'} = \emptyset$.
Let $\underline{O} = (O_\omega)_{\omega \in \Events{\Theta}}$ and $\underline{O}' = (O_\omega)_{\omega \in \Events{\Theta'}}$ be families of non-empty output sets.
The set of causal functions for the parallel composition space $\Theta \cup \Theta'$ factors into the product of the causal functions on the individual spaces:
\begin{equation}
     \CausFun{\Theta \cup \Theta', \underline{O}\vee\underline{O}'}
     \cong
     \CausFun{\Theta, \underline{O}}
     \times
     \CausFun{\Theta', \underline{O}'}
\end{equation}
More explicitly, causal functions in $\CausFun{\Theta \cup \Theta', \underline{O}\vee\underline{O}'}$ take the following form, for an arbitrary choice of causal functions $f \in \CausFun{\Theta, \underline{O}}$ and $f' \in \CausFun{\Theta', \underline{O}'}$:
\[
h
\mapsto
\left\{
\begin{array}{l}
f(h) \text{ if } h \in \Theta\\
f'(h) \text{ if } h \in \Theta'
\end{array}
\right.
\]
\end{theorem}
\begin{proof}
See \ref{proof:theorem:caus-fun-parallel-composition-factorisation}
\end{proof}

\begin{theorem}
\label{theorem:caus-fun-sequential-composition-factorisation}
Let $\Theta, \Theta'$ be spaces of input histories such that $\Events{\Theta} \cap \Events{\Theta'} = \emptyset$.
Let $\underline{O} = (O_\omega)_{\omega \in \Events{\Theta}}$ and $\underline{O}' = (O_\omega)_{\omega \in \Events{\Theta'}}$ be families of non-empty output sets. 
The set of causal functions for the sequential composition space $\Theta \seqcomposeSym \Theta'$ factors into the product of the causal functions on $\Theta$ and the families of causal functions on $\Theta'$ indexed by the maximal extended input histories of $\Theta$:
\begin{equation}
     \CausFun{\Theta \seqcomposeSym \Theta', \underline{O}\vee\underline{O}'}
     \cong
     \CausFun{\Theta, \underline{O}}
     \times
     \CausFun{\Theta', \underline{O}'}^{\max\Ext{\Theta}}
\end{equation}
More explicitly, causal functions in $\CausFun{\Theta \seqcomposeSym \Theta', \underline{O}\vee\underline{O}'}$ take the following form, for an arbitrary choice of causal function $f \in \CausFun{\Theta, \underline{O}}$ and family of causal functions $f'_k \in \CausFun{\Theta', \underline{O}'}$:
\[
h
\mapsto
\left\{
\begin{array}{l}
f(h) \text{ if } h \in \Theta\\
f'_k(h') \text{ if } h=k\vee h' \in \max\Ext{\Theta}\allJoinsSym\Theta'
\end{array}
\right.
\]
\end{theorem}
\begin{proof}
See \ref{proof:theorem:caus-fun-sequential-composition-factorisation}
\end{proof}

\begin{theorem}
\label{theorem:caus-fun-conditional-sequential-composition-factorisation}
Let $\Theta$ be a causally complete space of input histories.
Let $(\Theta'_k)_{k \in \max\Ext{\Theta}}$ be a family of causally complete spaces of input histories, with $\Events{\Theta} \cap \Events{\Theta'_k} = \emptyset$ for all $k \in \max\Ext{\Theta}$.
Let $\underline{O} = (O_\omega)_{\omega \in \Events{\Theta}}$ and $\underline{O}' = (O_{\omega})_{\omega \in \bigcup_{k \in \max\Ext{\Theta}}\Events{\Theta'_k}}$ be families of non-empty output sets, and define $\underline{O}'_k := (O_\omega)_{\omega \in \Events{\Theta'_k}}$ to be the restriction of $\underline{O}'$ to the events in $\Theta'_k$.
The set of causal functions for the conditional sequential composition space $\Theta \seqcomposeSym \underline{\Theta'}$ factors into the product of the causal functions on $\Theta$ and the families of causal functions on the spaces $\Theta'_k$ indexed by the maximal extended input histories $k \in \max\Ext{\Theta}$:
\begin{equation}
     \CausFun{\Theta \seqcomposeSym \underline{\Theta'}, \underline{O}\vee\underline{O}'}
     \cong
     \CausFun{\Theta, \underline{O}}
     \times
     \hspace{-5mm}
     \prod_{k \in \max\Ext{\Theta}}
     \hspace{-5mm}
     \CausFun{\Theta', \underline{O}'_k}
\end{equation}
More explicitly, causal functions in $\CausFun{\Theta \seqcomposeSym \underline{\Theta'}, \underline{O}\vee\underline{O}'}$ take the following form, for an arbitrary choice of causal function $f \in \CausFun{\Theta, \underline{O}}$ and family of causal functions $f'_k \in \CausFun{\Theta', \underline{O}'_k}$:
\[
h
\mapsto
\left\{
\begin{array}{l}
f(h) \text{ if } h \in \Theta\\
f'_k(h') \text{ if } h=k\vee h' \in \bigcup_{k \in \max\Ext{\Theta}}\{k\}\allJoinsSym\Theta'_k
\end{array}
\right.
\]
\end{theorem}
\begin{proof}
See \ref{proof:theorem:caus-fun-conditional-sequential-composition-factorisation}
\end{proof}

\subsection{The presheaf of (extended) causal functions}
\label{subsection:topology-causality-psheaf}

The (extended) causal functions defined in the previous subsections are an example of ``causal data'', that is, data defined on a space of input histories and respecting the associated causal constraints.
However, (extended) causal functions are a rather special case: as the correspondence with joint IO functions shows, they can be obtained from objects defined globally on the space by applying certain constraints, making them an example of non-contextual data.
This means that we can't use causal functions directly as a blueprint to define causal data in theories, such as quantum theory, where contextuality is a feature---both expected and desirable.
Instead, we must first shift from the global perspective to a more general contextual perspective: this is where topology finally comes into the picture.

The definition of compatible contextual data and the process that ``glues'' it together into global data is the purview of \emph{sheaf theory}, a complex mathematical discipline at the intersection of topology, geometry, algebra and category theory.
From the perspective of sheaf theory, ``contextual data'' is data associated to the open sets of some topological space, the ``contexts'', together with an appropriate definition of what it means to ``restrict'' data to open subsets.
The possible values taken by data on the open sets of a topological space, together with a specification for its restriction, defines what's known as a ``presheaf''.
Contextual data specified on different open sets is ``compatible'' if it has the same restrictions to all common open subsets: a ``sheaf'' is a presheaf where compatible contextual data can always be ``glued'' together, in a unique way.

\begin{definition}
Let $X$ be a topological space and let $\mathcal{T}(X)\subseteq \Subsets{X}$ be it collection of open sets, which we also refer to as the \emph{contexts}.
A \emph{presheaf} $P$ on $X$ is an association of:
\begin{itemize}
    \item a set $P(U)$ to each $U \in \mathcal{T}(X)$, specifying the possible values for \emph{contextual data} on $U$;
    \item a \emph{restriction} $P(U,V): P(U) \rightarrow P(V)$ for each open set $U$ and each open subset $V \subseteq U$, restricting contextual data on $U$ to corresponding contextual data on $V$.
\end{itemize}
The restrictions are required to satisfy the following conditions: 
\begin{enumerate}
    \item $P(U,U) = id_{P(U)}$, i.e. the trivial restriction from $U$ to $U$ is the identity on $P(U)$;
    \item $P(V, W) \circ P(U, V) = P(U, W)$, i.e. restrictions are stable under function composition.
\end{enumerate}
When the presheaf $P$ is clear from context, we adopt the following lightweight notation for restriction of contextual data $a \in P(U)$ to some open subset $V \subseteq U$:
\[
    \restrict{a}{V} := P(U,V)(a)
\]
The two restriction conditions can then be rewritten as follows, for all $a \in P(U)$
\begin{equation}
\text{(i)}\hspace{2mm}
\restrict{a}{U} = a
\hspace{30mm}
\text{(ii)}\hspace{2mm}
\restrict{\left(\restrict{a}{V}\right)\!}{W} = \restrict{a}{W}    
\end{equation}
\end{definition}

\begin{remark}
In categorical terms, a presheaf is a functor $P: \mathcal{T}(X) \rightarrow \text{Set}^{\text{op}}$, where the partial order $(\mathcal{T}(X), \subseteq)$ of open sets of $X$ is seen as a posetal category.
Note how the op-category $\text{Set}^{\text{op}}$ is used to make the functor ``contravariant'', from the point of view of sets and functions: to a morphism $V \subseteq U$ in the posetal category the functor $P$ associates a function $P(U) \rightarrow P(V)$.
\end{remark}

\begin{definition}
Let $X$ be a topological space and let $P$ be a presheaf on $X$.
Let $\mathfrak{U} \subseteq \mathcal{T}(X)$ be a family of open sets in $X$ and let $a = (a_U)_{U \in \mathfrak{U}}$ be a family specifying contextual data on the open sets in $\mathfrak{U}$.
We say that $a$ is a \emph{compatible family} (for $P$) if for every $U, U' \in \mathfrak{U}$ we have:
\begin{equation}
    \restrict{\left(a_U\right)\!}{U \cap U'} = \restrict{\left(a_{U'}\right)\!}{U \cap U'}
\end{equation}
that is, if the restriction of data on $U$ to the intersection $U\cap U'$ coincides with the restriction of data on $U'$ to the same intersection.
\end{definition}

\begin{definition}
Let $X$ be a topological space and let $P$ be a presheaf on $X$.
Let $\mathfrak{U} \subseteq \mathcal{T}(X)$ be a family of open sets in $X$ and let $a = (a_U)_{U \in \mathfrak{U}}$ be a family specifying contextual data on the open sets in $\mathfrak{U}$.
We say that some contextual data $\hat{a} \in P\left(\bigcup \mathfrak{U}\right)$ is a \emph{gluing} for the family $a$ if it restricts to $a_U$ for all $U \in \mathfrak{U}$:
\begin{equation}
    \restrict{\hat{a}}{U} = a_U
\end{equation}
We say that $P$ is a \emph{separated} presheaf if every compatible family has at most one gluing.
We say that $P$ is a \emph{sheaf} if every compatible family has exactly one gluing.
\end{definition}

Typically, data can be specified in many isomorphic ways, corresponding to different sets of values connected by suitable bijections.
A similar notion of isomorphism holds for contextual data specified by presheaves: in this ``natural'' isomorphism, the individual sets of data values assigned by two presheaf to each context are put in bijection, in such a way as to respect restrictions.

\begin{definition}
Let $X$ be a topological space and let $\mathcal{T}(X)\subseteq \Subsets{X}$ be it collection of open sets. 
We say that two presheaves $P$ and $P'$ on $X$ are \emph{naturally isomorphic}, written $P \cong P'$, if there is a family $\phi = \left(\phi_U\right)_{U \in \mathcal{T}(X)}$ of bijections $\phi_U: P(U) \rightarrow P'(U)$ such that for all inclusions $V \leq U$ of open sets we have:
\begin{equation}
\phi_V \circ P(U, V)
=
P'(U, V) \circ \phi_U    
\end{equation}
The commutation condition can be written more succinctly as follows, for all $a \in P(U)$:
\[
\phi_V(\restrict{a}{V})
=
\restrict{\phi_U(a)}{V}
\]
If we wish to specify a specific \emph{natural isomorphism} $\phi$, we can also write $\phi: P \cong P'$.
\end{definition}

\begin{observation}
\label{observation:natural-isomorphism-compatible-families}
Let $\phi: P \cong P'$ be a natural isomorphism between presheaves $P, P'$ on the same topological space $X$. Then $a = \left(a_U\right)_{U \in \mathfrak{U}}$ is a compatible family for $P$ if and only if $\phi\left(a\right)$ is a compatible family for $P'$, where we have defined:
\[
    \phi\left(a\right)
    :=
    \left(\phi_U(a_U)\right)_{U \in \mathfrak{U}}
\]
Also, $\hat{a}$ is a gluing of $a$ for $P$ if and only if $\phi\left(\hat{a}\right)$ is a gluing of $\phi\left(a\right)$ in $P'$.
Hence, $P$ is separated if and only if $P'$ is separated, and $P$ is a sheaf if and only if $P'$ is a sheaf.
\end{observation}

For the basic purpose of defining (pre)sheaves, it is enough to think of the collection $\mathcal{T}(X)$ of open sets as a partial order under subset inclusion.
However, additional properties possessed by $\mathcal{T}(X)$ play an important role in sheaf theory, and are thus worth an explicit introduction: in a word, $\mathcal{T}(X)$ is what's known as a ``locale''.

\begin{definition}
A \emph{locale} is a partially ordered set $(\mathcal{L}, \leq)$ satisfying the following properties:
\begin{enumerate}
    \item $(\mathcal{L}, \leq)$ has all finite meets, i.e. for all finite $F \subseteq \mathcal{L}$ there is a $\bigwedge F \in \mathcal{L}$ such that:
    \begin{itemize}
        \item $f \geq \bigwedge F$ for all $f \in F$
        \item if $f \geq g$ for all $f \in F$ then $\bigwedge F \geq g$
    \end{itemize}
    \item $(\mathcal{L}, \leq)$ has all joins, i.e. for all $F \subseteq \mathcal{L}$ there is a $\bigvee F \in \mathcal{L}$ such that:
    \begin{itemize}
        \item $f \leq \bigvee F$ for all $f \in F$
        \item if $f \leq g$ for all $f \in F$ then $\bigvee F \leq g$
    \end{itemize}
\end{enumerate}
The meets and joins are unique when they exist (because of the second property).
The empty meet $\bigwedge \emptyset$ is the unique global maximum, while the empty join $\bigvee \emptyset$ is the unique global minimum.
\end{definition}

\begin{observation}
Let $X$ be a topological space.
The axioms for the topology $\mathcal{T}(X) \subseteq \Subsets{X}$, the set of open sets in $X$, are exactly equivalent to asking that $\left(\mathcal{T}(X), \subseteq\right)$ be a locale:
\begin{enumerate}
    \item Finite intersections of open sets are open, including the empty intersection $X$.
    \item Arbitrary unions of open sets are open, including the empty union $\emptyset$.
\end{enumerate}
Note that for the inclusion order $\subseteq$, intersections are the meets and unions are the joins.
\end{observation}

In order for the machinery of sheaf theory to become available to us, we must first endow our spaces of input histories with a suitable topology.
Because a space of input histories $\Theta$ is a partial order, a natural choice of topology is given by taking its lower sets $\Lsets{\Theta}$ to be the open sets.
This is the dual of the Alexandrov topology, where the upper sets are taken to be the open sets, and all techniques applicable to Alexandrov topologies naturally dualise to lowerset topologies.
In particular, we make the following standard observations:
\begin{itemize}
    \item The points of $\Theta$, i.e. the input histories $h \in \Theta$, can be identified with downsets $\downset{h} \in \Lsets{\Theta}$.
    \item The downsets of input histories are exactly the lowersets $U \in \Lsets{\Theta}$ which are $\cup$-prime, i.e. those which cannot be written as $U=V\cup W$ for some lowersets $V,W \neq U$.
    \item The order on $\Theta$ can be reconstructed from the inclusion order on $\Lsets{\Theta}$, by observing that $h \leq h'$ if and only if $\downset{h} \subseteq \downset{h'\!}$.
\end{itemize}
But there's more!
The extended input histories $k \in \Ext{\Theta}$ can themselves be identified with certain lowersets, namely with the intersection $\downset{k\!}\!\cap\;\Theta$ of their downset in $\Ext{\Theta}$ with the space $\Theta$.
This identification is both injective and order-preserving, generalising the previous identification of $h \in \Theta$ with $\downset{h}$:
\[
\begin{array}{rcl}
\left(\Ext{\Theta}, \leq\right)
&\lhook\joinrel\longrightarrow&
\left(\Lsets{\Theta}, \subseteq\right)
\\
k
&\mapsto&
\downset{k\!}\!\cap\;\Theta
\end{array}
\]
Clearly, the locale of lowersets $\Lsets{\Theta}$ provides an equivalent way to talk about input histories, extended input histories and their order.
We adopt lowersets as our default topology for spaces of (extended) input histories other spaces derived from partial functions, and proceed to show that causality for extended functions is the same as continuity.

\begin{definition}
When talking about spaces of (extended) input histories as topological spaces, we take them endowed with the lowerset topology.
Unless otherwise specified, when talking about subsets $S \subseteq \PFun{\underline{Y}}$ we take them endowed with the partial order of $\PFun{\underline{Y}}$ and the lowerset topology.
\end{definition}


We start with a couple of useful---and well-known---facts about the lowerset topology on partial orders, which we prove for the convenience of readers who might not have previously encountered them.

\begin{proposition}
\label{proposition:lowerset-topology-continuity}
Let $X, Y$ be partial orders endowed with the lowerset topology.
A function $f: X \rightarrow Y$ is continuous if and only if it is order-preserving.
\end{proposition}
\begin{proof}
See \ref{proof:proposition:lowerset-topology-continuity}
\end{proof}

\begin{proposition}
\label{proposition:lowerset-topology-subset}
Let $Y$ be a partial order endowed with the lowerset topology and $X \subseteq Y$ be a subset.
The subspace topology on $X$ is the lowerset topology for the partial order on $X$ induced by $Y$.
\end{proposition}
\begin{proof}
See \ref{proof:proposition:lowerset-topology-subset}
\end{proof}

\begin{definition}
Let $X$ be a topological space.
The \emph{lowerset specialisation preorder} on $X$ is the preorder induced on $X$ by its topology $\mathcal{T}(X)$ in the following way:
\begin{equation}
    x' \leq x
    \stackrel{def}{\Leftrightarrow}
    \forall U \in \mathcal{T}(X).\;
    x \in U \Rightarrow x' \in U
\end{equation}
\end{definition}

\begin{remark}
We added ``lowerset'' to the above definition to distinguish it from the traditional definition for the ``specialisation preorder'', which follows the opposite convention:
\begin{equation}
    x' \leq x
    \stackrel{def}{\Leftrightarrow}
    \forall U \in \mathcal{T}(X).\;
    x' \in U \Rightarrow x \in U
\end{equation}
This is in line with our adoption of the lowerset topology instead of the Alexandrov topology: the latter uses upper sets as its open, instead of lowersets.
\end{remark}

\begin{proposition}
\label{proposition:lowerset-topology-specialisation-preorder}
Let $X$ be a partial order and let $\Lsets{X}$ be its lowerset topology.
The lowerset specialisation preorder induced by the topology $\Lsets{X}$ is exactly the partial order on $X$.
\end{proposition}
\begin{proof}
See \ref{proof:proposition:lowerset-topology-specialisation-preorder}
\end{proof}

We now show that continuity with respect to the lowerset topology provides an alternative, mathematically elegant characterisation for extended causal functions.
This provides the main motivation for our topological choice.

\begin{observation}
Let $\Theta$ be a space of input histories and let $\underline{O} = (O_\omega)_{\omega \in \Events{\Theta}}$ be a family of non-empty output sets.
The extended functions on $\Theta$ are a subset of the functions from $\Ext{\Theta}$ to $\PFun{\underline{O}}$:
\[
    \ExtFun{\Theta, \underline{O}}
    =
    \prod_{k \in \Ext{\Theta}}\prod_{\omega \in \dom{k}}\hspace{-2mm}O_\omega\hspace{2mm}
    \subseteq\hspace{2mm}
    \Ext{\Theta} \rightarrow \PFun{\underline{O}}
\]
In fact, they are exactly the functions $\hat{F}: \Ext{\Theta} \rightarrow \PFun{\underline{O}}$ which commute with the domain map, in the following sense:
\[
\dom{\hat{F}(k)} = \dom{k}
\]
\end{observation}

\begin{remark}
Commutation with the domain map is the same as saying that extended functions are morphisms $(\Ext{\Theta}, \domSym) \rightarrow (\PFun{\underline{O}}, \domSym)$ in the slice category $\text{Set}/\Subsets{\Events{\Theta}}$.
\end{remark}

\begin{theorem}
\label{theorem:function-causal-is-continuous}
Let $\Theta$ be a space of input histories and let $\underline{O} = (O_\omega)_{\omega \in \Events{\Theta}}$ be a family of non-empty output sets.
An extended function $\hat{F} \in \ExtFun{\Theta, \underline{O}}$ is causal if and only if it is continuous as a function $\hat{F}:\Ext{\Theta} \rightarrow \PFun{\underline{O}}$ where both $\Ext{\Theta}$ and $\PFun{\underline{O}}$ are equipped with the lowerset topology.
\end{theorem}
\begin{proof}
See \ref{proof:theorem:function-causal-is-continuous}
\end{proof}

Having drawn a clear connection between causality and lowerset topology, we now wish to formulate (extended) causal functions as a special case of causal data, potentially patched together from contextual parts.
In other words, we now proceed to construct a ``presheaf of (extended) causal functions''.
As our first step, we show that the lowersets of spaces of input histories, i.e. the open sets in their topologies, are themselves spaces of input histories, and that they inherit tightness and tip events from their parent spaces.
Despite inheriting tips, open sets don't necessarily inherit causal completeness, because they might fail to satisfy the free-choice condition.

\begin{proposition}
\label{proposition:lowersets-subspaces}
Let $\Theta$ be a space of input histories and let $\underline{O} = (O_\omega)_{\omega \in \Events{\Theta}}$ be a family of non-empty sets of outputs.
Any lowerset $\lambda \in \Lsets{\Theta}$ is a space of input histories, with $\Ext{\lambda} \in \Lsets{\Ext{\Theta}}$.
\end{proposition}
\begin{proof}
See \ref{proof:proposition:lowersets-subspaces}
\end{proof}

\begin{proposition}
\label{proposition:lowerset-cc-tight}
Let $\Theta$ be a space of input histories and let $\underline{O} = (O_\omega)_{\omega \in \Events{\Theta}}$ be a family of non-empty sets of outputs.
Let $\lambda \in \Lsets{\Theta}$ be a lowerset:
\begin{itemize}
    \item For all $h \in \lambda$, we have $\tips{\lambda}{h} = \tips{\Theta}{h}$.
    \item For all $h, h' \in \lambda$, we have that $\histconstr{\omega}{h}{h'}$ in $\lambda$ implies $\histconstr{\omega}{h}{h'}$ in $\Theta$. Hence, if $\Theta$ is tight then so is $\lambda$.
\end{itemize}
\end{proposition}
\begin{proof}
See \ref{proof:proposition:lowerset-cc-tight}
\end{proof}

Taken together, preservation of tips and constraints for input histories immediately imply that causal functions restrict to causal functions, providing the impetus for a very straightforward definition of the presheaf of causal functions.

\begin{corollary}
\label{corollary:caus-fun-restriction-is-caus}
Let $\Theta$ be a space of input histories and let $\underline{O} = (O_\omega)_{\omega \in \Events{\Theta}}$ be a family of non-empty sets of outputs.
Let $f \in \CausFun{\Theta, \underline{O}}$ be a causal function for $\Theta$, and $\lambda \in \Lsets{\Theta}$ be a lowerset.
Then $\restrict{f}{\lambda} \in \CausFun{\lambda, \underline{O}}$.
\end{corollary}
\begin{proof}
See \ref{proof:corollary:caus-fun-restriction-is-caus}
\end{proof}

\begin{definition}
\label{definition:causal-function-presheaf}
Let $\Theta$ be a space of input histories and let $\underline{O} = (O_\omega)_{\omega \in \Events{\Theta}}$ be a family of non-empty sets of outputs.
The \emph{presheaf of causal functions} $\CausFun{\Lsets{\Theta}, \underline{O}}$ for space $\Theta$ and outputs $\underline{O}$ is the presheaf on the topological space $\Theta$ defined as follows:
\begin{itemize}
    \item to lowersets $\lambda \in \Lsets{\Theta}$, the open sets of $\Theta$, it associates the causal functions for $\lambda$:
    \begin{equation}
        \CausFun{\Lsets{\Theta}, \underline{O}}\left(\lambda\right)
        :=
        \CausFun{\lambda, \underline{O}}
    \end{equation}
    \item to inclusions $\lambda' \subseteq \lambda$ of lowersets, it associates ordinary function restriction:
    \begin{equation}
        \CausFun{\Lsets{\Theta}, \underline{O}}\left(\lambda, \lambda'\right)
        :=
        f \mapsto \restrict{f}{\lambda'}
    \end{equation}
\end{itemize}
\end{definition}

\begin{theorem}
\label{theorem:compatibility-caus-fun-restrictions}
Let $\Theta$ be a space of input histories and let $\underline{O} = (O_\omega)_{\omega \in \Events{\Theta}}$ be a family of non-empty sets of outputs.
$\CausFun{\Lsets{\Theta}, \underline{O}}$ is a well-defined separated presheaf.
Compatible families are families of functions which are compatible in the sense of \cite{gogioso2022combinatorics} Definition~3.3 (p.22).
Their gluing is given by their compatible join in the sense of \cite{gogioso2022combinatorics} Definition 3.4 (p.22) whenever the compatible join is causal, and no gluing exists otherwise.
\end{theorem}
\begin{proof}
See \ref{proof:theorem:compatibility-caus-fun-restrictions}
\end{proof}


Previous work on non-locality \cite{abramsky2011sheaf} shows that the presheaf of causal functions for the discrete/no-signalling space---therein called the ``sheaf of events''---is a sheaf.
We show that tight spaces of input histories always lead to a sheaf of causal functions, but this is no longer guaranteed once the requirement for tightness is dropped.
We explore this peculiar phenomenon in the simple case of non-tight causally complete space $\Theta_3$, from the previous section.
You might remember this space as the meet of the order-induced spaces for causal orders $\Omega = \total{\ev{A},\ev{B}}\vee\discrete{\ev{C}}$ and $\Omega' = \discrete{A}\vee\total{\ev{C},\ev{B}}$:
\begin{center}
    \begin{tabular}{ccc}
    \includegraphics[height=2.5cm]{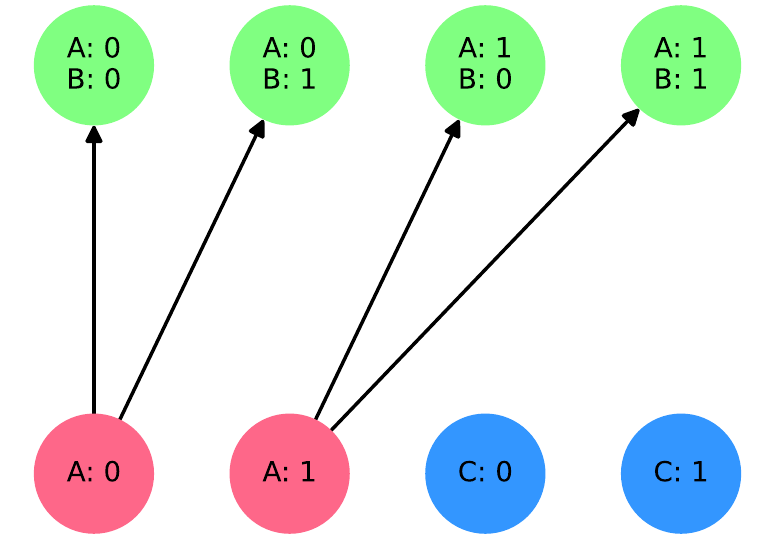}
    &
    \includegraphics[height=2.5cm]{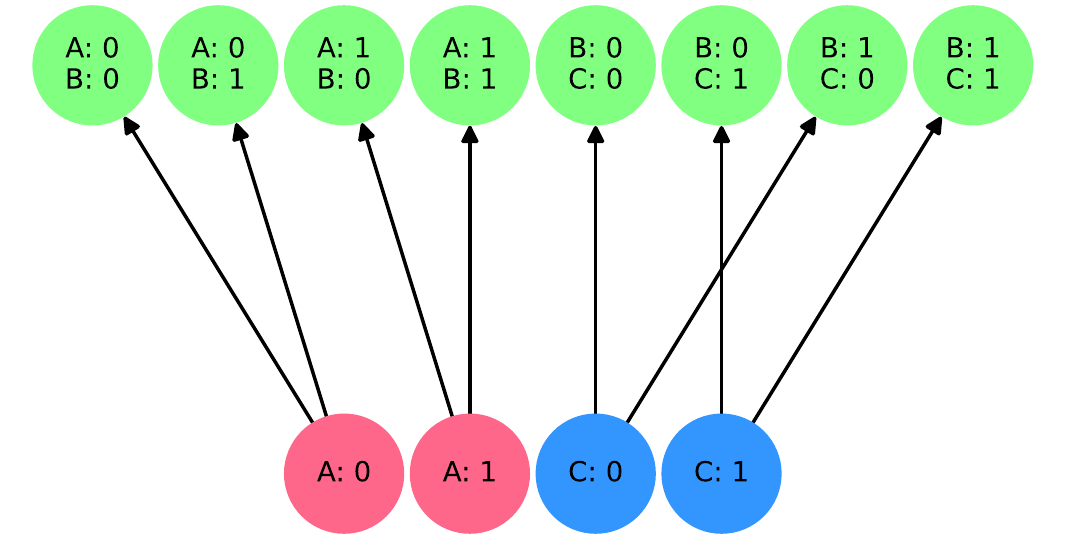}
    &
    \includegraphics[height=2.5cm]{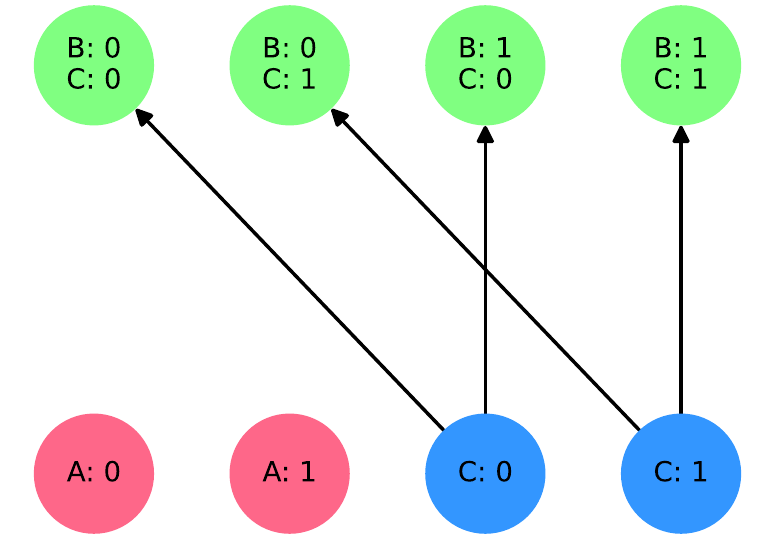}
    \\
    $\Hist{\Omega,\{0,1\}}$
    &
    \hspace{2mm}
    $\Theta_3 = \Hist{\Omega',\{0,1\}}\wedge\Hist{\Omega',\{0,1\}}$
    \hspace{2mm}
    &
    $\Hist{\Omega',\{0,1\}}$
    \end{tabular}
\end{center}
Space $\Theta_3$ is non-tight, because the downsets of event $\ev{B}$ are incomparable in the two spaces (cf. Theorem 3.33 p.50 of ``The Combinatorics of Causality'' \cite{gogioso2022combinatorics}):
\[
\begin{array}{rcl}
    \downset{\ev{B}\!}_{\Omega}
    &=&
    \evset{A,B}
    \\
    \downset{\ev{B}\!}_{\Omega'}
    &=&
    \evset{B,C}
\end{array}
\]
Since $\ev{B}$ is a maximum in both $\Omega$ and $\Omega'$, space $\Theta_3$ has conflicting histories with tip event $\ev{B}$ for all maximum input histories:
\begin{center}
    \begin{tabular}{cc}
    \includegraphics[height=3cm]{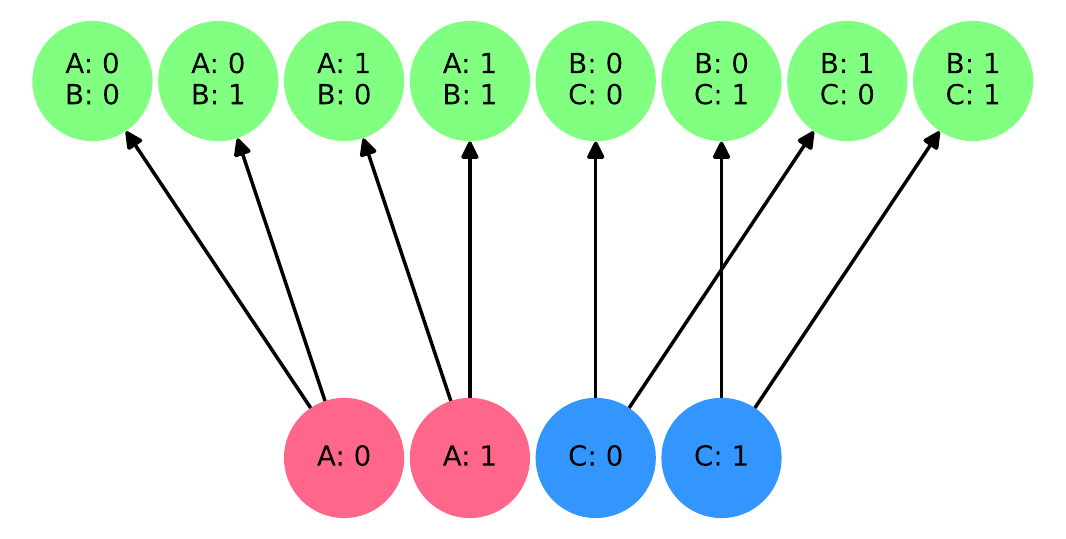}
    &
    \hspace{-0.7cm}
    \includegraphics[height=3.5cm]{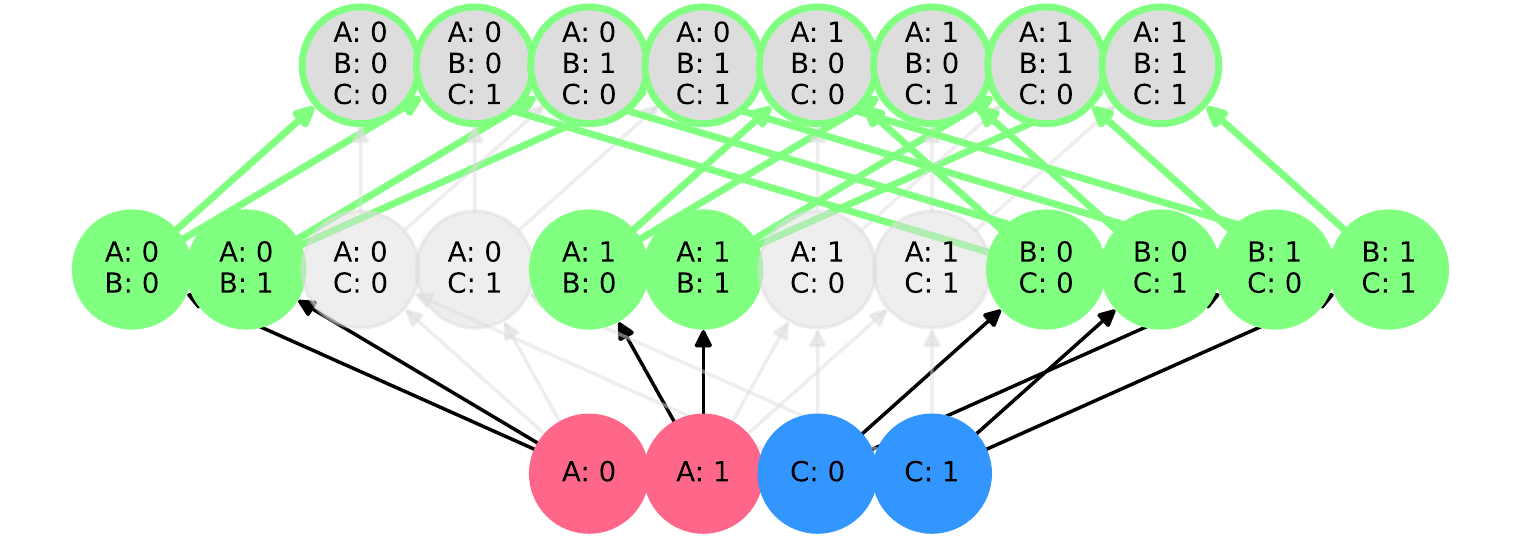}
    \\
    $\Theta_{3}$
    &
    \hspace{-0.7cm}
    $\Ext{\Theta_{3}}$ with highlights
    \end{tabular}
\end{center}
Specifically, every maximal input history $k=\hist{A/a,B/b,C/c}$ has exactly two input histories $h=\hist{A/a,B/b}$ and $h'=\hist{B/b,C/c}$ featuring tip event $\omega=\ev{B}$, i.e. such that $\histconstr{\omega}{h}{h'}$.
For one such maximal input history $k=\hist{A/a,B/b,C/c}$, and associated $h, h' \in \Theta_3$, we can consider the following lowersets:
\[
\begin{array}{rcccl}
\lambda
&:=&
\downset{h}
&=&
\left\{
\hist{A/a,B/b},
\hist{A/a}
\right\}
\\
\lambda'
&:=&
\downset{h'}
&=&
\left\{
\hist{B/b,C/c},
\hist{C/c}
\right\}
\end{array}
\]
Operationally, the lowerset $\lambda$ defines a context---arising from causal order $\Omega$---in which event $\ev{A}$ causally precedes event $\ev{B}$, but $\ev{C}$ does not.
Analogously, the lowerset $\lambda'$ defines a context---arising from causal order $\Omega$---in which event $\ev{C}$ causally precedes event $\ev{B}$, but $\ev{A}$ does not.
In the context of $\lambda$, we can define the following causal function $f$:
\[
\begin{array}{rcl}
f\left(\hist{A/a,B/b}\right)_{\ev{B}}
&:=&
1
\\
f\left(\hist{A/a}\right)_{\ev{A}}
&:=&
0
\end{array}
\]
In the context of $\lambda'$, we can define the following causal function $f'$:
\[
\begin{array}{rcl}
f'\left(\hist{B/b,C/c}\right)_{\ev{B}}
&:=&
0
\\
f'\left(\hist{C/c}\right)_{\ev{C}}
&:=&
0
\end{array}
\]
The functions $(f, f')$ form a compatible family over $\lambda, \lambda'$: we have $\lambda \cap \lambda' = \emptyset$, so that $\restrict{f}{\lambda \cap \lambda'} = \emptyset = \restrict{f'}{\lambda \cap \lambda'}$.
By Theorem \ref{theorem:compatibility-caus-fun-restrictions}, the gluing of $f$ and $f'$ exists exactly when their compatible join $f \vee f'$ is causal.
Unfortunately, the compatible join $f \vee f'$ is not causal in this example, because it satisfies $f(h)_{\ev{B}} \neq f'(h')_{\ev{B}}$:
\[
\begin{array}{rcl}
(f \vee f')\left(\hist{A/a,B/b}\right)_{\ev{B}}
&=&
1
\\
(f \vee f')\left(\hist{B/b,C/c}\right)_{\ev{B}}
&=&
0
\\
(f \vee f')\left(\hist{A/a}\right)_{\ev{A}}
&=&
0
\\
(f \vee f')\left(\hist{C/c}\right)_{\ev{C}}
&=&
0
\end{array}
\]
Hence, the presheaf of causal functions on space $\Theta_3$ is not a sheaf.
We refer to this peculiar behaviour as ``solipsistic contextuality'': certain spaces of input histories defined contexts with incompatible causal structure, allowing the contextual definition of causal functions which cannot be glued to a global causal function over the entire space.
We provide a general definition for this property, and prove that it exactly characterises the spaces of input histories for which the presheaf of causal functions fails to be a sheaf.

\begin{definition}
\label{definition:solipsitic-contextuality-witness}
Let $\Theta$ be a space of input histories.
A \emph{solipsistic contextuality witness} for $\Theta$ is a quadruple $(k, \omega, h, h')$ where:
\begin{enumerate}
    \item $k \in \Ext{\Theta}$ is an extended input history
    \item $\omega \in \dom{k}$ is an event
    \item $h, h' \in \Theta$ are distinct input histories such that $\histconstr{\omega}{h}{h'}$
    \item there is no $h'' \in \Theta$ such that $h, h' \leq h''$ and $h'' \leq k$
\end{enumerate}
We say that $\Theta$ admits \emph{solipsistic contextuality} if there is a solipsistic contextuality witness for $\Theta$.
\end{definition}

\begin{observation}
If $(k, \omega, h, h')$ is a solipsistic contextuality witness for a space of input histories $\Theta$, then by property (iv) we must have $k \notin \Theta$.
\end{observation}

\begin{lemma}
\label{lemma:deterministic-causal-contextuality-witness-sub-ext-hist}
Let $\Theta$ be a space of input histories.
If $(k, \omega, h, h')$ is a solipsistic contextuality witness for $\Theta$ and $k' \in \Ext{\Theta}$ is such that $h, h' \leq k' \leq k$, then $(k', \omega, h, h')$ is also a solipsistic contextuality witness for $\Theta$.
\end{lemma}
\begin{proof}
See \ref{proof:lemma:deterministic-causal-contextuality-witness-sub-ext-hist}
\end{proof}

\begin{proposition}
\label{proposition:causal-contextuality-characterisation}
A space of input histories $\Theta$ admits solipsistic contextuality if and only if there are distinct input histories $h, h' \in \Theta$ such that $\histconstr{\omega}{h}{h'}$ for some event $\omega$ and the following condition holds:
\[
\min\suchthat{k \in \Ext{\Theta}}{h, h' \leq k} \not\subseteq \Theta
\]
\end{proposition}
\begin{proof}
See \ref{proof:proposition:causal-contextuality-characterisation}
\end{proof}

\begin{theorem}
\label{theorem:causal-contextuality-sheaf-condition}
Let $\Theta$ be a space of input histories and let $\underline{O} = (O_\omega)_{\omega \in \Events{\Theta}}$ be a family of sets of outputs such that $O_\omega$ has at least two elements for all $\omega \in \Events{\Theta}$.
If $\Theta$ is tight, then $\CausFun{\Lsets{\Theta}, \underline{O}}$ is a sheaf.
More generally, $\CausFun{\Lsets{\Theta}, \underline{O}}$ is a sheaf if and only if $\Theta$ does not admit solipsistic contextuality.
\end{theorem}
\begin{proof}
See \ref{proof:theorem:causal-contextuality-sheaf-condition}
\end{proof}

The extended causal functions on a space of input histories can also be arranged into a presheaf, which is naturally isomorphic to the presheaf of causal functions via the $\ExtSym$ bijection.
The reason for explicitly defining such a presheaf is that the output data of extended causal functions is already ``glued together'', providing outputs for all events in the domain of any extended input history.
This will prove more convenient when connecting distributions over (extended) causal functions to the conditional probability distributions used by other works on causality.

\begin{definition}
\label{definition:ext-causal-function-presheaf}
Let $\Theta$ be a space of input histories and let $\underline{O} = (O_\omega)_{\omega \in \Events{\Theta}}$ be a family of non-empty sets of outputs.
The \emph{presheaf of extended causal functions} $\ExtCausFun{\Lsets{\Theta}, \underline{O}}$ for space $\Theta$ and outputs $\underline{O}$ is the presheaf on the topological space $\Theta$ defined as follows:
\begin{itemize}
    \item to lowersets $\lambda \in \Lsets{\Theta}$, it associates the extended causal functions for $\lambda$:
    \begin{equation}
        \ExtCausFun{\Lsets{\Theta}, \underline{O}}\left(\lambda\right)
        :=
        \ExtCausFun{\lambda, \underline{O}}
    \end{equation}
    \item to inclusions $\lambda' \subseteq \lambda$ of lowersets, it associates the following restrictions:
    \begin{equation}
        \ExtCausFun{\Lsets{\Theta}, \underline{O}}\left(\lambda, \lambda'\right)
        :=
        \Ext{f} \mapsto \restrict{\Ext{f}}{\Ext{\lambda}}
    \end{equation}
\end{itemize}
\end{definition}

\begin{proposition}
\label{proposition:presheaves-caus-fun-isomorphic}
Let $\Theta$ be a space of input histories and let $\underline{O} = (O_\omega)_{\omega \in \Events{\Theta}}$ be a family of non-empty sets of outputs.
$\ExtCausFun{\Lsets{\Theta}, \underline{O}}$ is a well-defined presheaf.
The family of bijections $\ExtSym = \left(\ExtSym: \CausFun{\lambda, \underline{O}} \rightarrow \ExtCausFun{\lambda, \underline{O}}\right)_{\lambda \in \Lsets{\Theta}}$ defines a natural isomorphism of presheaves $\ExtSym: \CausFun{\Lsets{\Theta}, \underline{O}} \cong \ExtCausFun{\Lsets{\Theta}, \underline{O}}$:
\begin{equation}
\label{equation:presheaves-caus-fun-isomorphic}
\restrict{\Ext{f}}{\Ext{\lambda}}
=
\Ext{\restrict{f}{\lambda}}
\end{equation}
As a consequence, $\ExtCausFun{\Lsets{\Theta}, \underline{O}}$ is always a separated presheaf, and it is a sheaf if and only if $\Theta$ does not admit solipsistic contextuality.
\end{proposition}
\begin{proof}
See \ref{proof:proposition:presheaves-caus-fun-isomorphic}
\end{proof}

\subsection{Empirical models}
\label{subsection:topology-causality-empmod}

Having recast causal functions into contextual objects, we are finally ready to step away from determinism and enter the world of probabilities and empirical models.
We proceed as follows:
\begin{enumerate}
    \item We define a ``presheaf of (probability) distributions'' over causal functions.
    \item We show how the causal conditional distributions used by other literature, known to us as ``empirical models'', arise as compatible families in the presheaf of distributions.
    \item We generalise empirical models to compatible families over arbitrary open covers of the space of input histories, with applications to the study of contextuality in causal settings.
\end{enumerate}
In this work, we only deal with topological and sheaf-theoretic aspects of empirical models.
The geometric picture, where empirical models are identified with the points of certain polytopes, is entirely the purview of the companion work ``The Geometry of Causality'' \cite{gogioso2022geometry}.

As our first ingredient, we turn our presheaves of (extended) causal functions into presheaves of distributions over (extended) causal functions, where restrictions are given by distribution marginalisation.
These presheaves are obtained from causal functions by using a mapping known as the ``distribution monad''.
The distribution monad sends a set $X$ to the set $\Dist{X}$ of finitely supported probability distributions over $X$, and it linearly extends functions $f: X \rightarrow Y$ between sets to functions $\Dist{f}: \Dist{X} \rightarrow \Dist{Y}$ defined on probability distributions.

\begin{definition}
The \emph{distribution monad} $\DistSym$ is the following mapping on sets and functions:
\begin{itemize}
    \item If $X$ is a set, $\Dist{X}$ is the set of probability distributions over $X$ with finite support:
    \begin{equation}
        \Dist{X}
        :=
        \suchthat{d: X \rightarrow \mathbb{R}^+}{\sum_{x \in X}d(x) = 1, \supp{d}\text{ is finite}}
    \end{equation}
    where the support of a distribution is the set of points over which it is non-zero:
    \begin{equation}
        \supp{d} := \suchthat{x \in X}{d(x) \neq 0}
    \end{equation}
    \item If $f: X \rightarrow Y$ is a function between sets, $\Dist{f}$ is the function $\Dist{X} \rightarrow \Dist{Y}$ defined as the linear extension of $f$ to probability distributions with finite support:
    \begin{equation}
        \Dist{f}
        :=
        d
        \mapsto
        \sum_{x \in X} d(x) \delta_{f(x)}
    \end{equation}
    where $\delta_{y} \in \Dist{Y}$ is the delta distribution at $y$:
    \begin{equation}
        \delta_{y}
        :=
        y' \mapsto
        \left\{
        \begin{array}{l}
        1 \text{ if } y' = y\\
        0 \text{ otherwise}
        \end{array}
        \right.
    \end{equation}
\end{itemize}
\end{definition}

\begin{observation}
The distribution monad satisfies the properties of a \emph{functor}:
\begin{enumerate}
    \item If $f: X \rightarrow Y$, then $\Dist{f}: \Dist{X} \rightarrow \Dist{Y}$.
    \item It acts as $\Dist{\id{X}} = \id{\Dist{X}}$ on the identity functions $\id{X}: X \rightarrow X$.
    \item It respects composition: if $f: X \rightarrow Y$ and $g: Y \rightarrow Z$, then $\Dist{g \circ f} = \Dist{g} \circ \Dist{f}$.
\end{enumerate}
\end{observation}

\begin{remark}
The term ``monad'' comes from category theory, where it defines a functor with specific additional structure.
We don't need this additional structure in our work, but we have preserved the name for compatibility with other sheaf-theoretic work.
\end{remark}

\begin{remark}
In this work, we are only interested in probability distributions, valued in $\mathbb{R}^+$.
However, the same definition extends from $R=\mathbb{R}^+$ to an arbitrary commutative semiring $R$.
This includes the \emph{possibilistic} case of $R=(\{0,1\}, \vee, 0, \wedge, 1)$ and the \emph{quasi-probabilistic} case of $R=\mathbb{R}$, previously studied in the sheaf-theoretic literature on non-locality and contextuality \cite{abramsky2011sheaf,abramsky2015contextuality,abramsky2016possibilities}, as well as the more exotic cases of $p$-adic probabilities and modalities, arising in some quantum-like theories \cite{gogioso2017fantastic}.
\end{remark}

As it turns our, we can ``compose'' the distribution monad $\DistSym$ with a presheaf $P$ to form a new presheaf $\DistSym P$, where the contextual data defined by $\DistSym P$ consists of probability distributions over the contextual data originally defined by $P$.
Composition is simple to define, and it conveniently respects natural isomorphism between presheaves.

\begin{definition}
Let $X$ be a topological space and let $\mathcal{T}(X)\subseteq \Subsets{X}$ be it collection of open sets.
For a presheaf $P$ on $X$, let $\DistSym P$ be the following mapping:
\begin{itemize}
    \item On an open set $U \in \mathcal{T}(X)$, define $\DistSym P(U) := \Dist{P(U)}$.
    \item On a subset inclusion $V \subseteq U$, define $\DistSym P(U, V) := \Dist{P(U, V)}$.
\end{itemize}
\end{definition}

\begin{lemma}
\label{lemma:distr-monad-bij}
Let $\phi: X \rightarrow Y$ be a function between sets.
If $\psi$ is a left (resp. right) inverse of $\phi$, then $\Dist{\psi}$ is a left (resp. right) inverse of $\Dist{\phi}$.
Hence, if $\phi$ is injective (resp. surjective), then $\Dist{\phi}$ is injective (resp. surjective).
\end{lemma}
\begin{proof}
See \ref{proof:lemma:distr-monad-bij}
\end{proof}

\begin{proposition}
\label{proposition:presheaf-dist-presheaf}
Let $X$ be a topological space.
If $P$ is a presheaf on $X$, then $\DistSym P$ is also a presheaf on $X$.
\end{proposition}
\begin{proof}
See \ref{proof:proposition:presheaf-dist-presheaf}
\end{proof}

\begin{proposition}
\label{proposition:presheaf-dist-presheaf-natural-isomorphism}
Let $X$ be a topological space and let $\mathcal{T}(X)\subseteq \Subsets{X}$ be it collection of open sets.
Let $P, P'$ be presheaves on some topological space $X$.
If $\phi: P \cong P'$ are naturally isomorphic, then $\DistSym\phi: \DistSym P \cong \DistSym P'$ are also naturally isomorphic, where we defined:
\[
\DistSym\phi
:=
\left(
\Dist{\phi_U}
\right)_{U \in \mathcal{T}(X)}
\]
\end{proposition}
\begin{proof}
See \ref{proof:proposition:presheaf-dist-presheaf-natural-isomorphism}
\end{proof}

Because composition by the distribution monad respects natural isomorphisms, we could define our causal distributions equivalently using causal functions or extended causal functions, since the two presheaves are isomorphic by Proposition \ref{proposition:presheaves-caus-fun-isomorphic}.
To simplify our upcoming definition of empirical models, we choose extended causal functions as our base for causal distributions.

\begin{definition}
Let $\Theta$ be a space of input histories and let $\underline{O} = (O_\omega)_{\omega \in \Events{\Theta}}$ be a family of non-empty output sets.
The \emph{presheaf of causal distributions} for $\Theta$ is defined as follows:
\begin{equation}
    \CausDist{\Lsets{\Theta},\underline{O}}
    :=
    \DistSym \,\ExtCausFun{\Lsets{\Theta},\underline{O}}
\end{equation}
We also define the following notation for the individual sets of distributions:
\begin{equation}
    \CausDist{\lambda,\underline{O}}
    :=
    \Dist{\ExtCausFun{\lambda, \underline{O}}}
\end{equation}
\end{definition}

\begin{proposition}
\label{proposition:causal-dist-marginal}
Let $\Theta$ be a space of input histories and let $\underline{O} = (O_\omega)_{\omega \in \Events{\Theta}}$ be a family of non-empty output sets.
The restrictions of the presheaf $\CausDist{\Lsets{\Theta},\underline{O}}$ act by marginalisation on probability distributions $d \in \CausDist{\lambda,\underline{O}}$:
\begin{equation}
\label{equation:causal-dist-marginal}
    \restrict{d}{\lambda'}
    =
    \Ext{f'}
    \mapsto
    \hspace{-3mm}
    \sum_{f \text{ s.t. } \restrict{f}{\lambda'}=f'}
    \hspace{-3mm}
    d\left(\Ext{f}\right)
\end{equation}
In words, the probability assigned by the marginal $\restrict{d}{\lambda'} \in \CausDist{\lambda',\underline{O}}$ to a generic extended causal function $\Ext{f'} \in \ExtCausFun{\lambda', \underline{O}}$ is the sum of the probabilities assigned by $d$ to all extended causal functions $\Ext{f} \in \ExtCausFun{\lambda, \underline{O}}$ which restrict to $\Ext{f'}$.
\end{proposition}
\begin{proof}
See \ref{proof:proposition:causal-dist-marginal}
\end{proof}


\begin{definition}
\label{definition:standard-emp-model}
Let $\Theta$ be a space of input histories and let $\underline{O} = (O_\omega)_{\omega \in \Events{\Theta}}$ be a family of non-empty sets of outputs.
A \emph{standard empirical model} $e$ is a compatible family $e = (e_{\downset{k}})_{k \in \max\Ext{\Theta}}$ for the presheaf of causal distributions:
\[
e_{\downset{k}} \in \CausDist{\downset{k\!}, \underline{O}}
\]
\end{definition}

In the companion work ``The Geometry of Causality'' \cite{gogioso2022geometry}, we prove that standard empirical models admit a description in terms of distributions on extended output histories conditional to maximal extended input histories.
For example, on p.\pageref{subsubsection:fork-empirical-model} we describe how to obtain the following empirical model for the causal fork space $\Theta = \Hist{\discrete{C}\seqcomposeSym\discrete{A,B}, \{0,1\}}$:
\begin{center}
\begin{tabular}{l|rrrrrrrr}
\hfill
ABC & 000 & 001 & 010 & 011 & 100 & 101 & 110 & 111\\
\hline
000 & $1/4$ & $1/4$ & $0$ & $0$ & $0$ & $0$ & $1/4$ & $1/4$\\
001 & $0$ & $0$ & $1/4$ & $1/4$ & $1/4$ & $1/4$ & $0$ & $0$\\
010 & $1/8$ & $1/8$ & $1/8$ & $1/8$ & $1/8$ & $1/8$ & $1/8$ & $1/8$\\
011 & $1/8$ & $1/8$ & $1/8$ & $1/8$ & $1/8$ & $1/8$ & $1/8$ & $1/8$\\
100 & $1/8$ & $1/8$ & $1/8$ & $1/8$ & $1/8$ & $1/8$ & $1/8$ & $1/8$\\
101 & $1/8$ & $1/8$ & $1/8$ & $1/8$ & $1/8$ & $1/8$ & $1/8$ & $1/8$\\
110 & $1/4$ & $0$ & $0$ & $1/4$ & $0$ & $1/4$ & $1/4$ & $0$\\
111 & $1/4$ & $0$ & $0$ & $1/4$ & $0$ & $1/4$ & $1/4$ & $0$\\
\end{tabular}
\end{center}
\label{page:fork-empirical-model-descr}
In this tabular form, each row $i_\ev{A}i_\ev{B}i_\ev{C}$ corresponds to a maximal extended input history $\hist{A/i_\ev{A}, B/i_\ev{B}, C/i_\ev{C}} \in \StdCov{\Theta}$, while each column $o_\ev{A}o_\ev{B}o_\ev{C}$ corresponds to an associated extended output history $\hist{A/o_\ev{A}, B/o_\ev{B}, C/o_\ev{C}}$.
Formally, however, each line $i_\ev{A}i_\ev{B}i_\ev{C}$ of the same empirical model has to be defined explicitly as a distribution on extended causal functions:
\[
\Dist{\ExtCausFun{\downset{\hist{A/i_\ev{A}, B/i_\ev{B}, C/i_\ev{C}}}, \{0,1\}}}
\]
Equivalently, we look at distributions on causal functions, which are freely characterised:
\[
\Dist{\CausFun{\downset{\hist{A/i_\ev{A}, B/i_\ev{B}, C/i_\ev{C}}}, \{0,1\}}}
\]
Because the space $\Theta$ is both tight and causally complete, the causal functions on the downset $\downset{\hist{A/i_\ev{A}, B/i_\ev{B}, C/i_\ev{C}}}$ take the following form:
\[
    \prod_{h \leq \hist{A/i_\ev{A}, B/i_\ev{B}, C/i_\ev{C}}}
    O_{\tip{\Theta}{h}}
\]
where we used the fact that $\tip{\downset{\hist{A/i_\ev{A}, B/i_\ev{B}, C/i_\ev{C}}}}{h}=\tip{\Theta}{h}$ to simplify the expression.
The input histories $h \leq \hist{A/i_\ev{A}, B/i_\ev{B}, C/i_\ev{C}}$ are exactly:
\begin{itemize}
    \item $h = \hist{C/i_\ev{C}}$ with tip event \ev{C}
    \item $h = \hist{C/i_\ev{C}, A/i_\ev{A}}$ with tip event \ev{A}
    \item $h = \hist{C/i_\ev{C}, B/i_\ev{B}}$ with tip event \ev{B}
\end{itemize}
Hence the causal functions in $\CausFun{\downset{\hist{A/i_\ev{A}, B/i_\ev{B}, C/i_\ev{C}}}, \{0,1\}}$ are:
\[
f_{o_\ev{A}o_\ev{B}o_\ev{C}|i_\ev{A}i_\ev{B}i_\ev{C}}:=
\left\{
\begin{array}{rl}
    \hist{C/i_\ev{C}}       &\mapsto o_\ev{C}\\
    \hist{C/i_\ev{C},A/i_\ev{A}} &\mapsto o_\ev{A}\\
    \hist{C/i_\ev{C},B/i_\ev{B}} &\mapsto o_\ev{B}
\end{array}
\right.
\]
Using these functions, we can reconstruct the desired distribution for each row of the empirical model above.
For example, the second row is indexed by the maximal extended input history $\hist{A/0, B/0, C/1}$ and it corresponds to the following distribution on extended causal functions:
\[
\frac{1}{4}\delta_{\Ext{f_{010|001}}}
+\frac{1}{4}\delta_{\Ext{f_{011|001}}}
+\frac{1}{4}\delta_{\Ext{f_{100|001}}}
+\frac{1}{4}\delta_{\Ext{f_{101|001}}}
\]
Doing this for all rows yields the following standard empirical model:
\[
e_{\downset{\hist{A/i_\ev{A}, B/i_\ev{B}, C/i_\ev{C}}}}
:=
\left\{
\begin{array}{rl}
\frac{1}{4}\sum\limits_{o_\ev{A}\oplus o_\ev{B} = i_\ev{C}}\sum\limits_{o_\ev{C}} \delta_{\Ext{f_{o_\ev{A}o_\ev{B}o_\ev{C}|i_\ev{A}i_\ev{B}i_\ev{C}}}}
&\text{ if } i_\ev{A} = i_\ev{B}\\
\frac{1}{8}\sum\limits_{o_\ev{A}}\sum\limits_{o_\ev{B}}\sum\limits_{o_\ev{C}} \delta_{\Ext{f_{o_\ev{A}o_\ev{B}o_\ev{C}|i_\ev{A}i_\ev{B}i_\ev{C}}}}
&\text{ if } i_\ev{A} \neq i_\ev{B}
\end{array}
\right.
\]

In Definition \ref{definition:standard-emp-model}, we referred to our empirical models as ``standard''.
As we discuss in the companion work ``The Geometry of Causality'' \cite{gogioso2022geometry}, standard empirical models encompass the probability distributions on joint outputs conditional to joint inputs which are considered in previous literature on causality, further extended to the very many new causal constraints that can be expressed using arbitrary (e.g. non-tight) spaces of input histories.
As a very special case---corresponding to the discrete spaces at the bottom of our causal hierarchies---standard empirical models capture all non-locality scenarios previously described by the sheaf-theoretic literature.
However, standard empirical models are too specific to cover more general examples of contextuality, such as those studied in previous sheaf-theoretic literature or the solipsistic contextuality examples we previously explored.
Luckily, it is straightforward to extend our definition to encompass all such examples.
The key is to observe that $\suchthat{\downset{k}}{k \in \max\ExtHist{\Theta, \underline{O}}}$ is an example of an ``open cover'' for the topological space $\Theta$, and to explore what happens if we define empirical models on other such open covers.
What we derive is a ``hierarchy of contextuality'', corresponding to different operational assumptions. Three covers are of particular interest.
\begin{itemize}
    \item The ``standard cover'' $\suchthat{\downset{k}}{k \in \max\ExtHist{\Theta, \underline{O}}}$, accommodating generic causal distributions on joint outputs conditional to the maximal extended input histories.
    It models settings where it is, at the very least, possible to define conditional distributions when all events are taken together.
    Empirical models on the standard cover are the standard empirical models defined above.
    \item The ``classical cover'' $\{\Theta\}$ is the ``coarsest'' cover, lying at the top of the hierarchy.
    It models settings admitting a deterministic causal hidden variable explanation.
    Empirical models on the classical cover can be restricted to every other open cover: the empirical models arising this way are known as ``non-contextual''.
    \item The ``fully solipsistic cover'' $\suchthat{\downset{h}}{h \in \max\Hist{\Theta, \underline{O}}}$ is the ``finest'' cover, lying at the bottom of the hierarchy.
    It models settings more restrictive than those modelled by the standard cover, where it might only be possible to define distributions over the events in the past of some event---``witnessing'' the existence of the events before it, so to speak.
    That is, the fully solipsistic cover accommodates all causal distributions on joint outputs conditional to the maximal input histories.
    Every empirical model can be restricted to the fully solipsistic cover, and if an empirical model is contextual then so is its restriction to the fully solipsistic cover.
    In particular, the fully solipsistic cover contains the empirical models witnessing solipsistic contextuality.
\end{itemize}
We adopt different nomenclature to distinguish different sources of contextuality: we say that an empirical model displays ``solipsistic contextuality'' if it doesn't arise by restriction from an empirical model defined on a cover above the standard cover, and that an empirical model is ``contextual'' if it doesn't arise by restriction from an empirical model on the classical cover.

\begin{definition}
Let $X$ be a topological space and $\mathcal{T}(X) \subseteq \Subsets{X}$ be its topology.
An \emph{open cover}, or simply a \emph{cover}, for $X$ is a maximal antichain in the partial order $\mathcal{T}(X)$, i.e. a collection $\mathcal{C} \subseteq \mathcal{T}(X)$ of open sets which are incomparable and which cover the whole space $X$:
\[
    \forall U, V \!\in \mathcal{C}.\; U \!\neq\! V \Rightarrow
        \left[\, V \!\not\subseteq\! U \text{ and } U \!\not\subseteq V\! \,\right]
    \hspace{2cm}
    X = \bigcup_{U \in \mathcal{C}} U
\]
If $\mathcal{C}$ and $\mathcal{C}'$ are covers on $X$, we say that $\mathcal{C}'$ is \emph{finer} than $\mathcal{C}$, written $\mathcal{C}' \preceq \mathcal{C}$, if the following holds:
\begin{equation}
    \mathcal{C}' \preceq \mathcal{C}
    \Leftrightarrow
    \forall V \in \mathcal{C}'.\,
    \exists U \in \mathcal{C}.\;
    \text{ s.t. } V \subseteq U
\end{equation}
Equivalently, we say that $\mathcal{C}$ is \emph{coarser} than $\mathcal{C}'$.
Note that $\preceq$ is a partial order on covers for $X$, known as the \emph{refinement order}.
\end{definition}

\begin{remark}
The definition of ``open cover'' given above is more specific than the definition typically used in point-set topology, where an open cover is not required to be an antichain:
\begin{itemize}
    \item every cover according to our definition is a cover in the sense of point-set topology;
    \item to go from a cover in the sense of point-set topology to one according to our definition, one has to restrict the cover to its maximal open sets (under inclusion).
\end{itemize}
The reason for our definition of covers as antichains is that the main use for covers in this work---and in other sheaf-theoretic works---is to define compatible families for a presheaf: data for a compatible family \textit{must} be specified on the maximal open sets of the cover, while data on all other open sets can always be derived by restriction.
The choice to define covers as antichains is then a ``minimalistic'' one: the open sets at the top are the only ones we have to care about when specifying data.
It also has the advantage of creating a kind of ``normal form'' for the spaces of empirical models: two covers with the same maximal open sets would have the same compatible families, and hence would give rise to the same empirical models.
\end{remark}

\begin{observation}
Let $X$ be a topological space and let $P$ be a presheaf on $X$.
Let $\mathcal{C}$ be a cover for $X$ and $a = (a_U)_{U \in \mathcal{C}}$ be a compatible family over $\mathcal{C}$.
If $\mathcal{C}' \preceq \mathcal{C}$ is a finer cover for $X$, then the following is a compatible family over $\mathcal{C}'$, known as the \emph{restriction} of $a$ to $\mathcal{C}'$:
\begin{equation}
    \restrict{a}{\mathcal{C}'}
    :=
    (\restrict{a_{U_V}}{V})_{V \in \mathcal{C'}}
\end{equation}
where $U_V \in \mathcal{C}$ is any open such that $V \subseteq U_V$.
\end{observation}

\begin{definition}
Let $\Theta$ be a space of input histories.
The \emph{standard cover} on $\Theta$ is the following open cover:
\begin{equation}
    \StdCov{\Theta}
    :=
    \suchthat{\downset{k}}{k \in \max\ExtHist{\Theta, \underline{O}}}
\end{equation}
The \emph{fully solipsistic cover} on $\Theta$ is the following open cover:
\begin{equation}
    \SolCov{\Theta}
    :=
    \suchthat{\downset{h}}{h \in \max\Hist{\Theta, \underline{O}}}
\end{equation}
The \emph{classical cover} on $\Theta$ is the following open cover:
\begin{equation}
    \ClsCov{\Theta}
    :=
    \{\Theta\}
\end{equation}
The \emph{hierarchy of covers} for $\Theta$ is the set $\Covers{\Theta}$ of open covers ordered by refinement $\preceq$.
We refer to covers $\mathcal{C}$ such that $\mathcal{C} \not\preceq \StdCov{\Theta}$ as \emph{solipsistic covers}.
\end{definition}

\begin{proposition}
\label{proposition:solipsistic-standard-global-hierarchy}
Let $\Theta$ be a space of input histories.
The partial order $\Covers{\Theta}$ is a lattice, with the fully solipsistic cover $\SolCov{\Theta}$ as its unique minimum and the classical cover $\ClsCov{\Theta}$ as its unique maximum.
In particular:
\[
\SolCov{\Theta} \preceq \StdCov{\Theta} \preceq \ClsCov{\Theta}
\]
\end{proposition}
\begin{proof}
See \ref{proof:proposition:solipsistic-standard-global-hierarchy}
\end{proof}

A our first and simplest example, we look at the open covers on the discrete space with 1 event and ternary inputs $\Hist{\ev{A}, \{0,1,2\}}$: we chose this particular example because it is simple enough that all covers can be enumerated explicitly, but at the same time supports an interesting contextual empirical model (on cover \#7 below).
There are 9 open covers for this space, arranged in the following hierarchy.
\begin{center}
\includegraphics[height=3cm]{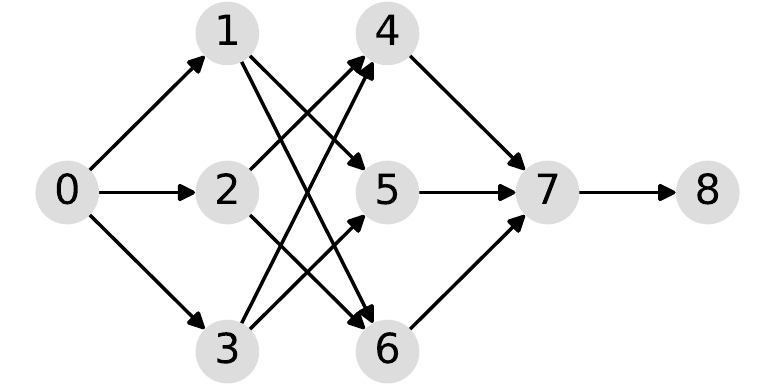}
\end{center}
Because $\Hist{\ev{A}, \{0,1,2\}}=\ExtHist{\ev{A}, \{0,1,2\}}$, the standard cover and fully solipsistic cover coincide for this example.
\begin{itemize}
\item Cover \#0 (standard/fully solipsistic cover) contains the following lowersets:
\[
    \left\{\pigl\{\hist{A/0}\pigr\}, \pigl\{\hist{A/1}\pigr\}, \pigl\{\hist{A/2}\pigr\}\right\}
\]
\item Cover \#1 contains the following lowersets:
\[
    \left\{\pigl\{\hist{A/0}\pigr\}, \pigl\{\hist{A/1}, \hist{A/2}\pigr\}\right\}
\]
\item Cover \#2 contains the following lowersets:
\[
    \left\{\pigl\{\hist{A/1}\pigr\}, \pigl\{\hist{A/0}, \hist{A/2}\pigr\}\right\}
\]
\item Cover \#3 contains the following lowersets:
\[
    \left\{\pigl\{\hist{A/2}\pigr\}, \pigl\{\hist{A/0}, \hist{A/1}\pigr\}\right\}
\]
\item Cover \#4 contains the following lowersets:
\[
    \left\{\pigl\{\hist{A/0}, \hist{A/1}\pigr\}, \pigl\{\hist{A/0}, \hist{A/2}\pigr\}\right\}
\]
\item Cover \#5 contains the following lowersets:
\[
    \left\{\pigl\{\hist{A/0}, \hist{A/1}\pigr\}, \pigl\{\hist{A/1}, \hist{A/2}\pigr\}\right\}
\]
\item Cover \#6 contains the following lowersets:
\[
    \left\{\pigl\{\hist{A/0}, \hist{A/2}\pigr\}, \pigl\{\hist{A/1}, \hist{A/2}\pigr\}\right\}
\]
\item Cover \#7 contains the following lowersets:
\[
    \left\{\pigl\{\hist{A/0}, \hist{A/1}\pigr\}, \pigl\{\hist{A/0}, \hist{A/2}\pigr\}, \pigl\{\hist{A/1}, \hist{A/2}\pigr\}\right\}
\]
\item Cover \#8 (global cover) contains the following lowersets:
\[
    \left\{\pigl\{\hist{A/0}, \hist{A/1}, \hist{A/2}\pigr\}\right\}
\]
\end{itemize}

As our second example, we look at the no-signalling space $\Hist{\discrete{A,B}, \{0,1\}}$ on 2 events with binary inputs. 
This space has 114 open covers, arranged into the following hierarchy.
\begin{center}
\includegraphics[width=0.8\textwidth]{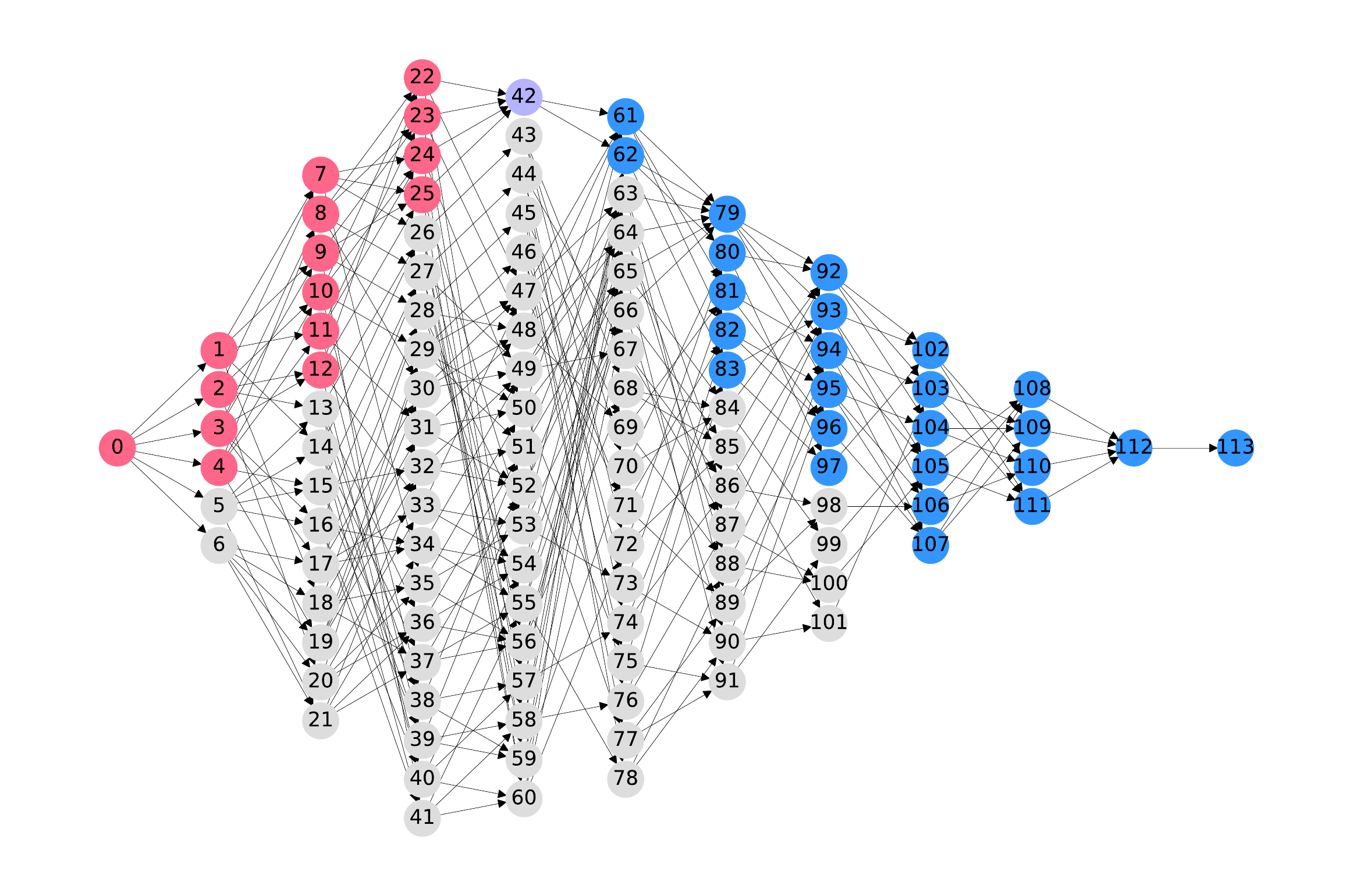}
\end{center}
The standard cover \#42 is coloured violet in the hierarchy and it takes the following form:
\[
\scalebox{0.95}{$\left\{\pigl\{\hist{A/0}, \hist{B/0}\pigr\}, \pigl\{\hist{A/0}, \hist{B/1}\pigr\}, \pigl\{\hist{A/1}, \hist{B/0}\pigr\}\right\}$}
\]
The refinements of the standard cover are coloured red in the hierarchy above.
They include the fully solipsistic cover \#0, which takes the following form:
\[
\left\{\pigl\{\hist{A/0}\pigr\}, \pigl\{\hist{A/1}\pigr\}, \pigl\{\hist{B/0}\pigr\}, \pigl\{\hist{B/1}\pigr\}\right\}
\]
The closest refinements of the standard cover are obtained by removing one of its 4 open sets.
For example, cover \#22 takes the following form:
\[
\left\{\pigl\{\hist{A/0}, \hist{B/0}\pigr\}, \pigl\{\hist{A/0}, \hist{B/1}\pigr\}, \pigl\{\hist{A/1}, \hist{B/0}\pigr\}\right\}
\]
The coarsenings of the standard cover are coloured blue in the hierarchy above.
They include the classical cover \#113, which takes the following form:
\[
\left\{\pigl\{\hist{A/0}, \hist{A/1}, \hist{B/0}, \hist{B/1}\pigr\}\right\}
\]
The closest coarsenings of the standard cover are obtained by adding both input histories for either event \ev{A} (cover \#61) or event \ev{B} (cover \#62).
For example, cover \#61 takes the following form:
\[
\scalebox{0.78}{$\left\{\pigl\{\hist{A/0}, \hist{A/1}\pigr\}, \pigl\{\hist{A/0}, \hist{B/0}\pigr\}, \pigl\{\hist{A/0}, \hist{B/1}\pigr\}, \pigl\{\hist{A/1}, \hist{B/0}\pigr\}, \pigl\{\hist{A/1}, \hist{B/1}\pigr\}\right\}$}
\]
Finally, there are covers which don't lie either below or above the standard cover.
The minimal covers unrelated to the standard cover are \#5 and \#6: these covers add both input histories for either event \ev{A} (cover \#6) or event \ev{B} (cover \#5) to the fully solipsistic cover, much as covers \#61 and \#62 did for the standard cover.
For example, cover \#5 takes the following form:
\[
\left\{\pigl\{\hist{A/0}\pigr\}, \pigl\{\hist{A/1}\pigr\}, \pigl\{\hist{B/0}, \hist{B/1}\pigr\}\right\}
\]
The maximal covers unrelated to the standard cover are \#98, \#99, \#100 and \#101.
The take the following form, for all $i_\ev{A}, i_\ev{B} \in \{0, 1\}$:
\[
\left\{\pigl\{\hist{A/0}, \hist{A/1}, \hist{B/i_\ev{B}}\pigr\}, \pigl\{\hist{A/i_\ev{A}}, \hist{B/0}, \hist{B/1}\pigr\}\right\}
\]

As our third example, we look at the following space, one of the four spaces lying in the middle layer of the hierarchy of causally complete spaces on 2 events with binary inputs:
\begin{center}
    \begin{tabular}{cc}
    \includegraphics[height=2.5cm]{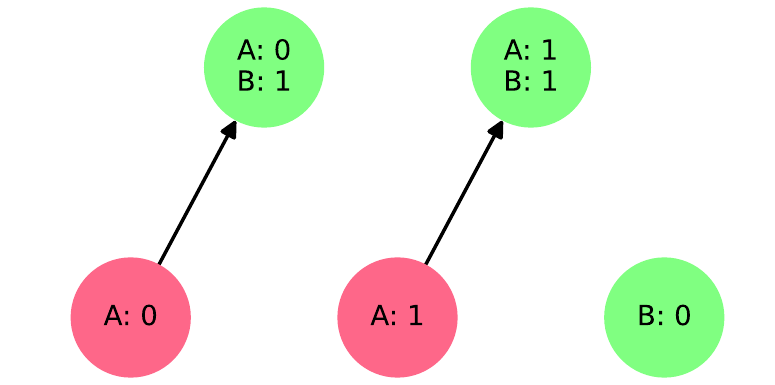}
    &
    \includegraphics[height=2.5cm]{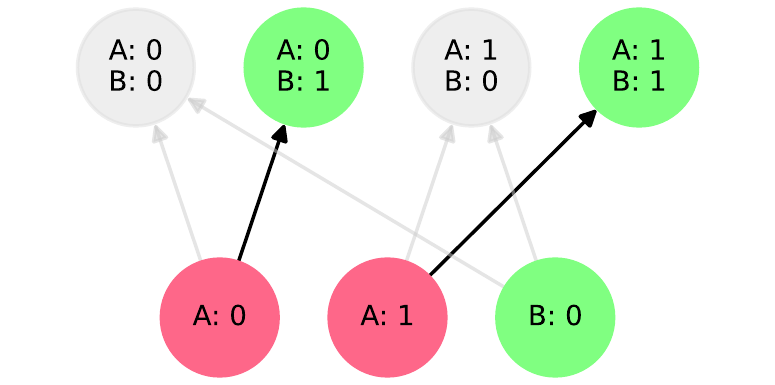}
    \\
    $\Theta$
    &
    $\Ext{\Theta}$
    \end{tabular}
\end{center}
This space has 80 open covers, arranged into the following hierarchy.
\begin{center}
\includegraphics[width=0.7\textwidth]{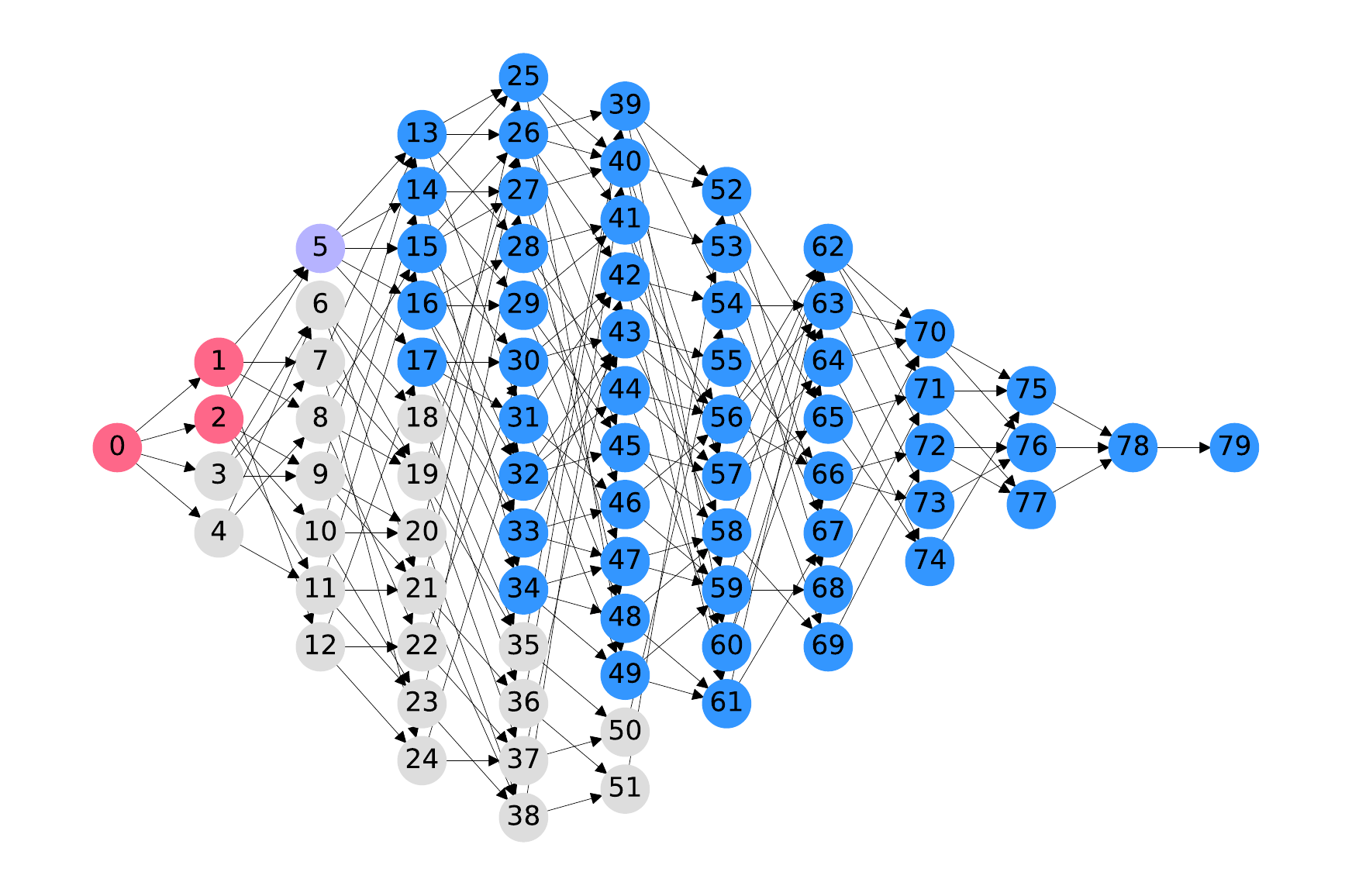}
\end{center}
The standard cover \#5 is coloured violet in the hierarchy and takes the following form:
\[
\scalebox{0.80}{$\left\{\pigl\{\hist{A/0}, \hist{B/0}\pigr\}, \pigl\{\hist{A/0}, \hist{B/1,A/0}\pigr\}, \pigl\{\hist{A/1}, \hist{B/0}\pigr\}, \pigl\{\hist{A/1}, \hist{B/1,A/1}\pigr\}\right\}$}
\]
The refinements of the standard cover are coloured red in the hierarchy above.
They include the fully solipsistic cover \#0, which takes the following form:
\[
\left\{\pigl\{\hist{B/0}\pigr\}, \pigl\{\hist{A/0}, \hist{B/1,A/0}\pigr\}, \pigl\{\hist{A/1}, \hist{B/1,A/1}\pigr\}\right\}
\]
The two covers \#1 and \#2 lying between the solipsistic and standard cover take the following form, for $i_\ev{A} \in \{0,1\}$
\[
\left\{\pigl\{\hist{A/i_\ev{A}}, \hist{B/0}\pigr\}, \pigl\{\hist{A/0}, \hist{B/1,A/0}\pigr\}, \pigl\{\hist{A/1}, \hist{B/1,A/1}\pigr\}\right\}
\]
The coarsenings of the standard cover are coloured blue in the hierarchy above.
They include the classical cover \#79, which takes the following form:
\[
\left\{\pigl\{\hist{A/0}, \hist{A/1}, \hist{B/0}, \hist{B/1,A/0}, \hist{B/1,A/1}\pigr\}\right\}
\]

As our fourth and final example, we look at the space $\Hist{\total{A,B}, \{0,1\}}$ on 2 events with binary inputs.
This space has 380 open covers, arranged into the following hierarchy.
\begin{center}
\includegraphics[width=\textwidth]{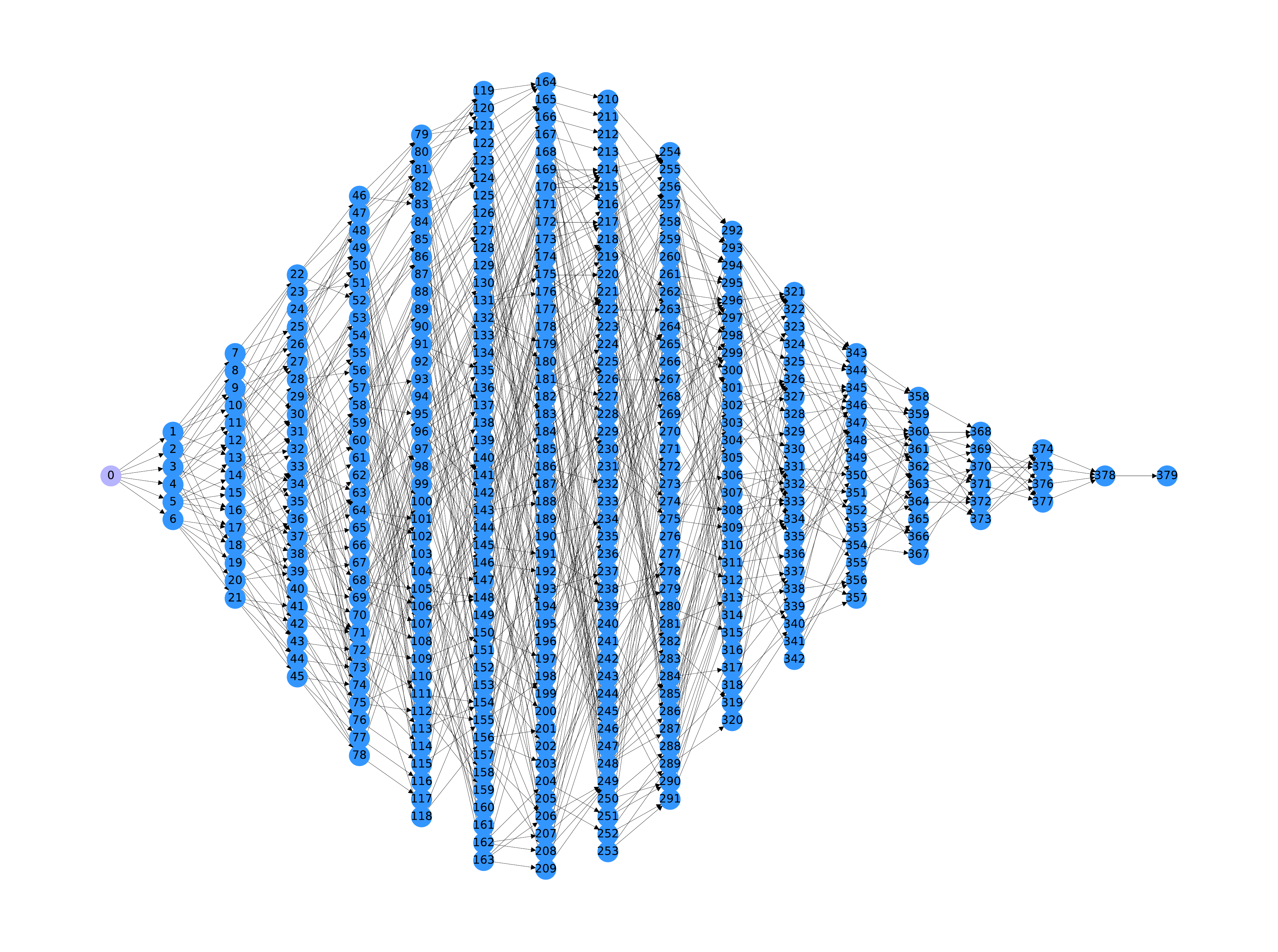}
\end{center}
Because $\Hist{\total{A,B}, \{0,1\}} = \ExtHist{\total{A,B}, \{0,1\}}$, the standard and fully solipsistic covers coincide in this example:
\[
\scalebox{0.74}{$\left\{\pigl\{\hist{A/0}, \hist{B/0,A/0}\pigr\}, \pigl\{\hist{A/0}, \hist{B/1,A/0}\pigr\}, \pigl\{\hist{A/1}, \hist{B/0,A/1}\pigr\}, \pigl\{\hist{A/1}, \hist{B/1,A/1}\pigr\}\right\}$}
\]
The classical cover takes the following form:
\[
\scalebox{0.9}{$\left\{\pigl\{\hist{A/0}, \hist{A/1}, \hist{B/0,A/0}, \hist{B/1,A/0}, \hist{B/0,A/1}, \hist{B/1,A/1}\pigr\}\right\}$}
\]
The immediate refinements of the standard cover take one of two possible forms.
Covers \#1, \#2, \#4 and \#5 take the following form, for $i_\ev{A}, i_\ev{B} \in \{0,1\}$:
\[
\scalebox{0.65}{$\left\{\pigl\{\hist{A/0}, \hist{B/0,A/0}\pigr\}, \pigl\{\hist{A/0}, \hist{B/1,A/0}\pigr\}, \pigl\{\hist{A/1}, \hist{B/0,A/1}\pigr\}, \pigl\{\hist{A/0}, \hist{A/1}, \hist{A/i_\ev{A},B/i_\ev{B}}\pigr\}\right\}$}
\]
Covers \#3 and \#6 are bit-flips of each other, taking the following form:
\[
\scalebox{0.80}{$\left\{\pigl\{\hist{A/0}, \hist{B/0,A/0}\pigr\}, \pigl\{\hist{A/0}, \hist{B/1,A/0}\pigr\}, \pigl\{\hist{A/1}, \hist{B/0,A/1}, \hist{B/1,A/1}\pigr\}\right\}$}
\]
\[
\scalebox{0.80}{$\left\{\pigl\{\hist{A/1}, \hist{B/0,A/1}\pigr\}, \pigl\{\hist{A/1}, \hist{B/1,A/1}\pigr\}, \pigl\{\hist{A/0}, \hist{B/0,A/0}, \hist{B/1,A/0}\pigr\}\right\}$}
\]

Having seen a few examples of covers, we now move to the definition of empirical models for an arbitrary cover $\mathcal{C}$.
These are a straightforward generalisation of those for the standard cover: they are simply families of distributions on causal functions for each lowerset $\lambda \in \mathcal{C}$ in the cover.

\begin{definition}
\label{definition:emp-model}
Let $\Theta$ be a space of input histories and let $\underline{O} = (O_\omega)_{\omega \in \Events{\Theta}}$ be a family of non-empty sets of outputs.
If $\mathcal{C}$ is a cover of $\Theta$, an \emph{empirical model} $e$ on $\mathcal{C}$ is a compatible family $e = (e_{\lambda})_{\lambda \in \mathcal{C}}$ for the presheaf of causal distributions $\CausDist{\Lsets{\Theta}, \underline{O}}$.
A \emph{standard empirical model} is an empirical model on the standard cover, a \emph{solipsistic empirical model} is an empirical model on some solipsistic cover, and a \emph{classical empirical model} is an empirical model on the classical cover.
We write $\EmpModels{\mathcal{C}, \underline{O}}$ for the empirical models on a cover $\mathcal{C}$ of $\Theta$, with outputs valued in $\underline{O}$.
\end{definition}

\begin{observation}
The cover can always be recovered from the empirical model as $\mathcal{C} = \dom{e}$, so it is not strictly necessary to explicitly state ``on $\mathcal{C}$'' when talking about an empirical model $e$.
Similarly, $\Theta$ can be recovered from any one of its covers $\mathcal{C} \in \Covers{\Theta}$, as the union $\bigcup\mathcal{C}$ of the open sets in the cover.
\end{observation}

For a familiar example of an empirical model on a cover different from the standard one, we look at cover \#7 for the space $\Theta := \Hist{\ev{A}, \{0,1,2\}}$ previously discussed:
\[
    \mathcal{C} := \left\{\pigl\{\hist{A/0}, \hist{A/1}\pigr\}, \pigl\{\hist{A/0}, \hist{A/2}\pigr\}, \pigl\{\hist{A/1}, \hist{A/2}\pigr\}\right\}
\]
Each lowerset in this cover takes the form $\lambda_{i, i'} := \pigl\{\hist{A/i}, \hist{A/i'}\pigr\}$ for distinct $i, i' \in \{0,1,2\}$ and it has the following binary-valued causal functions, for all $o, o' \in \{0,1\}$:
\[
f_{oo'|ii'}:=
\left\{
\begin{array}{rl}
    \hist{A/i}        &\mapsto o\\
    \hist{A/i'}       &\mapsto o'\\
\end{array}
\right.
\]
We define the following empirical model $e^{tri} \in \EmpModels{\mathcal{C}, \{0,1\}}$, sometimes known as the ``contextual triangle'' and originally due to \cite{Specker1960,Liang2011}:
\[
\begin{array}{rcl}
e^{tri}_{\lambda_{01}}
&:=& \frac{1}{2}\delta_{\Ext{f_{01|01}}}+\frac{1}{2}\delta_{\Ext{f_{10|01}}}
\\
e^{tri}_{\lambda_{02}}
&:=& \frac{1}{2}\delta_{\Ext{f_{01|02}}}+\frac{1}{2}\delta_{\Ext{f_{10|02}}}
\\
e^{tri}_{\lambda_{12}}
&:=& \frac{1}{2}\delta_{\Ext{f_{01|12}}}+\frac{1}{2}\delta_{\Ext{f_{10|12}}}
\end{array}
\]
We can represent this empirical model in tabular form, with rows indexed by $i\,i'$ and columns indexed by $o\,o'$:
\begin{center}
\begin{tabular}{l|rrrr}
\hfill
   & 00  & 01    & 10    & 11\\
\hline
01 & $0$ & $1/2$ & $1/2$ & $0$ \\
02 & $0$ & $1/2$ & $1/2$ & $0$ \\
12 & $0$ & $1/2$ & $1/2$ & $0$
\end{tabular}
\end{center}
As the name suggests, this empirical model is an example of a ``contextual'' empirical model: these are the models which cannot be explained ``classically''.
Classical empirical models for a space $\Theta$ are, by definition, probability distributions on extended causal functions defined on the entire space:
\[
\CausDist{\ClsCov{\Theta}, \underline{O}}
=\Dist{\ExtCausFun{\Theta, \underline{O}}}
\]
As a consequence, any empirical model $e \in \EmpModels{\mathcal{C}, \underline{O}}$ which arises as restriction $\restrict{\hat{e}}{\dom{e}}$ of some classical empirical model $\hat{e} \in \EmpModels{\ClsCov{\Theta}, \underline{O}}$ admits a \emph{deterministic causal hidden variable model (HVM)}: the observed probabilities are fully explained by some probabilistic mixture of causal functions defined globally on $\Theta$.
Empirical model admitting such a deterministic causal HVM are known as ``non-contextual'' (or ``local'', in the special case of the standard cover).

\begin{definition}
Let $\Theta$ be a space of input histories and let $\underline{O} = (O_\omega)_{\omega \in \Events{\Theta}}$ be a family of non-empty sets of outputs.
Let $e \in \EmpModels{\mathcal{C}, \underline{O}}$ be an empirical model for some cover $\mathcal{C}$ on $\Theta$.
We say that $e$ is \emph{non-contextual} if it arises as restriction $e = \restrict{\hat{e}}{\dom{e}}$ of a classical empirical model $\hat{e} \in \EmpModels{\ClsCov{\Theta}, \underline{O}}$; otherwise, we say that $e$ is \emph{contextual}.
If $e$ is a standard empirical model, we adopt \emph{local} as a synonym of non-contextual, and \emph{non-local} as a synonym of contextual.
\end{definition}

\begin{definition}
Let $\Theta$ be a space of input histories and let $\underline{O} = (O_\omega)_{\omega \in \Events{\Theta}}$ be a family of non-empty sets of outputs.
Let $e$ be a solipsistic empirical model.
We say that $e$ is \emph{solipsistically non-contextual} if it arises as restriction $e = \restrict{\hat{e}}{\dom{e}}$ of an empirical model $\hat{e} \in \EmpModels{\mathcal{C}, \underline{O}}$ for some cover $\mathcal{C} \succeq \StdCov{\Theta}$; otherwise, we say that $e$ is \emph{solipsistically contextual}.
\end{definition}

\begin{definition}
Let $\Theta$ be a space of input histories and let $\underline{O} = (O_\omega)_{\omega \in \Events{\Theta}}$ be a family of non-empty sets of outputs.
Let $e \in \EmpModels{\mathcal{C}, \underline{O}}$ be an empirical model for some cover $\mathcal{C}$ on $\Theta$.
The \emph{non-contextual fraction} of $e$ is defined to be the largest $p \in [0,1]$ such that $e$ can be written as $e = p e^{NC} + (1-p) e'$, where $e^{NC}, e' \in \EmpModels{\mathcal{C}, \underline{O}}$ and $e^{NC}$ is a non-contextual empirical model; the \emph{contextual fraction} of $e$ is defined to be 1 minus its non-contextual fraction.
When $e$ is a standard empirical model, we adopt \emph{local fraction} as a synonym of non-contextual fraction, and \emph{non-local fraction} as a synonym of contextual fraction.
\end{definition}

\begin{observation}
Let $\Theta$ be a space of input histories and let $\underline{O} = (O_\omega)_{\omega \in \Events{\Theta}}$ be a family of non-empty sets of outputs.
Let $e$ be an empirical model on a cover $\mathcal{C}$, let $\mathcal{C}'$ be a finer cover and let $e' := \restrict{e}{\mathcal{C}'}$ be the restriction of $e$ to $\mathcal{C}'$.
If $e = \restrict{\hat{e}}{\mathcal{C}}$ is non-contextual, then $e' = \restrict{\hat{e}}{\mathcal{C}'}$ is non-contextual.
Hence, if $\restrict{e}{\mathcal{C}'}$ is contextual, then $e$ is contextual.
\end{observation}

The contextual triangle empirical model $e^{tri}$ previously defined on cover \#7 of space $\Theta := \Hist{\ev{A}, \{0,1,2\}}$ is a known example of a contextual empirical model.
The causal functions in $\CausFun{\Theta, \{0,1\}}$ form the following set:
\[
\prod_{h \in \Theta} O_{\tip{\Theta}{h}}
\]
Specifically, there are 8 causal functions, taking the following form for $(o_{0},o_{1},o_{2}) \in \{0,1\}^3$:
\[
g_{o_{0} o_{1} o_{2}}:=
\left\{
\begin{array}{rl}
    \hist{A/0}        &\mapsto o_{0}\\
    \hist{A/1}        &\mapsto o_{1}\\
    \hist{A/2}        &\mapsto o_{2}
\end{array}
\right.
\]
Classical empirical models for $\Theta$ then take the following form, for probability distributions $d \in \Dist{\{0,1\}^3}$:
\[
e^{(d)}
:= \sum_{o_0}\sum_{o_1}\sum_{o_2} d(o_0,o_1,o_2) \delta_{\Ext{g_{o_{0} o_{1} o_{2}}}}
\]
The restrictions of the 8 causal functions for $\Theta$ to the lowersets in cover \#7 take the following form:
\[
\restrict{g_{o_{0} o_{1} o_{2}}}{\lambda_{i,i'}}
= f_{o_{i}o_{i'}|ii'}
\]
Hence, the restrictions of the classical empirical models to cover \#7 take the following form:
\[
\restrict{e^{(d)}}{\lambda_{i,i'}}
= \sum_{o_0}\sum_{o_1}\sum_{o_2} d(o_0,o_1,o_2) \delta_{\Ext{f_{o_{i}o_{i'}|ii'}}}
\]
We can represent the generic restriction of a classical model in tabular form:
\begin{center}
\begin{tabular}{l|rrrr}
\hfill
   & 00  & 01    & 10    & 11\\
\hline
01 & $d(000)+d(001)$ & $d(010)+d(011)$ & $d(100)+d(101)$ & $d(110)+d(111)$ \\
02 & $d(000)+d(010)$ & $d(001)+d(011)$ & $d(100)+d(110)$ & $d(101)+d(111)$ \\
12 & $d(000)+d(100)$ & $d(001)+d(101)$ & $d(010)+d(110)$ & $d(011)+d(111)$
\end{tabular}
\end{center}
For a non-contextual empirical model for cover \#7 of $\Hist{\ev{A}, \{0,1,2\}}$, the table above shows that the difference between the sum of the elements in the first and fourth columns and the sum of the elements in the second and third column---that is, the sum of the output correlation coefficients over the 3 input contexts---is bounded below by -1:
\[
\begin{array}{rrl}
&&\left(d(000)+d(001)\right)
+\left(d(110)+d(111)\right)
+\left(d(000)+d(010)\right)
\\
&+&\left(d(101)+d(111)\right)
+\left(d(000)+d(100)\right)
+\left(d(011)+d(111)\right)
\\
&-&\left(d(010)+d(011)\right)
-\left(d(100)+d(101)\right)
-\left(d(001)+d(011)\right)
\\
&-&\left(d(100)+d(110)\right)
-\left(d(001)+d(101)\right)
-\left(d(010)+d(110)\right)
\\
= && 4(d(000)+d(111))-1 \geq -1
\end{array}
\]
For the contextual triangle empirical model $e^{tri}$, the same number comes to $-3$ instead, proving that the empirical model is contextual.

\subsection{Contextuality and Causality}
\label{subsection:contextuality-and-causality}

Contextuality in the case of no-signalling spaces has been extensively studied by previous literature, with the contextual triangle example chosen for its familiarity.
In this Subsection, we will focus our attention on some interactions between contextuality and causality:
\begin{enumerate}
    \item We will investigate ``solipsistic'' contextuality, arising in solipsistic empirical models for spaces where the presheaf of causal functions fails to be a sheaf.
    \item We will prove a no-go result for non-locality on causal switch spaces, related to results on causal separability by \cite{abbott2016multipartite}.
    \item We will consider the implication of causal inseparability of functions for non-contextual empirical models on causally incomplete spaces.
\end{enumerate}
In the next Subsection, we will remark on the correlation between contextuality and causal inseparability more broadly, a notion which is introduced and thoroughly investigated in the companion work ``The Geometry of Causality'' \cite{gogioso2022geometry}.

\subsubsection{Causally-induced contextuality.}
As our first example of interaction between causality and contextuality, we consider examples of contextuality which we can safely deem to be purely due to the additional constraints imposed on causal functions by non-tight spaces.
Since maximal extended input histories witness the output consistency for all histories below them, such examples of contextuality can only manifest below the standard cover, prompting the following definition.
In particular, if empirical models displaying such solipsistic contextuality exist at all, then some necessarily exist on the fully solipsistic cover.

\begin{definition}
Let $\Theta$ be a space of input histories and let $\underline{O} = (O_\omega)_{\omega \in \Events{\Theta}}$ be a family of non-empty sets of outputs.
Let $e$ be an empirical model.
We say that $e$ displays \emph{solipsistic contextuality} if it does not arise as restriction $e = \restrict{\hat{e}}{\dom{e}}$ of a standard empirical model $\hat{e}$.
\end{definition}

\begin{observation}
Let $\Theta$ be a space of input histories and let $\underline{O} = (O_\omega)_{\omega \in \Events{\Theta}}$ be a family of non-empty sets of outputs.
Let $e$ be an empirical model on a cover $\mathcal{C}$, let $\mathcal{C}'$ be a finer cover and let $e' := \restrict{e}{\mathcal{C}'}$ be the restriction of $e$ to $\mathcal{C}'$.
If $\restrict{e}{\mathcal{C}'}$ displays solipsistic contextuality, then $e$ displays solipsistic contextuality.
\end{observation}

The solipsistic contextuality displayed by these spaces is surprisingly ``strong'': in the examples we explore, it manifests as the impossibility of gluing certain ``deterministic'' empirical models to causal functions defined on the whole space.
Since deterministic empirical models correspond to causal functions on certain lowersets, such examples arise only for non-tight spaces where the presheaf of causal functions is not a sheaf.

\begin{definition}
Let $\Theta$ be a space of input histories and let $\underline{O} = (O_\omega)_{\omega \in \Events{\Theta}}$ be a family of non-empty sets of outputs.
We say that an empirical model $e=\left(e_\lambda\right)_{\lambda \in \mathcal{C}}$ is \emph{deterministic} if for every $\lambda \in \mathcal{C}$ there is some $f_{\lambda} \in \CausFun{\lambda, \underline{O}}$, necessarily unique, such that:
\[
e_\lambda = \delta_{\Ext{f_\lambda}}
\]
We say that an empirical model is \emph{globally deterministic} if it is the restriction of a deterministic empirical model on the classical cover.
\end{definition}

\begin{proposition}
\label{proposition:deterministic-empirical-models}
Let $\Theta$ be a space of input histories and let $\underline{O} = (O_\omega)_{\omega \in \Events{\Theta}}$ be a family of non-empty sets of outputs.
Let $\mathcal{C}$ be a cover for $\Theta$.
Deterministic empirical models $e=\left(e_\lambda\right)_{\lambda \in \mathcal{C}}$ correspond exactly to compatible families $f=\left(f_\lambda\right)_{\lambda \in \mathcal{C}}$ for $\CausFun{\Lsets{\Theta}, \underline{O}}$:
\[
e_\lambda = \delta_{\Ext{f_\lambda}}
\]
This correspondence respects restrictions:
\[
\restrict{e_\lambda}{\lambda'}
=
\delta_{\Ext{\restrict{f_\lambda}{\lambda'}}}
\]
In particular, globally deterministic empirical models correspond exactly to causal functions in $\CausFun{\Theta, \underline{O}}$.
\end{proposition}
\begin{proof}
See \ref{proof:proposition:deterministic-empirical-models}
\end{proof}

\begin{corollary}
\label{corollary:deterministic-empirical-models-noncontextual}
Let $\Theta$ be a space of input histories and let $\underline{O} = (O_\omega)_{\omega \in \Events{\Theta}}$ be a family of non-empty sets of outputs.
If $\CausFun{\Lsets{\Theta}, \underline{O}}$ is a sheaf, then every deterministic empirical model is globally deterministic---and, in particular, it does not display solipsistic contextuality.
\end{corollary}
\begin{proof}
See \ref{proof:corollary:deterministic-empirical-models-noncontextual}
\end{proof}

Recall from Definition \ref{definition:solipsitic-contextuality-witness} that a solipsistic contextuality witness $(k, \omega, h, h')$ is a quadruple where:
\begin{enumerate}
    \item $k \in \Ext{\Theta}$ is an extended input history
    \item $\omega \in \dom{k}$ is an event in the domain of $k$
    \item $h, h' \in \Theta$ are distinct input histories such that $\histconstr{\omega}{h}{h'}$, i.e. ones for which causal function outputs at $\omega$ must coincide
    \item there is no $h'' \in \Theta$ such that $h, h' \leq h''$ and $h'' \leq k$, whose existence would force consistency of outputs for $h, h'$ on the fully solipsistic cover.
\end{enumerate}
Recall from Theorem \ref{theorem:causal-contextuality-sheaf-condition} that the existence of such witnesses in a space $\Theta$ is equivalent to the failure for the presheaf of causal functions on $\Theta$ to be a sheaf.
Using such a witness $w := (k, \omega, h, h')$ for a space $\Theta$, we can construct a solipsistic empirical model $e^{(w)} \in \EmpModels{\SolCov{\Theta}, \{0,1\}}$ which is deterministic and displays solipsistic contextuality.
For a simple example, we look again at space $\Theta_{17}$ from \cite{gogioso2022combinatorics} Figure~5 (p.~42).
This space admits a single solipsistic contextuality witness $w := (k, \omega, h, h')$:
\[
\begin{array}{rcl}
k &:=& \hist{A/1,B/1,C/2}
\\
\omega &:=& \ev{C}
\\
h &:=& \hist{A/1, C/1}
\\
h' &:=& \hist{B/1, C/1}
\end{array}
\]
Below, the extended input history $k:=\hist{A/1,B/1,C/2}$ has been circled in blue, the same colour used to indicate that event $\omega := \ev{C}$ appears as a tip event in the input histories $h:=\hist{A/1, C/1}$ and $h':=\hist{B/1, C/1}$; edges from $h, h'$ to $k$ have also been highlighted blue, for clarity.
Note that there is no input history $h''$ such that $h, h' \leq h''$ in this space.
\begin{center}
    \begin{tabular}{cc}
    \includegraphics[height=3.5cm]{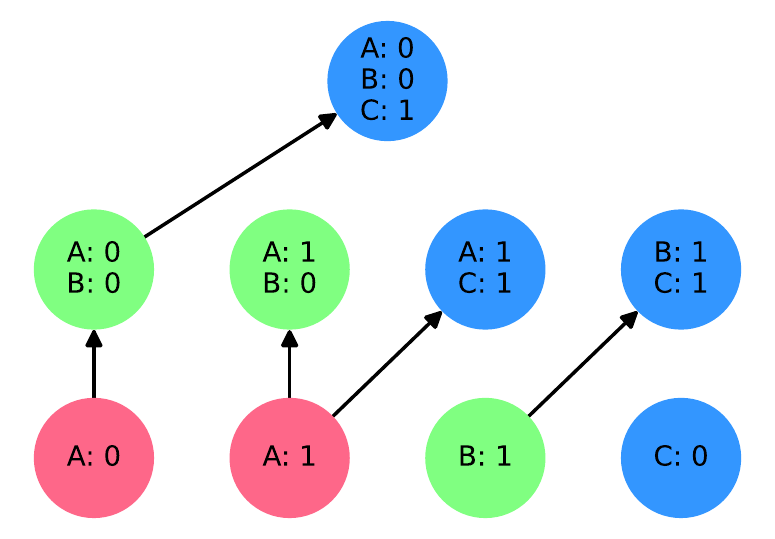}
    &
    \includegraphics[height=3.5cm]{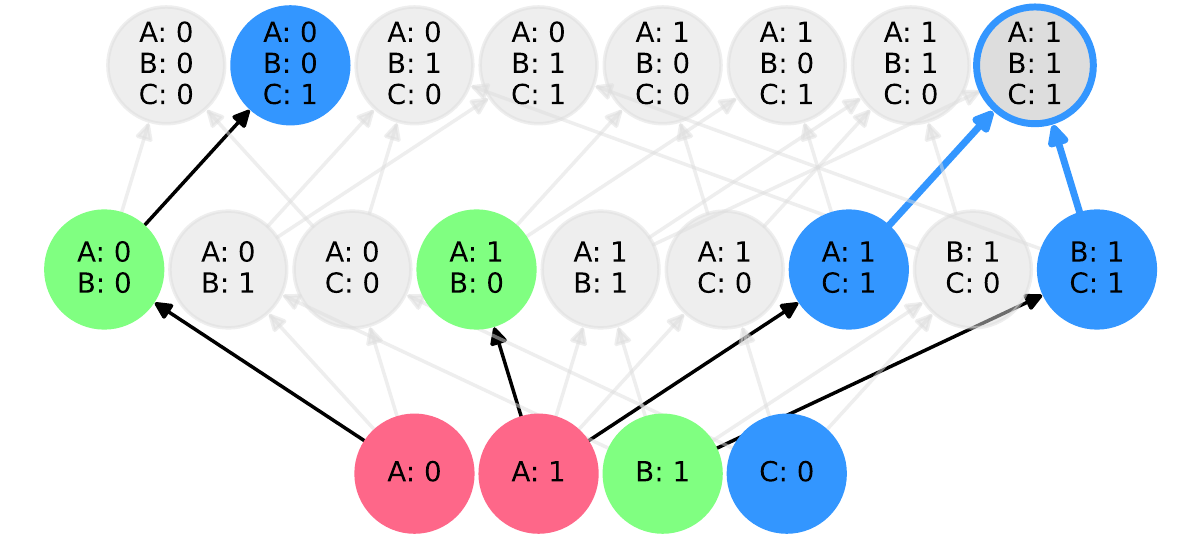}
    \\
    $\Theta_{17}$
    &
    $\Ext{\Theta_{17}}$ with highlights
    \end{tabular}
\end{center}
For all $t \in \max\Theta_{17}$ we consider the following ``zero functions'', defined on all $h'' \in \downset{t}$:
\[
f_{0|t} := h'' \mapsto 0
\]
For $h = \hist{A/1, C/1}$ we consider the following ``indicator function'', defined on all $h'' \in \downset{h}$:
\[
f_{\delta_{\ev{C}}|h}
:= h'' \mapsto
\left\{
\begin{array}{rl}
1 &\text{ if } \tip{\Theta_{17}}{h''} = \ev{C}\\
0 &\text{ otherwise}
\end{array}
\right.
\]
Explicitly, the indicator function takes the following form:
\[
f_{\delta_{\ev{C}}|\hist{A/1, C/1}}
:=
\left\{
\begin{array}{rl}
\hist{A/1} &\mapsto 0\\
\hist{A/1, C/1} &\mapsto 1
\end{array}
\right.
\]
We then consider the following solipsistic empirical model, where components are indexed by the downsets $\downset{t}$ of the 5 maximal input histories $t \in \max \Theta_{17}$:
\[
e^{(w)}_{\downset{t}}
:=
\left\{
\begin{array}{rl}
f_{\delta_{\ev{C}}|h} & \text{ if } t = h\\
f_{0|t} & \text{ otherwise}
\end{array}
\right.
\]
The tabular representation previously considered for standard empirical models (on spaces satisfying the free-choice condition) can be extended to solipsistic empirical models, by explicitly accounting for the fact that different rows/columns involve different subsets of events.
The table below corresponds to empirical model $e^{(w)}$ above: because the empirical model is deterministic, the entries are only 0 and 1, with a single 1 in each row.
\begin{center}
\scalebox{0.6}{
\begin{tabular}{l|rrrrrrrrrrrrrrrrrrrrrr}
\hfill
ABC
& \_\_0 & \_\_1
& \_00 & \_01 & \_10 & \_11
& 0\_0 & 0\_1 & 1\_0 & 1\_1 
& 00\_ & 01\_ & 10\_ & 11\_
& 000 & 001 & 010 & 011 & 100 & 101 & 110 & 111
\\
\hline
\_\_0
& 1 & 0
&   &   &   &
&   &   &   &
&   &   &   &
&   &   &   &   &   &   &   &
\\
\_11
&   &
& 1 & 0 & 0 & 0 
&   &   &   &
&   &   &   &
&   &   &   &   &   &   &   &
\\
1\_1
&   &
&   &   &   &
& 0 & 1 & 0 & 0 
&   &   &   &
&   &   &   &   &   &   &   &
\\ 
10\_
&   &
&   &   &   &
&   &   &   &
& 1 & 0 & 0 & 0 
&   &   &   &   &   &   &   &
\\
001
&   &
&   &   &   &
&   &   &   &
&   &   &   &
& 1 & 0 & 0 & 0 & 0 & 0 & 0 & 0
\\
\end{tabular}
}
\end{center}
Each row of the table is indexed by a maximal input history $t \in \max\Theta_{17}$: we indicate such a history by the inputs it assigns to events in its domain, using an underscore $\_$ to indicate that an event is not in the domain of the input history.
For example, row $1\_1$ corresponds to the maximal input history $\hist{A/1,C/1}$, while row $10\_$ corresponds to the maximal input history $\hist{A/1,B/0}$.
Each column of the table is similarly indexed by an output history: for the row corresponding to a given $t \in \max \Theta_{17}$, the relevant columns are those for output histories $\prod_{h'' \in \downset{t}} O_{\tip{\Theta_{17}}{h''}}$; entries for other columns are omitted.
The same indexing scheme as rows can be adopted for columns, thanks to the following observations:
\begin{itemize}
    \item $\Theta_{17}$ is causally complete, so each $h'' \in \downset{t}$ has a single tip event;
    \item there is exactly one $h'' \in \downset{t}$ with $\tip{\downset{t}}{h''} = \omega$ for each $\omega \in \dom{t}$.
\end{itemize}
For example, column $0\_1$ for row $1\_1$ indicates the causal function $f_{\delta_{\ev{C}}|h}$: the row is input history $h = \hist{A:1/C:1}$, while the column indicates output history $\hist{A/0, C/1}$, which corresponds to $\hist{A/1} \mapsto 0$ (because $\tip{\Theta_{17}}{\hist{A/1}} = \ev{A}$) and $\hist{A/1, C/1} \mapsto 1$ (because $\tip{\Theta_{17}}{\hist{A/1, C/1}} = \ev{C}$).

To prove that the solipsistic empirical model $e^{(w)}$ cannot arise by restriction of some standard empirical model $e$, it suffices to focus on the rows corresponding to the two maximal input histories $h = \hist{A/1,C/1}$ and $h' = \hist{B/1,C/1}$ used by witness $w$:
\begin{center}
\scalebox{1}{
\begin{tabular}{l|rrrrrrrr}
\hfill
ABC
& \_00 & \_01 & \_10 & \_11
& 0\_0 & 0\_1 & 1\_0 & 1\_1
\\
\hline
\_11
& 1 & 0 & 0 & 0 
&   &   &   &
\\
1\_1
&   &   &   &
& 0 & 1 & 0 & 0
\\
\end{tabular}
}
\end{center}
For $e$, it suffices to focus on the row corresponding to the maximal extended input history $k = \hist{A/1,B/1,C/1}$:
\begin{center}
\begin{tabular}{l|rrrrrrrr}
\hfill
ABC
& 000 & 001 & 010 & 011
& 100 & 101 & 110 & 111
\\
\hline
111
& $p_{000}$ & $p_{001}$ & $p_{010}$ & $p_{011}$
& $p_{100}$ & $p_{101}$ & $p_{110}$ & $p_{111}$
\end{tabular}
\end{center}
If $e$ restricts to $e^{(w)}$, then the zero components on row $\_11$ of $e^{(w)}$ restrict the possible support for $e$, forcing the following components of row $111$ to be zero:
\begin{center}
\begin{tabular}{l|rrrrrrrr}
\hfill
ABC
& 000 & 001 & 010 & 011
& 100 & 101 & 110 & 111
\\
\hline
111
& $p_{000}$ & 0 & 0 & 0
& $p_{100}$ & 0 & 0 & 0
\end{tabular}
\end{center}
If $e$ restricts to $e^{(w)}$, then the zero components on row $1\_1$ of $e^{(w)}$ alternatively restrict the possible support for $e$, forcing the following components of row $111$ to be zero:
\begin{center}
\begin{tabular}{l|rrrrrrrr}
\hfill
ABC
& 000 & 001 & 010 & 011
& 100 & 101 & 110 & 111
\\
\hline
111
& 0 & $p_{001}$ & 0 & $p_{011}$
& 0 & 0 & 0 & 0
\end{tabular}
\end{center}
Putting the two restrictions together forces all components of row $111$ to be zero:
\begin{center}
\begin{tabular}{l|rrrrrrrr}
\hfill
ABC
& 000 & 001 & 010 & 011
& 100 & 101 & 110 & 111
\\
\hline
111
& 0 & 0 & 0 & 0
& 0 & 0 & 0 & 0
\end{tabular}
\end{center}
This is inconsistent with the requirement that $e_{\downset{\hist{A/1,B/1,C/1}}}$ be a probability distribution: hence no standard model $e$ can restrict to the solipsistic empirical model $e^{(w)}$ defined above, proving that the latter displays solipsistic contextuality.
The following proposition takes the construction above and generalises it to arbitrary spaces and witnesses.

\begin{theorem}
\label{theorem:deterministic-empirical-models-2}
Let $\Theta$ be a space of input histories and let $\underline{O} = (O_\omega)_{\omega \in \Events{\Theta}}$ be a family of sets of outputs such that $O_\omega$ has at least two elements for all $\omega \in \Events{\Theta}$.
If $(k, \omega, h, h')$ is a solipsistic contextuality witness for $\Theta$, then the deterministic solipsistic empirical model $e=\left(e_{\downset{t}}\right)_{t \in \max\Theta}$ defined as follows displays solipsistic contextuality:
\begin{equation}
\begin{array}{rcl}
e_{\downset{t}}
&:=&
\delta_{\Ext{f_t}}
\in
\CausDist{\downset{t}, \underline{O}}
\\
f_t(h'')_\xi
&:=&
\left\{
\begin{array}{rl}
o_{\xi,1} &\text{ if $h''=h$ and $\xi=\omega$}\\
o_{\xi,0} &\text{ otherwise}
\end{array}
\right.
\end{array}
\end{equation}
where $o_{\omega,0}, o_{\omega,1} \in O_\omega$ are any distinct elements for each $\omega \in \Events{\Theta}$.
Hence, if $\Theta$ admits deterministic casual contextuality, then there exist deterministic solipsistic empirical models for $\Theta$ which display solipsistic contextuality---and, in particular, which are not globally deterministic.
\end{theorem}
\begin{proof}
See \ref{proof:theorem:deterministic-empirical-models-2}
\end{proof}

Finally, the results below prove that solipsistic contextuality cannot happen on or above the standard cover, making the additional covers introduced in this work---such as the fully solipsistic cover---fundamentally necessary to describe it.

\begin{proposition}
\label{proposition:gluing-standard-function-compat-family}
Let $\Theta$ be a space of input histories and let $\underline{O} = (O_\omega)_{\omega \in \Events{\Theta}}$ be a family of non-empty sets of outputs.
Let $f = (f_{\downset{k}})_{k \in \max\Ext{\Theta}}$ be a compatible family in \CausFun{\Lsets{\Theta}, \underline{O}}.
Then $f$ admits a gluing $\hat{f} \in \CausFun{\Theta, \underline{O}}$.
\end{proposition}
\begin{proof}
See \ref{proof:proposition:gluing-standard-function-compat-family}
\end{proof}

\begin{corollary}
\label{corollary:standard-emp-model-deterministic-globally-deterministic}
Let $\Theta$ be a space of input histories and let $\underline{O} = (O_\omega)_{\omega \in \Events{\Theta}}$ be a family of non-empty sets of outputs.
Let $e = (e_{\downset{k}})_{k \in \max\Ext{\Theta}}$ be a standard empirical model.
If $e$ is deterministic, then $e$ is globally deterministic.
\end{corollary}
\begin{proof}
See \ref{proof:corollary:standard-emp-model-deterministic-globally-deterministic}
\end{proof}

\subsubsection{No non-locality on causal switch spaces.}
\label{prologue:nogo-nonlocality-switch-spaces}
As our second example of interaction between causality and contextuality, a no-go result for non-locality on causal switch spaces, with spaces induced by total orders as a special case.
More generally, we consider non-empty spaces $\Theta$ such that $\Theta = \Ext{\Theta}$ and where each input history $h \in \Theta$ has a single tip event $\tips{\Theta}{h} = \{\tip{\Theta}{h}\}$.
Theorem~3.34 and Corollary~3.35 from \cite{gogioso2022combinatorics} characterise such spaces as inductively defined by conditional sequential composition:
\[
\Theta = \Hist{\{\omega_1\}, \restrict{\underline{\Inputs{\Theta}}}{\{\omega_1\}}}
\seqcomposeSym \underline{\Theta'}
\]
where $\omega_1 \in \Events{\Theta}$ is some event which ``comes before all others'', and either $\Events{\Theta} = \{\omega_1\}$ or every $\Theta_{i_1}$ for $i_1 \in \Inputs{\Theta}_{\omega_1}$ is itself a causal switch space on events $\Events{\Theta} \backslash \{\omega_1\}$ and inputs $\restrict{\Inputs{\Theta}}{\Events{\Theta} \backslash \{\omega_1\}}$.
In particular, we have:
\begin{itemize}
    \item $\omega \notin \Events{\Theta_{i_1}}$
    \item $\Theta_{i_1} = \Ext{\Theta_{i_1}}$
    \item for all $h \in \Theta'_{i_1}$ we have $\tips{\Theta'_{i_1}}{h} = \tip{\Theta}{\{\omega_1:i_1\}\vee h}$
    \item $\max\Ext{\Theta'_{i_1}} = \prod_{\omega \in \Events{\Theta} \backslash \{\omega_1\}} \Inputs{\Theta}_\omega$
\end{itemize}
Theorem \ref{theorem:caus-fun-conditional-sequential-composition-factorisation} provides a recursive factorisation result for causal functions on such spaces:
\[
\CausFun{\Theta, \underline{O}}
\cong
O^{\Inputs{\Theta}_{\omega_1}}
\times
\prod_{i_1 \in \Inputs{\Theta}_{\omega_1}}
\CausFun{\Theta'_{i_1}, \restrict{\underline{O}}{\Events{\Theta}\backslash\{\omega_1\}}}
\]
More explicitly, causal functions in $\CausFun{\Theta, \underline{O}}$ take the following form, for an arbitrary choice of outputs $\underline{o} \in O_{\omega_1}^{\Inputs{\Theta}_{\omega_1}}$ for the inputs at the first event $\omega_1$ and a family $\underline{f'}$ of causal functions $f'_{i_1} \in \CausFun{\Theta'_{i_1}, \restrict{\underline{O}}{\Events{\Theta}\backslash\{\omega_1\}}}$:
\[
g_{\underline{o}, \underline{f'}}
:=
h
\mapsto
\left\{
\begin{array}{rl}
o_{i_1} &\text{ if } h=\{\omega_1:i_1\} \in \Hist{\{\omega_1\}, \restrict{\underline{\Inputs{\Theta}}}{\{\omega_1\}}}\\
f'_{i_1}(h') &\text{ if } h=\{\omega_1:i_1\}\vee h' \text{ and } h' \in \Theta'_{i_1}
\end{array}
\right.
\]
The factorisation result for causal functions allows us to prove that every standard empirical model $e$ on such a causal switch space $\Theta$ is necessarily local, i.e that it arises by restriction of a classical empirical model $\hat{e}$ on $\Theta$.
The proof proceeds by induction on the number of events in the causal switch space.
In the ``base case'' where $\Events{\Theta} = \{\omega_1\}$, this is obvious: $\hat{e}$ is just the product probability distribution, where the output values for different inputs are sampled independently.
In the ``inductive case'' where $|\Events{\Theta}| \geq 2$, the proof proceeds as follows:
\begin{enumerate} 
    \item We consider the probability $p_{i_1, o_1}$ assigned by empirical model $e$ to output $o_1$ on input $i_1$:
    \[
    p_{i_1, o_1} := \restrict{e}{\downset{\{\omega_1:i_1\}}}\left(\{\omega_1:o_1\}\right)
    = \sum_{\underline{o}' \in \underline{O}'}
    e_{\downset{\left(\{\omega_1:i_1\}\vee h'\right)}}\left(\{\omega_1:o_1\}\vee\underline{o}'\right)
    \]
    \item For all $i_1, o_1$ for which $p_{i_1, o_1} > 0$, we define empirical models $e^{(i_1, o_1)}$ on the spaces $\Theta'_{i_1}$:
    \[
    e^{(i_1, o_1)}_{\downset{h'}}\left(\underline{o}'\right)
    :=
    \frac{1}{p_{i_1,o_1}}
    e_{\downset{\left(\{\omega_1:i_1\}\vee h'\right)}}\left(\{\omega_1:o_1\}\vee\underline{o}'\right)
    \]
    \item By inductive hypothesis, the standard empirical models $e^{(i_1, o_1)}$ arise by restriction of classical empirical models $\hat{e}^{(i_1, o_1)}$ on the spaces $\Theta'_{i_1}$.
    \item We collate the classical empirical models $\hat{e}^{(i_1, o_1)}$ together into a classical empirical model $\hat{e}$ on $\Theta$, weighted by the probabilities $p_{i_1, o_1}$
    \[
        \hat{e}_{\Theta}\left(\Ext{g_{\underline{o}, \underline{f'}}}\right)
        :=
        \prod\limits_{i_1 \in \Inputs{\Theta}_{\omega_1}}
        p_{i_1, o_1}
        \hat{e}^{(i_1, o_1)}_{\Theta'_{i_1}}\left(\Ext{f'_{i_1}}\right)
    \]
\end{enumerate}
The result below formalises the above procedure and proves that the classical empirical model $\hat{e}$ restricts to the standard empirical model $e$ on the standard cover for $\Theta$.

\begin{theorem}
\label{theorem:nogo-nonlocality-switch-spaces}
Let $\Theta \in \CSwitchSpaces{\underline{I}}$ be a non-empty causal switch space and let $e \in \EmpModels{\StdCov{\Theta}, \underline{O}}$ be a standard empirical model on $\Theta$.
Then $e$ is local, i.e. it arises as a restriction of a classical empirical model $\hat{e} \in \EmpModels{\ClsCov{\Theta}, \underline{O}}$ to the standard cover $\StdCov{\Theta}$.
\end{theorem}
\begin{proof}
See \ref{proof:theorem:nogo-nonlocality-switch-spaces}
\end{proof}

The no-go result does not generalise to arbitrary covers on arbitrary causal switch spaces, or even on arbitrary spaces induced by total orders: the contextual triangle provides a simple counterexample.
However, we can easily generalise it so spaces obtained by sequential conditional composition of ``indiscrete'' spaces, ones where all extended input histories are unrelated: causal switch spaces are a special case (each indiscrete space involved in the composition has a single event), as are indiscrete spaces for any given set of events and choice of inputs.

\begin{corollary}
\label{corollary:nogo-nonlocality-switch-spaces-indef}
Let $\Theta \in \CSwitchSpaces{\underline{I}}$ be a space of input histories which is obtained by repeated conditional sequential composition of ``indiscrete'' spaces, i.e. spaces such that $\Xi = \max\Ext{\Xi}$.
If $e \in \EmpModels{\StdCov{\Theta}, \underline{O}}$ is a standard empirical model on $\Theta$, then $e$ is local.
\end{corollary}
\begin{proof}
See \ref{proof:corollary:nogo-nonlocality-switch-spaces-indef}
\end{proof}

\subsubsection{Contextuality and inseparable functions.}

As our third example of interaction between causality and contextuality, we now look at the implications of causal incompleteness on contextuality.
Consider an empirical model $e$ on some cover of a causally incomplete space $\Theta$: for $e$ to be non-contextual means that $e$ can be expressed as a convex combination of causal functions on $\Theta$, but says nothing about the causal separability of the latter.
This is so by design: we wished indefinite causal orders to accommodate the possibility of event delocalisation, so we included all causal functions---rather than just the separable ones---in our presheaves.

However, it is sometimes interesting to ask whether a non-contextual empirical model can be decomposed classically in a separable way, or whether every non-contextual decomposition necessarily involves an inseparable function.
As an example of the former, consider the following empirical model on the causally incomplete space $\Hist{\indiscrete{A,B}, [0,1]}$:
\begin{center}
\begin{tabular}{l|rrrr}
\hfill
   & 00  & 01    & 10    & 11\\
\hline
00 & $1/2$ & $0$ & $0$ & $1/2$ \\
01 & $0$ & $1/2$ & $1/2$ & $0$ \\
10 & $0$ & $1/2$ & $1/2$ & $0$ \\
11 & $1/2$ & $0$ & $0$ & $1/2$
\end{tabular}
\end{center}
The empirical model can be decomposed as a 50\%-50\% mixture of the swap function and function obtained by post-composing a swap with a bit-flip on each output, both of which are inseparable:
\begin{center}
\raisebox{0mm}{$0.5\;\times$}
\scalebox{1}{$
\begin{tabular}{l|rrrr}
\hfill
   & 00  & 01    & 10    & 11\\
\hline
00 & $1$ & $0$ & $0$ & $0$ \\
01 & $0$ & $0$ & $1$ & $0$ \\
10 & $0$ & $1$ & $0$ & $0$ \\
11 & $0$ & $0$ & $0$ & $1$
\end{tabular}
$}
\raisebox{0mm}{$\;+\;\;0.5\;\times$}
\scalebox{1}{$
\begin{tabular}{l|rrrr}
\hfill
   & 00  & 01    & 10    & 11\\
\hline
00 & $0$ & $0$ & $0$ & $1$ \\
01 & $0$ & $1$ & $0$ & $0$ \\
10 & $0$ & $0$ & $1$ & $0$ \\
11 & $1$ & $0$ & $0$ & $0$
\end{tabular}
$}
\end{center}
However, the same empirical model can be also decomposed as a 50\%-50\% mixture of the identity function and the bit-flip function on each output, both of which are separable:
\begin{center}
\raisebox{0mm}{$0.5\;\times$}
\scalebox{1}{$
\begin{tabular}{l|rrrr}
\hfill
   & 00  & 01    & 10    & 11\\
\hline
00 & $1$ & $0$ & $0$ & $0$ \\
01 & $0$ & $1$ & $0$ & $0$ \\
10 & $0$ & $0$ & $1$ & $0$ \\
11 & $0$ & $0$ & $0$ & $1$
\end{tabular}
$}
\raisebox{0mm}{$\;+\;\;0.5\;\times$}
\scalebox{1}{$
\begin{tabular}{l|rrrr}
\hfill
   & 00  & 01    & 10    & 11\\
\hline
00 & $0$ & $0$ & $0$ & $1$ \\
01 & $0$ & $0$ & $1$ & $0$ \\
10 & $0$ & $1$ & $0$ & $0$ \\
11 & $1$ & $0$ & $0$ & $0$
\end{tabular}
$}
\end{center}
There are obvious cases where non-contextual empirical models can only be decomposed in terms of separable functions, e.g. on every causally complete space.
There are also obvious cases where non-contextual empirical models can only be decomposed in terms of inseparable functions, e.g. for the deterministic empirical models associated to the inseparable functions themselves.
As for the intermediate cases, there are some interesting questions.

For example, we might wonder whether a non-contextual empirical model $e$ on a causally incomplete space $\Theta$ which ``decomposes'' as a convex combination $e = \sum_{j=1}^{n} p_j e^{(j)}$ of empirical models for some causally complete sub-spaces $\Theta_1,...,\Theta_n$---where each $e^{(j)}$ is identified as an empirical model for both $\Theta$ and $\Theta_j$ in some meaningful sense---is non-contextual also for the causally complete sub-spaces $\Theta_1,...,\Theta_n$, and hence can be expressed as a convex combination of separable functions on the original, causally incomplete space $\Theta$.
This turns out to be a complicated questions, with answers either way depending on the specific circumstances:
\begin{enumerate}
    \item If $\Theta_1,...,\Theta_n$ are selected amongst the causal completions of $\Theta$ (not necessarily all of them), then the above holds for standard empirical models when $\Theta$ takes a certain form (see Proposition \ref{proposition:std-empmodel-local-par-seq-spaces-separability}) for details).
    \item If $\Theta_1,...,\Theta_n$ are not restricted to causal completions of $\Theta$, then the above is does not hold in general, not even for standard empirical models.
\end{enumerate}
We now proceed to substantiate the two points above, by proving a result for the first and providing a counterexample for the second.

\begin{definition}
    Let $\Theta$ be a space of input histories and let $\underline{O} = (O_\omega)_{\omega \in \Events{\Theta}}$ be a family of non-empty sets of outputs.
    Let $e \in \EmpModels{\mathcal{C}, \underline{O}}$ be a non-contextual (resp. local) empirical model.
    We say that $e$ is \emph{separably} non-contextual (resp. local) if it arises as restriction $e = \restrict{\hat{e}}{\dom{e}}$ of a classical empirical model $\hat{e} \in \EmpModels{\ClsCov{\Theta}, \underline{O}}$ which is entirely supported by causal functions which are separable.
\end{definition}

\begin{definition}
Let $\Theta$ be a space of input histories and let $\underline{O} = (O_\omega)_{\omega \in \Events{\Theta}}$ be a family of non-empty sets of outputs.
Let $e \in \EmpModels{\mathcal{C}, \underline{O}}$ be an empirical model for some cover $\mathcal{C}$ on $\Theta$.
The \emph{separable non-contextual fraction} of $e$ is defined to be the largest $p \in [0,1]$ such that $e$ can be written as $e = p e^{NC} + (1-p) e'$, where $e^{NC}, e' \in \EmpModels{\mathcal{C}, \underline{O}}$ and $e^{NC}$ is an empirical model which is separably non-contextual.
When $e$ is a standard empirical model, we adopt \emph{separable local fraction} as a synonym of separable non-contextual fraction.
\end{definition}

\begin{observation}
Let $\Theta \leq \Theta'$ are spaces of input histories such that $\Events{\Theta'} = \Events{\Theta}$ and $\Inputs{\Theta'} = \Inputs{\Theta}$ and $\max\Ext{\Theta}=\max\Ext{\Theta'}$; this is, for example, the case when $\Theta$ is a causal completion of $\Theta'$.
It is a fact (cf. Proposition 5.26 Section 5.4 of ``The Geometry of Causality'' \cite{gogioso2022geometry}) that standard empirical models for a space $\Theta$ can be canonically identified with standard empirical models for $\Theta'$.
It is therefore legitimate to ask whether a standard empirical $e$ for $\Theta'$ ``lifts'' to one for $\Theta$, i.e. whether it corresponds to a standard empirical model for $\Theta$ under the above identification.
\end{observation}

\begin{proposition}
\label{proposition:std-empmodel-local-par-seq-spaces-separability}
Let $\Theta$ be any space of input histories which can be obtained by repeated applications of parallel and (conditional) sequential composition, starting from indiscrete spaces (possibly on a single event).
Let $\Theta^{(1)},...,\Theta^{(n)}$ be causal completions of $\Theta$ (possibly not all of them).
Let $e$ be a local standard empirical model for $\Theta$.
If $e$ can be written as a convex combination $e = \sum_{j=1}^{n} p_j e^{(j)}$ of standard empirical models $e^{(1)},...,e^{(n)}$ for $\Theta$, where each $e^{(j)}$ lifts to an empirical model for the causally complete space $\Theta^{(j)}$, then $e$ is necessarily local in a separable way.
\end{proposition}
\begin{proof}
See \ref{proof:proposition:std-empmodel-local-par-seq-spaces-separability}
\end{proof}

It is an open question to which extent this result can be generalised beyond its current formulation, e.g. to empirical models on covers other than the standard cover, or to a broader class of spaces.
While pursuing such generalisations, there are some factors to consider:
\begin{itemize}
    \item The proof makes use of factorisation results for causal completions (cf. Theorems 3.26 and 3.27 p.38 of ``The Combinatorics of Causality'' \cite{gogioso2022combinatorics}) and the characterisation of causal switch spaces as causal completions for indiscrete spaces (cf. Theorem 3.36 p.51 of ``The Combinatorics of Causality'' \cite{gogioso2022combinatorics}), but a more general characterisation of causal completions would be required for a generalisation in this direction.
    \item The proof makes use of the no-go result for locality on causal switch spaces (cf. Theorem \ref{theorem:nogo-nonlocality-switch-spaces}), but there is an open question as to the extent to which this result can be generalised beyond the standard cover (cf. discussion after the result).
    \item The proof makes use of factorisation results for causal functions (cf. Theorems \ref{theorem:caus-fun-parallel-composition-factorisation}, \ref{theorem:caus-fun-sequential-composition-factorisation} and \ref{theorem:caus-fun-conditional-sequential-composition-factorisation}), and hence it only applies to non-contextual empirical models. Generalised factorisation results for empirical models would be required to strengthen the result from the qualitative statement of ``separable non-contextuality'' to quantitative bounds on the ``separable non-contextual fraction''.
\end{itemize}
That said, we already know that the result does not generalise to the case where $\Theta^{(1)},...,\Theta^{(n)}$ are not constrained to be causal completions of $\Theta$.
As a counterexample, we can consider the causal fork empirical model from p.\pageref{page:fork-empirical-model-descr}, discussed in further detail in the next Subsection.
The causal fork empirical model is a standard empirical model for the space $\Theta = \Hist{\ev{C} \seqcomposeSym (\ev{A}\vee\ev{B}), [0,1]}$, induced by the following fork causal order:
\begin{center}
\includegraphics[height=1.8cm]{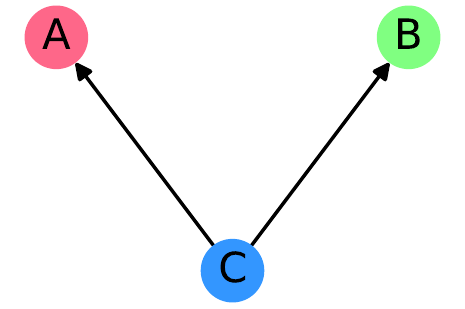}
\end{center}
The space $\Theta$ can be obtained by parallel and sequential composition from single-event (in)discrete spaces:
\[
\Theta = \Hist{\ev{C}, [0,1]} \seqcomposeSym (\Hist{\ev{A}, [0,1]} \cup \Hist{\ev{B}, [0,1]})
\]
Let $n=1$ and let $\Theta^{(1)} \leq \Theta$ be the following causally complete sub-space 
\begin{center}
    \includegraphics[height=2cm]{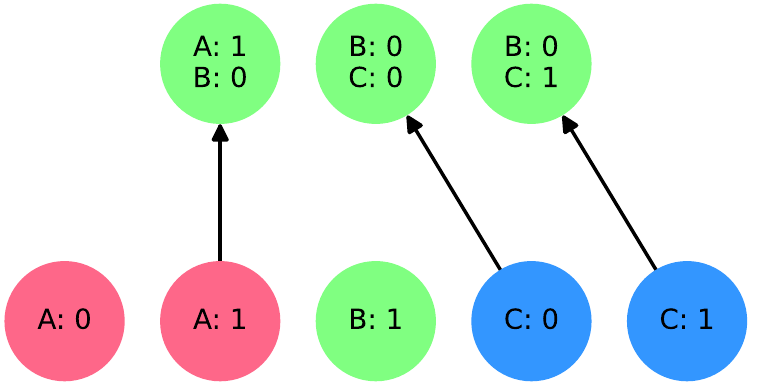}
\end{center}
The sub-space above is non-tight, it falls into equivalence class 2 from Figure~5 (p.42) of \cite{gogioso2022combinatorics} and it has the same causal functions as the discrete space in equivalence class 0.
It turns out that the causal fork empirical model lifts to an empirical model on $\Theta^{(1)}$, but also that it has a 0\% no-signalling fraction (cf. Subsubsection 2.5.1 of \cite{gogioso2022geometry}): as a consequence, it is (maximally) non-local on $\Theta^{(1)}$, providing our desired counterexample.

\subsection{Further Examples of Empirical Models}
\label{subsection:topology-causality-examples}

In this subsection, we present empirical models for a selection of examples of interest.
All empirical models are for the standard cover, so that any non-classicality arises from non-locality rather than other forms of contextuality (solipsistic or ordinary).

All models have binary inputs and outputs $I_\omega = O_\omega = \{0, 1\}$ at each event, unless otherwise specified.
For convenience, we will describe our scenarios in terms of agents performing operations at the events, always following the same convention: Alice acts at event \ev{A}, Bob acts at event \ev{B}, Charlie acts at event \ev{C}, Diane acts at event \ev{D}, Eve acts at event \ev{E} and Felix acts at event \ev{F}.

\subsubsection{A Causal Fork Empirical Model.}
\label{subsubsection:fork-empirical-model}

In this example, Charlie produces one of the four Bell basis states and forwards one qubit each to Alice and Bob, who measure it in either the Z or X basis:
\begin{enumerate}
    \item On input $c \in \{0,1\}$ Charlie prepares the 2-qubit state $|0c\rangle$. He then performs a XX parity measurement, resulting in one of $|\Phi^{\pm}\rangle$ states (if his input was 0) or one of $|\Psi^{\pm}\rangle$ states (if his input was 1), all with 50\% probability. He forwards this state to Alice and Bob, one qubit each.
    \item Alice and Bob perform a Z basis measurement on input 0 and an X basis measurement on input 1, and use the measurement outcome as their output.
\end{enumerate}
Below is the ``fork'' causal order which naturally supports this example, together with a figure summarising the experiment:
\begin{center}
\raisebox{-10mm}{
\includegraphics[height=3cm]{svg-inkscape/vee-C-AB-color_svg-tex.pdf}
}
\hspace{2cm}
\scalebox{1.25}{    
\begin{tikzpicture}
	\begin{pgfonlayer}{nodelayer}
		\node [style=none] (0) at (-1, -2) {};
		\node [style=none] (1) at (1, -2) {};
		\node [style=none] (3) at (-2.25, 3.25) {};
		\node [style=none] (5) at (-2.25, 3.25) {};
		\node [style=none] (6) at (-2.25, 1) {};
		\node [style=none] (10) at (-4.025, 4.5) {};
		\node [style=none] (11) at (-0.525, 4.5) {};
		\node [style=none] (12) at (-0.525, 1.75) {};
		\node [style=none] (13) at (-4.025, 1.75) {};
		\node [style=red label] (14) at (-4.75, 3.25) {$A$};
		\node [style=none] (16) at (2.225, 3.25) {};
		\node [style=none] (18) at (2.225, 3.25) {};
		\node [style=none] (19) at (2.225, 1) {};
		\node [style=none] (23) at (4, 4.5) {};
		\node [style=none] (24) at (0.5, 4.5) {};
		\node [style=none] (25) at (0.5, 1.75) {};
		\node [style=none] (26) at (4, 1.75) {};
		\node [style=red label] (27) at (4.725, 3.25) {$B$};
		\node [style=small box] (29) at (-2.25, 3.25) {$X/Z$};
		\node [style=small box] (30) at (2.25, 3.25) {$X/Z$};
		\node [style=none] (31) at (-2.5, 3) {};
		\node [style=none] (32) at (-2.25, 3.5) {};
		\node [style=none] (33) at (2.5, 3) {};
		\node [style=none] (34) at (2.25, 3.5) {};
		\node [style=none] (35) at (-2.25, 5.25) {};
		\node [style=none] (36) at (2.25, 5.25) {};
		\node [style=none] (37) at (-3.25, 1.25) {};
		\node [style=none] (38) at (3.25, 1.25) {};
		\node [style=none] (39) at (2.725, -1.5) {};
		\node [style=none] (40) at (-1.775, -1.5) {};
		\node [style=none] (41) at (-1.775, -4) {};
		\node [style=none] (42) at (2.725, -4) {};
		\node [style=small box] (43) at (1, -2.5) {$\Phi^{\pm}/\Psi^{\pm}$};
		\node [style=none] (44) at (1, -3) {};
		\node [style=none] (45) at (-1, -3) {};
		\node [style=none] (46) at (1.25, -2.75) {};
		\node [style=eslabel] (47) at (1.25, -4.75) {$\{ \Phi,\Psi \}$};
		\node [style=none] (48) at (0.75, -2.25) {};
		\node [style=none] (49) at (0, -0.75) {$$};
		\node [style=red label] (50) at (-2.5, -2.75) {$C$};
		\node [style=eslabel] (51) at (-2.25, 5.5) {$\{0,1\}$};
		\node [style=eslabel] (53) at (-3.25, 1) {$\{X,Z\}$};
		\node [style=eslabel] (54) at (3.25, 1) {$\{X,Z\}$};
		\node [style=eslabel] (55) at (2.225, 5.5) {$\{0,1\}$};
		\node [style=none] (56) at (1.25, -4.25) {};
		\node [style=eslabel] (57) at (0, -0.5) {$\{+,-\}$};
	\end{pgfonlayer}
	\begin{pgfonlayer}{edgelayer}
		\draw [style=carta da zucchero thin] (12.center)
			 to (13.center)
			 to (10.center)
			 to (11.center)
			 to cycle;
		\draw [style=carta da zucchero thin] (25.center)
			 to (26.center)
			 to (23.center)
			 to (24.center)
			 to cycle;
		\draw [style=classical, in=-90, out=90] (32.center) to (35.center);
		\draw [style=classical, in=90, out=-90] (31.center) to (37.center);
		\draw [style=classical, in=-90, out=90] (34.center) to (36.center);
		\draw [style=classical, in=90, out=-90] (33.center) to (38.center);
		\draw [style=carta da zucchero thin] (41.center)
			 to (42.center)
			 to (39.center)
			 to (40.center)
			 to cycle;
		\draw [style=bold black] (45.center) to (0.center);
		\draw (44.center) to (43);
		\draw (43) to (1.center);
		\draw [style=bold black, bend right=90, looseness=1.25] (45.center) to (44.center);
		\draw [style=classical, in=270, out=90] (48.center) to (49.center);
		\draw [style=bold black] (6.center) to (5.center);
		\draw [style=bold black] (19.center) to (18.center);
		\draw [style=bold black, in=270, out=90] (0.center) to (6.center);
		\draw [style=bold black, in=270, out=90] (1.center) to (19.center);
		\draw [style=classical] (56.center) to (46.center);
	\end{pgfonlayer}
\end{tikzpicture}
}
\end{center}
The description above results in the following empirical model on 3 events:
\begin{center}
\scalebox{0.9}{
\begin{tabular}{l|rrrrrrrr}
\hfill
ABC & 000 & 001 & 010 & 011 & 100 & 101 & 110 & 111\\
\hline
000 & $1/4$ & $1/4$ & $0$ & $0$ & $0$ & $0$ & $1/4$ & $1/4$\\
001 & $0$ & $0$ & $1/4$ & $1/4$ & $1/4$ & $1/4$ & $0$ & $0$\\
010 & $1/8$ & $1/8$ & $1/8$ & $1/8$ & $1/8$ & $1/8$ & $1/8$ & $1/8$\\
011 & $1/8$ & $1/8$ & $1/8$ & $1/8$ & $1/8$ & $1/8$ & $1/8$ & $1/8$\\
100 & $1/8$ & $1/8$ & $1/8$ & $1/8$ & $1/8$ & $1/8$ & $1/8$ & $1/8$\\
101 & $1/8$ & $1/8$ & $1/8$ & $1/8$ & $1/8$ & $1/8$ & $1/8$ & $1/8$\\
110 & $1/4$ & $0$ & $0$ & $1/4$ & $0$ & $1/4$ & $1/4$ & $0$\\
111 & $1/4$ & $0$ & $0$ & $1/4$ & $0$ & $1/4$ & $1/4$ & $0$\\
\end{tabular}
}
\end{center}
To better understand the process, we restrict our attention to the rows where Charlie has input 0, corresponding to Alice and Bob receiving the Bell basis states $|\Phi^\pm\rangle$:
\begin{center}
\scalebox{0.9}{
\begin{tabular}{l|rrrrrrrr}
\hfill
ABC & 000 & 001 & 010 & 011 & 100 & 101 & 110 & 111\\
\hline
000 & $1/4$ & $1/4$ & $0$ & $0$ & $0$ & $0$ & $1/4$ & $1/4$\\
010 & $1/8$ & $1/8$ & $1/8$ & $1/8$ & $1/8$ & $1/8$ & $1/8$ & $1/8$\\
100 & $1/8$ & $1/8$ & $1/8$ & $1/8$ & $1/8$ & $1/8$ & $1/8$ & $1/8$\\
110 & $1/4$ & $0$ & $0$ & $1/4$ & $0$ & $1/4$ & $1/4$ & $0$\\
\end{tabular}
}
\end{center}
When Charlie's output is 0 (left below) Alice and Bob receive the Bell basis state $|\Phi^+\rangle$: they get perfectly correlated outputs when they both measure in Z or both measure in X, and uncorrelated uniformly distributed outputs otherwise.
When Charlie's output is 1 (right below) Alice and Bob receive the Bell basis state $|\Phi^-\rangle$: they get perfectly correlated outputs when they both measure in Z, perfectly anti-correlated outputs when they both measure in X, and uncorrelated uniformly distributed outputs otherwise. 
\begin{center}
\scalebox{0.9}{
\begin{tabular}{l|rrrr}
\hfill
ABC & 000 & 010 & 100 & 110\\
\hline
000 & $1/4$ & $0$ & $0$ & $1/4$\\
010 & $1/8$ & $1/8$ & $1/8$ & $1/8$\\
100 & $1/8$ & $1/8$ & $1/8$ & $1/8$\\
110 & $1/4$ & $0$ & $0$ & $1/4$\\
\end{tabular}
}
\hspace{2cm}
\scalebox{0.9}{
\begin{tabular}{l|rrrr}
\hfill
ABC & 001 & 011 & 101 & 111\\
\hline
000 & $1/4$ & $0$ & $0$ & $1/4$\\
010 & $1/8$ & $1/8$ & $1/8$ & $1/8$\\
100 & $1/8$ & $1/8$ & $1/8$ & $1/8$\\
110 & $0$ & $1/4$ & $1/4$ & $0$\\
\end{tabular}
}
\end{center}
The description for the associated empirical model was already worked out on p.\pageref{page:fork-empirical-model-descr} as an early example, but it is repeated here for completeness.
In tabular form, each row $i_\ev{A}i_\ev{B}i_\ev{C}$ corresponds to a maximal extended input history $\hist{A/i_\ev{A}, B/i_\ev{B}, C/i_\ev{C}} \in \StdCov{\Theta}$, while each column $o_\ev{A}o_\ev{B}o_\ev{C}$ corresponds to an associated extended output history $\hist{A/o_\ev{A}, B/o_\ev{B}, C/o_\ev{C}}$.
Formally, however, each line $i_\ev{A}i_\ev{B}i_\ev{C}$ of the same empirical model has to be defined explicitly as a distribution on extended causal functions:
\[
\Dist{\ExtCausFun{\downset{\hist{A/i_\ev{A}, B/i_\ev{B}, C/i_\ev{C}}}, \{0,1\}}}
\]
Equivalently, we look at distributions on causal functions, which are freely characterised:
\[
\Dist{\CausFun{\downset{\hist{A/i_\ev{A}, B/i_\ev{B}, C/i_\ev{C}}}, \{0,1\}}}
\]
Because the space $\Theta$ is both tight and causally complete, the causal functions on the downset $\downset{\hist{A/i_\ev{A}, B/i_\ev{B}, C/i_\ev{C}}}$ take the following form:
\[
    \prod_{h \leq \hist{A/i_\ev{A}, B/i_\ev{B}, C/i_\ev{C}}}
    O_{\tip{\Theta}{h}}
\]
where we used the fact that $\tip{\downset{\hist{A/i_\ev{A}, B/i_\ev{B}, C/i_\ev{C}}}}{h}=\tip{\Theta}{h}$ to simplify the expression.
The input histories $h \leq \hist{A/i_\ev{A}, B/i_\ev{B}, C/i_\ev{C}}$ are exactly:
\begin{itemize}
    \item $h = \hist{C/i_\ev{C}}$ with tip event \ev{C}
    \item $h = \hist{C/i_\ev{C}, A/i_\ev{A}}$ with tip event \ev{A}
    \item $h = \hist{C/i_\ev{C}, B/i_\ev{B}}$ with tip event \ev{B}
\end{itemize}
Hence the causal functions in $\CausFun{\downset{\hist{A/i_\ev{A}, B/i_\ev{B}, C/i_\ev{C}}}, \{0,1\}}$ are:
\[
f_{o_\ev{A}o_\ev{B}o_\ev{C}|i_\ev{A}i_\ev{B}i_\ev{C}}:=
\left\{
\begin{array}{rl}
    \hist{C/i_\ev{C}}       &\mapsto o_\ev{C}\\
    \hist{C/i_\ev{C},A/i_\ev{A}} &\mapsto o_\ev{A}\\
    \hist{C/i_\ev{C},B/i_\ev{B}} &\mapsto o_\ev{B}
\end{array}
\right.
\]
Using these functions, we can reconstruct the desired distribution for each row of the empirical model above.
For example, the second row is indexed by the maximal extended input history $\hist{A/0, B/0, C/1}$ and it corresponds to the following distribution on extended causal functions:
\[
\frac{1}{4}\delta_{\Ext{f_{010|001}}}
+\frac{1}{4}\delta_{\Ext{f_{011|001}}}
+\frac{1}{4}\delta_{\Ext{f_{100|001}}}
+\frac{1}{4}\delta_{\Ext{f_{101|001}}}
\]
Doing this for all rows yields the following standard empirical model:
\[
e_{\downset{\hist{A/i_\ev{A}, B/i_\ev{B}, C/i_\ev{C}}}}
:=
\left\{
\begin{array}{rl}
\frac{1}{4}\sum\limits_{o_\ev{A}\oplus o_\ev{B} = i_\ev{C}}\sum\limits_{o_\ev{C}} \delta_{\Ext{f_{o_\ev{A}o_\ev{B}o_\ev{C}|i_\ev{A}i_\ev{B}i_\ev{C}}}}
&\text{ if } i_\ev{A} = i_\ev{B}\\
\frac{1}{8}\sum\limits_{o_\ev{A}}\sum\limits_{o_\ev{B}}\sum\limits_{o_\ev{C}} \delta_{\Ext{f_{o_\ev{A}o_\ev{B}o_\ev{C}|i_\ev{A}i_\ev{B}i_\ev{C}}}}
&\text{ if } i_\ev{A} \neq i_\ev{B}
\end{array}
\right.
\]

Rather interestingly, the empirical model for this experiment lifts to an empirical model on two unrelated sub-spaces of the causal fork space, both in equivalence class 33 from Figure~5 (p.42) of \cite{gogioso2022combinatorics}: the space $\Theta_{\ev{A}\vee (\ev{C}\seqcomposeSym \ev{B})}$ induced by causal order $\ev{A}\vee (\ev{C}\seqcomposeSym \ev{B})$ (left below) an the space $\Theta_{\ev{B}\vee (\ev{C}\seqcomposeSym \ev{A})}$ induced by causal order $\ev{B}\vee (\ev{C}\seqcomposeSym \ev{A})$. In other words, the empirical data is compatible both with absence of signalling from \ev{C} to \ev{A} (left below) and with absence of signalling from \ev{C} to \ev{B} (right below).
\begin{center}
    \begin{tabular}{ccc}
    \includegraphics[height=3cm]{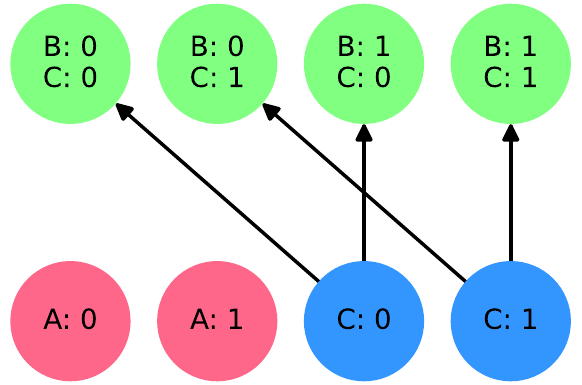}
    & \hspace{2cm} &
    \includegraphics[height=3cm]{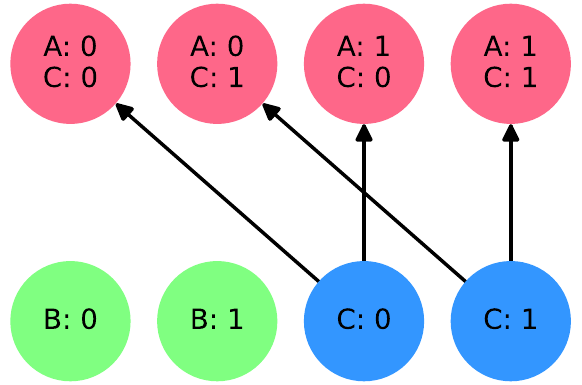}
    \end{tabular}
\end{center}
For example, below is a decomposition of the empirical model as a uniform mixture of 8 causal functions for $\Theta_{\ev{B}\vee (\ev{C}\seqcomposeSym \ev{A})}$.
The black dots are classical copies, the $\oplus$ dots are classical XORs and the $\wedge$ dots classical ANDs. 
\[
    \frac{1}{8} \sum_{(x,y,z) \in \{0,1\}^3}
    \scalebox{0.5}{
        \begin{tikzpicture}[scale=2.7]
	\begin{pgfonlayer}{nodelayer}
		\node [style=black dot] (0) at (-2, -3.75) {};
		\node [style=none] (3) at (-1.25, -3) {};
		\node [style=none] (4) at (-1.25, -1) {};
		\node [style=none] (5) at (0.75, -1) {};
		\node [style=none] (6) at (0.75, -3) {};
		\node [style=none] (7) at (1.5, 2.5) {};
		\node [style=none] (8) at (1.5, 5) {};
		\node [style=none] (9) at (4, 5) {};
		\node [style=none] (10) at (4, 2.5) {};
		\node [style=none] (11) at (-5, 2.5) {};
		\node [style=none] (12) at (-5, 5.75) {};
		\node [style=none] (13) at (-1.5, 5.75) {};
		\node [style=none] (14) at (-1.5, 2.5) {};
		\node [style=none] (15) at (-0.5, -3) {};
		\node [style=none] (16) at (-4.25, 2.5) {};
		\node [style=point] (17) at (-2, -4.75) {$z$};
		\node [style=none] (18) at (0, -1) {};
		\node [style=none] (19) at (0, -3) {};
		\node [style=none] (20) at (-0.5, -1) {};
		\node [style=none] (21) at (-2.5, 2.5) {};
		\node [style=none] (22) at (-3.5, 2.5) {};
		\node [style=black dot] (23) at (-2.25, 0.5) {};
		\node [style=none] (24) at (1.75, 2.5) {};
		\node [style=none] (25) at (2, 2.5) {};
		\node [style=point] (26) at (-2.25, -0.5) {$x$};
		\node [style=black dot] (27) at (0.5, 0.5) {};
		\node [style=point] (28) at (0.5, -0.5) {$y$};
		\node [style=none] (29) at (-2, 2.5) {};
		\node [style=none] (30) at (0, -5) {};
		\node [style=none] (32) at (0, 7.25) {};
		\node [style=white dot] (33) at (2.75, 3.25) {$\wedge$};
		\node [style=none] (34) at (2.75, 4.25) {$\oplus$};
		\node [style=none] (35) at (3.5, 2.5) {};
		\node [style=none] (36) at (3.5, -5) {};
		\node [style=none] (37) at (2.75, 5) {};
		\node [style=none] (38) at (2.75, 7.25) {};
		\node [style=none] (39) at (-3.25, 7.25) {};
		\node [style=none] (40) at (-4.5, 2.5) {};
		\node [style=none] (41) at (-4.5, -5) {};
		\node [style=white dot] (42) at (-4.5, 4.5) {$\wedge$};
		\node [style=black dot] (45) at (-2.5, 3) {};
		\node [style=none] (46) at (-3.25, 5.75) {};
		\node [style=none] (47) at (-3.25, 5.25) {$\oplus$};
		\node [style=none] (48) at (-3.5, 3.5) {$\oplus$};
		\node [style=label] (49) at (-2, -1.75) {$C$};
		\node [style=label] (50) at (-5.75, 4) {$A$};
		\node [style=label] (51) at (0.75, 3.5) {$B$};
	\end{pgfonlayer}
	\begin{pgfonlayer}{edgelayer}
		\draw [style=carta da zucchero thin] (5.center)
			 to (4.center)
			 to (3.center)
			 to (6.center)
			 to cycle;
		\draw [style=carta da zucchero thin] (9.center)
			 to (8.center)
			 to (7.center)
			 to (10.center)
			 to cycle;
		\draw [style=carta da zucchero thin] (13.center)
			 to (12.center)
			 to (11.center)
			 to (14.center)
			 to cycle;
		\draw [style=bold black, in=-90, out=0, looseness=1.25] (0) to (15.center);
		\draw [style=bold black, in=-90, out=180, looseness=0.75] (0) to (16.center);
		\draw [style=bold black] (17) to (0);
		\draw [style=bold black, in=-90, out=90] (15.center) to (18.center);
		\draw [style=bold black, in=-90, out=90] (19.center) to (20.center);
		\draw [style=bold black, in=-90, out=90, looseness=0.50] (20.center) to (21.center);
		\draw [style=bold black, in=-90, out=180] (23) to (22.center);
		\draw [style=bold black, in=-90, out=0, looseness=0.75] (23) to (25.center);
		\draw [style=bold black] (23) to (26);
		\draw [style=bold black] (28) to (27);
		\draw [style=bold black, in=180, out=-90] (29.center) to (27);
		\draw [style=bold black, in=-90, out=0] (27) to (24.center);
		\draw [style=bold black] (32.center) to (18.center);
		\draw [style=bold black] (30.center) to (19.center);
		\draw [style=bold black] (36.center) to (35.center);
		\draw [style=bold black, bend left=45] (25.center) to (33);
		\draw [style=bold black, bend left=45] (33) to (35.center);
		\draw [style=bold black] (33) to (34.center);
		\draw [style=bold black, in=180, out=90, looseness=1.50] (24.center) to (34.center);
		\draw [style=bold black] (34.center) to (37.center);
		\draw [style=bold black] (37.center) to (38.center);
		\draw [style=bold black] (41.center) to (40.center);
		\draw [style=bold black] (40.center) to (42);
		\draw [style=bold black] (21.center) to (45);
		\draw [style=bold black] (46.center) to (39.center);
		\draw [style=bold black, in=-90, out=15] (45) to (47.center);
		\draw [style=bold black, in=180, out=90] (16.center) to (48.center);
		\draw [style=bold black] (22.center) to (48.center);
		\draw [style=bold black, in=90, out=0] (42) to (48.center);
		\draw [style=bold black, in=0, out=-180, looseness=1.25] (45) to (48.center);
		\draw [style=bold black, in=-180, out=90] (42) to (47.center);
		\draw [style=bold black, in=90, out=0, looseness=1.25] (47.center) to (29.center);
		\draw [style=bold black] (46.center) to (47.center);
	\end{pgfonlayer}
\end{tikzpicture}
    }
\]
What makes this empirical model even more interesting is that its no-signalling fraction is 0\% (cf. Subsubsection 2.5.1 of \cite{gogioso2022geometry}): no part of it can be explained without signalling from \ev{C} to at least one of \ev{A} or \ev{B}.
\begin{center}
    \begin{tabular}{c}
    \includegraphics[height=3cm]{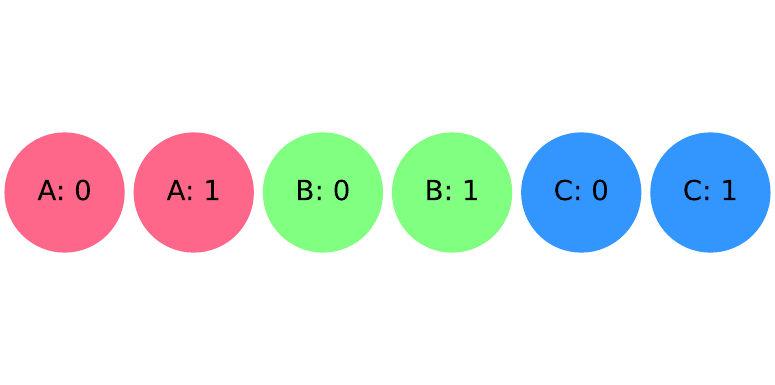}
    \end{tabular}
\end{center}
It is, however, the case (cf. Subsubsection 2.5.1 of \cite{gogioso2022geometry}) that the empirical model lifts to an empirical model for the following two unrelated non-tight subspaces of $\Theta_{\ev{A}\vee (\ev{C}\seqcomposeSym \ev{B})}$ and $\Theta_{\ev{B}\vee (\ev{C}\seqcomposeSym \ev{A})}$ respectively, both falling into equivalence class 2 from from Figure~5 (p.42) of \cite{gogioso2022combinatorics}:
\begin{center}
    \begin{tabular}{ccc}
    \includegraphics[height=3cm]{svg-inkscape/example-2-space-4_svg-tex.pdf}
    & \hspace{2cm} &
    \includegraphics[height=3cm]{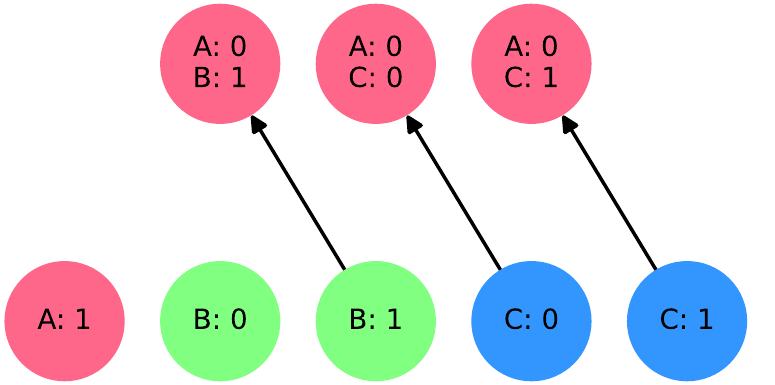}
    \end{tabular}
\end{center}
From Figure~5 (p.42) of \cite{gogioso2022combinatorics}, we see that spaces in equivalence class 2 have exactly the same causal functions as the discrete space, in equivalence class 0, which is also their meet.
Since the empirical model has a no-signalling fraction of 0\%, it immediately follows that it is maximally non-local there for both spaces above. To recap, this example bears three separate gifts:
\begin{itemize}
    \item It shows that there are empirical models can be fully supported by multiple spaces but 0\% supported by their meet.
    \item It provides an empirical model whose minimally supporting spaces are non-tight, providing additional evidence for the importance of non-tight spaces in the study of non-locality and causality.
    \item It shows that the notions of non-locality and contextuality depend on a specific choice of causal constraints, by providing an empirical model which is local in for a space and maximally non-local for a sub-space, while being fully supported by both.
\end{itemize}

\subsubsection{A Classical Switch Empirical Model.}

In this example, Alice classically controls the order of Bob and Charlie, as follows:
\begin{itemize}
    \item Alice flips one of two biased coins, depending on her input: when her input is 0, her output is 75\% 0 and 25\% 1; when her input is 1, her output is 25\% 0 and 75\% 1 instead.
    \item Bob and Charlie are in a quantum switch, controlled in the Z basis and with $|0\rangle$ as a fixed input: on output $a \in \{0, 1\}$, Alice feeds state $|a\rangle$ into the control system of the switch, determining the relative causal order of Bob and Charlie.
    \item Bob and Charlie both apply the same quantum instrument: they measure the incoming qubit they receive in the Z basis, obtaining their output, and then encode their input into the Z basis of the outgoing qubit.
    \item Both the control qubit and the outgoing qubit of the switch are discarded: even without Alice controlling the switch in the Z basis, discarding the control qubit would be enough to make the control classical.
\end{itemize}
The description above results in the following empirical model on 3 events:
\begin{center}
\scalebox{0.9}{
\begin{tabular}{l|rrrrrrrr}
\hfill
ABC & 000 & 001 & 010 & 011 & 100 & 101 & 110 & 111\\
\hline
000 & $3/4$ & $0$ & $0$ & $0$ & $1/4$ & $0$ & $0$ & $0$\\
001 & $3/4$ & $0$ & $0$ & $0$ & $0$ & $0$ & $1/4$ & $0$\\
010 & $0$ & $3/4$ & $0$ & $0$ & $1/4$ & $0$ & $0$ & $0$\\
011 & $0$ & $3/4$ & $0$ & $0$ & $0$ & $0$ & $1/4$ & $0$\\
100 & $1/4$ & $0$ & $0$ & $0$ & $3/4$ & $0$ & $0$ & $0$\\
101 & $1/4$ & $0$ & $0$ & $0$ & $0$ & $0$ & $3/4$ & $0$\\
110 & $0$ & $1/4$ & $0$ & $0$ & $3/4$ & $0$ & $0$ & $0$\\
111 & $0$ & $1/4$ & $0$ & $0$ & $0$ & $0$ & $3/4$ & $0$\\
\end{tabular}
}
\end{center}
Below is the indefinite causal order which naturally supports this example:
\begin{center}
\includegraphics[height=3cm]{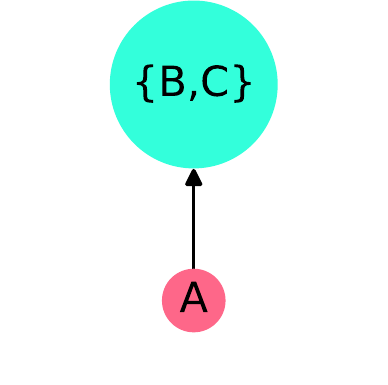}
\end{center}
To better understand the table above, we focus on the second row, corresponding to input 001:
\begin{enumerate}
    \item Alice's input is 0, so her output is 75\% 0 and 25\% 1. This means that the probabilities of outputs $0\_\_$ in row 001 of the empirical model must sum to 75\%, and the probabilities of output $1\_\_$ must sum to 25\%.
    \item Conditional to Alice's output being 0, the output is $000$ with 100\% probability:
    \begin{enumerate}
        \item Bob goes first and receives the input state $|0\rangle$ for the switch: he measures the state in the Z basis, obtaining output 0 with 100\% probability.
        Because his input is 0, he then prepares the state $|0\rangle$, which he forwards into the switch.
        \item Charlie goes second and receives the state $|0\rangle$ prepared by Bob: he measures the state in the Z basis, obtaining output 0 with 100\% probability.
        Because his input is 1, he then prepares the state $|1\rangle$, which he forwards into the switch.
        \item Charlie's state $|1\rangle$ comes out of the switch, and is discarded.
    \end{enumerate}
    \item Conditional to Alice's output being 1, the output is $110$ with 100\% probability:
    \begin{enumerate}
        \item Charlie goes first and receives the input state $|0\rangle$ for the switch: he measures the state in the Z basis, obtaining output 0 with 100\% probability.
        Because his input is 1, he then prepares the state $|1\rangle$, which he forwards into the switch.
        \item Bob goes second and receives the state $|1\rangle$ prepared by Charlie: he measures the state in the Z basis, obtaining output 1 with 100\% probability.
        Because his input is 0, he then prepares the state $|0\rangle$, which he forwards into the switch.
        \item Bob's state $|0\rangle$ comes out of the switch, and is discarded.
    \end{enumerate}
\end{enumerate}
The space $\Theta$ is tight but not causally complete, with the following input histories:
\begin{itemize}
    \item $h = \hist{A/i_\ev{A}}$ with tip event \ev{A}
    \item $h = \hist{A/i_\ev{A}, B/i_\ev{B}, C/i_\ev{C}}$ with tip events \evset{A,B}
\end{itemize}
The causal functions in $\CausFun{\downset{\hist{A/i_\ev{A}, B/i_\ev{B}, C/i_\ev{C}}}, \{0,1\}}$ are:
\[
f_{o_\ev{A}o_\ev{B}o_\ev{C}|i_\ev{A}i_\ev{B}i_\ev{C}}:=
\left\{
\begin{array}{rl}
    \hist{A/i_\ev{A}}       &\mapsto o_{\ev{A}}\\
    \hist{A/i_\ev{A}, B/i_\ev{B}, C/i_\ev{C}} &\mapsto \left(o_\omega\right)_{\omega \in \evset{B,C}}\\
\end{array}
\right.
\]
Using these functions, we can reconstruct the desired distribution for each row of the empirical model above.
For example, the second row is indexed by the maximal extended input history $\hist{A/0, B/0, C/1}$ and it corresponds to the following distribution on extended causal functions:
\[
\frac{3}{4}\delta_{\Ext{f_{000|001}}}
+\frac{1}{4}\delta_{\Ext{f_{110|001}}}
\]
Doing this for all rows yields the following standard empirical model:
\[
e_{\downset{\hist{A/i_\ev{A}, B/i_\ev{B}, C/i_\ev{C}}}}
:=
\left\{
\begin{array}{rl}
\frac{3}{4}\delta_{\Ext{f_{00i_\ev{B}|i_\ev{A}i_\ev{B}i_\ev{C}}}}
+\frac{1}{4}\delta_{\Ext{f_{1i_\ev{C}0|i_\ev{A}i_\ev{B}i_\ev{C}}}}
&\text{ if } i_\ev{A} = 0
\\
\frac{1}{4}\delta_{\Ext{f_{00i_\ev{B}|i_\ev{A}i_\ev{B}i_\ev{C}}}}
+\frac{3}{4}\delta_{\Ext{f_{1i_\ev{C}0|i_\ev{A}i_\ev{B}i_\ev{C}}}}
&\text{ if } i_\ev{A} = 1
\end{array}
\right.
\]

\subsubsection{A Causal Cross Empirical Model.}

In this example, Charlie receives qubits from Alice and Bob and forwards them to Diane and Eve, choosing whether to forward the qubits as $A\rightarrow D, B\rightarrow E$ or as $A\rightarrow E, B\rightarrow D$.
Below is the ``cross'' causal order which naturally supports this example:
\begin{center}
\includegraphics[height=3cm]{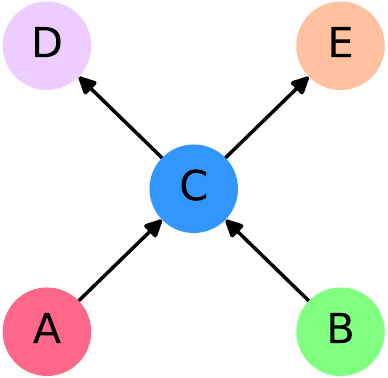}
\end{center}
More specifically, the parties act as follows:
\begin{enumerate}
    \item Alice and Bob encode their input into the Z basis of one qubit each, which they then forward to Charlie. Their output is trivial, constantly set to 0.
    \item Charlie receives the two qubits from Alice and Bob, measures the ZZ parity and uses the outcome of the parity measurement as his output. He decides how to forward the two qubits based on his input:
    \begin{itemize}
         \item On input 0, Charlie forwards Alice's qubit to Diane and Bob's qubit to Eve.
         \item On input 1, Charlie forwards Alice's qubit to Eve and Bob's qubit to Diane.
     \end{itemize} 
    \item Diane and Eve have trivial input, with only 0 as an option. They measure the qubit they receive in the Z basis and use the outcome as their output.
\end{enumerate}
The following figure summarises the experiment. The $P$ box is a ZZ parity measurement:
\begin{center}
    \scalebox{1.2}{
        \begin{tikzpicture}
	\begin{pgfonlayer}{nodelayer}
		\node [style=none] (0) at (-3.5, 5.5) {};
		\node [style=white dot] (3) at (-2.5, 3.75) {};
		\node [style=none] (7) at (-3.5, 6.5) {};
		\node [style=none] (8) at (3.475, 5.5) {};
		\node [style=white dot] (11) at (2.475, 3.75) {};
		\node [style=none] (15) at (3.475, 6.5) {};
		\node [style=none] (20) at (-3.5, -5.5) {};
		\node [style=white dot] (23) at (-2.5, -3.75) {};
		\node [style=none] (27) at (-3.5, -6.5) {};
		\node [style=none] (28) at (3.475, -5.5) {};
		\node [style=white dot] (31) at (2.475, -3.75) {};
		\node [style=none] (35) at (3.475, -6.5) {};
		\node [style=none] (39) at (0, -6.5) {};
		\node [style=none] (41) at (-0.25, 3) {};
		\node [style=none] (46) at (1.975, 6) {};
		\node [style=none] (47) at (3.975, 6) {};
		\node [style=none] (48) at (3.975, 3.25) {};
		\node [style=none] (49) at (1.975, 3.25) {};
		\node [style=none] (50) at (1.975, -3.25) {};
		\node [style=none] (51) at (3.975, -3.25) {};
		\node [style=none] (52) at (3.975, -6) {};
		\node [style=none] (53) at (1.975, -6) {};
		\node [style=none] (54) at (-4.025, 6) {};
		\node [style=none] (55) at (-2.025, 6) {};
		\node [style=none] (56) at (-2.025, 3.25) {};
		\node [style=none] (57) at (-4.025, 3.25) {};
		\node [style=none] (58) at (-4.025, -3.25) {};
		\node [style=none] (59) at (-2.025, -3.25) {};
		\node [style=none] (60) at (-2.025, -6) {};
		\node [style=none] (61) at (-4.025, -6) {};
		\node [style=red label] (62) at (4.75, 4.75) {$E$};
		\node [style=red label] (64) at (-4.75, -4.5) {$A$};
		\node [style=red label] (65) at (4.75, -4.5) {$B$};
		\node [style=red label] (66) at (-4.75, 4.75) {$D$};
		\node [style=none] (67) at (-2.5, 2.5) {};
		\node [style=none] (68) at (2.475, 2.5) {};
		\node [style=none] (69) at (-1, -1) {};
		\node [style=none] (70) at (-0.5, 1) {};
		\node [style=none] (71) at (-1, 1) {};
		\node [style=none] (72) at (-0.5, -1) {};
		\node [style=none] (73) at (-2.5, -2.5) {};
		\node [style=none] (74) at (2.475, -2.5) {};
		\node [style=none] (75) at (-0.5, -1) {};
		\node [style=none] (76) at (-1, -1) {};
		\node [style=copoint] (77) at (0.75, -1) {$0$};
		\node [style=none] (78) at (-1.75, 1.75) {};
		\node [style=none] (79) at (1.75, 1.75) {};
		\node [style=none] (80) at (1.75, -1.75) {};
		\node [style=none] (81) at (-1.75, -1.75) {};
		\node [style=red label] (82) at (2.5, 0) {$C$};
		\node [style=none] (83) at (-1, 0) {};
		\node [style=none] (84) at (-0.5, 0) {};
		\node [style=box] (85) at (-0.75, -0.5) {$P$};
		\node [style=none] (86) at (-0.25, 0) {};
	\end{pgfonlayer}
	\begin{pgfonlayer}{edgelayer}
		\draw [style=classical, in=-90, out=90] (3) to (0.center);
		\draw [style=classical] (0.center) to (7.center);
		\draw [style=classical, in=-90, out=90] (11) to (8.center);
		\draw [style=classical] (8.center) to (15.center);
		\draw [style=classical, in=90, out=-90] (23) to (20.center);
		\draw [style=classical] (20.center) to (27.center);
		\draw [style=classical, in=90, out=-90] (31) to (28.center);
		\draw [style=classical] (28.center) to (35.center);
		\draw [style=carta da zucchero thin] (46.center)
			 to (47.center)
			 to (48.center)
			 to (49.center)
			 to cycle;
		\draw [style=carta da zucchero thin] (53.center)
			 to (50.center)
			 to (51.center)
			 to (52.center)
			 to cycle;
		\draw [style=carta da zucchero thin] (54.center)
			 to (55.center)
			 to (56.center)
			 to (57.center)
			 to cycle;
		\draw [style=carta da zucchero thin] (61.center)
			 to (58.center)
			 to (59.center)
			 to (60.center)
			 to cycle;
		\draw [style=carta da zucchero thin] (78.center)
			 to (79.center)
			 to (80.center)
			 to (81.center)
			 to cycle;
		\draw [style=bold black, in=90, out=-90] (67.center) to (71.center);
		\draw [style=bold black, in=-90, out=90] (70.center) to (68.center);
		\draw [style=bold black, in=-90, out=90] (73.center) to (76.center);
		\draw [style=bold black, in=90, out=-90] (75.center) to (74.center);
		\draw [style=bold black] (76.center) to (83.center);
		\draw [style=bold black] (75.center) to (84.center);
		\draw [style=bold black] (68.center) to (11);
		\draw [style=bold black] (67.center) to (3);
		\draw [style=bold black] (23) to (73.center);
		\draw [style=bold black] (31) to (74.center);
		\draw [style=bold black] (83.center) to (71.center);
		\draw [style=bold black] (84.center) to (70.center);
		\draw [style=classical, in=-90, out=90] (39.center) to (77);
		\draw [style=classical] (86.center) to (41.center);
	\end{pgfonlayer}
\end{tikzpicture}
    }
    \hspace{0.6cm}
    \scalebox{1.2}{
        \begin{tikzpicture}
	\begin{pgfonlayer}{nodelayer}
		\node [style=none] (0) at (-3.5, 5.5) {};
		\node [style=white dot] (3) at (-2.5, 3.75) {};
		\node [style=none] (4) at (-2.5, 2.5) {};
		\node [style=none] (7) at (-3.5, 6.5) {};
		\node [style=none] (8) at (3.475, 5.5) {};
		\node [style=none] (10) at (3.475, 3.75) {};
		\node [style=white dot] (11) at (2.475, 3.75) {};
		\node [style=none] (12) at (2.475, 2.5) {};
		\node [style=none] (15) at (3.475, 6.5) {};
		\node [style=none] (16) at (-1, -1) {};
		\node [style=none] (17) at (-0.5, 1) {};
		\node [style=none] (18) at (-1, 1) {};
		\node [style=none] (19) at (-0.5, -1) {};
		\node [style=none] (20) at (-3.5, -5.5) {};
		\node [style=white dot] (23) at (-2.5, -3.75) {};
		\node [style=none] (24) at (-2.5, -2.5) {};
		\node [style=none] (27) at (-3.5, -6.5) {};
		\node [style=none] (28) at (3.475, -5.5) {};
		\node [style=white dot] (31) at (2.475, -3.75) {};
		\node [style=none] (32) at (2.475, -2.5) {};
		\node [style=none] (35) at (3.475, -6.5) {};
		\node [style=none] (36) at (-0.5, -1) {};
		\node [style=none] (37) at (-1, -1) {};
		\node [style=copoint] (38) at (0.75, -1) {$1$};
		\node [style=none] (39) at (0, -6.5) {};
		\node [style=none] (41) at (-0.25, 3) {};
		\node [style=none] (42) at (-1.75, 1.75) {};
		\node [style=none] (43) at (1.75, 1.75) {};
		\node [style=none] (44) at (1.75, -1.75) {};
		\node [style=none] (45) at (-1.75, -1.75) {};
		\node [style=none] (46) at (1.975, 6) {};
		\node [style=none] (47) at (3.975, 6) {};
		\node [style=none] (48) at (3.975, 3.25) {};
		\node [style=none] (49) at (1.975, 3.25) {};
		\node [style=none] (50) at (1.975, -3.25) {};
		\node [style=none] (51) at (3.975, -3.25) {};
		\node [style=none] (52) at (3.975, -6) {};
		\node [style=none] (53) at (1.975, -6) {};
		\node [style=none] (54) at (-4.025, 6) {};
		\node [style=none] (55) at (-2.025, 6) {};
		\node [style=none] (56) at (-2.025, 3.25) {};
		\node [style=none] (57) at (-4.025, 3.25) {};
		\node [style=none] (58) at (-4.025, -3.25) {};
		\node [style=none] (59) at (-2.025, -3.25) {};
		\node [style=none] (60) at (-2.025, -6) {};
		\node [style=none] (61) at (-4.025, -6) {};
		\node [style=red label] (62) at (4.75, 4.75) {$E$};
		\node [style=red label] (63) at (2.5, 0) {$C$};
		\node [style=red label] (64) at (-4.75, -4.5) {$A$};
		\node [style=red label] (65) at (4.75, -4.5) {$B$};
		\node [style=red label] (66) at (-4.75, 4.75) {$D$};
		\node [style=none] (67) at (-1, 0) {};
		\node [style=none] (68) at (-0.5, 0) {};
		\node [style=box] (69) at (-0.75, -0.5) {$P$};
		\node [style=none] (71) at (-0.25, 0) {};
	\end{pgfonlayer}
	\begin{pgfonlayer}{edgelayer}
		\draw [style=classical, in=-90, out=90] (3) to (0.center);
		\draw [style=bold black] (4.center) to (3);
		\draw [style=classical] (0.center) to (7.center);
		\draw [style=classical, in=-90, out=90] (11) to (8.center);
		\draw [style=bold black] (12.center) to (11);
		\draw [style=classical] (8.center) to (15.center);
		\draw [style=classical, in=90, out=-90] (23) to (20.center);
		\draw [style=bold black] (24.center) to (23);
		\draw [style=classical] (20.center) to (27.center);
		\draw [style=classical, in=90, out=-90] (31) to (28.center);
		\draw [style=bold black] (32.center) to (31);
		\draw [style=classical] (28.center) to (35.center);
		\draw [style=classical, in=90, out=-90] (38) to (39.center);
		\draw [style=carta da zucchero thin] (42.center)
			 to (43.center)
			 to (44.center)
			 to (45.center)
			 to cycle;
		\draw [style=carta da zucchero thin] (46.center)
			 to (47.center)
			 to (48.center)
			 to (49.center)
			 to cycle;
		\draw [style=carta da zucchero thin] (53.center)
			 to (50.center)
			 to (51.center)
			 to (52.center)
			 to cycle;
		\draw [style=carta da zucchero thin] (57.center)
			 to (54.center)
			 to (55.center)
			 to (56.center)
			 to cycle;
		\draw [style=carta da zucchero thin] (61.center)
			 to (58.center)
			 to (59.center)
			 to (60.center)
			 to cycle;
		\draw [style=bold black, in=90, out=-90] (4.center) to (18.center);
		\draw [style=bold black, in=-90, out=90] (17.center) to (12.center);
		\draw [style=bold black, in=-90, out=90] (24.center) to (37.center);
		\draw [style=bold black, in=90, out=-90] (36.center) to (32.center);
		\draw [style=bold black, in=-90, out=90] (67.center) to (17.center);
		\draw [style=bold black, in=-90, out=90] (68.center) to (18.center);
		\draw [style=bold black] (37.center) to (67.center);
		\draw [style=bold black] (36.center) to (68.center);
		\draw [style=classical] (71.center) to (41.center);
	\end{pgfonlayer}
\end{tikzpicture}
    }
\end{center}
The description above results in the following empirical model on 5 events; note that the outputs of Alice and Bob, as well as the inputs of Diane and Eve, are fixed to 0. 
\begin{center}
\scalebox{0.9}{
\begin{tabular}{l|rrrrrrrr}
\hfill
ABCDE & 00000 & 00001 & 00010 & 00011 & 00100 & 00101 & 00110 & 00111\\
\hline
00000 & $1$ & $0$ & $0$ & $0$ & $0$ & $0$ & $0$ & $0$\\
00100 & $1$ & $0$ & $0$ & $0$ & $0$ & $0$ & $0$ & $0$\\
01000 & $0$ & $0$ & $0$ & $0$ & $0$ & $1$ & $0$ & $0$\\
01100 & $0$ & $0$ & $0$ & $0$ & $0$ & $0$ & $1$ & $0$\\
10000 & $0$ & $0$ & $0$ & $0$ & $0$ & $0$ & $1$ & $0$\\
10100 & $0$ & $0$ & $0$ & $0$ & $0$ & $1$ & $0$ & $0$\\
11000 & $0$ & $0$ & $0$ & $1$ & $0$ & $0$ & $0$ & $0$\\
11100 & $0$ & $0$ & $0$ & $1$ & $0$ & $0$ & $0$ & $0$\\
\end{tabular}
}
\end{center}
Because of the fixed inputs/outputs, the table above defines a standard empirical model for the following tight, causally complete space $\Theta$:
\begin{center}
\includegraphics[height=3.5cm]{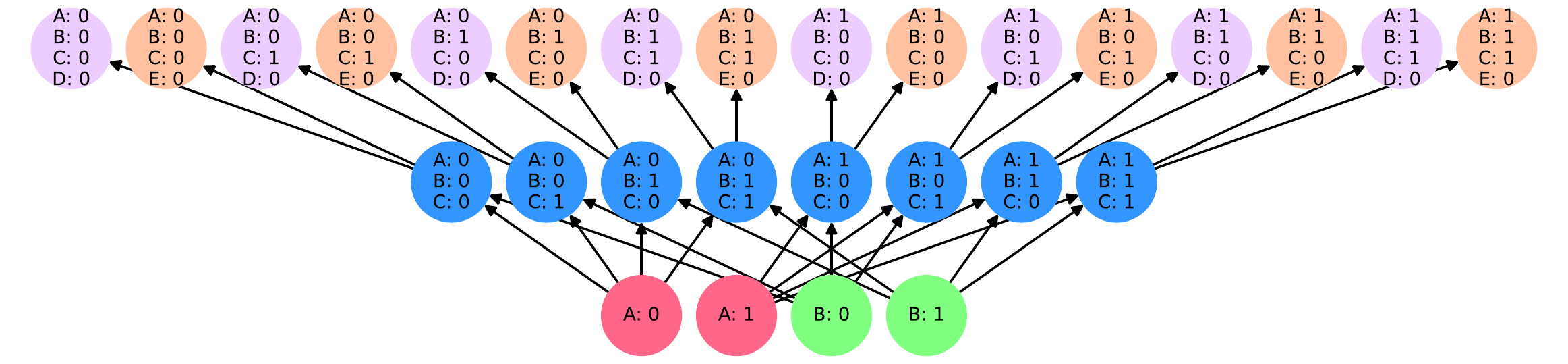}
\end{center}
The causal functions in $\CausFun{\downset{\hist{A/i_\ev{A}, B/i_\ev{B}, C/i_\ev{C}, D/0, E/0}}, \{0,1\}}$ are as follows:
\[
f_{00o_\ev{C}o_\ev{D}o_\ev{E}|i_\ev{A}i_\ev{B}i_\ev{C}00}:=
\left\{
\begin{array}{rl}
    \hist{A/i_\ev{A}}                                           &\mapsto 0\\
    \hist{B/i_\ev{B}}                                           &\mapsto 0\\
    \hist{A/i_\ev{A}, B/i_\ev{B}, C/i_\ev{C}}                   &\mapsto o_{\ev{C}}\\
    \hist{A/i_\ev{A}, B/i_\ev{B}, C/i_\ev{C}, D/0}       &\mapsto o_{\ev{D}}\\
    \hist{A/i_\ev{A}, B/i_\ev{B}, C/i_\ev{C}, E/0}       &\mapsto o_{\ev{E}}\\
\end{array}
\right.
\]
The empirical model is deterministic, consisting of a single causal function:
\[
e_{\downset{\hist{A/i_\ev{A}, B/i_\ev{B}, C/i_\ev{C}, D/0, E/0}}}
:=
\left\{
\begin{array}{rl}
\delta_{\Ext{
    f_{0,0,i_\ev{A}\oplus i_\ev{B},i_\ev{A},i_\ev{B}|i_\ev{A}i_\ev{B}i_\ev{C}00}
}}
&\text{ if } i_\ev{C} = 0
\\
\delta_{\Ext{
    f_{0,0,i_\ev{A}\oplus i_\ev{B},i_\ev{B},i_\ev{A}|i_\ev{A}i_\ev{B}i_\ev{C}00}
}}
&\text{ if } i_\ev{C} = 1
\end{array}
\right.
\]

\subsubsection{The Leggett-Garg Empirical Model.}

To disprove macro-realistic explanations of quantum mechanical phenomena, the authors of \cite{leggett1985quantum} propose the following experiment on a 2-level quantum system, i.e. a qubit, which evolves in time by rotating about the Y axis at a constant angular rate.
Writing $\Delta t > 0$ for the minimum time over which qubit evolution performs a $\frac{2\pi}{3}$ Y rotation, the experiment proceeds as follows:
\begin{enumerate}
    \item The qubit is prepared in the $|+\rangle$ state at time $t_0$ and left alone to evolve.
    \item At time $t_1 := t_0 + \Delta t$, known to us as event $\ev{A}$, the qubit is either left alone (input 0 at $\ev{A}$, with output fixed to $0$) or a non-demolition measurement in the Z basis is performed on it (input 1 at $\ev{A}$, with meas. outcome as output). The qubit is again left alone to evolve.
    \item At time $t_2 := t_1 + \Delta t$, known to us as event $\ev{B}$, the qubit is either left alone (input 0 at $\ev{B}$, with output fixed to $0$) or a non-demolition measurement in the Z basis is performed on it (input 1 at $\ev{B}$, with meas. outcome as output). The qubit is again left alone to evolve.
    \item At time $t_3 := t_2 + \Delta t$, known to us as event $\ev{C}$, the qubit is either discarded (input 0 at $\ev{C}$, with output fixed to $0$) or a demolition measurement in the Z basis is performed on it (input 1 at $\ev{C}$, with meas. outcome as output).
\end{enumerate}
The figure below exemplifies the scenario we have just described:
\begin{center}
\scalebox{1.3}{
    \begin{tikzpicture}
	\begin{pgfonlayer}{nodelayer}
		\node [style=small box] (0) at (-3, -7.25) {$R_y(\frac{2\pi}{3})$};
		\node [style=small box] (1) at (-3, -5.5) {$M$};
		\node [style=small box] (2) at (-3, -3.5) {$R_y(\frac{2\pi}{3})$};
		\node [style=point] (4) at (-3, -8.75) {$+$};
		\node [style=small box] (5) at (-3, 0.5) {$R_y(\frac{2\pi}{3})$};
		\node [style=small box] (6) at (-3, -1.5) {$M$};
		\node [style=small box] (7) at (-3, 2.5) {$M$};
		\node [style=upground] (8) at (-3, 4) {};
		\node [style=none] (9) at (-2.75, 2.25) {};
		\node [style=none] (10) at (-2.75, 2.75) {};
		\node [style=none] (11) at (-1, 4) {};
		\node [style=none] (12) at (-1, 1) {};
		\node [style=none] (13) at (-5, 0) {};
		\node [style=none] (14) at (-5, -3) {};
		\node [style=none] (15) at (-3.25, -1.25) {};
		\node [style=none] (16) at (-3.25, -1.75) {};
		\node [style=none] (17) at (-5, 1) {};
		\node [style=none] (18) at (-5, -4) {};
		\node [style=none] (19) at (-5, -3) {};
		\node [style=none] (20) at (-1, 0) {};
		\node [style=none] (21) at (-1, 5) {};
		\node [style=none] (22) at (-1, -4) {};
		\node [style=none] (23) at (-1, -7) {};
		\node [style=none] (24) at (-2.5, -5.25) {};
		\node [style=none] (25) at (-2.5, -5.75) {};
		\node [style=none] (26) at (-1, -3) {};
		\node [style=none] (27) at (-1, -8) {};
		\node [style=none] (28) at (-1, -7) {};
		\node [style=none] (29) at (-3.75, 3.5) {};
		\node [style=none] (30) at (-2, 3.5) {};
		\node [style=none] (31) at (-2, 1.5) {};
		\node [style=none] (32) at (-3.75, 1.5) {};
		\node [style=none] (33) at (-3.75, -0.5) {};
		\node [style=none] (34) at (-2, -0.5) {};
		\node [style=none] (35) at (-2, -2.5) {};
		\node [style=none] (36) at (-3.75, -2.5) {};
		\node [style=none] (37) at (-3.75, -4.5) {};
		\node [style=none] (38) at (-2, -4.5) {};
		\node [style=none] (39) at (-2, -6.5) {};
		\node [style=none] (40) at (-3.75, -6.5) {};
		\node [style=small box] (41) at (1.5, 0.5) {$M$};
		\node [style=none] (42) at (1.75, 0.25) {};
		\node [style=none] (43) at (1.75, 0.75) {};
		\node [style=none] (44) at (2.25, 2) {};
		\node [style=point] (45) at (2.75, -1) {$0$};
		\node [style=none] (46) at (1.5, -1) {};
		\node [style=none] (47) at (1.5, 2) {};
		\node [style=small box] (48) at (1.5, -5.5) {$M$};
		\node [style=none] (49) at (1.75, -5.75) {};
		\node [style=none] (50) at (1.75, -5.25) {};
		\node [style=none] (51) at (2.25, -4) {};
		\node [style=point] (52) at (2.75, -7) {$1$};
		\node [style=none] (53) at (1.5, -7) {};
		\node [style=none] (54) at (1.5, -4) {};
		\node [style=none] (55) at (3.5, 0.5) {$=$};
		\node [style=none] (56) at (3.5, -5.5) {$=$};
		\node [style=none] (57) at (5, 0.5) {};
		\node [style=point] (59) at (5.75, 0.75) {$0$};
		\node [style=none] (60) at (5.75, 2) {};
		\node [style=none] (61) at (5, -1) {};
		\node [style=none] (62) at (5, 2) {};
		\node [style=white dot] (63) at (5, -5.5) {};
		\node [style=none] (65) at (5.75, -5) {};
		\node [style=none] (66) at (5.75, -4) {};
		\node [style=none] (67) at (5, -7) {};
		\node [style=none] (68) at (5, -4) {};
	\end{pgfonlayer}
	\begin{pgfonlayer}{edgelayer}
		\draw [style=bold black] (0) to (1);
		\draw [style=bold black] (1) to (2);
		\draw [style=bold black] (4) to (0);
		\draw [style=bold black] (2) to (6);
		\draw [style=bold black] (6) to (5);
		\draw [style=bold black] (5) to (7);
		\draw [style=bold black] (7) to (8);
		\draw [style=classical, in=270, out=90] (10.center) to (11.center);
		\draw [style=classical, in=90, out=-90] (9.center) to (12.center);
		\draw [style=classical, in=-90, out=90] (15.center) to (13.center);
		\draw [style=classical, in=90, out=-90, looseness=1.25] (16.center) to (14.center);
		\draw [style=classical] (18.center) to (19.center);
		\draw [style=classical] (13.center) to (17.center);
		\draw [style=classical] (20.center) to (12.center);
		\draw [style=classical] (11.center) to (21.center);
		\draw [style=classical, in=-90, out=90] (24.center) to (22.center);
		\draw [style=classical, in=90, out=-90, looseness=1.25] (25.center) to (23.center);
		\draw [style=classical] (27.center) to (28.center);
		\draw [style=classical] (22.center) to (26.center);
		\draw [style=carta da zucchero thin] (29.center)
			 to (30.center)
			 to (31.center)
			 to (32.center)
			 to cycle;
		\draw [style=carta da zucchero thin] (33.center)
			 to (34.center)
			 to (35.center)
			 to (36.center)
			 to cycle;
		\draw [style=carta da zucchero thin] (37.center)
			 to (38.center)
			 to (39.center)
			 to (40.center)
			 to cycle;
		\draw [style=bold black] (41) to (47.center);
		\draw [style=bold black] (41) to (46.center);
		\draw [style=classical, in=-90, out=90] (43.center) to (44.center);
		\draw [style=classical, in=90, out=-90] (42.center) to (45);
		\draw [style=bold black] (48) to (54.center);
		\draw [style=bold black] (48) to (53.center);
		\draw [style=classical, in=-90, out=90] (50.center) to (51.center);
		\draw [style=classical, in=90, out=-90] (49.center) to (52);
		\draw [style=bold black] (57.center) to (62.center);
		\draw [style=bold black] (57.center) to (61.center);
		\draw [style=classical, in=-90, out=90] (59) to (60.center);
		\draw [style=bold black] (63) to (68.center);
		\draw [style=bold black] (63) to (67.center);
		\draw [style=classical, in=-90, out=90] (65.center) to (66.center);
		\draw [style=classical, in=-90, out=0, looseness=0.75] (63) to (65.center);
	\end{pgfonlayer}
\end{tikzpicture}
}
\end{center}
The description above results in the following empirical model on 3 events:
\begin{center}
\scalebox{0.9}{
\begin{tabular}{l|rrrrrrrr}
\hfill
ABC & 000 & 001 & 010 & 011 & 100 & 101 & 110 & 111\\
\hline
000 & 1.000 & 0.000 & 0.000 & 0.000 & 0.000 & 0.000 & 0.000 & 0.000\\
001 & 0.933 & 0.067 & 0.000 & 0.000 & 0.000 & 0.000 & 0.000 & 0.000\\
010 & 0.067 & 0.000 & 0.933 & 0.000 & 0.000 & 0.000 & 0.000 & 0.000\\
011 & 0.017 & 0.050 & 0.700 & 0.233 & 0.000 & 0.000 & 0.000 & 0.000\\
100 & 0.500 & 0.000 & 0.000 & 0.000 & 0.500 & 0.000 & 0.000 & 0.000\\
101 & 0.125 & 0.375 & 0.000 & 0.000 & 0.375 & 0.125 & 0.000 & 0.000\\
110 & 0.125 & 0.000 & 0.375 & 0.000 & 0.375 & 0.000 & 0.125 & 0.000\\
111 & 0.031 & 0.094 & 0.281 & 0.094 & 0.094 & 0.281 & 0.094 & 0.031\\
\end{tabular}
}
\end{center}
The Leggett-Garg inequalities provides bounds, valid in macro-realistic interpretations, for the sum of the expected $\pm1$-valued parity of outputs when the $\pm1$-valued parity of inputs is $+1$:
\[
-1 \leq 
\mathbb{E}(-1^{o_\ev{A}\!\oplus\! o_\ev{B} \!\oplus\! o_\ev{C}}|011)
+\mathbb{E}(-1^{o_\ev{A}\!\oplus\! o_\ev{B} \!\oplus\! o_\ev{C}}|101)
+\mathbb{E}(-1^{o_\ev{A}\!\oplus\! o_\ev{B} \!\oplus\! o_\ev{C}}|110)
\leq 3
\]
The authors then observe that, in the experiment they propose, the sum of such expected parities is $-\frac{3}{2}$, violating the lower bound and thus excluding a macro-realistic explanation.
Indeed, we can restrict ourselves to the relevant rows of the empirical model:
\begin{center}
\begin{tabular}{l|rrrrrrr}
\hfill
ABC & 000 & 001 & 010 & 011 & 100 & 101 & 110\\
\hline
011 & 0.017 & 0.050 & 0.700 & 0.233 & 0.000 & 0.000 & 0.000\\
101 & 0.125 & 0.375 & 0.000 & 0.000 & 0.375 & 0.125 & 0.000\\
110 & 0.125 & 0.000 & 0.375 & 0.000 & 0.375 & 0.000 & 0.125\\
\end{tabular}
\end{center}
The sum of the expected parity of outputs is then computed as follows: 
\[
\begin{array}{rl}
&\mathbb{E}(-1^{o_\ev{A}\!\oplus\! o_\ev{B} \!\oplus\! o_\ev{C}}|011)
+\mathbb{E}(-1^{o_\ev{A}\!\oplus\! o_\ev{B} \!\oplus\! o_\ev{C}}|101)
+\mathbb{E}(-1^{o_\ev{A}\!\oplus\! o_\ev{B} \!\oplus\! o_\ev{C}}|110)
\\
=& (0.017-0.050-0.700+0.233)
\\
 &+(0.125-0.375-0.375+0.125)
\\
 &+(0.125-0.375-0.375+0.125)
\\
=& \frac{1}{2}+\frac{1}{2}+\frac{1}{2} = \frac{3}{2}
\end{array}
\]
By construction, this is a standard empirical model for (the space of inputs histories induced by) the total order $\total{A, B, C}$.

Because of Theorem \ref{theorem:nogo-nonlocality-switch-spaces}, the Leggett-Garg empirical model is necessarily non-contextual/local for this space: it can be obtained as mixture of 12 causal functions, with weights and empirical models enumerated below.
The following causal functions explain 1.674\% (below left) and 1.451\% (below right) of the empirical model:
\begin{center}
\scalebox{0.6}{
\begin{tabular}{l|rrrrrrrr}
\hfill
ABC & 000 & 001 & 010 & 011 & 100 & 101 & 110 & 111\\
\hline
000 & $1$ & $0$ & $0$ & $0$ & $0$ & $0$ & $0$ & $0$\\
001 & $1$ & $0$ & $0$ & $0$ & $0$ & $0$ & $0$ & $0$\\
010 & $1$ & $0$ & $0$ & $0$ & $0$ & $0$ & $0$ & $0$\\
011 & $1$ & $0$ & $0$ & $0$ & $0$ & $0$ & $0$ & $0$\\
100 & $1$ & $0$ & $0$ & $0$ & $0$ & $0$ & $0$ & $0$\\
101 & $1$ & $0$ & $0$ & $0$ & $0$ & $0$ & $0$ & $0$\\
110 & $1$ & $0$ & $0$ & $0$ & $0$ & $0$ & $0$ & $0$\\
111 & $1$ & $0$ & $0$ & $0$ & $0$ & $0$ & $0$ & $0$
\end{tabular}
}
\hspace{1cm}
\scalebox{0.6}{
\begin{tabular}{l|rrrrrrrr}
\hfill
ABC & 000 & 001 & 010 & 011 & 100 & 101 & 110 & 111\\
\hline
000 & $1$ & $0$ & $0$ & $0$ & $0$ & $0$ & $0$ & $0$\\
001 & $1$ & $0$ & $0$ & $0$ & $0$ & $0$ & $0$ & $0$\\
010 & $1$ & $0$ & $0$ & $0$ & $0$ & $0$ & $0$ & $0$\\
011 & $0$ & $1$ & $0$ & $0$ & $0$ & $0$ & $0$ & $0$\\
100 & $1$ & $0$ & $0$ & $0$ & $0$ & $0$ & $0$ & $0$\\
101 & $1$ & $0$ & $0$ & $0$ & $0$ & $0$ & $0$ & $0$\\
110 & $1$ & $0$ & $0$ & $0$ & $0$ & $0$ & $0$ & $0$\\
111 & $1$ & $0$ & $0$ & $0$ & $0$ & $0$ & $0$ & $0$
\end{tabular}
}
\end{center}
The following causal functions explain 3.573\% (below left) and 5.801\% (below right) of the empirical model:
\begin{center}
\scalebox{0.6}{
\begin{tabular}{l|rrrrrrrr}
\hfill
ABC & 000 & 001 & 010 & 011 & 100 & 101 & 110 & 111\\
\hline
000 & $1$ & $0$ & $0$ & $0$ & $0$ & $0$ & $0$ & $0$\\
001 & $1$ & $0$ & $0$ & $0$ & $0$ & $0$ & $0$ & $0$\\
010 & $1$ & $0$ & $0$ & $0$ & $0$ & $0$ & $0$ & $0$\\
011 & $0$ & $1$ & $0$ & $0$ & $0$ & $0$ & $0$ & $0$\\
100 & $1$ & $0$ & $0$ & $0$ & $0$ & $0$ & $0$ & $0$\\
101 & $1$ & $0$ & $0$ & $0$ & $0$ & $0$ & $0$ & $0$\\
110 & $1$ & $0$ & $0$ & $0$ & $0$ & $0$ & $0$ & $0$\\
111 & $0$ & $1$ & $0$ & $0$ & $0$ & $0$ & $0$ & $0$
\end{tabular}
}
\hspace{1cm}
\scalebox{0.6}{
\begin{tabular}{l|rrrrrrrr}
\hfill
ABC & 000 & 001 & 010 & 011 & 100 & 101 & 110 & 111\\
\hline
000 & $1$ & $0$ & $0$ & $0$ & $0$ & $0$ & $0$ & $0$\\
001 & $1$ & $0$ & $0$ & $0$ & $0$ & $0$ & $0$ & $0$\\
010 & $0$ & $0$ & $1$ & $0$ & $0$ & $0$ & $0$ & $0$\\
011 & $0$ & $0$ & $1$ & $0$ & $0$ & $0$ & $0$ & $0$\\
100 & $1$ & $0$ & $0$ & $0$ & $0$ & $0$ & $0$ & $0$\\
101 & $1$ & $0$ & $0$ & $0$ & $0$ & $0$ & $0$ & $0$\\
110 & $1$ & $0$ & $0$ & $0$ & $0$ & $0$ & $0$ & $0$\\
111 & $0$ & $1$ & $0$ & $0$ & $0$ & $0$ & $0$ & $0$
\end{tabular}
}
\end{center}
The following causal functions explain 28.125\% (below left) and 9.375\% (below right) of the empirical model:
\begin{center}
\scalebox{0.6}{
\begin{tabular}{l|rrrrrrrr}
\hfill
ABC & 000 & 001 & 010 & 011 & 100 & 101 & 110 & 111\\
\hline
000 & $1$ & $0$ & $0$ & $0$ & $0$ & $0$ & $0$ & $0$\\
001 & $1$ & $0$ & $0$ & $0$ & $0$ & $0$ & $0$ & $0$\\
010 & $0$ & $0$ & $1$ & $0$ & $0$ & $0$ & $0$ & $0$\\
011 & $0$ & $0$ & $1$ & $0$ & $0$ & $0$ & $0$ & $0$\\
100 & $1$ & $0$ & $0$ & $0$ & $0$ & $0$ & $0$ & $0$\\
101 & $0$ & $1$ & $0$ & $0$ & $0$ & $0$ & $0$ & $0$\\
110 & $0$ & $0$ & $1$ & $0$ & $0$ & $0$ & $0$ & $0$\\
111 & $0$ & $0$ & $1$ & $0$ & $0$ & $0$ & $0$ & $0$
\end{tabular}
}
\hspace{1cm}
\scalebox{0.6}{
\begin{tabular}{l|rrrrrrrr}
\hfill
ABC & 000 & 001 & 010 & 011 & 100 & 101 & 110 & 111\\
\hline
000 & $1$ & $0$ & $0$ & $0$ & $0$ & $0$ & $0$ & $0$\\
001 & $1$ & $0$ & $0$ & $0$ & $0$ & $0$ & $0$ & $0$\\
010 & $0$ & $0$ & $1$ & $0$ & $0$ & $0$ & $0$ & $0$\\
011 & $0$ & $0$ & $1$ & $0$ & $0$ & $0$ & $0$ & $0$\\
100 & $1$ & $0$ & $0$ & $0$ & $0$ & $0$ & $0$ & $0$\\
101 & $0$ & $1$ & $0$ & $0$ & $0$ & $0$ & $0$ & $0$\\
110 & $0$ & $0$ & $1$ & $0$ & $0$ & $0$ & $0$ & $0$\\
111 & $0$ & $0$ & $0$ & $1$ & $0$ & $0$ & $0$ & $0$
\end{tabular}
}
\end{center}
The following causal functions explain 9.375\% (below left) and 17.300\% (below right) of the empirical model:
\begin{center}
\scalebox{0.6}{
\begin{tabular}{l|rrrrrrrr}
\hfill
ABC & 000 & 001 & 010 & 011 & 100 & 101 & 110 & 111\\
\hline
000 & $1$ & $0$ & $0$ & $0$ & $0$ & $0$ & $0$ & $0$\\
001 & $1$ & $0$ & $0$ & $0$ & $0$ & $0$ & $0$ & $0$\\
010 & $0$ & $0$ & $1$ & $0$ & $0$ & $0$ & $0$ & $0$\\
011 & $0$ & $0$ & $1$ & $0$ & $0$ & $0$ & $0$ & $0$\\
100 & $0$ & $0$ & $0$ & $0$ & $1$ & $0$ & $0$ & $0$\\
101 & $0$ & $0$ & $0$ & $0$ & $1$ & $0$ & $0$ & $0$\\
110 & $0$ & $0$ & $0$ & $0$ & $1$ & $0$ & $0$ & $0$\\
111 & $0$ & $0$ & $0$ & $0$ & $1$ & $0$ & $0$ & $0$
\end{tabular}
}
\hspace{1cm}
\scalebox{0.6}{
\begin{tabular}{l|rrrrrrrr}
\hfill
ABC & 000 & 001 & 010 & 011 & 100 & 101 & 110 & 111\\
\hline
000 & $1$ & $0$ & $0$ & $0$ & $0$ & $0$ & $0$ & $0$\\
001 & $1$ & $0$ & $0$ & $0$ & $0$ & $0$ & $0$ & $0$\\
010 & $0$ & $0$ & $1$ & $0$ & $0$ & $0$ & $0$ & $0$\\
011 & $0$ & $0$ & $1$ & $0$ & $0$ & $0$ & $0$ & $0$\\
100 & $0$ & $0$ & $0$ & $0$ & $1$ & $0$ & $0$ & $0$\\
101 & $0$ & $0$ & $0$ & $0$ & $1$ & $0$ & $0$ & $0$\\
110 & $0$ & $0$ & $0$ & $0$ & $1$ & $0$ & $0$ & $0$\\
111 & $0$ & $0$ & $0$ & $0$ & $0$ & $1$ & $0$ & $0$
\end{tabular}
}
\end{center}
The following causal functions explain 10.825\% (below left) and 5.802\% (below right) of the empirical model:
\begin{center}
\scalebox{0.6}{
\begin{tabular}{l|rrrrrrrr}
\hfill
ABC & 000 & 001 & 010 & 011 & 100 & 101 & 110 & 111\\
\hline
000 & $1$ & $0$ & $0$ & $0$ & $0$ & $0$ & $0$ & $0$\\
001 & $1$ & $0$ & $0$ & $0$ & $0$ & $0$ & $0$ & $0$\\
010 & $0$ & $0$ & $1$ & $0$ & $0$ & $0$ & $0$ & $0$\\
011 & $0$ & $0$ & $0$ & $1$ & $0$ & $0$ & $0$ & $0$\\
100 & $0$ & $0$ & $0$ & $0$ & $1$ & $0$ & $0$ & $0$\\
101 & $0$ & $0$ & $0$ & $0$ & $1$ & $0$ & $0$ & $0$\\
110 & $0$ & $0$ & $0$ & $0$ & $1$ & $0$ & $0$ & $0$\\
111 & $0$ & $0$ & $0$ & $0$ & $0$ & $1$ & $0$ & $0$
\end{tabular}
}
\hspace{1cm}
\scalebox{0.6}{
\begin{tabular}{l|rrrrrrrr}
\hfill
ABC & 000 & 001 & 010 & 011 & 100 & 101 & 110 & 111\\
\hline
000 & $1$ & $0$ & $0$ & $0$ & $0$ & $0$ & $0$ & $0$\\
001 & $1$ & $0$ & $0$ & $0$ & $0$ & $0$ & $0$ & $0$\\
010 & $0$ & $0$ & $1$ & $0$ & $0$ & $0$ & $0$ & $0$\\
011 & $0$ & $0$ & $0$ & $1$ & $0$ & $0$ & $0$ & $0$\\
100 & $0$ & $0$ & $0$ & $0$ & $1$ & $0$ & $0$ & $0$\\
101 & $0$ & $0$ & $0$ & $0$ & $0$ & $1$ & $0$ & $0$\\
110 & $0$ & $0$ & $0$ & $0$ & $0$ & $0$ & $1$ & $0$\\
111 & $0$ & $0$ & $0$ & $0$ & $0$ & $0$ & $1$ & $0$
\end{tabular}
}
\end{center}
The following causal functions explain 3.572\% (below left) and 3.125\% (below right) of the empirical model:
\begin{center}
\scalebox{0.6}{
\begin{tabular}{l|rrrrrrrr}
\hfill
ABC & 000 & 001 & 010 & 011 & 100 & 101 & 110 & 111\\
\hline
000 & $1$ & $0$ & $0$ & $0$ & $0$ & $0$ & $0$ & $0$\\
001 & $0$ & $1$ & $0$ & $0$ & $0$ & $0$ & $0$ & $0$\\
010 & $0$ & $0$ & $1$ & $0$ & $0$ & $0$ & $0$ & $0$\\
011 & $0$ & $0$ & $0$ & $1$ & $0$ & $0$ & $0$ & $0$\\
100 & $0$ & $0$ & $0$ & $0$ & $1$ & $0$ & $0$ & $0$\\
101 & $0$ & $0$ & $0$ & $0$ & $0$ & $1$ & $0$ & $0$\\
110 & $0$ & $0$ & $0$ & $0$ & $0$ & $0$ & $1$ & $0$\\
111 & $0$ & $0$ & $0$ & $0$ & $0$ & $0$ & $1$ & $0$
\end{tabular}
}
\hspace{1cm}
\scalebox{0.6}{
\begin{tabular}{l|rrrrrrrr}
\hfill
ABC & 000 & 001 & 010 & 011 & 100 & 101 & 110 & 111\\
\hline
000 & $1$ & $0$ & $0$ & $0$ & $0$ & $0$ & $0$ & $0$\\
001 & $0$ & $1$ & $0$ & $0$ & $0$ & $0$ & $0$ & $0$\\
010 & $0$ & $0$ & $1$ & $0$ & $0$ & $0$ & $0$ & $0$\\
011 & $0$ & $0$ & $0$ & $1$ & $0$ & $0$ & $0$ & $0$\\
100 & $0$ & $0$ & $0$ & $0$ & $1$ & $0$ & $0$ & $0$\\
101 & $0$ & $0$ & $0$ & $0$ & $0$ & $1$ & $0$ & $0$\\
110 & $0$ & $0$ & $0$ & $0$ & $0$ & $0$ & $1$ & $0$\\
111 & $0$ & $0$ & $0$ & $0$ & $0$ & $0$ & $0$ & $1$
\end{tabular}
}
\end{center}
Taking a convex combination of the 12 deterministic empirical models above with the corresponding weights yields the Leggett-Garg empirical model.

Interestingly, the macro-realist constraints specified by Leggett and Garg in Equation (1) of \cite{leggett1985quantum} are causal constraints, stating that a macro-realist model has to be an empirical model for the following 3 indefinite causal orders.
\begin{center}
    \begin{tabular}{ccc}
    \includegraphics[height=3.5cm]{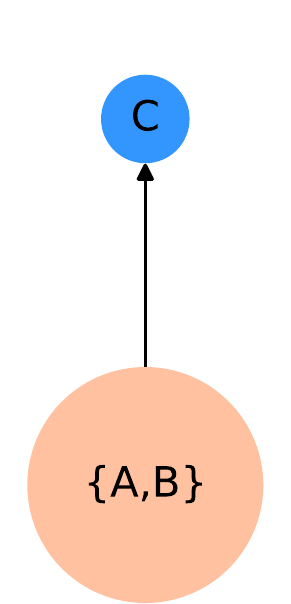}
    &\hspace{2cm}
    \includegraphics[height=3.5cm]{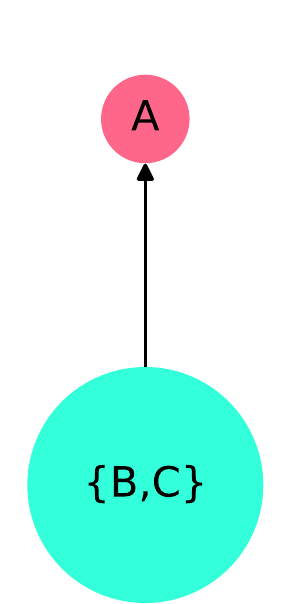}
    &\hspace{2cm}
    \includegraphics[height=3.5cm]{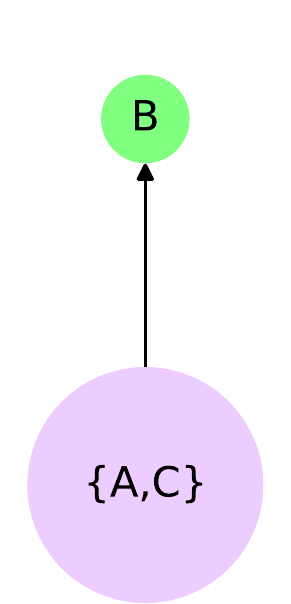}
    \end{tabular}
\end{center}
The Leggett-Garg empirical model is an empirical model for the leftmost causal order, because \total{A, B, C} is a sub-order, but not for the middle or right one (cf. Subsubsection 2.5.4 of \cite{gogioso2022geometry}).
It can be shown that Leggett-Garg empirical model lifts to an empirical model for the following space:
\begin{center}
\includegraphics[height=3cm]{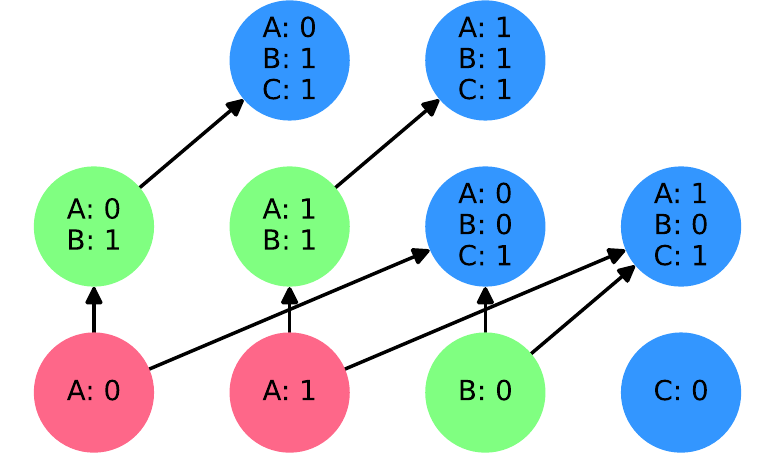}
\end{center}
This space captures the causal constraints---additional to the ones from the total order \total{A,B,C}---associated with the absence of measurement on input 0:
\begin{itemize}
    \item The presence of history \hist{C/0} states that there is no signalling from \ev{A} nor \ev{B} to \ev{C} when no measurement is performed at \ev{C}.
    \item The presence of history \hist{B/0} states that there is no signalling from \ev{A} to \ev{B} when no measurement is performed at \ev{B}.
\end{itemize}
Indeed, it is easy to check that the 12 causal functions involved in the deterministic causal HVM for the empirical model over \total{A, B, C} are also causal functions for the causally complete space shown above: hence the Leggett-Garg empirical model is lifts to an empirical model on that space, and it is non-contextual/local there.

\subsubsection{The BFW Empirical Model.}

We now look at the empirical model introduced by Baumeler, Feix and Wolf in \cite{baumeler2014maximal,baumeler2014perfect}, described by the authors as the 50\%-50\% mixture of a ``circular identity'' classical process and a ``circular bitflip'' classical process, for three agents Alice, Bob and Charlie.
\begin{center}
\scalebox{0.9}{
\begin{tabular}{l|rrrrrrrr}
\hfill
ABC & 000 & 001 & 010 & 011 & 100 & 101 & 110 & 111\\
\hline
000 & $1/2$ & $0$ & $0$ & $0$ & $0$ & $0$ & $0$ & $1/2$\\
001 & $0$ & $0$ & $0$ & $1/2$ & $1/2$ & $0$ & $0$ & $0$\\
010 & $0$ & $1/2$ & $0$ & $0$ & $0$ & $0$ & $1/2$ & $0$\\
011 & $0$ & $0$ & $1/2$ & $0$ & $0$ & $1/2$ & $0$ & $0$\\
100 & $0$ & $0$ & $1/2$ & $0$ & $0$ & $1/2$ & $0$ & $0$\\
101 & $0$ & $1/2$ & $0$ & $0$ & $0$ & $0$ & $1/2$ & $0$\\
110 & $0$ & $0$ & $0$ & $1/2$ & $1/2$ & $0$ & $0$ & $0$\\
111 & $1/2$ & $0$ & $0$ & $0$ & $0$ & $0$ & $0$ & $1/2$\\
\end{tabular}
}
\end{center}
Specifically, the empirical model above is the 50\%-50\% mixture of the following inseparable functions for the causally incomplete space induced by the causal order \indiscrete{A, B, C}:
\begin{center}
\scalebox{0.7}{
    \begin{tabular}{l|rrrrrrrr}
    \hfill
    ABC & 000 & 001 & 010 & 011 & 100 & 101 & 110 & 111\\
    \hline
    000 & $1$ & $0$ & $0$ & $0$ & $0$ & $0$ & $0$ & $0$\\
    001 & $0$ & $0$ & $0$ & $0$ & $1$ & $0$ & $0$ & $0$\\
    010 & $0$ & $1$ & $0$ & $0$ & $0$ & $0$ & $0$ & $0$\\
    011 & $0$ & $0$ & $0$ & $0$ & $0$ & $1$ & $0$ & $0$\\
    100 & $0$ & $0$ & $1$ & $0$ & $0$ & $0$ & $0$ & $0$\\
    101 & $0$ & $0$ & $0$ & $0$ & $0$ & $0$ & $1$ & $0$\\
    110 & $0$ & $0$ & $0$ & $1$ & $0$ & $0$ & $0$ & $0$\\
    111 & $0$ & $0$ & $0$ & $0$ & $0$ & $0$ & $0$ & $1$\\
    \end{tabular}
}
\hspace{5mm}
\scalebox{0.7}{
    \begin{tabular}{l|rrrrrrrr}
    \hfill
    ABC & 000 & 001 & 010 & 011 & 100 & 101 & 110 & 111\\
    \hline
    000 & $0$ & $0$ & $0$ & $0$ & $0$ & $0$ & $0$ & $1$\\
    001 & $0$ & $0$ & $0$ & $1$ & $0$ & $0$ & $0$ & $0$\\
    010 & $0$ & $0$ & $0$ & $0$ & $0$ & $0$ & $1$ & $0$\\
    011 & $0$ & $0$ & $1$ & $0$ & $0$ & $0$ & $0$ & $0$\\
    100 & $0$ & $0$ & $0$ & $0$ & $0$ & $1$ & $0$ & $0$\\
    101 & $0$ & $1$ & $0$ & $0$ & $0$ & $0$ & $0$ & $0$\\
    110 & $0$ & $0$ & $0$ & $0$ & $1$ & $0$ & $0$ & $0$\\
    111 & $1$ & $0$ & $0$ & $0$ & $0$ & $0$ & $0$ & $0$\\
    \end{tabular}
}
\end{center}
Recall that an inseparability witness for a causal function $f \in \CausFun{\Theta, \underline{O}}$ is a triple $(k, (k'_\omega)_{\omega \in \dom{k}}, (\xi_\omega)_{\omega \in \dom{k}})$ where: 
\begin{itemize}
    \item $k \in \Ext{\Theta}$ is an extended input history;
    \item for every $\omega \in \dom{k}$, $k'_\omega \in \Ext{\Theta}$ is such that $\restrict{k}{\dom{k}\backslash\{\omega\}} \leq k'_\omega$;
    \item for every $\omega \in \dom{k}$, $\xi_\omega \in \dom{k}\backslash\{\omega\}$ is such that $\Ext{f}(k)_{\xi_{\omega}} \neq \Ext{f}(k'_\omega)_{\xi_{\omega}}$.
\end{itemize}
Both the circular identity function (on the left) and the circular bitflip function (on the right) admits the following inseparability witness:
\begin{itemize}
    \item $k := \hist{A/0,B/0,C/0}$
    \item $k'_\ev{A} := \hist{A/1,B/0,C/0}$, $k'_\ev{B} := \hist{A/0,B/1,C/0}$, $k'_\ev{C} := \hist{A/0,B/0,C/1}$
    \item $\xi_\ev{A} := \ev{B}$, $\xi_\ev{B} := \ev{C}$, $\xi_\ev{C} := \ev{A}$
\end{itemize}

\newpage

\subsection{Proofs for Section \ref{section:topology-causality}}

\subsubsection{Proof of Proposition \ref{proposition:causal-function-traditional}}
\label{proof:proposition:causal-function-traditional}
\begin{proof}
For each given $\omega \in \Omega$, the histories $h \in \Theta = \Hist{\Omega, \underline{I}}$ with $\omega \in \tips{\Theta}{h}$ are exactly the partial functions assigning a choice of input to each event $\xi \in \downset{\omega}$:
\[
\omega \in \tips{\Theta}{h}
\Leftrightarrow
h \in \prod_{\xi \in \downset{\omega}} I_\xi
\]
Fixing a $\omega \in \Omega$ and an $h$ such that $\omega \in \tips{\Theta}{h}$, the $k \in \prod_{\xi \in \Omega} I_\omega$ such that $h \leq k$ are determined by extending $h$ through an arbitrary assignment of values to all $\xi \in \Omega\backslash\downset{\omega}$:
\[
k = h \vee \hat{k}
\text{ for all choices of }
\hat{k} \in \prod_{\xi \in \Omega\backslash\downset{\omega}} I_\omega
\]
Causality of $F$ for $\Theta$ then means that $F_\omega\left(h\vee\hat{k}\right)$ is independent of the choice of $\hat{k}$, i.e. that it takes the form $F_\omega\left(h\vee\hat{k}\right) = G_\omega\left(h\right)$ for some $G_\omega: \{0,1\}^{\downset{\omega}} \rightarrow \{0,1\}$.
\end{proof}

\subsubsection{Proof of Proposition \ref{proposition:consistency-gluing-conditions}}
\label{proof:proposition:consistency-gluing-conditions}
\begin{proof}
In one direction, observe that $k' \leq k$ is a special case of compatibility between $k$ and $k'$: if $\hat{F}$ satisfies the gluing condition, then $\hat{F}(k')$ and $\hat{F}(k)$ must be compatible and $\hat{F}(k') = \hat{F}(k\vee k') = \hat{F}(k) \vee \hat{F}(k')$, which is the same as $\hat{F}(k') \leq \hat{F}(k)$.
In the other direction, observe that $k \leq k \vee k'$ and $k' \leq k \vee k'$:
if $\hat{F}$ satisfies the consistency condition, then $\hat{F}(k) \leq \hat{F}(k \vee k')$ and $\hat{F}(k') \leq \hat{F}(k \vee k')$, which jointly with $\dom{k\vee k'} = \dom{k} \cup \dom{k'}$ implies that $\hat{F}(k\vee k') = \hat{F}(k) \vee \hat{F}(k')$.
\end{proof}

\subsubsection{Proof of Theorem \ref{theorem:ext-f-characterisation-tight-cc}}
\label{proof:theorem:ext-f-characterisation-tight-cc}
\begin{proof}
In one direction, we show that $\Ext{f}$ satisfies the consistency condition for any $f \in \CausFun{\Theta, \underline{O}}$.
We start by observing that tightness implies $h_{k',\omega}=h_{k,\omega}$ for all $\omega \in \dom{k'} \subseteq \dom{k}$, where $k'\leq k$ are two extended input histories.
The output value at each $\omega \in \dom{k'}$ is then the same for both extended output histories $\Ext{f}(k')$ and $\Ext{f}(k)$, proving that that $\Ext{f}(k') \leq \Ext{f}(k)$:
\[
\Ext{f}(k)_\omega
=
f\left(h_{k, \omega}\right)
=
f\left(h_{k', \omega}\right)
=
\Ext{f}(k')_\omega
\]
In the other direction, we consider an extended function $\hat{F}$ satisfying the consistency condition.
Causal functions have no restriction on them other, than conditions on their domain and the type of their outputs: both conditions are satisfied by the definition of $\Prime{\hat{F}}$, which is therefore a causal function $\Prime{\hat{F}} \in \CausFun{\Theta, \underline{O}}$.
For any $k \in \Ext{\Theta}$ and $\omega \in \dom{k}$, the input history $h_{k, \omega} \in \Theta$ satisfies $h_{k, \omega} \leq k$ and $\tip{\Theta}{h_{k,\omega}} = \omega$.
By the consistency condition, we must have $\hat{F}(k)_\omega = \hat{F}(h_{k,\omega})_\omega$.
By the definition of $\Prime{\hat{F}}$, we must also have $\hat{F}(h_{k,\omega})_\omega = \Prime{\hat{F}}(h_{k,\omega}) = \Ext{\Prime{\hat{F}}}(k)_\omega$.
Putting the two together, we conclude that $\hat{F}(k)=\Ext{\Prime{\hat{F}}}(k)$ for all $k \in \Ext{\Theta}$.
Also, the definition of $\Ext{f}$ implies that $\Ext{f}(h)_{\tip{\Theta}{h}} = f(h)$, proving the uniqueness claim.
\end{proof}

\subsubsection{Proof of Proposition \ref{proposition:causal-function-joint-function-tight-cc}}
\label{proof:proposition:causal-function-joint-function-tight-cc}
\begin{proof}
Because $\Theta$ satisfies the free-choice condition, its maximal extended input histories are:
\[
\prod_{\omega \in \Events{\Theta}} \Inputs{\Theta}_\omega
\subseteq
\Ext{\Theta}
\]
Restricting $\Ext{f}$ to the extended input histories clearly yields a joint IO function for the operational scenario $(\Events{\Theta}, \underline{\Inputs{\Theta}}, \underline{O})$.
This joint IO function is causal for $\Theta$ if for every $h \in \Theta$ the output value $\Ext{f}_{\tip{\Theta}{h}}(k)$ is the same for all maximal extended input histories $k$ such that $h \leq k$.
Because $\Theta$ is tight, we necessarily have $h = h_{k, \omega}$, so that $\Ext{f}_{\tip{\Theta}{h}}(k) = f(h)$ is indeed the same for all such $k$.

Conversely, consider a joint IO function $F$ which is causal for $\Theta$.
We intend to extend $F$ to a function $\hat{F} \in \ExtFun{\Theta, \underline{O}}$ as follows:
\[
\hat{F}(k)
:=
\restrict{F(\hat{k})}{\dom{k}}
\text{ for any maximal ext. input history $\hat{k}$ s.t. } k \leq \hat{k}
\]
We must show that the function $\hat{F}$ is well-defined.
Given a $k \in \Ext{\Theta}$ and an $\omega \in \dom{k}$, we have $h_{k, \omega} \leq k \leq \hat{k}$ for all choices of $\hat{k}$ in the definition of $\hat{F}(k)$ above.
Causality of the joint IO function $F$ for $\Theta$ then implies that the value $F_\omega(\hat{k})$ be the same for all such choices of $\hat{k}$, making $\hat{F}(k)_\omega$ well-defined for all $\omega \in \dom{k}$; as a consequence, $\hat{F}$ is well defined as a whole.

If $f$ is defined as in Equation \ref{equation:causal-fun-from-extended-fun}, we have $\hat{F}(h)_{\tip{\Theta}{h}} = f(h)$, so that $\hat{F}$ satisfies the first condition in Theorem \ref{theorem:ext-f-characterisation-tight-cc}.
Furthermore, for any two compatible $k, k' \in \Ext{\Theta}$ and any maximal extended input history $\hat{k}$ such that $k \vee k' \leq \hat{k}$, we have that:
\[
\begin{array}{rcl}
\hat{F}(k\vee k')&=&\restrict{F(\hat{k})}{\dom{k \vee k'}}\\
\hat{F}(k)&=&\restrict{F(\hat{k})}{\dom{k}}\\
\hat{F}(k')&=&\restrict{F(\hat{k})}{\dom{k'}}\\
\end{array}
\]
From the above, it immediately follows that $\hat{F}(k\vee k') = \hat{F}(k) \vee \hat{F}(k')$: this means that $\hat{F}$ also satisfies the second condition in Theorem \ref{theorem:ext-f-characterisation-tight-cc}, allowing us to conclude that $\hat{F} = \Ext{f}$.
Restricting $\hat{F} = \Ext{f}$ to the maximal extended input histories yields back $F$, because on such histories $k$ we have $\hat{F}(k) = F(k)$ by definition of $\hat{F}$.
\end{proof}

\subsubsection{Proof of Proposition \ref{proposition:caus-fun-subspaces}}
\label{proof:proposition:caus-fun-subspaces}
\begin{proof}
By Theorem \ref{theorem:ext-f-characterisation-tight-cc}:
\[
\begin{array}{rl}
&\Prime{\restrict{\Ext{f'}}{\Ext{\Theta}}}=\Prime{\restrict{\Ext{g'}}{\Ext{\Theta}}}
\\
\Rightarrow&
\restrict{\Ext{f'}}{\Ext{\Theta}} = \restrict{\Ext{g'}}{\Ext{\Theta}}
\end{array}
\]
Because $\Theta', \Theta \in \SpacesFC{\underline{I}}$, the two spaces have the same maximal extended input histories, the above further implies $\Ext{f'}(k)=\Ext{g'}(k)$ for all $k \in \prod_{\omega \in \Events{\Theta}} I_\omega$.
Since every $h \in \Theta'$ satisfies $h \leq k$ for some maximal extended input history $k$, the consistency condition in turn implies that $f'(h) = g'(h)$ for all $h \in \Theta'$, proving that $\CausFunInj{\Theta'}{\underline{O}'}{\Theta}{\underline{O}}$ is an injection.
These injections are stable under composition:
\[
\begin{array}{rl}
&\Prime{\restrict{\Ext{
    \Prime{\restrict{
    \Ext{f''}
    }{\Ext{\Theta'}}}
}}{\Ext{\Theta}}}
\\
=&
\Prime{\restrict{\left(
    \restrict{
    \Ext{f''}
    }{\Ext{\Theta'}}
\right)}{\Ext{\Theta}}} 
\\
=&
\Prime{\restrict{
    \Ext{f''}
}{\Ext{\Theta}}}
\end{array}
\]
This completes our proof.
\end{proof}


\subsubsection{Proof of Lemma \ref{lemma:inseparability-witness-1}}
\label{proof:lemma:inseparability-witness-1}
\begin{proof}
We necessarily have $|\dom{k}| \geq 1$, and for any $\omega \in \dom{k}$ we also have $\xi_\omega \in \dom{k}\backslash\{\omega\}$, so necessarily $|\dom{k}| \geq 2$.
The partial function $h_\omega$ cannot be a an extended input history for $\Theta$, because otherwise the consistency condition on $\Ext{f}$ would imply the following, contradicting the definition of $\xi_\omega$:
\[
\begin{array}{rcl}
\Ext{f}_{\xi_{\omega}}(h_\omega) &=& \Ext{f}_{\xi_{\omega}}(k)\\
\Ext{f}_{\xi_{\omega}}(h_\omega) &=& \Ext{f}_{\xi_{\omega}}(k'_\omega)\\
\end{array}
\]
\end{proof}

\subsubsection{Proof of Lemma \ref{lemma:inseparability-witness-2}}
\label{proof:lemma:inseparability-witness-2}
\begin{proof}
Be definition, $\Theta' \leq \Theta$ is the same as $\Ext{\Theta'} \supseteq \Ext{\Theta}$, so that $k$ and all $k'_\omega$ are extended histories for $\Theta'$.
Also by definition, $f$ arising from $f'$ is the same as $\restrict{\Ext{f'}}{\Ext{\Theta}} = \Ext{f}$, so that:
\[
\Ext{f'}_{\xi_\omega}(k)
=
\Ext{f}_{\xi_\omega}(k)
\neq
\Ext{f}_{\xi_\omega}(k'_\omega)
=
\Ext{f'}_{\xi_\omega}(k'_\omega)
\]
Hence $(k, \underline{k'}, \underline{\xi})$ is an inseparability witness for $f'$.
\end{proof}

\subsubsection{Proof of Proposition \ref{proposition:inseparability-witness}}
\label{proof:proposition:inseparability-witness}
\begin{proof}
In one direction, let $(k, \underline{k'}, \underline{\xi})$ be an inseparability witness for $f$.
If $\Theta' \leq \Theta$ is causally complete, the by Theorem~3.37 (p.~51) of \cite{gogioso2022combinatorics} there exists an $\omega_k \in \dom{k}$ such that $h_{\omega_k}:=\restrict{k}{\dom{k}\backslash\{\omega_k\}} \in \Ext{\Theta'}$, because Lemma \ref{lemma:inseparability-witness-1} forces $|\dom{k}| \geq 2$.
Then $f$ cannot arise from some $f' \in \CausFun{\Theta', \underline{O}}$: if it did, Lemma \ref{lemma:inseparability-witness-2} would imply that $(k, \underline{k'}, \underline{\xi})$ is an inseparability witness for $f'$, and Lemma \ref{lemma:inseparability-witness-1} would in turn force $h_{\omega_k} \notin \Ext{\Theta'}$.

In the other direction, assume that $f \in \CausFun{\Theta, \underline{O}}$ does not have an inseparability witness.
For each $k \in \Ext{\Theta}$ and each $\omega \in \dom{k}$, we define $h_{k, \omega} := \restrict{k}{\dom{k}\backslash\{\omega\}}$.
The absence of an inseparability witness for $f$ means that for all $k \in \Ext{\Theta}$ there is a $\omega_k \in \dom{k}$ such that for all $k' \in \Ext{\Theta}$ with $h_{k,\omega_k} \leq k'$ and all $\xi \in \dom{k}\backslash\{\omega\}$ we have $\Ext{f}_\xi(k) = \Ext{f}_\xi(k')$.
This allows us to consistently extend the definition of $\Ext{f}$ to $h_{k, \omega_k}$ for all $k$:
\[
\Ext{f'}(h_{k, \omega_k})
:=
\bigwedge\suchthat{\Ext{f}(k')}{k' \in \Ext{\Theta}, h_{k, \omega_k} \leq k'}
\]
The $\vee$-closure of $\Theta\cup\suchthat{h_{k, \omega_k}}{k \in \Ext{\Theta}}$ is the space of extended input histories $\Ext{\Theta'}$ for a causally complete sub-space $\Theta' \leq \Theta$.
Furthermore, the definition of $\Ext{f'}(h_{k, \omega_k})$ above can be completed---by gluing over compatible joins---to that of an extended causal function $f' \in \CausFun{\Theta', \underline{O}}$ such that $f$ arises from $f'$, proving that $f$ is separable.
\end{proof}

\subsubsection{Proof of Proposition \ref{proposition:histconstr-eqrel}}
\label{proof:proposition:histconstr-eqrel}
\begin{proof}
The definition of $\histconstr{\omega}{h}{h'}$ makes it immediate to see that the binary relation $\histconstrSym{\omega}$ is reflexive and symmetric.
It is also easy to see that it is transitive:
\[
    \hat{F}(h)_\omega = \hat{F}(h')_\omega
    \text{ and }
    \hat{F}(h')_\omega = \hat{F}(h'')_\omega
    \Rightarrow
    \hat{F}(h)_\omega = \hat{F}(h'')_\omega
\]
Given two $h, h' \in \Theta$ with $\omega$ as a tip event, the existence of a $k \in \Ext{\Theta}$ such that $h, h' \leq k$ forces all extended functions $\hat{F}$ satisfying the consistency condition to have $\hat{F}(h)_\omega = \hat{F}(h')_\omega$:
\[
\hat{F}(h)_\omega
\stackrel{cons. cond.}{=}
\hat{F}(k)_\omega
\stackrel{cons. cond.}{=}
\hat{F}(h')_\omega
\]
This shows that $R_\omega\!\subseteq\, \histconstrSym{\omega}$, and hence that $\bar{R}_\omega\!\subseteq\, \histconstrSym{\omega}$.

We now wish to show that $\histconstrSym{\omega}\subseteq\bar{R}_\omega$; equivalently, for all $h,h' \in \TipHists{\Theta}{\omega}$, we wish to show that $(h, h') \notin \bar{R}_\omega$ implies $\nothistconstr{\omega}{h}{h'}$.
Consider any event $\omega \in \Events{\Theta}$ and an input history $h\in \TipHists{\Theta}{\omega}$.
Let $\hat{F} \in \ExtFun{\Theta, \{0,1\}}$ be the extended function defined as follows:
\[
\hat{F}(k)_\xi
:=
\left\{
\begin{array}{l}
1 \text{ if } \xi=\omega \text{ and } \exists h' \text{ s.t. }\!\left[(h, h') \in \bar{R}_\omega \text{ and } h' \leq k\right]\\
0 \text{ otherwise}
\end{array}
\right.
\]
We show that $\hat{F}$ satisfies the consistency condition.
By construction, any violation $\hat{F}(k') \not\leq \hat{F}(k)$ for $k' \leq k$ of the consistency condition would necessarily stem from an inequality $0=\hat{F}(k')_\omega \neq \hat{F}(k)_\omega=1$ where:
\[
\begin{array}{l}
\exists h' \text{ s.t. }\!\left[(h, h') \in \bar{R}_\omega \text{ and } h' \leq k\right]\\
\not\exists h' \text{ s.t. }\!\left[(h, h') \in \bar{R}_\omega \text{ and } h' \leq k'\right]
\end{array}
\]
However, $\omega \in \dom{k'}$ would imply the existence of some $h'' \in \TipHists{\Theta}{\omega}$ such that $h'' \leq k'$: $k' \leq k$ would in turn imply that $h'' \leq k$, hence $(h', h'') \in \bar{R}_\omega$, hence $(h, h'') \in \bar{R}_\omega$, contradicting our previous statement about $k'$.
Since $\hat{F}$ satisfies the consistency condition, it provides a witness that $\nothistconstr{\omega}{h}{h'}$ for all $h' \in \TipHists{\Theta}{\omega}$ such that $(h, h') \notin \bar{R}_\omega$: this proves that $\histconstrSym{\omega}\subseteq\bar{R}_\omega$, and hence that $\histconstrSym{\omega}=\bar{R}_\omega$.
\end{proof}

\subsubsection{Proof of Corollary \ref{corollary:tightness-constrained-histories}}
\label{proof:corollary:tightness-constrained-histories}
\begin{proof}
For each $k \in \Ext{\Theta}$ and each $\omega \in \dom{k}$, the tightness requirement is exactly that there is a unique $h \in \TipHists{\Theta}{\omega}$ with $h \leq k$.
\end{proof}

\subsubsection{Proof of Theorem \ref{theorem:ext-f-characterisation}}
\label{proof:theorem:ext-f-characterisation}
\begin{proof}
The proof is essentially the same as for Theorem \ref{theorem:ext-f-characterisation-tight-cc}: we only need to check that nothing goes wrong when dropping causal completeness and tightness.
The definition of extended causal function has been amended to accommodate the more general definition of causal functions on causally incomplete spaces, potentially involving outputs at multiple tip events: the proof of Theorem \ref{theorem:ext-f-characterisation-tight-cc} straightforwardly extends to this more general setting (by changing $\tip{\Theta}{h_{k,\omega}} = \omega$ to $\omega \in \tips{\Theta}{h_{k, \omega}}$).
The definition of extended causal function has also been amended to accommodate the case of non-tight spaces, by letting $h_{k,\omega}$ be any one of the $h \in \TipHists{\Theta}{\omega}$ such that $h \leq k$: since $f(h_{k, \omega})$ has the same value for all choices, the original proof of Theorem \ref{theorem:ext-f-characterisation-tight-cc} goes through unaltered in this respect.
\end{proof}

\subsubsection{Proof of Proposition \ref{proposition:ext-f-gluing}}
\label{proof:proposition:ext-f-gluing}
\begin{proof}
By Proposition \ref{proposition:consistency-gluing-conditions}, the consistency condition is equivalent to the gluing condition: for consistent $k, k' \in \Ext{\Theta}$, we have $\Ext{f}(k\vee k') = \Ext{f}(k) \vee \Ext{f}(k')$.
Writing an arbitrary extended input history $k$ as the compatible join of the input histories below it yields our desired expression for $\Ext{f}$.
\end{proof}

\subsubsection{Proof of Proposition \ref{proposition:causal-function-joint-function}}
\label{proof:proposition:causal-function-joint-function}
\begin{proof}
The proof is essentially the same as for Proposition \ref{proposition:causal-function-joint-function-tight-cc}: we only need to check that nothing goes wrong when dropping causal completeness and tightness.
The proof for the first part of the statement only makes use of the free-choice condition, which we have now explicitly required, so it goes through unaltered.
The second part of the statement was modified to account for the more general definition of causal functions on causally incomplete spaces, potentially involving outputs at multiple tip events.
Aside from this modification, and the explicit assumption of the free-choice condition, the proof for the second part of the statement also goes through unaltered.
\end{proof}

\subsubsection{Proof of Theorem \ref{theorem:caus-fun-parallel-composition-factorisation}}
\label{proof:theorem:caus-fun-parallel-composition-factorisation}
\begin{proof}
Because event sets are disjoint, we have that $\histconstr{\omega}{h}{h'}$ implies either $h,h' \in \Theta$ or $h,h' \in \Theta'$.
This allows any constraints associated with lack of tightness to be factored over the two spaces, so we simply indicate them by $\text{...c'ts...}$ below.
We get:
\[
\begin{array}{rcl}
    &&\CausFun{\Theta \cup \Theta', \underline{O}\vee\underline{O}'}
    \\
    &=&
    \suchthat{\!\!
            f \in \prod\limits_{h \in \Theta\cup\Theta'}
            \prod\limits_{\omega \in \tips{\Theta\cup\Theta'}{h}}
            \!\!\!O_{\omega}
        }
        {\text{...c'ts...}}
    \\
    &=&
    \suchthat{\!\!
            f
            \in
            \left(
            \prod\limits_{h \in \Theta}
            \prod\limits_{\omega \in \tips{\Theta}{h}}
            \!\!\!O_{\omega}
            \right)
            \times
            \left(
            \prod\limits_{h \in \Theta'}
            \prod\limits_{\omega \in \tips{\Theta'}{h}}
            \!\!\!O_{\omega}
            \right)
        }
        {\text{...c'ts...}}
    \\
    &=&
    \suchthat{\!\!
            f
            \in
            \prod\limits_{h \in \Theta}
            \prod\limits_{\omega \in \tips{\Theta}{h}}
            \!\!\!O_{\omega}
        }
        {\text{...c'ts...}}
    \times
    \suchthat{\!\!
            f
            \in
            \prod\limits_{h \in \Theta'}
            \prod\limits_{\omega \in \tips{\Theta'}{h}}
            \!\!\!O_{\omega}
        }
        {\text{...c'ts...}}
    \\
    &=&
    \CausFun{\Theta, \underline{O}}
    \times
    \CausFun{\Theta', \underline{O}'}
\end{array}
\]
This concludes our proof.
\end{proof}

\subsubsection{Proof of Theorem \ref{theorem:caus-fun-sequential-composition-factorisation}}
\label{proof:theorem:caus-fun-sequential-composition-factorisation}
\begin{proof}
Because event sets are disjoint, we have that $\histconstr{\omega}{h}{h'}$ implies either that $h,h' \in \Theta$ or that $h=k\vee h''$ and $h' = k' \vee h'''$, where $h'',h''' \in \Theta'$ and $k,k' \in \max\Ext{\Theta}$, and the following two observations further imply that $k=k'$:
\begin{itemize}
    \item We can never have $\histconstr{\omega}{a \vee b}{c}$ for $a \in \max\Ext{\Theta}$, $b \in \Theta'$ and $c \in \Theta$;
    \item any $a \vee b$ and $a' \vee b'$ for distinct $a, a' \in \max\Ext{\Theta}$ and any $b, b' \in \Theta'$ are necessarily incompatible, and hence they cannot share a common $c \in \Ext{\Theta \seqcomposeSym \Theta'}$ such that $a \vee b \leq c$ and $a' \vee b' \leq c$.
\end{itemize}
This allows any constraints associated with lack of tightness to be factored over the two spaces, so we simply indicate them by $\text{...c'ts...}$ below.
We get:
\[
    \fl\begin{array}{rcl}
    &&\CausFun{\Theta \seqcomposeSym \Theta', \underline{O}\vee\underline{O}'}
    \\
    &=&
    \suchthat{\!\!
            f \in \prod\limits_{h \in \Theta\cup\left(\max\Ext{\Theta}\vee\Theta'\right)}
            \prod\limits_{\omega \in \tips{\Theta\seqcomposeSym\Theta'}{h}}
            \!\!\!O_{\omega}
        }
        {\text{...c'ts...}}
    \\
    &=&
    \suchthat{\!\!
            f
            \in
            \left(
            \prod\limits_{h \in \Theta}
            \prod\limits_{\omega \in \tips{\Theta}{h}}
            \!\!\!O_{\omega}
            \right)
            \times
            \left(
            \prod\limits_{h \in \max\Ext{\Theta}\vee\Theta'}
            \prod\limits_{\omega \in \tips{\Theta\seqcomposeSym\Theta'}{h}}
            \!\!\!O_{\omega}
            \right)
        }
        {\text{...c'ts...}}
    \\
    &=&
    \suchthat{\!\!
            f
            \in
            \prod\limits_{h \in \Theta}
            \prod\limits_{\omega \in \tips{\Theta}{h}}
            \!\!\!O_{\omega}
        }
        {\text{...c'ts...}}
    \times
    \suchthat{\!\!
            f
            \in
            \prod\limits_{h \in \max\Ext{\Theta}\vee\Theta'}
            \prod\limits_{\omega \in \tips{\Theta\seqcomposeSym\Theta'}{h}}
            \!\!\!O_{\omega}
        }
        {\text{...c'ts...}}
    \\
    &=&
    \suchthat{\!\!
            f
            \in
            \prod\limits_{h \in \Theta}
            \prod\limits_{\omega \in \tips{\Theta}{h}}
            \!\!\!O_{\omega}
        }
        {\text{...c'ts...}}
    \times
    \prod\limits_{k \in {\max\Ext{\Theta}}}
    \suchthat{\!\!
            f
            \in
            \prod\limits_{h \in \Theta'}
            \prod\limits_{\omega \in \tips{\Theta'}{h}}
            \!\!\!O_{\omega}
        }
        {\text{...c'ts...}}
    \\
    &=&
    \CausFun{\Theta, \underline{O}}
    \times
    \CausFun{\Theta', \underline{O}'}^{\max\Ext{\Theta}}
\end{array}
\]
This concludes our proof.
\end{proof}

\subsubsection{Proof of Theorem \ref{theorem:caus-fun-conditional-sequential-composition-factorisation}}
\label{proof:theorem:caus-fun-conditional-sequential-composition-factorisation}
\begin{proof}
The proof is entirely analogous to that of the previous Proposition, which it generalises: replace $\Theta'$ with $\Theta'_k$ and replace $\max\Ext{\Theta}\vee\Theta'$ with $\bigcup_{k \in \max\Ext{\Theta}}\{k\}\vee\Theta'_k$.
\end{proof}

\subsubsection{Proof of Proposition \ref{proposition:lowerset-topology-continuity}}
\label{proof:proposition:lowerset-topology-continuity}
\begin{proof}
In one direction, let $f: X \rightarrow Y$ be order-preserving and consider $\lambda \in \Lsets{Y}$.
If $f(x) \in \lambda$ and $x' \leq x$, then $f(x') \leq f(x)$ and thus $f(x') \in \lambda$.
Hence, $f^{-1}(\lambda) \in \Lsets{X}$.
In the other direction, let $f: X \rightarrow Y$ be continuous and consider $x, x' \in X$ such that $x' \leq x$.
Because of continuity, $f^{-1}{\downset{f(x)}} \in \Lsets{X}$, so that $x' \in f^{-1}{\downset{f(x)}}$ and hence $f(x') \in \downset{f(x)}$, i.e. $f(x') \leq f(x)$.
\end{proof}

\subsubsection{Proof of Proposition \ref{proposition:lowerset-topology-subset}}
\label{proof:proposition:lowerset-topology-subset}
\begin{proof}
The subspace topology on $X$ consists of $\lambda \cap X$ for all open sets $\lambda$ in $Y$, i.e. for all $\lambda \in \Lsets{Y}$.
The induced order on $X$ is defined by setting $x' \leq x$ in $X$ if and only if $x' \leq x$ in $Y$, for all $x, x' \in X$.
The downset $\downset{x}_X$ in $X$ for $x \in X$ is exactly the intersection of $\downset{x}_Y$ with $X$:
\[
\downset{x}_X
=
\suchthat{x' \in X}{x' \leq_X x}
=
\suchthat{x' \in Y}{x' \leq_Y x, x' \in X}
=
\downset{x}_Y \cap X
\]
A lowerset $\nu \in \Lsets{X}$ is the union of the downsets of its elements, and hence it is the intersection with $X$ of a lowerset in $Y$
\[
\nu
=
\bigcup_{x \in \nu} \downset{x}_X
=
\bigcup_{x \in \nu} \left(\downset{x}_Y \cap X\right)
=
\left(\bigcup_{x \in \nu} \downset{x}_Y\right) \cap X
\]
Conversely, the intersection with $X$ of a lowerset $\lambda \in \Lsets{Y}$ is a lowerset in $X$:
\[
\lambda \cap X
=
\left(\bigcup_{x \in \lambda} \downset{x}_Y\right) \cap X
=
\bigcup_{x \in \lambda} \left(\downset{x}_Y \cap X\right)
=
\bigcup_{x \in \lambda \cap X} \left(\downset{x}_X\right)
\]
As a consequence, the lowersets in $X$ for the induced order are exactly the subsets in the subspace topology.
\end{proof}

\subsubsection{Proof of Proposition \ref{proposition:lowerset-topology-specialisation-preorder}}
\label{proof:proposition:lowerset-topology-specialisation-preorder}
\begin{proof}
Write $\preceq$ for the lowerset specialisation preorder induced by the topology $\Lsets{X}$.
\[
x' \leq x
\Leftrightarrow
\downset{x'} \subseteq \downset{x}
\Leftrightarrow
\left(\forall \lambda \in \Lsets{X}.\;x \in \lambda \Rightarrow x' \in \lambda\right)
\Leftrightarrow
x' \preceq x
\]
where the downsets are taken with respect to $\leq$ and the middle $\Leftrightarrow$ follows from the fact that $x \in \lambda \Leftrightarrow \downset{x} \subseteq \lambda$ for all lowersets $\lambda \in \Lsets{X}$.
\end{proof}

\subsubsection{Proof of Theorem \ref{theorem:function-causal-is-continuous}}
\label{proof:theorem:function-causal-is-continuous}
\begin{proof}
An extended function $\hat{F}$ is continuous if and only if it is order-preserving, i.e. if and only if $k' \leq k$ implies $\hat{F}(k') \leq \hat{F}(k)$. This is exactly the consistency condition from Definition \ref{definition:consistency-condition}, which is equivalent to causality of $\hat{F}$ by Theorem \ref{theorem:ext-f-characterisation}.
\end{proof}

\subsubsection{Proof of Proposition \ref{proposition:lowersets-subspaces}}
\label{proof:proposition:lowersets-subspaces}
\begin{proof}
Taking subsets does not introduce new input histories, so that any partial function in $\lambda \subseteq \Theta$ which was $\vee$-prime in $\Theta$ is also $\vee$-prime in $\lambda$.
We have that $\Ext{\lambda}\subseteq \Ext{\Theta}$ by definition of $\ExtSym$, and $\Ext{\lambda}$ is a lowerset because $\lambda$ is a lowerset.
\end{proof}

\subsubsection{Proof of Proposition \ref{proposition:lowerset-cc-tight}}
\label{proof:proposition:lowerset-cc-tight}
\begin{proof}
Because $\lambda$ is a lowerset, all sub-histories in $\Theta$ of a history $h \in \lambda$ are also sub-histories of $h$ in $\lambda$.
This means that the tip events of $h$ in $\Theta$ are the same as the tip events of $h$ in $\lambda$, and in particular that $\lambda$ is causally complete whenever $\Theta$ is.
In turn, this means that any $h \in \lambda$ such that $h \leq k$ and $\omega \in \tips{\lambda}{h}$ is also a $h \in \Theta$ such that $h \leq k$ and $\omega \in \tips{\Theta}{h}$, for every $k \in \Ext{\lambda}\subseteq\Ext{\Theta}$.
This immediately implies that $\lambda$ is tight whenever $\Theta$ is, and it also implies that $h, h' \in \lambda$ are constrained at $\omega$ in $\lambda$ only if they are in $\Theta$.
\end{proof}

\subsubsection{Proof of Corollary \ref{corollary:caus-fun-restriction-is-caus}}
\label{proof:corollary:caus-fun-restriction-is-caus}
\begin{proof}
If $f \in \CausFun{\Theta, \underline{O}}$, then we certainly have the following, where we used the fact that $\tips{\lambda}{h} = \tips{\Theta}{h}$ from Proposition \ref{proposition:lowerset-cc-tight}:
\[
\restrict{f}{\lambda}
\in
\prod_{h \in \lambda}
\prod_{\omega \in \tips{\lambda}{h}}
O_\omega
\]
Proposition \ref{proposition:lowerset-cc-tight} also states that $\histconstr{\omega}{h}{h'}$ in $\lambda$ implies $\histconstr{\omega}{h}{h'}$ in $\Omega$, which in turn implies that $f(h)_\omega=f(h')_\omega$, because $f \in \CausFun{\Theta, \underline{O}}$. Hence we have $\restrict{f}{\lambda} \in \CausFun{\lambda, \underline{O}}$.
\end{proof}

\subsubsection{Proof of Theorem \ref{theorem:compatibility-caus-fun-restrictions}}
\label{proof:theorem:compatibility-caus-fun-restrictions}
\begin{proof}
$\CausFun{\Lsets{\Theta}, \underline{O}}\left(\lambda\right)$ is well-defined by Proposition \ref{proposition:lowersets-subspaces}, while the restrictions are well-defined by Corollary \ref{corollary:caus-fun-restriction-is-caus}.
Because restrictions are ordinary function restrictions, a family of functions is compatible if any pair of functions in the family agree on the intersections of their domains, agreeing with Definition~3.3 (p.22) of \cite{gogioso2022combinatorics}.
A gluing for a family of compatible functions, if it exists, must be a function defined on the union of their domains, which agrees with each function on its domain: the only option is the compatible join according to Definition~3.4 (p.22) of \cite{gogioso2022combinatorics}.
If the compatible join is causal, then it is the gluing of the family; otherwise, no gluing can exist.
\end{proof}

\subsubsection{Proof of Lemma \ref{lemma:deterministic-causal-contextuality-witness-sub-ext-hist}}
\label{proof:lemma:deterministic-causal-contextuality-witness-sub-ext-hist}
\begin{proof}
Let $(k, \omega, h, h')$ is a solipsistic contextuality witness for $\Theta$ and let $k' \in \Ext{\Theta}$ be such that $h, h' \leq k' \leq k$.
Because $h \leq k'$ and $\omega \in \dom{h}$, we have $\omega \in \dom{k'}$.
If there existed $h, h' \leq h'' \leq k'$, then we'd also have $h, h' \leq h'' \leq k$, but this is impossible since $(k, \omega, h, h')$ is a solipsistic contextuality witness.
Hence $(k', \omega, h, h')$ is also a solipsistic contextuality witness.
\end{proof}

\subsubsection{Proof of Proposition \ref{proposition:causal-contextuality-characterisation}}
\label{proof:proposition:causal-contextuality-characterisation}
\begin{proof}
In one direction, assume that $\Theta$ admits solipsistic contextuality and let $(k, \omega, h, h')$ be a witness.
We have:
\[
\min\suchthat{k \in \Ext{\Theta}}{h, h' \leq k}\cap \downset{k}
\subseteq
\min\suchthat{k \in \Ext{\Theta}}{h, h' \leq k}
\]
If $k' \in \min\suchthat{k \in \Ext{\Theta}}{h, h' \leq k}\cap \downset{k}$, then by the previous Proposition we have that $(k', \omega, h, h')$ is a contextuality witness, which in turn implies that $k' \notin \Theta$ and hence that $\min\suchthat{k \in \Ext{\Theta}}{h, h' \leq k} \not\subseteq \Theta$.
In the other direction, $\Theta$ satisfy the condition above.
By the definition of $\histconstrSym{\omega}$, we can without loss of generality assume that $h, h' \leq k$ for some $k \in \Ext{\Theta}$.
We can also restrict our attention to a minimal such $k$ and pick one which is not in $\Theta$:
\[
k \in \min\suchthat{k' \in \Ext{\Theta}}{h, h' \leq k'}\backslash\Theta
\]
By minimality of $k$, and since $k \notin \Theta$, there can be no $h'' \in \Theta$ such that $h, h' \leq h'' \leq k$: this means that $(k, \omega, h, h')$ is a solipsistic contextuality witness.
\end{proof}

\subsubsection{Proof of Theorem \ref{theorem:causal-contextuality-sheaf-condition}}
\label{proof:theorem:causal-contextuality-sheaf-condition}
\begin{proof}
If $\Theta$ is tight then so is $\lambda$ for all $\lambda \in \Lsets{\Theta}$, by Proposition \ref{proposition:lowerset-cc-tight}: causal functions are unconstrained ordinary function, compatibility is the usual notion for partial functions and the unique gluing of a compatible family is obtained by compatible join.
If $\Theta$ is non-tight, what can go wrong is that for some family $\mathcal{F}$ the compatible join $\hat{f} := \bigvee \mathcal{F}$ has $\hat{f}(h)_\omega \neq \hat{f}(h')_\omega$, for some distinct $h, h' \in \dom{\hat{f}}$ such that $\histconstr{\omega}{h}{h'}$ on the lowerset $\dom{\hat{f}}$.
Because of the way $\histconstr{\omega}{h}{h'}$ is defined, we can restrict our attention to distinct $h, h'$ such that $\omega \in \tips{\downset{h}}{h} \cap \tips{\downset{h'}}{h'}$ and $h, h' \leq k$ for some $k \in \Ext{\dom{\hat{f}}}$.

If every such $k$ always has a $h'' \leq k$ with $h, h' \leq h'' \leq k$, then all compatible families must compatibly restrict to a causal function $\tilde{f}$ on $\downset{h''}$, the causality of which forces $\tilde{f}(h)_\omega=\tilde{f}(h')_\omega$; this is exactly the case when $\min\suchthat{k \in \Ext{\Theta}}{h, h' \leq k} \subseteq \Theta$.
If, on the other hand, we can find a $k \in \min\suchthat{k' \in \Ext{\Theta}}{h, h' \leq k'}$ with $k \notin \Theta$, then we can explicitly construct a compatible family $\mathcal{F}$ which admits no gluing, generalising the construction used in the example of non-tight space $\Theta_3$.
Specifically, consider consider the following two lowersets:
\[
\begin{array}{rcl}
\lambda
&:=&
\downset{h}
\\
\lambda'
&:=&
\suchthat{h'' \in \Theta}{h'' \leq k, h \nleq h''}
\end{array}
\]
We have $h \in \lambda \backslash \lambda'$, $h' \in \lambda'\backslash\lambda$ and $k \in \Ext{\lambda \cup \lambda'}$.

For each $\omega \in \Events{\Theta}$, fix distinct $o_{\omega,0}, o_{\omega,1} \in O_\omega$.
We define $f' \in \CausFun{\lambda', \underline{O}}$ to be the constant function which takes the value $o_{\omega,0}$ for all input histories in $\lambda'$ and all tip events $\omega$.
We define $f \in \CausFun{\lambda, \underline{O}}$ as follows:
\[
f(h'')_{\xi}
:=
\left\{
\begin{array}{rl}
o_{\xi,1}&\text{ if }h''=h\text{ and }\xi=\omega\\
o_{\xi,0}&\text{ otherwise}
\end{array}
\right.
\]
Then $(f, f')$ forms a compatible family on $(\lambda, \lambda')$: if their gluing exists, then it must be the join $\hat{f} := f \vee f'$ on $\lambda \cup \lambda'$.
However, $\hat{f}(h)_\omega \neq \hat{f}(h')_\omega$ and $\histconstr{\omega}{h}{h'}$ on $\lambda \cup \lambda'$, because $k \in \Ext{\lambda \cup \lambda'}$.
Hence $\hat{f} \notin \CausFun{\lambda, \{0,1\}}$ and the compatible family $(f, f')$ does not admit a gluing.
\end{proof}

\subsubsection{Proof of Proposition \ref{proposition:presheaves-caus-fun-isomorphic}}
\label{proof:proposition:presheaves-caus-fun-isomorphic}
\begin{proof}
It suffices to prove that the equality above holds for all $\lambda \in \Lsets{\Theta}$: both well-definition of restrictions and natural isomorphism between the presheaves follow from it, using Proposition \ref{proposition:lowersets-subspaces} to extend the equation to all inclusions $\lambda'$ for $\lambda' \subseteq \lambda$ (which is equivalent to $\lambda' \in \Lsets{\lambda}$).
The separated presheaf/sheaf claims follow in turn from natural isomorphism (by Observation \ref{observation:natural-isomorphism-compatible-families}).
To prove the equation, we start by observing that the restriction $\restrict{\Ext{f}}{\Ext{\lambda}}$ is well-defined, because $\Ext{\lambda} \subseteq \Ext{\Theta}$ by Proposition \ref{proposition:lowersets-subspaces}.
We show that the functions on the two sides of the equation give the same result when applied to a generic extended input history $k \in \Ext{\lambda}$.
On the left hand side of the equation, we have the following:
\[
\restrict{\Ext{f}}{\Ext{\lambda}}(k)
=
\Ext{f}(k)
\]
On the right hand side of the equation, we have the following:
\[
\Ext{\restrict{f}{\lambda}}(k)
=
\bigvee_{h \in \downset{k} \cap \lambda}
\restrict{f}{\lambda}(h)
=
\bigvee_{h \in \downset{k}}
f(h)
=
\Ext{f}(k)
\]
The first and last equalities are by Proposition \ref{proposition:ext-f-gluing}, while the middle equality follows from the fact that $\lambda$ is a lowerset, so that $k \in \lambda$ implies $\downset{k} \subseteq \lambda$. 
\end{proof}

\subsubsection{Proof of Lemma \ref{lemma:distr-monad-bij}}
\label{proof:lemma:distr-monad-bij}
\begin{proof}
If $\psi \circ \phi = \id{X}$, then we have:
\[
\Dist{\psi} \circ \Dist{\phi}
=
\Dist{\psi \circ \phi}
=
\Dist{\id{X}}
=
\id{\Dist{X}}
\]
If $\phi \circ \psi = \id{Y}$, then we have:
\[
\Dist{\phi} \circ \Dist{\psi}
=
\Dist{\phi \circ \psi}
=
\Dist{\id{Y}}
=
\id{\Dist{Y}}
\]
\end{proof}

\subsubsection{Proof of Proposition \ref{proposition:presheaf-dist-presheaf}}
\label{proof:proposition:presheaf-dist-presheaf}
\begin{proof}
There are three properties to check for $\DistSym P$ to be a presheaf, each one an immediate consequence of a corresponding property of $\DistSym$ as a functor:
\begin{enumerate}
    \item $\DistSym P(U, V): \DistSym P(U) \rightarrow \DistSym P(V)$
    \item $\DistSym P(U, U) = \id{\DistSym P(U)}$
    \item $\DistSym P(V, W) \circ \DistSym P(U, V) = \Dist{P(V, W)\circ P(U, V)} = \DistSym P(U, W)$
\end{enumerate}
\end{proof}

\subsubsection{Proof of Proposition \ref{proposition:presheaf-dist-presheaf-natural-isomorphism}}
\label{proof:proposition:presheaf-dist-presheaf-natural-isomorphism}
\begin{proof}
Because $\phi_U: P(U) \rightarrow P'(U)$ is a bijection, so is its image $\Dist{\phi_U}$ under the distribution monad:
\[
\begin{array}{rcl}
\Dist{\phi_U^{-1} \circ \phi_U}
&=&
\Dist{\id{P(U)}}
=
\id{\DistSym P(U)}
\\
\Dist{\phi_U \circ \phi_U^{-1}}
&=&
\Dist{\id{P'(U)}}
=
\id{\DistSym P'(U)}
\end{array}
\]
Furthermore, the bijections commute with restrictions:
\[
\begin{array}{rl}
\Dist{\phi_V} \circ \DistSym P(U, V)
&=
\Dist{\phi_V \circ P(U,V)}
\\
&=
\Dist{P(U,V) \circ \phi_U}
=
\DistSym P(U, V) \circ \Dist{\phi_U}
\end{array}
\]
We conclude that $\DistSym\phi$ is a natural isomorphism between $\DistSym P$ and $\DistSym P'$.
\end{proof}

\subsubsection{Proof of Proposition \ref{proposition:causal-dist-marginal}}
\label{proof:proposition:causal-dist-marginal}
\begin{proof}
From the definition of $\DistSym$ on functions, we get:
\[
\Dist{\Ext{f} \mapsto \restrict{\Ext{f}}{\Ext{\lambda'}}}
=
d \mapsto
\sum_{\Ext{f}}
d(\Ext{f})\;\delta_{\restrict{\Ext{f}}{\Ext{\lambda'}}}
\]
The sum is in terms of $\Ext{f} \in \ExtCausFun{\lambda, \underline{O}}$, but we can rewrite it in terms of $\Ext{f'} \in \ExtCausFun{\lambda', \underline{O}}$:
\[
\sum_{\Ext{f}}
d(\Ext{f})\;\delta_{\restrict{\Ext{f}}{\Ext{\lambda'}}}
=
\sum_{\Ext{f'}}
\left(
\sum_{\Ext{f} \text{ s.t. }\restrict{\Ext{f}}{\Ext{\lambda'}} = \Ext{f'}}
\hspace{-10mm}
d(\Ext{f})
\right)
\delta_{\Ext{f'}}
\]
To improve concision, we use $\restrict{f}{\lambda'} = f'$ as an equivalent condition for the restriction $\restrict{\Ext{f}}{\Ext{\lambda'}} = \Ext{f'}$, according to Equation \ref{equation:presheaves-caus-fun-isomorphic} from Proposition \ref{proposition:presheaves-caus-fun-isomorphic}:
\[
\sum_{\Ext{f'}}
\left(
\sum_{\Ext{f} \text{ s.t. }\restrict{\Ext{f}}{\Ext{\lambda'}} = \Ext{f'}}
\hspace{-15mm}
d(\Ext{f})
\right)
\delta_{\Ext{f'}}
=
\sum_{\Ext{f'}}
\left(
\sum_{f \text{ s.t. } \restrict{f}{\lambda'}=f'}
\hspace{-5mm}
d(\Ext{f})
\right)
\delta_{\Ext{f'}}
\]
\end{proof}

\subsubsection{Proof of Proposition \ref{proposition:solipsistic-standard-global-hierarchy}}
\label{proof:proposition:solipsistic-standard-global-hierarchy}
\begin{proof}
Every $\lambda \in \Lsets{\Theta}$ satisfies $\lambda \subseteq \Theta$, so we have $\mathcal{C} \preceq \ClsCov{\Theta}$ for all covers $\mathcal{C} \in \Covers{\Theta}$.
For every $\mathcal{C} \in \Covers{\Theta}$ and every $h \in \Theta$, there must exist some $\lambda_h \in \mathcal{C}$ such that $h \in \lambda$, i.e. $\downset{h} \subseteq \lambda$: this is in particular true for all maximal input histories $h$, hence $\SolCov{\Theta} \preceq \mathcal{C}$.
\end{proof}

\subsubsection{Proof of Proposition \ref{proposition:deterministic-empirical-models}}
\label{proof:proposition:deterministic-empirical-models}
\begin{proof}
Let $f_\lambda$ be defined by $e_\lambda = \delta_{\Ext{f_\lambda}}$ for all $\lambda \in \mathcal{C}$.
It suffices to prove that $\restrict{f_\lambda}{\lambda'} = f_{\lambda'}$.
Indeed, compatibility of $e$ implies that:
\[
\left(
\Ext{f'}
\mapsto
\sum_{\restrict{f}{\lambda'}=f'}
\left\{
\begin{array}{rl}
1 &\text{ if } f = f_\lambda\\
0 &\text{ otherwise}
\end{array}
\right\}
\right)
=
\restrict{e_\lambda}{\lambda'}
= e_{\lambda'} = \delta_{\Ext{f_{\lambda'}}}
\]
This implies that $f$ is a compatible family and that the correspondence respects restrictions.
Finally, the globally deterministic empirical models are exactly the restrictions of the deterministic empirical models on the classical cover, which correspond exactly to compatible family of causal functions on the classical cover, which are just the causal functions $f_\Theta \in \CausFun{\Theta, \underline{O}}$ (more precisely, they are the singleton families $(f_\lambda)_{\lambda \in \{\Theta\}}$, but this is an unnecessary technicality).
\end{proof}

\subsubsection{Proof of Corollary \ref{corollary:deterministic-empirical-models-noncontextual}}
\label{proof:corollary:deterministic-empirical-models-noncontextual}
\begin{proof}
In one direction, assume that $\CausFun{\Lsets{\Theta}, \underline{O}}$ is a sheaf and let $e=\left(e_{\downset{h}}\right)_{h \in \max\Theta}$ be a solipsistic empirical model for $\Theta$.
If $e=(e_\lambda)_{\lambda \in \mathcal{C}}$ is deterministic, then by Proposition \ref{proposition:deterministic-empirical-models} there exists a compatible family $f$ for $\CausFun{\Lsets{\Theta}, \underline{O}}$ such that $e_\lambda = \delta_{\Ext{f_\lambda}}$ for all $\lambda \in \mathcal{C}$.
Because $\CausFun{\Lsets{\Theta}, \underline{O}}$ is a sheaf, this family can be glued to a causal function $\hat{f} \in \CausFun{\Theta, \underline{O}}$ such that $\restrict{\hat{f}}{\lambda} = f_\lambda$ for all $\lambda \in \mathcal{C}$.
Again by Proposition \ref{proposition:deterministic-empirical-models}, the function $\hat{f}$ corresponds to an empirical model $\hat{e}$ on the classical cover, such that $\restrict{\hat{e}}{\lambda} = e_\lambda$.
Hence, $e$ is globally deterministic.
By restricting $\hat{e}$ to the a standard empirical model $\hat{e}'$, and observing that $\hat{e}'$ further restricts to $e$, we also conclude that $e$ does not display solipsistic contextuality.
\end{proof}

\subsubsection{Proof of Theorem \ref{theorem:deterministic-empirical-models-2}}
\label{proof:theorem:deterministic-empirical-models-2}
\begin{proof}
Let $(k, \omega, h, h')$ be a solipsistic contextuality witness for $\Theta$ and let $o_{\omega,0}, o_{\omega,1} \in O_\omega$, $f_t \in \CausFun{\downset{t}, \underline{O}}$ and $e=\left(e_{\downset{t}}\right)_{t \in \max\Theta}$ be defined as in the statement of this Proposition.
There cannot exist a standard empirical model $\hat{e}$ such that $e = \restrict{\hat{e}}{\SolCov{\Theta}}$, because the component $\hat{e}_{\downset{k}}$ alone would fail the consistency requirement.
Indeed, we get the following restriction from the side of $h$:
\[
\left(
\Ext{f}
\mapsto
\sum_{\restrict{f'}{\downset{h}}=f}
\hat{e}_{\downset{k}}\!\left(\Ext{f'}\right)
\right)
=
\restrict{\hat{e}_{\downset{k}}}{\downset{h}}
=
e_{\downset{h}}
=
\delta_{\Ext{f_h}}
\]
This means that $\hat{e}_{\downset{k}}$ must be entirely supported by the $\Ext{f} \in \ExtCausFun{\downset{k}, \underline{O}}$ such that $\restrict{f}{\downset{h}} = f_h$, which implies $f(h)_\omega = o_{\omega,1}$.
Analogously, we get the following restriction from the side of $h'$:
\[
\left(
\Ext{f}
\mapsto
\sum_{\restrict{f'}{\downset{h'}}=f}
\hat{e}_{\downset{k}}\!\left(\Ext{f'}\right)
\right)
=
\restrict{\hat{e}_{\downset{k}}}{\downset{h'}}
=
e_{\downset{h'}}
=
\delta_{\Ext{f_{h'}}}
\]
This means that $\hat{e}_{\downset{k}}$ must also be entirely supported by the $\Ext{f} \in \ExtCausFun{\downset{k}, \underline{O}}$ such that $\restrict{f}{\downset{h'}} = f_{h'}$, which implies $f(h')_\omega = o_{\omega,0}$.
Because $\histconstr{\omega}{h}{h'}$, no $f \in \CausFun{\downset{k}, \underline{O}}$ can exist such that $f(h)_\omega = o_{\omega,1} \neq o_{\omega,0} = f(h')_\omega$, and hence the support of $\hat{e}$ would have to be empty.
This is impossible, so we conclude that there cannot exist a standard empirical model $\hat{e}$ such that $e = \restrict{\hat{e}}{\SolCov{\Theta}}$.
\end{proof}

\subsubsection{Proof of Proposition \ref{proposition:gluing-standard-function-compat-family}}
\label{proof:proposition:gluing-standard-function-compat-family}
\begin{proof}
It suffices to show that if $\histconstr{\omega}{h}{h'}$ then $f_{\downset{k}}(h)_\omega = f_{\downset{k'}}(h')_\omega$ for all $k, k' \in \max\ExtHist{\Theta}$ such that $h \leq k$ and $h' \leq k'$.
Let $\histconstr{\omega}{h}{h'}$ and $h, h' \leq k$ for some $k \in \max\ExtHist{\Theta}$: because $f_{\downset{k}} \in \CausFun{\downset{k}, \underline{O}}$, we necessarily have $f_{\downset{k}}(h)_\omega = f_{\downset{k}}(h')_\omega$.
Now let $\histconstr{\omega}{h}{h'}$, without further requirements.
By definition of $\histconstr{\omega}{h}{h'}$, there exist some $N \geq 1$ and some families $(h_j)_{j=0}^N$ and $(k_j)_{j=1}^N$ such that:
\begin{itemize}
    \item $h_0=h$, $h_N=h'$, $h_j \in \Theta$ and $\histconstr{\omega}{h_{j-1}}{h_j}$ for all $j=1,...,N$;
    \item $k_j \in \max\ExtHist{\Theta}$ such that $h_{j-1}, h_{j} \leq k_j$ for all $j=1,...,N$.
\end{itemize}
By the previous observation, we have $f_{\downset{k_j}}(h_{j-1})_\omega = f_{\downset{k_j}}(h_j)_\omega$ for all $j=1,...,N$.
By compatibility of $f$, we also have $f_{\downset{k_j}}(h_j)_\omega=f_{\downset{k_{j+1}}}(h_j)_\omega$ for all $j=1,...,N-1$.
Putting the two together, we have $f_{\downset{k_1}}(h)_\omega=f_{\downset{k_N}}(h')_\omega$.
Finally, let $k, k' \in \max\ExtHist{\Theta}$ be any maximal extended input histories such that $h \leq k$ and $h' \leq k'$.
By compatibility, we have $f_{\downset{k}}(h)_\omega=f_{\downset{k_1}}(h)_\omega$ and $f_{\downset{k_N}}(h')_\omega = f_{\downset{k'}}(h')_\omega$: together with the previously established $f_{\downset{k_1}}(h)_\omega = f_{\downset{k_N}}(h')_\omega$, this implies our desired $f_{\downset{k}}(h)_\omega=f_{\downset{k'}}(h')_\omega$.
\end{proof}

\subsubsection{Proof of Corollary \ref{corollary:standard-emp-model-deterministic-globally-deterministic}}
\label{proof:corollary:standard-emp-model-deterministic-globally-deterministic}
\begin{proof}
Let $e$ be deterministic, with $e_{\downset{k}} = \delta_{f_{\downset{k}}}$ for a compatible family $f = (f_{\downset{k}})_{k \in \max\ExtHist{\Theta}}$ in $\CausFun{\Lsets{\Theta}, \underline{O}}$.
By Proposition \ref{proposition:gluing-standard-function-compat-family}, the compatible family $f$ can be glued to a causal function $\hat{f} \in \CausFun{\Theta, \underline{O}}$.
By Proposition \ref{proposition:deterministic-empirical-models}, $\hat{f}$ corresponds to a deterministic empirical model $\hat{e}$ on the classical cover which restricts to $e$.
Hence $e$ is globally deterministic.
\end{proof}

\subsubsection{Proof of Theorem \ref{theorem:nogo-nonlocality-switch-spaces}}
\label{proof:theorem:nogo-nonlocality-switch-spaces}
\begin{proof}
We will use the observations made in the prologue to Theorem \ref{theorem:nogo-nonlocality-switch-spaces}, on page \pageref{prologue:nogo-nonlocality-switch-spaces}:
\begin{enumerate}
    \item The switch space $\Theta$ factorises as $\Theta = \Hist{\{\omega_1\}, \restrict{\underline{\Inputs{\Theta}}}{\{\omega_1\}}} \seqcomposeSym \underline{\Theta'}$.
    \item Each $\Theta_{i_1}$ is a switch space $\Theta_{i_1} \in \CSwitchSpaces{\restrict{\underline{I}}{E\backslash\{\omega_1\}}}$, so we have $\Theta_{i_1} = \Ext{\Theta_{i_1}}$.
    \item Causal functions on $\Theta$ recursively factorise as follows, where we define $\underline{O}' := \restrict{\underline{O}}{\Events{\Theta}\backslash\{\omega_1\}}$:
    \[
    \CausFun{\Theta, \underline{O}}
    \cong
    O^{\Inputs{\Theta}_{\omega_1}}
    \times
    \prod_{i_1 \in \Inputs{\Theta}_{\omega_1}}
    \CausFun{\Theta'_{i_1}, \underline{O}'}
    \]
    \item Causal functions in $\CausFun{\Theta, \underline{O}}$ take the following form, for an arbitrary choice of outputs $\underline{o} \in O_{\omega_1}^{\Inputs{\Theta}_{\omega_1}}$ and a family $\underline{f'}$ of causal functions $f'_{i_1} \in \CausFun{\Theta'_{i_1}, \underline{O}'}$:
    \[
    g_{\underline{o}, \underline{f'}}
    :=
    h
    \mapsto
    \left\{
    \begin{array}{rl}
    o_{i_1} &\text{ if } h=\{\omega_1:i_1\} \in \Hist{\{\omega_1\}, \restrict{\underline{\Inputs{\Theta}}}{\{\omega_1\}}}\\
    f'_{i_1}(h') &\text{ if } h=\{\omega_1:i_1\}\vee h' \text{ and } h' \in \Theta'_{i_1}
    \end{array}
    \right.
    \]
\end{enumerate}
For all $i_1 \in \Inputs{\Theta}_{\omega_1}$ and all $o_1 \in O_{\omega_1}$, we define the probability assigned to output $o_1$ by the empirical model $e$ on input $i_1$:
\[
p_{i_1, o_1} := \restrict{e}{\downset{\{\omega_1:i_1\}}}\left(\{\omega_1:o_1\}\right)
= \sum_{\underline{o}' \in \underline{O}'}
e_{\downset{\left(\{\omega_1:i_1\}\vee h'\right)}}\left(\{\omega_1:o_1\}\vee\underline{o}'\right)
\]
where $h'$ is an arbitrary choice of input history $h' \in \max\Theta'_{i_1}$, since they all result in the same marginal $p_{i_1, o_1}$.
In the base case $\Events{\Theta} = \{\omega_1\}$, the causal functions take the form $g_{\underline{o}, \emptyset} := \{\omega_1: i_1\} \mapsto o_{i_1}$ and the standard empirical model satisfies:
\[
e_{\downset{\{\omega_1:i_1\}}}\left(\{\omega_1:o_1\}\right)
=
\restrict{e}{\downset{\{\omega_1:i_1\}}}\left(\{\omega_1:o_1\}\right)
= p_{i_1, o_1}
\]
We define a classical empirical model $\hat{e}$ corresponding to the outputs for different inputs having independent probabilities:
\[
\hat{e}_{\Theta}\left(\Ext{g_{\underline{o}}}\right)
:=
\prod_{i_1 \in \Inputs{\Theta}_{\omega_1}}
e_{\downset{\{\omega_1: i_1\}}}\left(\{\omega_1: o_{i_1}\}\right)
\]
The classical empirical model $\hat{e}$ straightforwardly restricts to $e$ on the standard cover:
\[
\begin{array}{rcl}
\restrict{\hat{e}}{\downset{\{\omega_1: i_1\}}}\left(\{\omega_1: o_1\}\right)
&=& \sum\limits_{\underline{p}}
\hat{e}_{\Theta}\left(\Ext{g_{\{i_1:o_1\}\vee\underline{p}}}\right)
\\
&=&
p_{i_1, o_1}
\sum\limits_{\underline{q}}
\prod\limits_{j_1 \in \Inputs{\Theta}_{\omega_1}}
e_{\downset{\{\omega_1: j_1\}}}\left(\{\omega_1: p_{j_1}\}\right)
\\
&=&
p_{i_1, o_1}
\prod\limits_{j_1 \in \Inputs{\Theta}_{\omega_1}}
\sum\limits_{q_{j_1} \in O_{\omega_1}}
p_{j_1, q_{j_1}}
\\
&=&
p_{i_1, o_1}
\end{array}
\]
where the sums $\sum_{\underline{q}}$ are taken over all families $\underline{q} \in O_{\omega_1}^{\Inputs{\Theta}_{\omega_1}\backslash\{i_1\}}$, and we observed $\sum\limits_{q_{j_1} \in O_{\omega_1}}p_{j_1, q_{j_1}} = 1$.

For the rest of the proof, we consider the inductive case where $|\Events{\Theta}| \geq 2$.
We define the set of outputs at $\omega_1$ which are produced with non-zero probability in the empirical model, conditional to each choice of input $i_1$ at $\omega_1$:
\[
O_{\omega_1, i_1}
:= \suchthat{
    o_1 \in O_{\omega_1}
}{
    p_{i_1,o_1} > 0
}
\]
We then define empirical models $e^{(i_1, o_1)} \in \EmpModels{\StdCov{\Theta'_{i_1}}, \underline{O}'}$, by conditioning on each input $i_1 \in \Inputs{\Theta}_{\omega_1}$ and each output $o_1 \in O_{\omega_1, i_1}$:
\[
e^{(i_1, o_1)}_{\downset{h'}}\left(\underline{o}'\right)
:=
\frac{1}{p_{i_1,o_1}}
e_{\downset{\left(\{\omega_1:i_1\}\vee h'\right)}}\left(\{\omega_1:o_1\}\vee\underline{o}'\right)
\]
where $\downset{h'}$ for $h' \in \max\Theta'_{i_1}$ are contexts in the standard cover of $\Theta'_{i_1}$.
By inductive hypothesis, each $e^{(i_1, o_1)}$ arises by restriction of a classical empirical model $\hat{e}^{(i_1, o_1)}$ on $\Theta'_{i_1}$.
We use these empirical model to define the following $\reals^+$-valued function $\hat{e}$ on the extended causal functions for $\Theta$:
\[
    \hat{e}_{\Theta}\left(\Ext{g_{\underline{o}, \underline{f'}}}\right)
    :=
    \prod\limits_{i_1 \in \Inputs{\Theta}_{\omega_1}}
    p_{i_1, o_1}
    \hat{e}^{(i_1, o_1)}_{\Theta'_{i_1}}\left(\Ext{f'_{i_1}}\right)
\]
We now prove that the above marginalises to $e_{\downset{h}}$ for all $h \in \max\Theta$, and hence that $\hat{e}$ is a well-defined classical empirical model on $\Theta$, restricting to $e$ on the standard cover.
We start by writing out the marginal explicitly:
\[
    \restrict{\hat{e}}{\{\omega_1:i_1\}\vee h'}
    \left(\{\omega_1: o_1\}\vee k'\right)
    = \sum_{f} \hat{e}_{\Theta}\left(\Ext{f}\right)
\]
where the sum $\sum_f$ is taken over all causal functions $f$ on $\Theta$ such that $\Ext{f}\left(\{\omega_1:i_1\}\vee h'\right) = \{\omega_1: o_1\}\vee k'$, and we have considered the output history $k' \in \prod_{\omega \neq \omega_1} O_\omega$.
The causal functions $f = g_{\underline{o}, \underline{f}'}$ satisfying the condition above are those where $o_{i_1} = o_1$ and $\Ext{f'_{i_1}}(h')=k'$.
We adopt the following notation for the ``freely chosen'' parts of $\underline{o}$ and $\underline{f'}$:
\[
\begin{array}{rcl}
J
&:=& \Inputs{\Theta}_{\omega_1}\backslash\{i_1\}
\\
\underline{o}' &:=& \restrict{\underline{o}}{J}
\\
\underline{f}'' &:=& \restrict{\underline{f}'}{J}
\end{array}
\]
Using the definition of $\hat{e}_{\Theta}$, the marginal can be rewritten as follows:
\[
\sum_{\underline{o}'}
\sum_{\underline{f}''}
\sum_{f'_{i_1}}
p_{i_1, o_1} \hat{e}^{(i_1,o_1)}_{\Theta'_{i_1}}\left(\Ext{f'_{i_1}}\right)
\prod_{j_1 \in J}
p_{j_1, o'_{j_1}} \hat{e}^{(j_1,o'_{j_1})}_{\Theta'_{j_1}}\left(\Ext{f''_{j_1}}\right)
\]
where the sum $\sum_{\underline{o}'}$ is taken over all $\underline{o}' \in O_{\omega_1}^J$, the sum $\sum_{\underline{f}''}$ is taken over all $\underline{f}'' \in \prod_{j_1 \in J} \CausFun{\Theta'_{j_1}, \underline{O}'}$, and the sum $\sum_{f'_{i_1}}$ is take over all $f'_{i_1}$ such that $\Ext{f'_{i_1}}(h')=k'$.
By regrouping the summations, the marginal can be rewritten as follows:
\[
\left(
p_{i_1, o_1}
\sum_{f'_{i_1}}
\hat{e}^{(i_1,o_1)}_{\Theta'_{i_1}}\left(\Ext{f'_{i_1}}\right)
\right)
\left(
\sum_{\underline{o}'}
\sum_{\underline{f}''}
\prod_{j_1 \in J}
p_{j_1, o'_{j_1}} \hat{e}^{(j_1,o'_{j_1})}_{\Theta'_{j_1}}\left(\Ext{f''_{j_1}}\right)
\right)
\]
Because the summations $\sum_{\underline{o}'}$ and $\sum_{\underline{f}''}$ are taken over products, the right factor can be rearranged as follows, making it evident that it is equal to 1:
\[
\prod_{j_1 \in J}
\sum_{o'_{j_1}}
p_{j_1, o'_{j_1}}
\sum_{f''_{j_1}}
\hat{e}^{(j_1,o'_{j_1})}_{\Theta'_{j_1}}\left(\Ext{f''_{j_1}}\right)
=
\prod_{j_1 \in J}
\sum_{o'_{j_1}}
p_{j_1, o'_{j_1}} \cdot 1
=
\prod_{j_1 \in J} 1 = 1
\]
Our marginal is then reduced to the left factor above:
\[
\restrict{\hat{e}}{\{\omega_1:i_1\}\vee h'}
\left(\{\omega_1: o_1\}\vee k'\right)
=
p_{i_1, o_1}
\sum_{f'_{i_1}}
\hat{e}^{(i_1,o_1)}_{\Theta'_{i_1}}\left(\Ext{f'_{i_1}}\right)
\]
Because the sum $\sum_{f'_{i_1}}$ is take over all $f'_{i_1}$ such that $\Ext{f'_{i_1}}(h')=k'$, the marginal further simplifies as follows:
\[
\restrict{\hat{e}}{\{\omega_1:i_1\}\vee h'}
\left(\{\omega_1: o_1\}\vee k'\right)
=
p_{i_1, o_1}
\restrict{\hat{e}^{(i_1, o_1)}}{h'}
\left(k'\right)
\]
By inductive hypothesis, the marginal on the right hand side is $e^{(i_1, o_1)}_{\downset{h'}}\left(\underline{o}'\right)$:
\[
\begin{array}{rcl}
\restrict{\hat{e}}{\{\omega_1:i_1\}\vee h'}
\left(\{\omega_1: o_1\}\vee k'\right)
&=&
p_{i_1, o_1}
e^{(i_1, o_1)}_{\downset{h'}}\left(\underline{o}'\right)
\\
&=&
e_{\downset{\left(\{\omega_1:i_1\}\vee h'\right)}}\left(\{\omega_1:o_1\}\vee\underline{o}'\right)
\end{array}
\]
For all $i_1$ and $h'$, we have:
\[
\sum_{o_1} \sum_{k'}
\restrict{\hat{e}}{\{\omega_1:i_1\}\vee h'}
\left(\{\omega_1: o_1\}\vee k'\right)
=
\sum_{o_1} \sum_{k'}
e_{\downset{\left(\{\omega_1:i_1\}\vee h'\right)}}\left(\{\omega_1:o_1\}\vee\underline{o}'\right)
=1
\]
Hence $\hat{e}$ is a well-defined classical empirical model on $\Theta$ which restricts to $e$ on the standard cover, proving that $e$ is local.
\end{proof}

\subsubsection{Proof of Corollary \ref{corollary:nogo-nonlocality-switch-spaces-indef}}
\label{proof:corollary:nogo-nonlocality-switch-spaces-indef}
\begin{proof}
Extending Theorem \ref{theorem:nogo-nonlocality-switch-spaces} to this more general case is entirely straightforward: if $\Xi$ is such that $\Xi = \max\Ext\Xi$, then the causal functions on $\Xi$ for any given family of output sets $\underline{O}$ are in bijection with the causal functions on the space of input histories $\Hist{\{*\}, \max\Ext\Xi}$, supported by a single event and having the histories of $\Xi$ as its input set, using $* \mapsto \prod_{\omega } O_\omega$ as its (singleton) family of output sets.
To prove that $e$ is local, one converts all ``indefinite'' spaces $\Xi$ into spaces with a single event, applies Theorem \ref{theorem:nogo-nonlocality-switch-spaces} and then converts the causal functions appearing in the resulting classical empirical model $\hat{e}$ back into causal functions for the original ``indefinite'' spaces.
\end{proof}

\subsubsection{Proof of Proposition \ref{proposition:std-empmodel-local-par-seq-spaces-separability}}
\label{proof:proposition:std-empmodel-local-par-seq-spaces-separability}
\begin{proof}
Let $e$ be a standard empirical model for $\Theta$ which can be written as a convex combination $e = \sum_{j=1}^{n} p_j e^{(j)}$ of empirical models $e^{(j)}$ for some causal completions $\Theta^{(j)}$ of $\Theta$.
Assume that $e$ is local, so that it can be written as a convex combination of causal functions $g_y$ for $\Theta$:
\[
e = \sum_{y \in Y} r_y \delta_{\Ext{g_y}}
\]
The proof proceeds by structural induction on the way in which $\Theta$ is constructed by parallel and (conditional) sequential composition.

In the base case, $\Theta$ is a single indiscrete space.
In this case, $\Theta^{(1)},...,\Theta^{(n)}$ are all causal switch spaces (cf. Theorem 3.36 p.51 of ``The Combinatorics of Causality'' \cite{gogioso2022combinatorics}).
By Theorem \ref{theorem:nogo-nonlocality-switch-spaces}, each $e^{(j)}$ is local---regardless of the assumption that $e$ is local---so it can be written as a convex combination of causal functions $f_x^{(j)}$ for $\Theta^{(j)}$:
\[
e^{(j)} = \sum_{x \in X^{(j)}} q_x^{(j)} \delta_{\Ext{f_x^{(j)}}}
\]
Because each $\Theta^{(j)}$ is causally complete, the functions $f_x^{(j)}$ are all separable, and hence the empirical model $e$ is separably local:
\[
e = \sum_{j=1}^n p_j \sum_{x \in \in X^{(j)}} q_x^{(j)} \delta_{\Ext{f_x^{(j)}}}
\]

Consider the inductive case where $\Theta = \bigcup_{k=1}^{m} \Xi^{(k)}$ is a parallel composition of spaces themselves satisfying the assumptions of this Proposition.
It is possible to define the restriction of $e$ to each $\Xi^{(k)}$, written $\restrict{e}{\Xi^{(k)}}$, by marginalising over the outputs of events not in $\Xi^{(k)}$ and exploiting the independence of outputs for events in $\Xi^{(k)}$ from inputs for events not in $\Xi^{(k)}$.
By Theorem \ref{theorem:caus-fun-parallel-composition-factorisation}, each causal function $g_y$ factors as a parallel composition of causal functions $g_y^{(1)},...,g_y^{(m)}$ for the individual spaces $\Xi^{(1)},...,\Xi^{(m)}$.
Marginalising $\delta_{\Ext{g_y}}$ to each $\Xi^{(k)}$ yields $\delta_{\Ext{g_y^{(k)}}}$, so the restrictions $\restrict{e}{\Xi^{(k)}}$ are themselves local:
\[
\restrict{e}{\Xi^{(k)}}
=\sum_{y \in Y} r_y \delta_{\Ext{g_y^{(k)}}}
\]
It is also a fact (cf. Theorems 3.26 p.38 ``The Combinatorics of Causality'' \cite{gogioso2022combinatorics}) that each causal completion $\Theta^{(j)}$ of $\Theta$ factorises as a parallel composition of causal completions $\Xi^{(1,j)},...,\Xi^{(m,j)}$ for the spaces $\Xi^{(1)},...,\Xi^{(m)}$, respectively:
\[
\Theta^{(j)} = \bigcup_{k=1}^{m} \Xi^{(k,j)}
\]
We can therefore apply the inductive hypothesis to each restricted empirical model $\restrict{e}{\Xi^{(k)}}$, with the corresponding space $\Xi^{(k)}$, to conclude that the restricted empirical models are all separably local: since parallel composition of separable functions is separable, the empirical model $e$ is also separably local.

Consider instead the inductive case where $\Theta = \Gamma \seqcomposeSym \underline{\Xi}$ is a conditional sequential composition of spaces themselves satisfying the assumptions of this Proposition.
It is possible to define the restriction of $e$ to $\Gamma$, written $\restrict{e}{\Gamma}$, by marginalising over the outputs of events not in $\Gamma$ and exploiting the independence of outputs for events in $\Gamma$ from inputs for events not in $\Gamma$.
It is also possible to define the restriction of $e$ to $\Xi^{(k)}$ for each $k \in \max\Ext{\Gamma}$, by conditioning on the input values in $k$, marginalising over the outputs of events not in $\Xi^{(k)}$, and exploiting the independence of outputs for events in $\Xi^{(k)}$ from inputs for events in $\Xi^{(k')}$, for any given $k' \neq k$.
By Theorem \ref{theorem:caus-fun-conditional-sequential-composition-factorisation}, each causal function $g_y$ factors as a sequential composition of a causal function $g_y^{(\Gamma)}$ for $\Gamma$ with a family of causal functions $g_y^{(k)}$ for each $\Xi^{(k)}$, each of the latter only applied conditional to the corresponding $k \in \max\Ext{\Gamma}$ being a sub-history of the given input history.
Marginalising $\delta_{\Ext{g_y}}$ to $\Gamma$ yields $g_y^{(\Gamma)}$, so the restriction $\restrict{e}{\Gamma}$ is itself local:
\[
\restrict{e}{\Gamma}
=\sum_{y \in Y} r_y \delta_{\Ext{g_y^{(\Gamma)}}}
\]
Marginalising $g_y$ to $\Xi^{(k)}$ yields $\delta_{\Ext{g_y^{(k)}}}$, so the restrictions $\restrict{e}{\Xi^{(k)}}$ are themselves local:
\[
\restrict{e}{\Xi^{(k)}}
=\sum_{y \in Y} r_y \delta_{\Ext{g_y^{(k)}}}
\]
It is also a fact (cf. Theorems 3.27 p.38 ``The Combinatorics of Causality'' \cite{gogioso2022combinatorics}) that each causal completion $\Theta^{(j)}$ of $\Theta$ factorises as a conditional sequential composition of causal completions $\Gamma^{(j)}, \Xi^{(1,j)},...,\Xi^{(m,j)}$ for the spaces $\Gamma, \Xi^{(1)},...,\Xi^{(m)}$, respectively:
\[
\Theta^{(j)} = \Gamma^{(j)} \seqcomposeSym \left(\Xi^{(k,j)}\right)_{k \in \max\Ext{\Gamma}}
\]
We can therefore apply the inductive hypothesis to the restricted empirical models $\restrict{e}{\Gamma}$ and $\restrict{e}{\Xi^{(k)}}$ (for all $k \in \max\Ext{\Gamma}$), with the corresponding spaces $\Gamma$ and $\Xi^{(k)}$ (for all $k \in \max\Ext{\Gamma}$), to conclude that the restricted empirical models are all separably local: since conditional sequential composition of separable functions is separable, the empirical model $e$ is also separably local.
\end{proof}






\ack
Financial support from EPSRC, the Pirie-Reid Scholarship and Hashberg Ltd is gratefully acknowledged.
This publication was made possible through the support of the ID\#62312 grant from the John Templeton Foundation, as part of the project `The Quantum Information Structure of Spacetime' (QISS), https://www.templeton.org/grant/the-quantum-information-structure-ofspacetime-qiss-second-phase.
The opinions expressed in this project/publication are those of the author(s) and do not necessarily reflect the views of the John Templeton Foundation.

\section*{Bibliography}

\bibliographystyle{unsrt}
\bibliography{biblio}

\end{document}